\documentclass[twoside,12pt]{article}
\usepackage{epsfig}
\usepackage{axodraw}
\usepackage{latexsym}
\usepackage{hyperref}
\usepackage{rotating}
\input amssym.def
\input amssym

\def \dd{{\rm d}}

\def \q{{\bf q}}
\def \p{{\bf p}}

\def \h{{\rm h}}

\def \xx{x}
\def \ra{\rangle}
\def \la{\langle}

\def \de{\delta}

\def \tr{\mbox{tr}}
\def \fr{\frac}

\def \M{{\mathcal{M}}}

\def \g{\gamma}
\def \Ga{\Gamma}
\def \oGa{{\overline \Gamma}}
\def \oW{{\overline W}}
\def \ba{\begin{array}}
\def \ea{\end{array}}

\def \s{\sigma}

\def \onu{{\overline \nu}}

\def\lsim{\lower .7ex\hbox{$\;\stackrel{\textstyle <}{\sim}\;$}} 
\def\gsim{\lower .7ex\hbox{$\;\stackrel{\textstyle >}{\sim}\;$}} 
\def \oq{{\overline q}}
\def \ou{{\overline u}}

\def \od{{\overline d}}

\def \oq{{\overline q}}

\def \GeV{\rm GeV}
\def \TeV{\rm TeV}

\def \half{{\textstyle{\fr{1}{2}}}}

\def \ts{\textstyle}
\def\Journal#1#2#3#4{{#1} {#2} (#4) #3 }

\def\NPA{{\em Nucl. Phys.} A}

\def\PLB{{\em Phys. Lett.} B}

\def\PRL{\em Phys. Rev. Lett.}

\def\PREP{\em Phys. Rep.}

\def\PRD{{\em Phys. Rev.} D}
\def\PRC{{\em Phys. Rev.} C}

\def\p{\vec p}

\def\q{\vec q}

\newcommand{\be}{\begin{equation}}
\newcommand{\ee}{\end{equation}}
\newcommand{\bea}{\begin{eqnarray}}
\newcommand{\eea}{\end{eqnarray}}

\begin{document}

\title{ \vspace{1cm} Structure Functions}
\author{A.\ De Roeck,$^{1,2,3}$ R.S.\ Thorne,$^4$ \\
$^1$CERN, Geneva, Switzerland.\\
$^2$Universiteit Antwerpen, Belgium.\\
$^3$University of California, Davis, Davis, California 95616, USA\\
$^4$Department of Physics and Astronomy,\\ University College London, Gower 
Street,\\ London, WC1E 6BT, United Kingdom\\
}
\maketitle
\begin{abstract}  
Structure functions are a measure of the partonic structure of hadrons, which is important 
for any process which involves colliding hadrons. They are a key ingredient for deriving partons distributions in nucleons. In recent years dramatic progress has been made in the understanding of the nucleon 
structure and the
precision of its partonic content, due to vast theoretical progress, and the availability of 
new high precision
measurements. This review gives
an overview on present structure function and related data, and on the most recent 
techniques used to extract parton distribution functions  to describe the structure of the proton. Special attention is given to the determination of the uncertainties  on the parton distributions.
\end{abstract}
%\eject
%\tableofcontents
\section{Introduction}
In order to obtain very high energies more easily many particle colliders have
hadrons, in particular protons and antiprotons, in the initial state.  
Since the 30th of March 2010, the LHC has started a two year operation period at the highest collision
energy ever produced in the laboratory, namely 7 TeV in the centre of mass of the  particle collisions.
The LHC accelerates proton beams in opposite directions, each beam 
with an energy of 3.5 TeV. With these collisions the LHC has entered the so called Terascale energy regime, 
which is expected to lead to new insights in the dynamics particle physics, and will possibly revolutionise
the field.  Hadrons are however composite particles, consisting of quarks and gluons, and these partons are the fundamental constituents that are involved in the collisions. The full and detailed understanding of the 
structure of the protons will be required to extract the most of the physics from the LHC data.

Hadrons are bound together by the strong force, described by Quantum Chromodynamics 
(QCD). The strong coupling constant $\alpha_S(\mu^2)$  
runs with the energy scale $\mu^2$ of a process, decreasing as $\mu^2$ increases,a phenomenon known as asymptotic freedom. 
Hence, $\alpha_S(\mu^2)$ is very large if $\mu^2$ is at
the scale of nonperturbative physics, about $1\GeV$, 
but $\alpha_S(\mu^2)\ll 1$ if $\mu^2 \gg 
1\GeV^2$, and perturbation theory can be used.     
Because of the strong force it is difficult to 
perform analytic calculations of scattering 
processes involving hadronic particles from first 
principles. However, the weakening of $\alpha_S(\mu^2)$ at higher scales 
leads to the Factorisation  Theorem which separates 
processes into nonperturbative parton distribution functions (PDFs)
which describe the composition of the proton 
and can be determined from 
experiment, and perturbative coefficient functions 
associated with higher scales 
which are calculated as a power-series in $\alpha_S(\mu^2)$. 
Thus in order to understand any of the results of these experiments one needs to 
understand how the incoming hadron is made up from the constituent quarks and
gluons, the interactions of which we then know how to calculate using perturbation 
theory  as long as there is a large scale in the process so that perturbation theory 
is applicable. 

The production of any particle --say a 
Higgs boson -- at a hadron collider can be 
determined by the cross section of the parton-parton collision to produce the Higgs, convoluted 
the probabilities to find these partons within the incoming hadrons.
We can use deep inelastic scattering (DIS) experiments to probe the
structure
of hadrons and the fundamental interactions of quarks, gluons, and
leptons. 
In DIS experiments 
a lepton probes a target nucleon or nucleus via exchange of an electroweak boson
In fact, DIS was the first method to directly detect quarks in
hadrons, in an experiment at SLAC in 1969\cite{Breidenbach:1969kd}.
In DIS an elementary  particle transfers large energy-momentum to a
hadron, which then breaks up inelastically. Essentially it knocks a 
quark out of the target hadron, which then hadronises.
The assumption is
that for high energy momentum transfers, corrections to these basic
processes from gluon exchange between quarks can be treated in 
QCD perturbation theory. The hadronisation process, where perturbation
theory cannot be used, takes place over much longer timescales and 
larger distance scales than the initial point-like electroweak
scattering. However, we also have to consider the nonperturbative 
initial state. 

The advent 
of the HERA collider in particular has led to significant progress in the last 10-15 years on the precise understanding of the
structure of the proton, especially in the kinematic region of small momentum fractions $x$ carried by the partons 
with respect to the  proton momentum. DIS experiments extract information  from the lepton scattering
cross sections to measure Structure Functions of the target, which are directly related to the PDFs.

Apart from DIS experimental data, results from jet production at hadron colliders, Drell-Yan, prompt 
photon and heavy vector boson experimental data can be used to constrain the partons in the hadrons.
For over 20 years several groups have used all the available experimental  data to make global fits to extract
PDFs which can be used in studies that involve colliding hadrons. Nowadays these datasets contain thousands of data points from over a dozen different experiments, and fits to these data are performed
with next to leading, and even next to next to leading order QCD tools.  
The original PDF extractions were relatively 
simple\cite{Eichten:1984eu,Morfin:1990ck,Martin:1988nk}
 but increased precision of the measurements, especially
 of the available DIS data  have lead to a steady progress 
to more sophisticated extractions of the PDFs, notably on the strange sea asymmetry, the treatment 
of heavy quarks, the constraints on the gluon --which is not directly probed in DIS as it is 
invisible to the electroweak boson-- determining the PDFs to higher perturbative orders in the strong
coupling constants, etc. Moreover, for the experimentalists it is often equally important to know what 
the uncertainties on the PDFs are. In recent years the two main groups providing general fits (CTEQ
and MRST/MSTW)
have developed a prescription for determining uncertainties, and have now been joined by several other groups
(NNPDF, ABKM, GJR). Also the HERA experimental collaborations have combined their data and produced PDFs with uncertainty bands (HERAPDF).

In this paper we will review the present understanding of the structure or the proton, and the uncertainties
on such determinations. In the remainder of this section we introduce the formalism, kinematics, the 
important factorisation hypothesis, the QCD corrections, and a short overview of the latest data relevant
for the extraction of the nucleon structure.
Section 2 discusses the determination of parton distribution functions. In Section 3 the present different
approaches of the 
parton distribution determinations are presented, together with determination of the uncertainties. 
Section 4 discusses some theoretical sources of uncertainties in more detail. In Section 5 some specific
PDFs suited for LO Monte Carlo generator programs are presented. Section 6 gives an outlook for 
future measurements and developments and Section 7 summarises the main points of the review.

\subsection{\it Structure Functions -- Kinematics}

\begin{figure}
\begin{center}
\begin{minipage}[t]{14 cm}
\epsfig{file=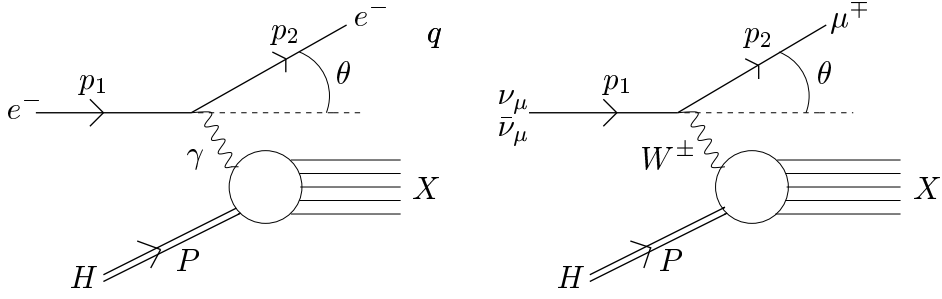,scale=0.7}
\end{minipage}
\begin{minipage}[t]{16.5 cm}
\caption{The kinematics for deep inelastic scattering. \label{dis1}}
\end{minipage}\end{center}
\end{figure}

We consider deep inelastic scattering  of electrons (or neutrinos) as a typical example.
Let us examine the former with scattering from a hadron $H$,
of mass $m_P$, we have in the first case, i.e. 
\be
e(p_1) + H(P) \longrightarrow e(p_2) + X \, ,
\label{deep} \ee
where $X$ is an arbitrary final state. To lowest order in 
the electromagnetic charge $e$, the electron
couples to the hadron through a virtual photon.
This can be seen in Fig. ~\ref{dis1} 
where we also show the corresponding diagram for neutrino interaction
via a $W$ boson. Concentrating first on the electromagnetic scattering process,
the lowest order  
QED amplitude is
\be
i\M = (ie)^2 \ou(p_2) \g^\mu u(p_1) \, i\frac{-g_{\mu\nu}}{q^2} \,
\la X | J_\h^\nu |H,P \ra \, , \quad q=p_1-p_2 \, .
\label{MH}
\ee
In the hadron rest frame $P = (m_P,{\bf 0})$, $p_1=(E_1,\p_1)$ and 
$p_2=(E_2,\p_2)$. The basic relativistic invariant variables are
\be
\nu \equiv P{\cdot q} = m_P(E_1-E_2) \, , \qquad Q^2 \equiv - q^2
= 2p_1{\cdot p_2} = 2E_1E_2(1-\cos \theta) \, ,
\label{invar}
\ee
where we have neglected the electron mass, so that $E_1=|\p_1|,\, E_2=|\p_2|$,
and $\theta$ is the electron scattering angle. Clearly $Q^2\ge 0$ and also
\be
m_X^{\, 2} = (P+q)^2 \ge m_P^2 \quad \Rightarrow \quad Q^2 \le 2\nu \, .
\ee
The standard expression for the differential cross section gives
\be
\dd \sigma = \frac{1}{F}\, \frac{\dd^3 p_2}{(2\pi)^3 2E_2} \,
\sum_X (2\pi)^4 \de^4(q+P-p_X) \, \half \!\sum_{e \ {\rm spins}} |\M |^2 \, ,
\label{S}
\ee
where $F$ is the flux factor, $F=  4E_1m_H$,
in the hadron rest frame. From Eq. (\ref{MH}) 
\be
\sum_{e \ {\rm spins}} |\M |^2 = \frac{e^4}{(q^2)^2} \, L_{\nu\mu} \,
\la H,P | J_\h^\nu |X \ra \, \la X | J_\h^\mu |H,P \ra \, ,
\ee
where, setting $m_e=0$,
\be
L_{\nu\mu} =  4(p_{1,\nu} p_{2,\mu} + p_{1,\mu} p_{2,\nu} - g_{\nu\mu} \,p_1{\cdot p_2}) \,.
\label{L}
\ee
If we define
\be
W_H^{\nu\mu}(q,P)=\frac{1}{4\pi} \sum_X \, (2\pi)^4 \de^4(q+P-p_X) \,
\la H,P | J_\h^\nu |X \ra \, \la X | J_\h^\mu |H,P \ra \, ,
\label{WH}
\ee
where we implicitly average over the hadron spin in this definition, 
then the cross section formula Eq. (\ref{S}) becomes
\be
 \frac{\dd \sigma}{\dd^3 p_2} = \frac{e^4}{8(2\pi)^2 E_1m_HE_2} \, \frac{1}
{(Q^2)^2} \,  L_{\nu\mu} W_H^{\nu\mu}(q,P) \, .
\label{S2}
\ee

By virtue of conservation of the electromagnetic current 
$(p_X-P)_\mu \la X | J_\h^\mu |H,P
\ra =0$,
we have
$q_\mu W_H^{\nu\mu}(q,P) = q_\nu  W_H^{\nu\mu}(q,P) =0$. The most
general Lorentz covariant form compatible with this is 
\be
W_H^{\nu\mu}(q,P) = \bigg ( -g^{\nu\mu} + \frac{q^\nu q^\mu}{q^2} \bigg ) W_1
+ \bigg ( P^\nu - \frac{P{\cdot q}}{q^2}\, q^\nu \bigg )
\bigg ( P^\mu - \frac{P{\cdot q}}{q^2}\, q^\mu \bigg ) W_2 \, ,
\label{W12}
\ee
where $W_{1,2}$ are Lorentz scalar functions characteristic of the
hadron $H$,  which depend on the two variables $Q^2$ and $\nu$. In writing
Eq. (\ref{W12}) we have neglected a possible term involving the $\epsilon$-tensor
but this can be excluded by using parity invariance. 
To calculate the contraction
in Eq. (\ref{S2}) we may use the fact that current conservation also leads to 
$L_{\nu\mu}q^\nu = L_{\nu\mu}q^\mu = 0$ so that
from Eq. (\ref{L}) and Eq. (\ref{W12}) we have, 
\begin{eqnarray}
L_{\nu\mu} W_H^{\nu\mu}(q,P) 
&=& 8p_2{\cdot p_1}\, W_1 + 4\big (2 p_1{\cdot P} \,
p_2 {\cdot P} - m_P^2 p_2{\cdot p_1} \big ) W_2 \nonumber \\
&=& 4 Q^2\, W_1 + 2m_H^2(4E_1E_2 - Q^2)\, W_2 
\label{LW}
\end{eqnarray}
where in the second line we have used $p_1{\cdot p_2}=-\half q^2$, if $m_e=0$,
together with $p_1{\cdot P} = m_HE_1, \, p_2{\cdot P} = m_HE_2$.
In the limit of large momentum transfer  
$Q^2 \sim{\rm O}(\nu) \to \infty$, we can define
dimensionless variables $\xx, y$ 
\be
\xx = \frac {Q^2}{2\nu} \, , \qquad y = \frac{\nu}{m_HE_1} = 1 - \frac{E_2}{E_1} \, ,
\label{xiy}
\ee
which stay fixed. It is easy to see that by definition 
\be
0\le \xx \le 1 \, , \qquad 0\le y \le 1 \, .
\ee
Then from Eq.(\ref{LW})
\be
L_{\nu\mu} W_H^{\nu\mu}(q,P) = 8E_1m_H 
\bigg ( \xx y\, W_1 + \frac{1-y}{y}\, \nu W_2 \bigg ) \left[ 
1 +
O\left( {m_P^2 \over Q^2} \right) \right] \ .
\ee
Since
\be
\dd^3 p_2 \to 2\pi\, E^{2}_2 \dd(\cos \theta) \, \dd E_2 = \pi E_2\, 
\dd Q^2 \, \dd y = 2\pi E_2\nu \, \dd\xx \, \dd y \, ,
\ee
we have 
\be
\frac{\dd \sigma}{\dd \xx \dd y} = \frac{4\pi\alpha^2}{Q^4} \, 2m_HE_1
\Big ( (1-y)F_2(\xx,Q^2) + \xx y^2 F_1(\xx,Q^2) \Big ) \left[ 
1 +
O\left( {m_P^2 \over Q^2} \right) \right]
\, ,
\label{S3}
\ee
where  $\alpha=e^2/4\pi$ and 
\be
F_2(\xx,Q^2) = \nu W_2 \, , \qquad F_1(\xx,Q^2) = W_1 \, ,
\ee
are dimensionless, frame invariant ``structure functions''. 
Clearly comparison of
cross section measurements with Eq. (\ref{S3}) allows
$F_{1,2}$ to be disentangled.

In the basic process Eq. (\ref{deep}) the electron $e$ may be replaced by a muon
without changing any of the subsequent results. A very similar analysis
also holds for inelastic scattering of neutrinos, or anti-neutrinos, where
\be
\nu_\mu(p_1) + H(P) \longrightarrow \mu^-(p_2) + X \  \ \mbox{or} \ \
\onu_\mu(p_1) + H(P) \longrightarrow \mu^+(p_2) + X  \, .
\ee
The scattering is now mediated by a virtual $W^+$ or $W^-$, instead of a
virtual $\gamma$, so that to first order in the weak interaction the amplitude
is similar to Eq. (\ref{MH}) but
\be
-\frac{e^2}{q^2} \longrightarrow \frac{1}{8}\, \frac{g^2}{m_W^{\, 2}-q^2}
= \frac{G_F}{\sqrt 2} \, \frac{m_W^{\, 2}}{m_W^{\, 2}+Q^2} \, .
\ee
If we assume $Q^2\ll m_W^{\, 2}$, then instead of Eq. (\ref{S3}) we have
\be
\frac{\dd \sigma_{\nu H,\onu H}}{\dd \xx \dd y} = 
\frac{G_F^{\, 2}}{2\pi} \, 2m_HE
\Big ( (1-y)F_2^\pm(\xx,Q^2) + \xx y^2 F_1^\pm(\xx,Q^2) \pm \xx y (1-\half y)
F_3^\pm(\xx,Q^2) \Big ) \, ,
\label{S5}
\ee
where now we have a parity violating structure function $F_3^\pm(\xx,Q^2)$.
More generally Eq. (\ref{S5}) should contain a factor $(1+Q^2/m_W^{\, 2})^{-2}$.

\subsection{\it Structure Functions -- Factorisation}

\begin{figure}
\begin{center}
\begin{minipage}[t]{14 cm}
\epsfig{file=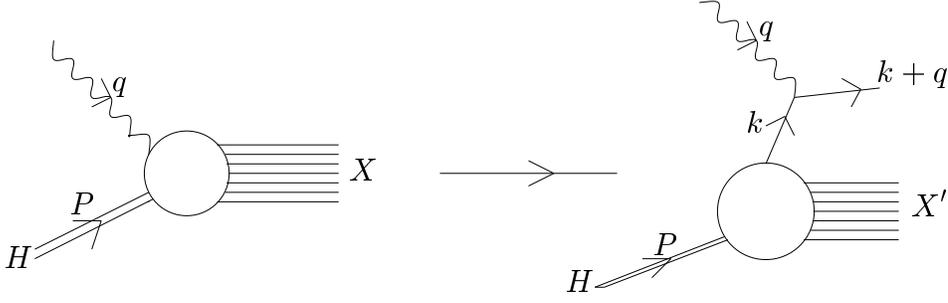,scale=0.7}
\end{minipage}
\begin{minipage}[t]{16.5 cm}
\caption{Deep inelastic scattering viewed in terms of scattering from a single parton. \label{dis2}}
\end{minipage}
\end{center}
\end{figure}

$W_H^{\mu \nu}(q,P)$ could be evaluated exactly if one knew the
wavefunctions of $|H\ra$ and $|X\ra$ in terms of quark and gluon Fock
states. 
In practice this is a difficult non-perturbative
problem. We can apply the assumption that since the creation of 
hadrons takes place on a time (and distance) scale 
${\cal O}(1/\Lambda_{QCD})$, while the creation of the final state 
in terms of quarks and gluons in the hard scattering 
happens over the short time (and distance) 
scale ${\cal O}(1/Q)$, we are able to sum over final state quarks and gluons
rather than hadrons up to corrections of ${\cal O}(\Lambda_{QCD}^2/Q^2)$. 
However, we also have the added complication that 
the target hadron momentum $P$ satisfies $P^2=m_P^2$ which is fixed and the
hadron wave function depends on low energy scales.
It is necessary to introduce a further factorisation assumption, which
can be derived to all orders in the perturbation expansion, in order to
justify using the ideas of asymptotic freedom.
We use similar physical reasoning to the hadronisation in the final state and 
assume that the large
momentum transfer from the virtual photon takes place to a single quark 
which has fluctuated out of the proton over the short time (and distance) 
scale ${\cal O}(1/Q)$,
and we can neglect the QCD interactions between hadron constituents
due to asymptotic freedom. On the larger time (and distance) scale 
${\cal O}(1/\Lambda_{QCD})$ the 
struck quark and the remaining quarks and gluons interact strongly
via QCD forces in order to hadronise in the final state, and this
processes is largely independent of the former so-called hard process. One can
prove in DIS scattering that this factorisation indeed holds up to corrections 
${\cal O}(\Lambda_{QCD}^2/Q^2)$. Within this framework
the leading term in the deep inelastic limit is then given in Eq. (\ref{WH})
by letting
$|X\ra \to |q_f,{\tilde k}\ra |X'\ra$, as illustrated in Fig. (\ref{dis2}), 
where $|q_f,{\tilde k}\ra$ denotes an on-shell 
`parton', either a single quark or anti-quark state with flavour index $f$ and 
4-momentum $\tilde k$, and $|X'\ra$ denotes the remnant of the scattered 
proton.

This allows us to rewrite Eq. (\ref{WH}) as 
\begin{eqnarray}
W_H^{\nu\mu}(q,P)&=&\frac{1}{4\pi} \sum_f
\sum_{X'} \, (2\pi)^4 \de^4(q+P-p_X) \,
\frac{1}{(2\pi)^3} \int \! \dd^4 {\tilde k} \,
\theta({\tilde k}^0) \delta({\tilde k}^2)  \sum_{q\ {\rm spins}}\,\nonumber\\
& &\biggl[Q_f^2\la H,P |\la 0 |\oq_f \g^\nu q_f |q_f,{\tilde k}\ra  |X \ra \, 
\la X | \la q_f,{\tilde k}|\oq_f \g^\mu q_f |0\ra|H,P \ra \biggr]\, .
\label{WHa}
\end{eqnarray}
The momentum of the on-shell struck quark (or antiquark) is 
$\tilde k = k+q$ and the 
proton momentum satisfies the momentum conservation constraint $P=k+p_{X'}$.
The second part of Eq. (\ref{WHa}) simplifies to 
\begin{eqnarray}
\sum_{q\ {\rm spins}}\,&Q_f^2&
\la H,P |\la 0 |\oq_f \g^\nu q_f |q_f,{\tilde k}\ra  |X \ra \, 
\la X | \la q_f,{\tilde k}|\oq_f \g^\mu q_f |0\ra|H,P \ra\nonumber \\
&=& Q_f^2
\la H,P |\oq_f  |X \ra \g^\nu \tilde k\cdot\g \g^\mu 
\la X | q_f |H,P \ra.
\end{eqnarray}
Summing over
both quarks and anti-quarks we can represent Eq. (\ref{WH}) as 
\be
W_H^{\nu\mu}(q,P) \to  \sum_f \int \! \dd^4 k \, \tr \Big (
W_f^{\nu\mu}(q,k) \Ga_{H,f}(P,k) + 
\oW{}_f^{\nu\mu}(q,k)\oGa_{H,f}(P,k)\Big ) \, ,
\label{parton2}
\ee
where $W_f^{\nu\mu}(q,k)$, $\oW{}_f^{\nu\mu}(q,k)$,
denotes the relevant contributions when
the virtual photon with momentum $q$ couples to a quark, anti-quark,
with flavour $f$ and momentum $k$,
\be
W_f^{\nu\mu}(q,k) = \oW{}_f^{\mu\nu}(q,k) =
\half Q_f^{\, 2} \, \g^\nu \g{\cdot (}k+q) \,\g^\mu \,
\delta ((k+q)^2 ) \theta((k+q)^0)\, ,
\label{Wq}
\ee
and where we define
\begin{eqnarray}
\Ga_{H,f}(P,k)_{\beta\alpha} = \sum_{X'} \delta^4(P-k-p_{X'})\,
\la H,P | \oq_{f\alpha} |X'\ra \, \la X' | q_{f\beta} | H,P \ra \, ,
\nonumber \\
\oGa_{H,f}(P,k)_{\beta\alpha} = \sum_{X'} \delta^4(P-k-p_{X'})\,
\la H,P | q_{f\beta} |X'\ra \, \la X' | \oq_{f\alpha} | H,P \ra \, .
\label{Gamma}
\end{eqnarray}
The average over the hadron spins is implicit in the above. 
The expression Eq. (\ref{parton2}) obtained assumes
that the quark, or anti-quark, does not interact with the state $X'$ after
it couples to the virtual photon. It may be proven in the deep inelastic 
limit, that this is true up to contributions suppressed by terms
${\cal O}(\Lambda_{QCD}^2/Q^2)$.

It may be shown that if we define the parton density functions
\begin{eqnarray}
\frac{1}{2P} \int \! \dd^4 k \, \delta \bigg(\frac{k}{P} - \xx \bigg) \, 
\tr \Big ( \g \Ga_{H,f}(P,k)\Big ) &=&  \, q_f(\xx) \, , \nonumber \\
\frac{1}{2P} \int \! \dd^4 k \, \delta \bigg(\frac{k}{P} - \xx \bigg) \, 
\tr \Big ( \g \oGa_{H,f}(P,k)\Big )&=&  \, \oq_f(\xx) \, ,
\label{qdist}
\end{eqnarray}
where strictly speaking it is the light-cone momenta $k^+=k^0+k^3$ and 
$P^+=P^0+P^3$ which are used in the delta function, and we are in a frame where
implicitly 
\be
P_\perp = \q_\perp = 0 \, ,
\label{perp}
\ee  
then we find     
\be
F_1(\xx,Q^2) \to \half \sum_f Q_f^{\, 2} \Big (q_f(\xx) + \oq_f(\xx) \Big )\,.
\label{Bj}
\ee

The result in Eq. (\ref{Bj}) demonstrates
that $F_1$  depends only on the dimensionless variable $\xx=Q^2/2\nu$ in
the deep inelastic limit, which is known as Bjorken scaling\cite{Bjorken:1968dy,Feynman}. The
experimental observation of this scaling was the first direct evidence
for point-like constituents in hadrons\cite{Miller:1971qb}. 
The quark
distribution functions $q_f(\xx), \, \oq_f(\xx)$ defined by Eq. (\ref{qdist})
for $\xx\ge 0$ are an intrinsic non-perturbative property of the
hadron $H$.
They may be interpreted as 
momentum distributions for quarks and  anti-quarks 
inside the  hadron and in principle (thought not yet in practice) 
they can be computed from a 
non-perturbative analysis in QCD. At present these distribution functions must simply
be determined experimentally  from (largely) DIS experiments.  
We also find that 
\be
F_2(\xx,Q^2)= 2\xx F_1(\xx,Q^2)
= \xx \sum_f Q_f^{\, 2} \Big (q_f(\xx) + \oq_f(\xx) \Big )\, .
\ee
The form of the relation between $F_1$ and $F_2$ 
is a consequence of the spin 1/2 nature of the
struck quark. The difference is proportional to 
the longitudinal structure function $F_L(x,Q^2)$, and is zero at lowest 
order due to helicity conservation\cite{Callan:1969uq}.  

Applying these results to deep inelastic scattering on a proton target 
the proton wavefunction  is dominated by 
$uud + \cdots$ where the dots
indicate $uud$ plus further quarks (including heavy flavours).
With notation $q_u(\xx)=u(\xx), \,
\oq_u(\xx)=\ou(\xx)$ etc, 
\be
F_{2,{\rm proton}}(\xx,Q^2) \sim \xx \Big ( {\ts\frac{4}{9}}(u(\xx)+\ou(\xx))
+ {\ts\frac{1}{9}}(d(\xx)+\od(\xx))+ 
\ {\rm heavy} \ {\rm flavours} \Big ) \, .
\label{proton}
\ee
We note that the derivation of Eq. (\ref{Bj}) is an approximation 
which relies on the assumption that $k$, being the the momentum
of a quark
(or antiquark) inside the proton, should have a very small probability
of having any momentum components greater than ${\cal O}(\Lambda_{QCD})$.
As such it also implies corrections of
${\cal O}(\Lambda_{QCD}^2/Q^2)$ corresponding to higher twist operators
(as discussed in\cite{Ellis:1982cd}). However, it also ignores rather more
important higher-order 
QCD corrections, which we consider next.

\subsection{\it QCD Corrections}

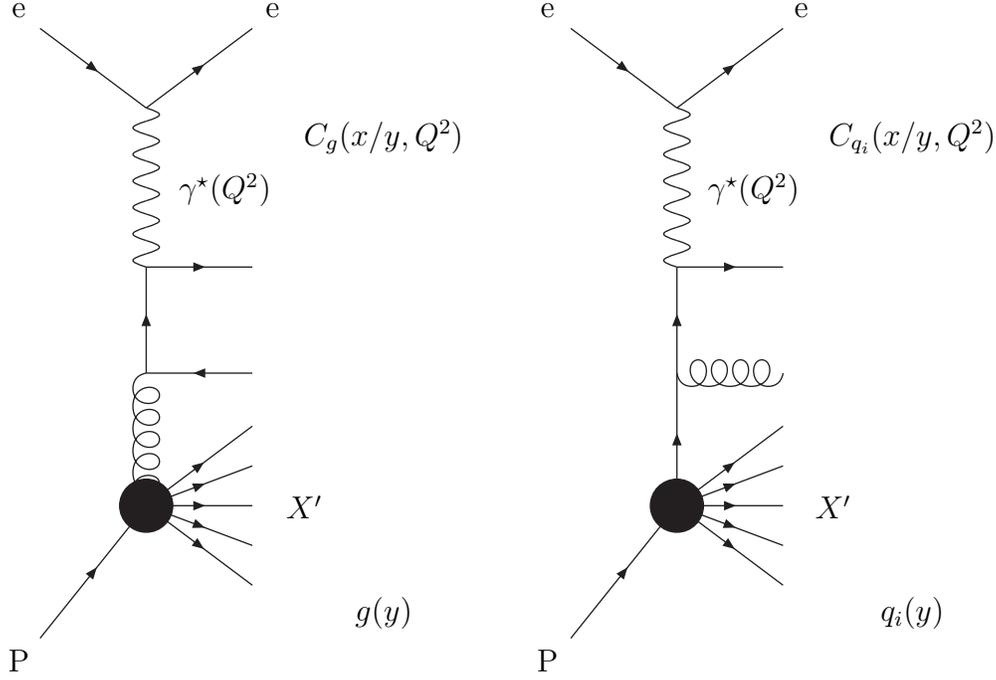
\begin{figure}
\begin{center}
\begin{minipage}[t]{14 cm}
\begin{picture}(400,270)(0,50)
\Text(12,298)[]{e}
\ArrowLine(20,290)(60,260)
\ArrowLine(60,260)(100,290)
\Text(108,298)[]{e}
\Photon(60,260)(60,200){5}{6}
\Text(90,230)[]{$\gamma^{\star}(Q^2)$}
\ArrowLine(60,200)(100,200)
\ArrowLine(60,160)(60,200)
\ArrowLine(100,160)(60,160)
\Gluon(60,110)(60,160){5}{5}
\ArrowLine(20,60)(60,110)
\Text(12,52)[]{P}
\ArrowLine(60,110)(100,140)
\ArrowLine(60,110)(100,125)
\ArrowLine(60,110)(100,110)
\ArrowLine(60,110)(100,95)
\ArrowLine(60,110)(100,80)
\CCirc(60,110){10}{Black}{Black}
\Text(120,110)[]{$X'$}
\Text(150,250)[]{$C_{g}(x/y,Q^2)$}
\Text(150,70)[]{$g(y)$}
\Text(212,298)[]{e}
\ArrowLine(220,290)(260,260)
\ArrowLine(260,260)(300,290)
\Text(308,298)[]{e}
\Photon(260,260)(260,200){5}{6}
\Text(290,230)[]{$\gamma^{\star}(Q^2)$}
\ArrowLine(260,200)(300,200)
\ArrowLine(260,110)(260,160)
\ArrowLine(260,160)(260,200)
\Gluon(300,160)(260,160){5}{4}
\ArrowLine(220,60)(260,110)
\Text(212,52)[]{P}
\ArrowLine(260,110)(300,140)
\ArrowLine(260,110)(300,125)
\ArrowLine(260,110)(300,110)
\ArrowLine(260,110)(300,95)
\ArrowLine(260,110)(300,80)
\CCirc(260,110){10}{Black}{Black}
\Text(320,110)[]{$X'$}
\Text(350,250)[]{$C_{q_i}(x/y,Q^2)$}
\Text(350,70)[]{$q_i(y)$}
\end{picture}
\caption{Factorisation with QCD corrections in deep inelastic scattering. \label{disfact}}
\end{minipage}
\end{center}
\end{figure}

In the previous section we have assumed
that the quark interacts with the virtual
$\g$ for large $Q^2$ with a point-like coupling, not
including any corrections due to QCD. In a field theory
approach the
quark fields in the currents are 
treated as if they were effectively free, disregarding QCD effects. This
is ultimately justified by asymptotic freedom but there 
are calculable perturbative QCD corrections to Bjorken
scaling. 
To simplify the discussion, we examine a generic structure 
function $F(\xx,Q^2)$, such as might be measured
in deep inelastic scattering. The dominant contributions for $Q^2\to \infty$
arise from the elementary particles of perturbative QCD, quarks and gluons,
but QCD corrections are no longer ignored and $F(\xx,Q^2)$ cannot any more be
represented in terms of solely point-like couplings to the quarks. 
Hence, we 
recognise that we can now create a number of quarks, 
antiquarks and gluons in the 
final state via a hard QCD perturbative process. 
The point-like
vertex is now also replaced by a ``coefficient function'' $C_i(q,k)$
representing this hard scattering process, where $i=q_f,\oq_f,G$ for an
incoming quark, 
antiquark or
gluon with 4-momentum $k$ coupling to a current $J$ carrying 4-momentum $q$,
$q^2=-Q^2$, and which includes all (perturbative) QCD
corrections. Some of the leading $\alpha_S$ corrections to the lowest 
order diagram are illustrated in Fig. \ref{disfact}

In the  relevant limit $Q^2=-q^2\to \infty$, used above,  
$\xx =Q^2/2\nu$ ($\nu=P{\cdot q}$)
fixed, $F(\xx,Q^2)$ is assumed to have the form of a sum over 
contributions for 
different $i=q_f,\oq_f,G$.
Taking into account these considerations the expression for the structure 
function reduces to a single 
variable integral
\be
F(\xx,Q^2) \sim \sum_{i=q_f,\oq_f,g} \int_\xx^1 \! \frac{\dd y}{y}
\, C_i \Big(\frac{\xx}{y}, \frac{Q^2}{\mu_F^2};\alpha_S\Big ) \, f_i(y,\mu_F^2) \, ,
\label{FQCD2}
\ee
where
\be
f_i(y,\mu^2) = \Big ( q_f (y,\mu^2) , \oq_f (y,\mu^2) , g(y,\mu^2) \Big ) \, ,
\quad i =q_f,\oq_f,g \, , 
\ee
and we now integrate over the possible values of the momentum fraction $y$. 

The definition of the parton distributions is 
the same as in the previous argument
except for three points. 

\begin{enumerate}

\item Now we also have a nonperturbative contribution 
corresponding to the possibility of scattering off a gluon in the hadron.

\item The momentum fraction of the parton leaving the hadron is denoted by
$y$, where $y \geq x$ since some of the original momentum may be lost by 
branching to other
particles before the scattering with the photon which defines the variable $x$.

\item The infrared singularities in the coefficient functions which have 
been regularised by $\mu_F$ must be absorbed into the nonperturbative
definition of $\Gamma(P,k)$ rendering it $\mu_F$ dependent when we include 
QCD corrections. This is natural because the singularities come from the 
infrared limit of the 
integral over $k$ where the coupling is strong and really we should be 
using nonperturbative physics. The divergences are determined entirely in terms of the 
incoming parton, and are independent of the particular scattering process as long
as it is one which sums over final states, though we can be slightly 
less inclusive and define e.g. final state jets.   

\end{enumerate}

\medskip

It is important to recognise that $F(\xx,Q^2)$ as a potentially measurable
physical quantity must be independent of $\mu_F$. In general for  vectors
$A_i,B_i$
\be
\mu_F\frac{\dd}{\dd \mu_F}\Big(A_i B_i\Big )=0  \ \ \Rightarrow \ \
\mu_F\frac{\dd}{\dd \mu_F} A_i = - A_jP_{ji} \, , \ \
\mu_F\frac{\dd}{\dd \mu_F} B_i =  P_{ij}B_j \, ,
\label{AB}
\ee
for some $P_{ij}$.
The integral convolution in Eq. (\ref{FQCD2}) can be regarded similarly as a form 
of matrix multiplication for two $\mu_F$-dependent factors. The analogous version
of the equations for $A,B$ in Eq. (\ref{AB}) become integral relations
\begin{eqnarray}
\mu_F\frac{\dd}{\dd \mu_F} 
C_i \Big(x, \frac{Q^2}{\mu_F^2};\alpha_S\Big )&=& - \!\!\!\!
\sum_{j=q_f,\oq_f,g} \int_x^1 \! \frac{\dd y}{y} \,
C_j \Big(y, \frac{Q^2}{\mu_F^2};\alpha_S\Big ) \, P_{ji}
\Big(\frac{x}{y};\alpha_S\Big ) \, , \label{AP1}\\
\mu_F\frac{\dd}{\dd \mu_F} f_i(y,\mu_F^2) &=& \!\!\! \sum_{j=q_f,\oq_f,g} 
\int_y^1 \! \frac{\dd z}{z} \, P_{ij}\Big(\frac{y}{z};\alpha_S\Big ) \, 
f_j(z,\mu_F^2) \, , \label{AP2}
\end{eqnarray}
where the $P_{ij}(y;\alpha_S)$ are determined by the form of the infrared divergences 
regularised by $\mu_f$ and absorbed into the nonperturbative
definition of the partons. As such they are
independent of $Q^2$, the particular current $J$
and the hadron $H$, and may be determined as an expansion in $\alpha_S$ from
Eq. (\ref{AP1}). In general all components of $P_{ij}(y;\alpha_S)$ are non
zero. 

The equations (\ref{AP1},\ref{AP2}), are referred to as the 
DGLAP equations\cite{Altarelli:1977zs,Lipatov:1974qm,Gribov:1972ri,Dokshitzer:1977sg}, 
and the perturbatively calculable $P_{ij}(y;\alpha_S)$
are known as splitting functions. 
They were effectively derived as anomalous dimensions of operators
within the context of the renormalisation group and operator product expansion\cite{Gross:1973ju,Georgi:1951sr}.
The coefficient functions and splitting functions were obtained at 
next to leading order (NLO) (${\cal O}(\alpha_S)$ for the coefficient functions
and ${\cal O}(\alpha_S^2)$ for splitting functions) within a few years\cite{Bardeen:1978yd,Floratos:1977au,Floratos:1978ny,GonzalezArroyo:1979df,Curci:1980uw,GonzalezArroyo:1979he,Furmanski:1980cm,Floratos:1981hs}.
In these above equations we should take $\alpha_S \to
\alpha_S(\mu_R^2)$ the running coupling.
It is important to note that $\alpha_S(\mu_R^2)$ is a function of the
renormalisation scale $\mu_R$ not the factorisation scale $\mu_F$ since its running is
determined by the renormalisation of the ultraviolet divergences in the theory, and is
nothing to do with the infrared regularisation which introduces $\mu_F$.

Since $\mu_R$ and $\mu_F$ are arbitrary we may choose their values independently.
However, it is natural, and very common to set $\mu_R^2=\mu_F^2=Q^2$ so that
Eq. (\ref{FQCD2}) becomes
\be
F(\xx,Q^2) \sim \sum_{i=q_f,\oq_f,g} \int_\xx^1 \! \frac{\dd y}{y}
\, C_i \Big(\frac{\xx}{y}, 1;\alpha_S(Q^2) \Big ) \, f_i(y,Q^2)
\equiv \sum_{i=q_f,\oq_f,g} \, C_i (\alpha_S(Q^2)) \otimes f_i(Q^2)\, ,
\label{FQCD3}
\ee
where from (\ref{AP2})
\be
Q\frac{\dd}{\dd Q} f_i(y,Q^2) = \! \sum_{j=q_f,\oq_f,g}
\int_y^1 \! \frac{\dd z}{z} \, P_{ij}\Big(\frac{y}{z};\alpha_S(Q^2)\Big )
\equiv \sum_{j=q_f,\oq_f,g} \,P_{ij}(\alpha_S(Q^2) ) \otimes
f_j(Q^2) \, .
\label{AP3}
\ee
The results Eq. (\ref{FQCD3}) and Eq. (\ref{AP3}) then provide the justification for
the claim that asymptotic freedom allows the $Q^2$ {\em evolution} 
of $F(\xx,Q^2)$
to be calculated perturbatively in the deep inelastic limit.
Hence, once we have measured the parton distributions at some low scale $Q_0^2$ 
we can calculate their evolution to higher scales perturbatively. Comparison 
of theory and data on structure functions and their scaling violations 
works extremely well, and is one of the best tests of QCD.  

\begin{figure}
\begin{center}
\begin{minipage}[t]{12 cm}
\begin{picture}(255,200)(0,0)
\Text(6,199)[]{P}
\ArrowLine(10,195)(50,160)
\ArrowLine(50,160)(75,130)
\ArrowLine(75,130)(100,100)
\ArrowLine(50,160)(90,175)
\ArrowLine(50,160)(90,165)
\ArrowLine(50,160)(90,155)
\ArrowLine(50,160)(90,145)
\CCirc(50,160){5}{Black}{Black}
\Text(100,160)[]{$X'$}
\Text(147.5,100)[]{$H$}
\DashLine(100,100)(137.5,100){5}
\Gluon(75,130)(115,130){5}{6}
\ArrowLine(75,70)(50,40)
\ArrowLine(100,100)(75,70)
\ArrowLine(10,5)(50,40)
\ArrowLine(50,40)(90,55)
\ArrowLine(50,40)(90,45)
\ArrowLine(50,40)(90,35)
\ArrowLine(50,40)(90,25)
\CCirc(50,40){5}{Black}{Black}
\Text(100,40)[]{$X'$}
\Gluon(75,70)(115,70){5}{6}
\Text(6,1)[]{P}
\Text(235,160)[]{$f_{i}(x_i,m_H^2,\alpha_S(m_H^2))$}
\Text(235,100)[]{$C^H_{ij}(x_i,x_j,\alpha_S(m_H^2))$}
\Text(235,40)[]{$f_{j}(x_j,m_H^2,\alpha_S(m_H^2))$}
\end{picture}
\caption{Factorisation in a hadron-hadron collision. \label{hadronfact}}
\end{minipage}
\end{center} 
\end{figure}
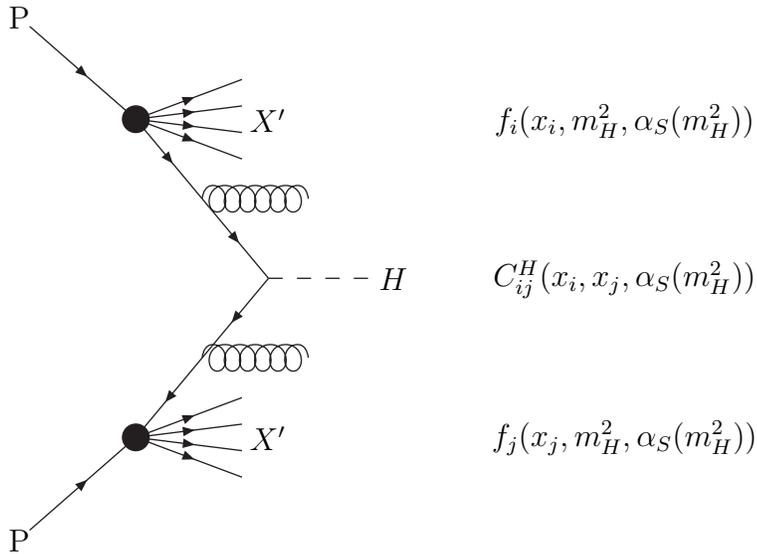

\medskip
We can apply the same sort of reasoning as above to hadron-hadron collisions.
The coefficient functions $C_{i}(x,\alpha_S(\mu^2))$ 
describing a particular hard 
scattering process involving incoming partons 
are process dependent but are calculable as a power-series
in the strong coupling constant $\alpha_S(\mu^2)$.
$$ 
C^P(x,\alpha_S(\mu^2))= \sum_k C^{P,k}(x)\alpha^k_s(\mu^2).
$$
The scale of the coupling will be set by the hard scale $q^2$ in the particular
process, e.g. if one produces a particle with large mass $m$ in the final state then 
$q^2=m^2$. If there is no hard scale in the perturbative scattering process, e.g.
if we simply have proton-proton scattering to hadrons with no identified hard final
state, perturbation theory cannot be reliably used.  
Since the parton distributions $f_{i}(x,q^2)$ 
are process-independent, i.e. {\it universal}, once they have been measured at 
one experiment, one can predict many other scattering processes. Consider for 
example the diagram for proton-proton scattering to form hadrons plus a Higgs 
boson, a contribution to which is shown in Fig. \ref{hadronfact}.  

The definition of the parton distributions is exactly the same for this diagram 
as it is in Deep Inelastic Scattering. Hence, once we calculate 
$C^H_{ij}(x_i,x_j,\alpha_S(m_H^2))$ we can calculate the cross section for Higgs
production at a proton-proton collider, i.e. the Tevatron and/or 
Large Hadron Collider (LHC). 
This is given simply by 
\be
\sigma_H(x_1,x_2,m_H^2) = \sum_{i,j=q_f,\oq_f,G} 
\int_{x_1}^1\int_{x_2}^1 \! \frac{\dd y_1}{y_1}
\frac{\dd y_2}{y_2}
\, C^H_{ij} \Big(\frac{x_1}{y_1},\frac{x_2}{y_2} \frac{m_H^2}{\mu^2};\alpha_S
\Big) \, f_i(y_1,\mu^2)f_j(y_2,\mu^2) \, ,
\label{FQCD4}
\ee
This general procedure can be applied to any process, so although parton 
distributions are essentially nonperturbative their determination in a small 
number of experiments then leads to huge predictive power. A comprehensive 
discussion of factorisation and its proof can be found in\cite{Collins:1989gx}. We finally note that the definition of the 
parton distributions can be altered so long as the coefficient functions are 
changed in a compensating manner such that all physical quantities are 
invariant. This is called a change of factorisation scheme. In the majority of 
cases we work with PDFs defined in the $\overline{\rm MS}$ scheme, since
calculations are done in manner where both ultraviolet and infrared divergences
are removed using dimensional regularisation and the  $\overline{\rm MS}$
procedure (the manner of dealing with ultraviolet divergences defining the 
renormalisation scheme and infrared divergences defining the factorisation 
scheme). An alternative factorisation scheme sometimes used is the DIS scheme\cite{Altarelli:1978id}, which is simply defined so that the LO relationship
between structure functions and quarks in Eq. (\ref{proton}) is true to all
orders.  

\subsection{\it Recent Progress on Experimental Data}
As will be shown  throughout this report, many different experiments have made measurements
 which can be used
to constrain the structure of the proton. DIS experiments 
constitute by far the most important
data input, and in the last two decades it has been especially the HERA electron proton collider that 
has led to spectacular progress in the understanding of the proton structure.

%\begin{figure}[tbp]
%%\vspace{-0.2cm} 
%%\vspace*{5pt}
%\centerline{
%\epsfig{figure=d09-158f20.eps  ,width=0.7\textwidth}}
%\caption {
%The parton distribution functions obtained using ACOT heavy-flavour scheme
%compared to  
%HERAPDF1.0 for $xu_v,xd_v,xS=2x(\bar{U}+\bar{D}),xg$, at $Q^2=10$~GeV$^2$. 
%The bands show total uncertainties of the HERAPDF1.0 fit.
%}
%\label{fig:acot}
%\end{figure}

HERA started to collect $ep$ collisions in 1992, at first at a centre of mass energy of 300 GeV, and 
later at 320 GeV. HERA collided 27.6 GeV electrons/positrons on 820 (920) GeV protons, which allowed
a measurement of structure functions at $x$ values down to $\sim 10^{-5}$ and to $Q^2$ values of up to 
$\sim 50,000$ GeV$^2$. HERA did not only open a new kinematic domain for DIS, but the collider experimental
environment also allowed the use of different type of detectors, as compared to the classical 
DIS fixed target experiments.
 In particular the hadronic final state is fully measurable for the majority of the $ep$ collisions in the
  collider experiments 
H1 and ZEUS,
allowing for either an excellent control of the systematics of the measurements, or for using hybrid methods 
based on scattered electron and hadronic final state reconstruction of the kinematical variables in 
each collision. Even the 
radiative corrections can be checked in part due the detection of the emitted photons in the direction 
of the electron beam.

\begin{figure}[tbp]
%\vspace{-0.5cm} 
%\vspace*{5pt}
\centerline{
\epsfig{figure=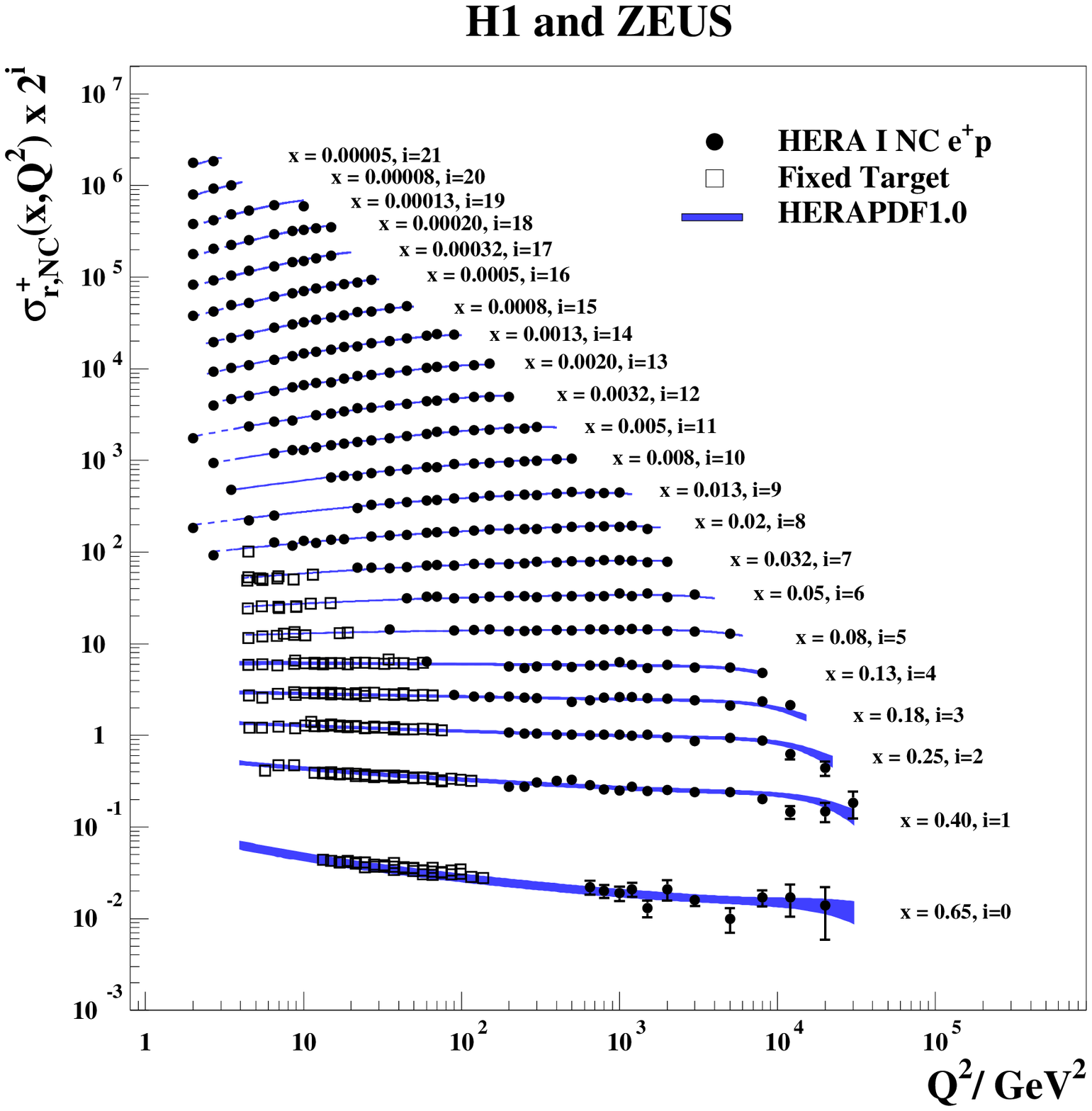,width=0.5\linewidth}
\epsfig{figure=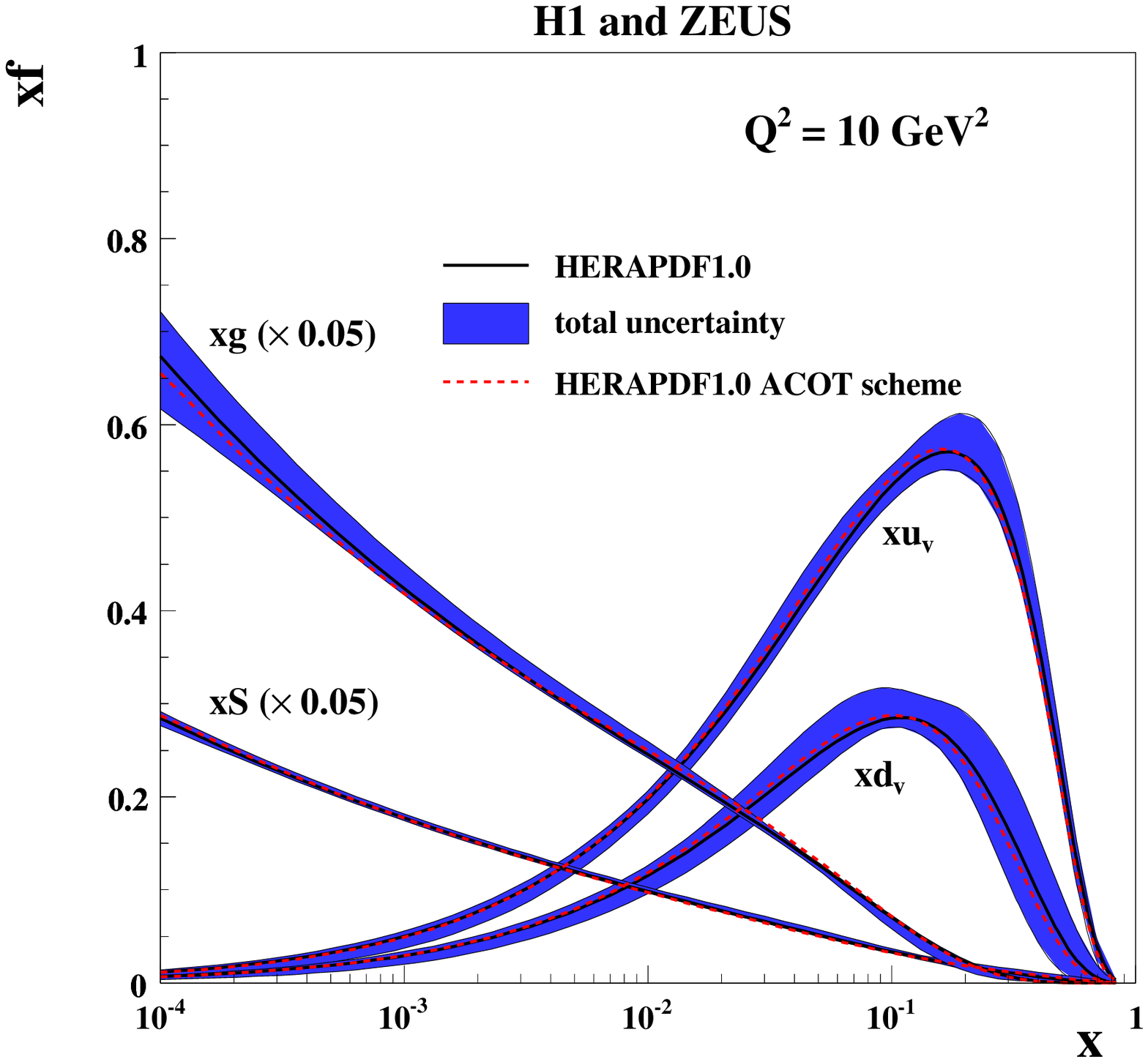  ,width=0.5\textwidth}}
\caption {On the left the HERA combined NC $e^+p$ reduced cross section
 and fixed-target data as a function of $Q^2$.
The error bars indicate the total experimental uncertainty.
The  HERAPDF1.0 fit is superimposed.
On the right the  bands represent the total uncertainty of the fit.
 Dashed lines are shown for $Q^2$ values not included in the QCD analysis.
The parton distribution functions obtained using ACOT heavy-flavour scheme
compared to  
HERAPDF1.0 for $xu_v,xd_v,xS=2x(\bar{U}+\bar{D}),xg$, at $Q^2=10$~GeV$^2$. 
The bands show total uncertainties of the HERAPDF1.0 fit.  
}
\label{fig:scal}
\end{figure}

\begin{figure}
\begin{center}
\begin{minipage}[t]{15.5 cm}
\epsfig{file=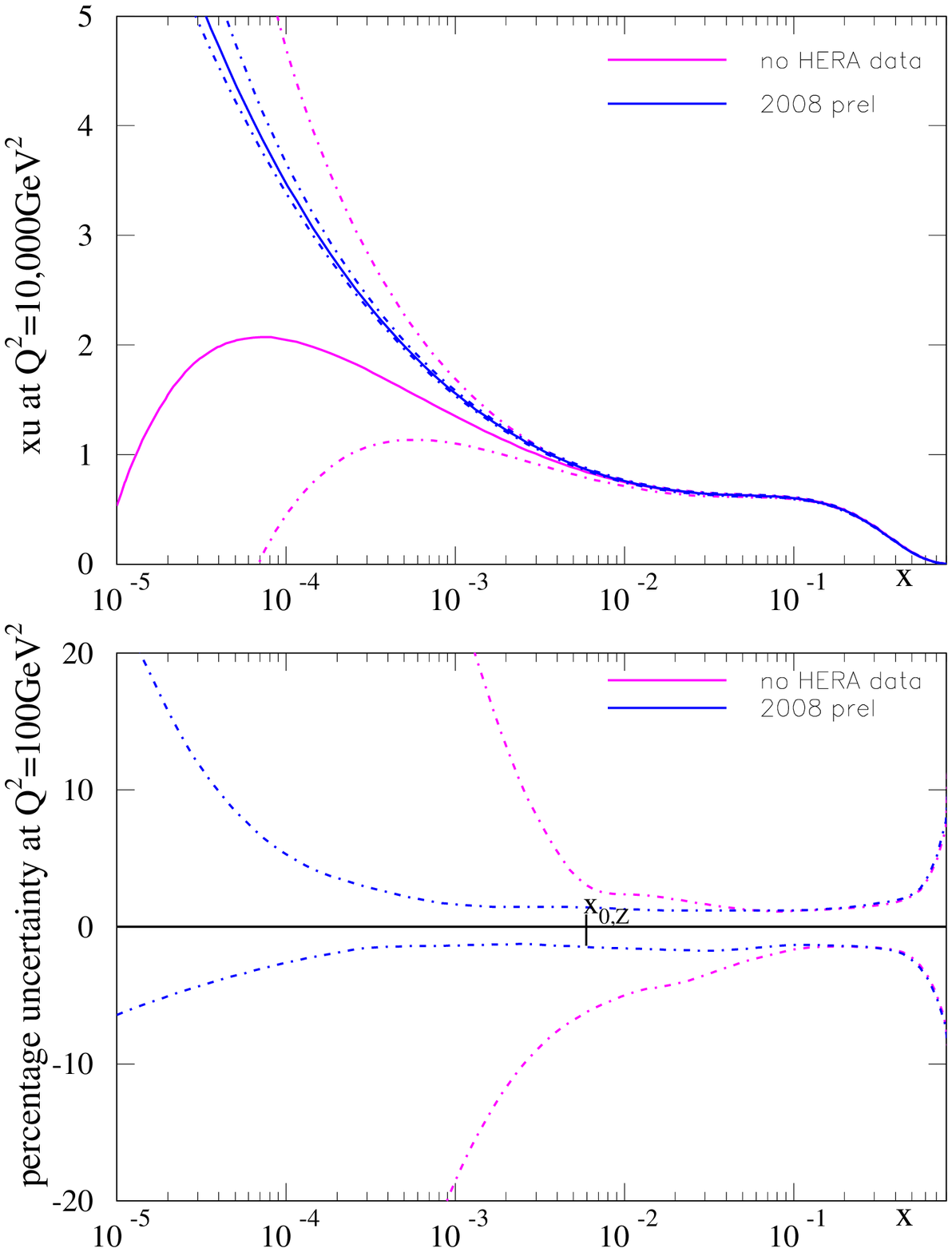,scale=0.4}
\hspace{1.5cm}
\epsfig{file=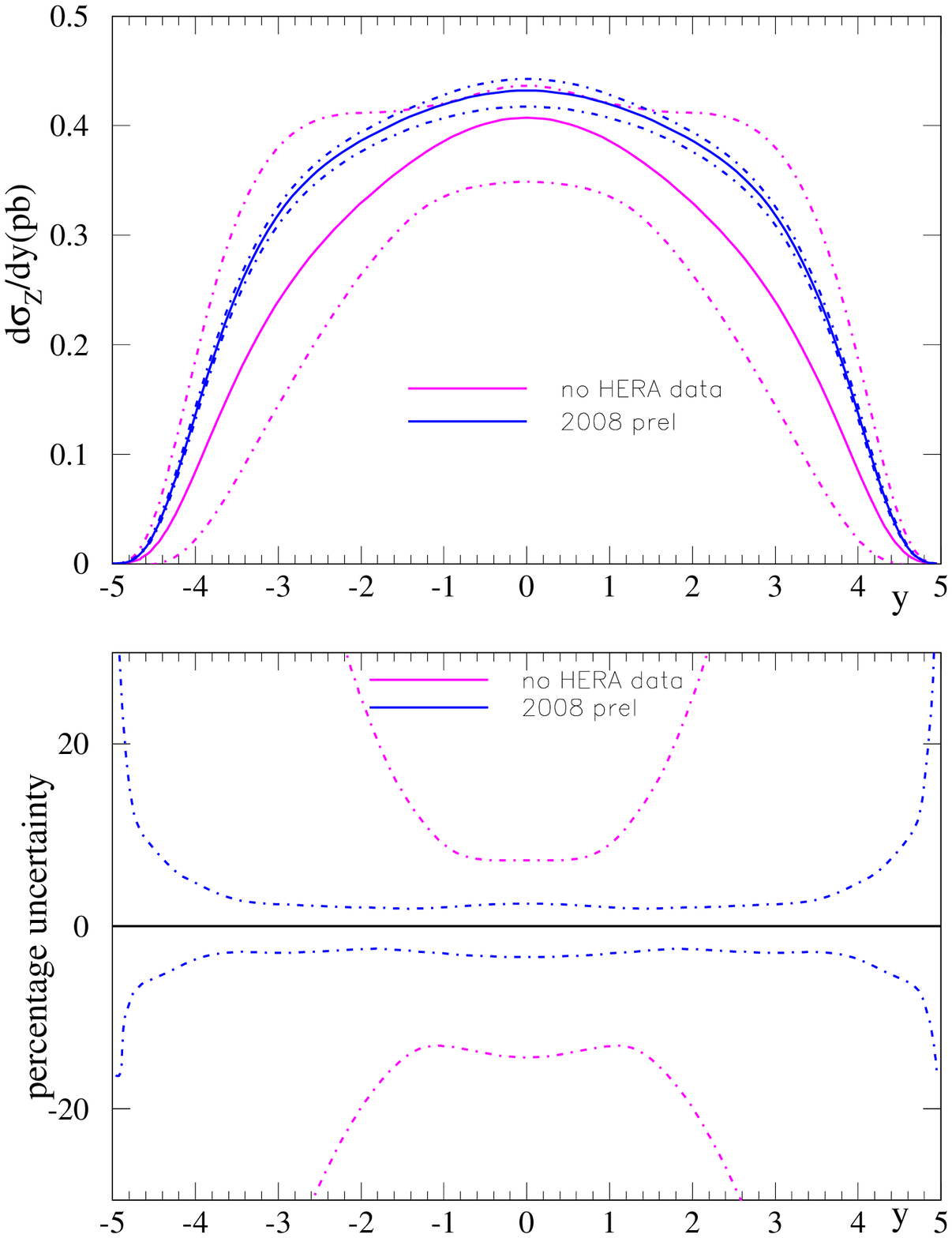,scale=0.4}
\end{minipage}
\begin{minipage}[t]{16.5 cm}
\caption{The up quark extracted from a global fit together with uncertainty
with and without HERA data (left) and the rapidity distribution for 
production of $Z$ at the LHC at 14 TeV, and the uncertainty for the same PDF 
fits with and without HERA data (right). 
\label{HERAdata}}
\end{minipage}
\end{center}
\end{figure}

As a result, in the low and medium $Q^2$ range, where the statistics is abundant, the
structure function measurements
have an accuracy of 1-2\%, and thus allow for a very precise determination of the quark content of the 
proton. The most recent structure function results, based on  the combined H1 and ZEUS data, 
and compared to lower energy fixed target experimental data, 
are shown in the left of Fig.~\ref{fig:scal}. These data have been used in QCD fits, as will be described below,
and the resulting parton distributions
 are shown in the right of Fig.\ref{fig:scal}. The blue bands show 
uncertainties on these parton distributions. Much of the discussion in this report will focus on how such
error bands can be determined.
The importance the HERA data have played for e.g. particle production at the LHC is shown
in Fig. \ref{HERAdata}: it shows the effect of the HERA data in the determination of 
the quark distribution and the precision of the
$Z$ production cross section at the LHC for collisions at a centre of mass energy of 14 TeV. The uncertainties are greatly reduced when  HERA data is included, compared to the result  
in a world without HERA data.

Apart from the fully inclusive measurements, the HERA measurements also allows for measurements
of final states including jets -which will help to constrain $\alpha_s$ in the fits- and more importantly, the 
presence of silicon vertex detectors in the H1 and ZEUS experiments allow for the tagging of collisions
with heavy flavours. The cross sections of events with either charm or bottom quarks in the final state
can be used to determine the $F_2^c(x,Q^2)$ and $F^b_2(x,Q^2)$ heavy flavour structure functions.
The present data analysed from run I at HERA (1992-2000) allows for measurements with a 
precision of the level of 
15\% for charm and 30\% for bottom tagged structure functions.
Just before its closure in 2007 the HERA machine was operated at reduced centre of mass energies, namely
575 and 460 GeV. Measuring the cross section for given $x, Q^2$ values at different centre of mass 
energies is equivalent to measuring at a different value of $y$. Thus the combination of measurements at
different energies allows to disentangle $F_2(x, Q^2) $ and $F_1(x, Q^2) (F_L(x,Q^2)) $ in eq. (17).
These measurements provide extra constraints in the QCD fits and the $F_L(x,Q^2) $ data in 
particular are directly sensitive to the gluon density distributions.

Besides the structure function data, important new data for constraining PDFs come from Tevatron 
measurements such as the di-jets, Z production and the W asymmetries at the collider 
experiments, as well as Drell-Yan measurements.
The usage and impact of these data sets will be discussed in the sections in the following.
In future similar data from the LHC itself will be useful input for further 
constraining the proton structure. The
experiments are gearing up to make such measurements.

\section{Fits to Determine Parton Distributions}

\subsection{\it General Procedure}

From the previous section we see that in order to make predictions for 
hard scattering processes at any collider which uses hadrons in the 
initial state we must first obtain an extraction of the parton distributions
of the hadron by fitting to the existing data which constrains these parton
distributions. The process for doing this is generally called a ``global
fit'', although there are a wide variety of definitions of what ``global''
actually means. 
However, in all cases the basic procedure is very similar.
Global fits\cite{Martin:2009iq,Nadolsky:2008zw,Ball:2010de,:2009wt,Alekhin:2009ni,Gluck:2007ck}
to determine parton distributions 
use available data, in all cases largely $ep \to eX$ (Structure Functions), and the most 
up-to-date QCD calculations to best determine the parton distributions and
their consequences. Currently the default is to use NLO--in--$\alpha_S(Q^2)$, i.e. for 
the coefficient functions and splitting functions this means
\begin{eqnarray} C^P_{i}(x,\alpha_S(Q^2))&=& \alpha_S^P C^{P,0}_{i}(x) + 
\alpha_S^{P+1}(Q^2)C^{P,1}_{i}(x).\nonumber \\
P_{ij}(x,\alpha_S(Q^2))&=& \alpha_S(Q^2)P^0_{ij}(x) + 
\alpha^2_S(Q^2)P^1_{ij}(x),
\label{nlodefinition}
\end{eqnarray}
where $P$ is process dependent, e.g. $P=0$ for deep inelastic scattering 
where the lowest order process is an electroweak boson scattering from
a quark, but $P=2$ for jet production in hadron-hadron collisions where
an example of a leading-order process is parton-parton annihilation
to form a new parton-parton pair. 
NNLO coefficient functions are known for some processes, e.g. 
structure functions\cite{vanNeerven:1991nn,Zijlstra:1991qc,Zijlstra:1992kj,Zijlstra:1992qd,Moch:2004xu,Vermaseren:2005qc}, and NNLO splitting functions have 
been completed\cite{Moch:2004pa,Vogt:2004mw}. Full NNLO 
fits\cite{Martin:2009iq,Alekhin:2009ni,Gluck:2007ck}
are now just possible using with some
(arguably) very good, and continually improving approximations, though the
precise form of the approximation depends on the group performing the fit. 
   
Perturbation theory is usually thought to be valid if $\alpha_S(Q^2)\lsim 0.4-0.5$. 
Since the running coupling constant $\alpha_S(Q^2)$ is roughly equal to 
\be
\alpha_S(Q^2) \approx 
\frac{4\pi}{(11-2/3N_f) \ln(Q^2/\Lambda^2_{QCD})},
\label{LOcoupling}
\ee
where $\Lambda_{QCD}$ is the scale of hadronic physics, i.e 
$\sim 150{\rm MeV}$ at LO in QCD, one can use perturbation theory if 
$Q^2 > 2 {\rm GeV^2}$. This cut should also remove the influence of 
higher twists. Hence, most global fits start evolution at
$Q_0^2$ in a range from about $1-5 \GeV^2$ (an exception is\cite{Gluck:2007ck} which start somewhat lower) and fit data with a minimum 
$Q^2$ of about $2-5 \GeV^2$. Additional cuts and/or higher twist
corrections are also generally applied, as discussed later. 

In principle there are 13 different parton distributions to consider
\be 
u, \bar u, \quad d, \bar d, \quad s, \bar s, \quad c, \bar c, 
\quad b, \bar b, \quad t, \bar t \quad g 
\label{pdflist}
\ee
However, $m_c, m_b, m_t \gg \Lambda_{{\rm QCD}}$ so these heavy 
flavour parton distributions 
are determined perturbatively. However, 
even at the LHC we are at energies not very 
far above the threshold for top production and it is normally most
useful to consider them as only final state particles. Hence, most PDF sets do not 
include the top quark and antiquark as a parton. 
This is consistent with the majority of cross section
calculations which use a renormalisation scheme where top is indeed
assumed to only be created in the final state. There are models of
nonperturbative ``intrinsic'' charm and bottom quark contributions\cite{Brodsky:1980pb}, and some fits investigate the importance of these,
but they are currently not part of the default fit for any group. 

Until recently it has been standard to assume that $s=\bar s$. This leaves 6 
independent combinations of partons. However, with the
most recent data there is some constraint on $s-\bar s$ (as discussed later)
and currently this is allowed to be nonzero at input in some sets\cite{Martin:2009iq,Ball:2010de}, though at NNLO a tiny asymmetry is generated 
by evolution even if it is zero at input\cite{Catani:2004nc}. 
Until the most recent sets it was also common to use $s(Q_0^2) = 
\kappa 1/2(\bar u(Q_0^2) + \bar d(Q_0^2))$, 
where in practice $\kappa\approx 0.4$, but this is also becoming more 
sophisticated with the most recent fits, and some shape as well as
normalisation difference is usually allowed when comparing the
strange and light quark sea. 

For the up and down quarks and antiquarks and the gluon there are then 
five degrees of freedom (which may overlap with the strange quark
parameterisation as explained above). These can be represented in a
variety of fashions, but some are more obviously useful than others. 
For example MSTW use
\be
u_V = u- \bar u, \quad d_V =d-\bar d, \quad \bar d - \bar u 
\quad {\rm sea}=2*(\bar u + \bar d + 
\bar s), \quad g,
\label{partonchoice}
\ee
where the first three combinations are all nonsinglet combinations. 
Even though it is a combination of the distributions already mentioned it 
is also often useful to define the singlet quark distribution
\be
\Sigma = u_V + d_V + {\rm sea} +(c+\bar c) +(b+\bar b).
\label{singletdefinition}
\ee

For each group the input partons are parametrised in some fashion (though 
for\cite{Ball:2010de} this involves a very large effective number of parameters).
For example in\cite{Martin:2009iq} a number of the input distributions
have the general form
\be
xf(x, Q_0^2) = (1-x)^{\eta}(1+\epsilon x^{0.5}+\gamma x)
x^{\delta}.
\label{inputparameterisation}
\ee
There is much variation, but all groups (including NNPDF) include the general 
feature of a power of $(1-x)$ as $x \to 1$ and a power (or possibly 
two powers) of $x$ as $x \to 0$. 
For non-singlet combinations, e.g. the valence quarks and 
$\bar d -\bar u$, 
$\delta$ is expected to be $\sim 0.5$. For singlet combinations, 
e.g. the sea and gluon, $\delta$ is expected to be $\sim 0$. 
The values extracted may vary significantly from these expectations,
and are rather dependent on the value of $Q_0^2$ taken, 
particularly for the gluon distribution.

Much of the structure function data is obtained from 
scattering off a deuterium target, so in practice one also needs to define
the parton distributions for the neutron. In the default fits
this is always done by assuming charge symmetry, i.e. a transformation
from $p \to n$ corresponds to 
\be
d(x) \to u(x) \qquad u(x) \to d(x),
\label{isotransform}
\ee
i.e. in this exact limit for a neutron target the LO expression 
equivalent to Eq. ({\ref{proton}}) is,
\be
F_{2,{\rm neutron}}(\xx,Q^2) \sim \xx \Big ( {\ts\frac{4}{9}}(d(\xx)+\od(\xx))
+ {\ts\frac{1}{9}}(u(\xx)+\ou(\xx))+  \ {\rm heavy} \ {\rm flavours} 
\Big ) \, .
\label{neutron}
\ee

In practice not all the parameters in the inputs for 
the parton distributions are free. There are a number of 
sum rules constraining the parton inputs and which are  
maintained in the evolution equations order by order in 
$\alpha_S$. There are the rules  
\be
\int_0^1 u_V(x)\, dx =2 \qquad \int_0^1 d_V(x)\, dx =1
\label{valencesum}
\ee
i.e. the conservation of the number of valence quarks. 
There is also the conservation of the momentum carried by the partons
\be
\int_0^1 x\Sigma(x) +x g(x) \, dx =1. 
\label{momentumsum}
\ee
This turns out to be an important constraint on the form of gluon 
distribution which is less directly and precisely constrained than 
the quarks. 

In determining the full sets of parton distributions we need to consider 
that not only are there at least 6 different combinations of partons, but 
there is also an extremely wide distribution of
$x$ both probed and needed which extends, in the former case from from 
$x=0.75$ to $x=0.00003$. Hence, in practice we need very many different types 
of experiment for a full and precise determination of all parton distributions.

\subsection{\it Large-$x$ Quarks}

\begin{figure}
\begin{center}
\begin{minipage}[t]{16 cm}
\epsfig{file=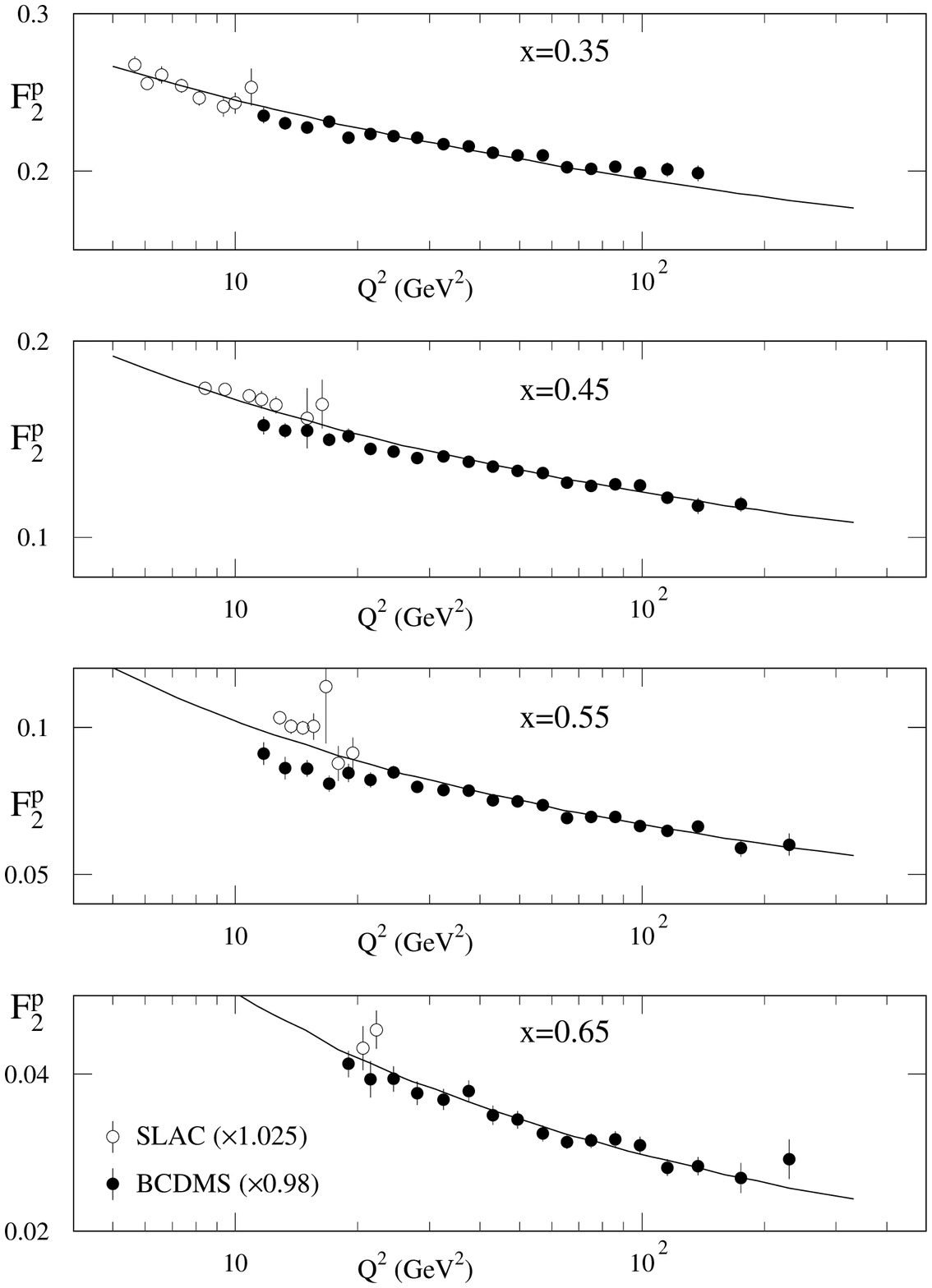,scale=0.4}
\hspace{1cm}
\epsfig{file=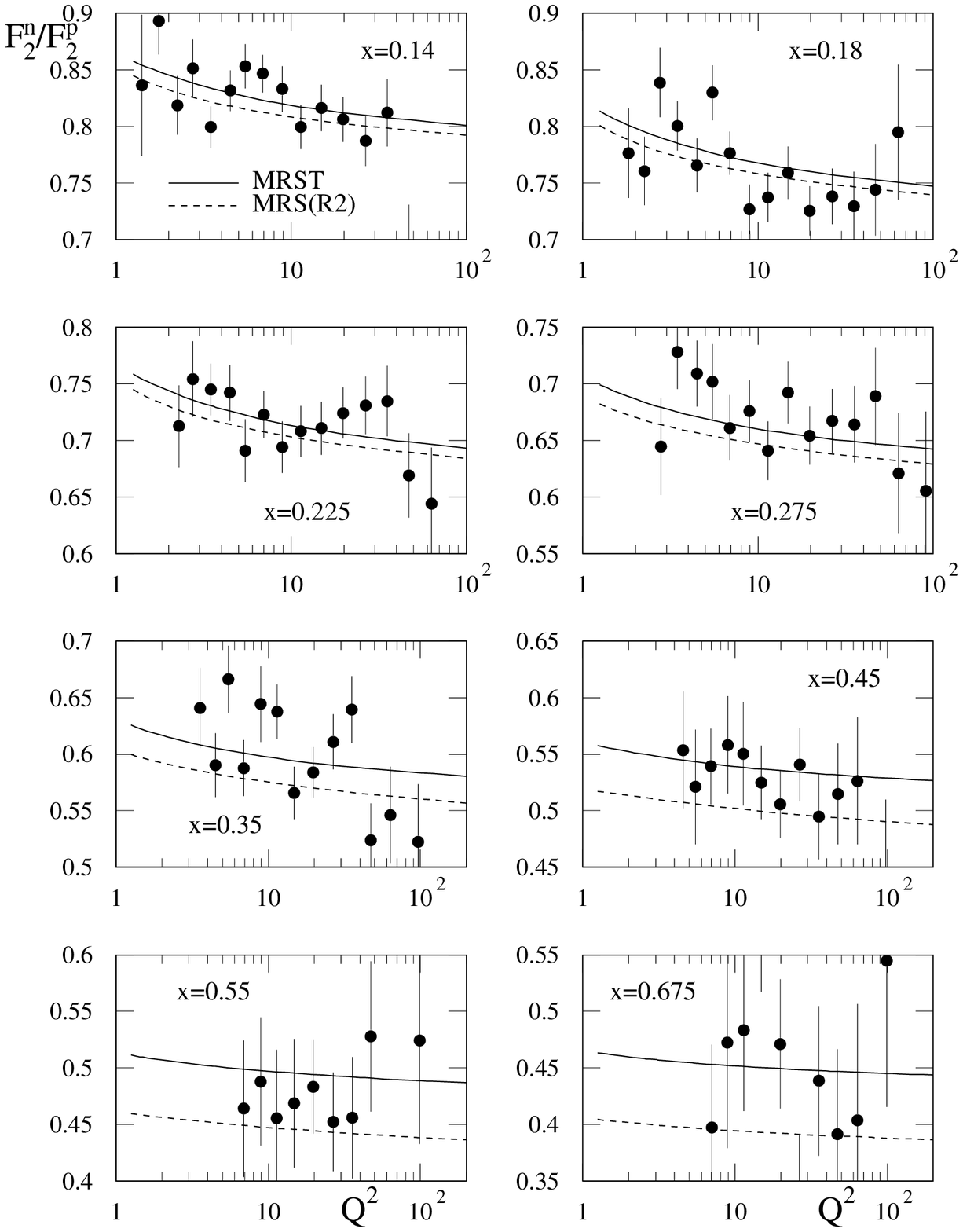,scale=0.4}
\end{minipage}
\begin{minipage}[t]{16.5 cm}
\caption{The description\cite{Martin:1998sq} of large $x$ 
BCDMS and SLAC measurements of $F_2^p$ (left) and of large $x$ NMC measurements of 
$F_2^n/F_2^p$ (right).  \label{BCDMS}}
\end{minipage}
\end{center}
\end{figure}

Let us consider how each type of parton distribution is determined
by a fit to experimental data. We start with probably the most
obvious example of the quark distributions at large $x$. In the
simplest parton model the parton distribution is dominated by
up and down valence quarks with $x \sim 0.3$. In detail the
picture is much more complicated, but the up and down 
valence distributions for $x>0.1$ are indeed a major constituent
of the proton and, until we approach $x=1$ are very precisely 
determined. For $x>0.1$ the quark distributions    
are determined almost entirely by comparison to structure 
function data. In this region they are dominated by the non-singlet 
valence distributions, i.e. one is unlikely to 
find sea quarks or gluons as $x \to 1$.  
The approximation of a non-singlet distribution leads to
a simple evolution of the parton distributions and conversion to 
the structure functions 
\begin{eqnarray} \frac{d\, f^{NS}(x,Q^2)}{d \ln Q^2} &=&  
P^{NS}(x,\alpha_S(Q^2))\otimes f^{NS}(x,Q^2) \nonumber \\
F_2^{NS}(x,Q^2)& = & C^{NS}(x,\alpha_S(Q^2))\otimes
f^{NS}(x,Q^2,\alpha_S(Q^2)) \nonumber 
\label{nonsingletevol}
\end{eqnarray}
This means that the evolution of the high $x$ structure functions 
is a good test of the theory of QCD and provides a 
direct measurement of  $\alpha_S(Q^2)$ which is 
the only parameter in Eq. (\ref{nonsingletevol}) other
than the parton distribution.
However - this very clean picture is disturbed somewhat by
the fact that perturbation theory involves contributions 
to coefficient functions $\sim \alpha_S^n(Q^2) \ln^{2n-1}(1-x)$.
Related to this reduced convergence of perturbation theory is the fact
that higher twist corrections of the form $(\Lambda_{\rm QCD}^2/Q^2)$
are known to be enhanced as $x \to 1$. Hence, it is common to 
 impose a cut of data and require $W^2 = Q^2(1/x-1)+m_P^2$ 
to be greater than $10-15 \GeV^2$ to avoid contamination of perturbation 
theory (or alternatively to put in a parameterisation of higher twists
to simultaneously fit these corrections\cite{Alekhin:2009ni}).
This leads to the precision of the extraction of the quarks
becoming more limited as $x$ increases above about $0.6$.  

%\begin{figure}
%\begin{center}
%\begin{minipage}[t]{8 cm}
%\epsfig{file=fig-NMCratio.ps,scale=0.4}
%\end{minipage}
%\begin{minipage}[t]{16.5 cm}
%\caption{The description\cite{Martin:1998sq} of large $x$ NMC measurements of 
%$F_2^n/F_2^p$. \label{NMCnp}}
%\end{minipage}
%\end{center}
%\end{figure}

There are various different types of structure function data
which constrain high-$x$ quarks. The most obvious is   
charged lepton proton scattering, for which the differential cross section is 
\be
\frac{d^2 \sigma}{dxdQ^2} = \frac{2 \pi \alpha^2}{Q^4}[
(1+(1-y)^2)F_2(x,Q^2) -y^2 F_L(x,Q^2)]
\label{neutralcurrent}
\ee
where we ignore $W$ and $Z$ exchange
(which is a small correction except for the highest-$Q^2$ HERA data), 
and where $y=Q^2/xs$. 
Both $F_L(x,Q^2)$ and $y$ are usually small so the cross section 
is effectively a measure of $F_2(x,Q^2)$. 
\begin{eqnarray} F^p_2(x) &\approx& x[4/9(u+\bar u +c + \bar c)
+1/9(d+\bar d+
s+\bar s)\bigr] \nonumber \\ F^d_2(x) &\approx& 
x[4/9(d+\bar d +c + \bar c)+1/9(u+\bar u+
s+\bar s)\bigr] 
\label{LOneutralcurrent}
\end{eqnarray}
This means that SLAC\cite{Whitlow:1991uw}, BCDMS\cite{Benvenuti:1989rh}, 
NMC\cite{Arneodo:1996qe} and E665\cite{Adams:1996gu} data on $F^p_2(x,Q^2)$ 
and $F^d_2(x,Q^2)$\cite{Whitlow:1990dr,Benvenuti:1989fm,Arneodo:1996qe,Adams:1996gu}
and  a dedicated measurement by NMC of of $F^d_2(x,Q^2)/F^p_2(x,Q^2)$\cite{Arneodo:1996kd}
help determine high $x$ parton distributions dominated by valence quarks. 
The fall of the structure functions, and implicitly parton distributions, 
at high $x$ is shown in the left of Fig. \ref{BCDMS} for SLAC and BCDMS $F_2^p(x,Q^2)$
data where high-$x$ partons evolve through splitting 
to smaller $x$ partons. The NMC data translated into the form $F^n_2(x,Q^2)/F^p_2(x,Q^2)$
is shown in the right of Fig. \ref{BCDMS} and compared to two PDF sets 
which postdate and predate this data. One sees that the ratio falls as $x$ approaches 1, 
leading to the clear conclusion that $d(x,Q^2)$ falls more quickly than $u(x,Q^2)$ 
in this limit. However, the behaviour as $x$ reaches 1 is not determined.

\begin{figure}
\begin{center}
\begin{minipage}[t]{16.5 cm}
\epsfig{file=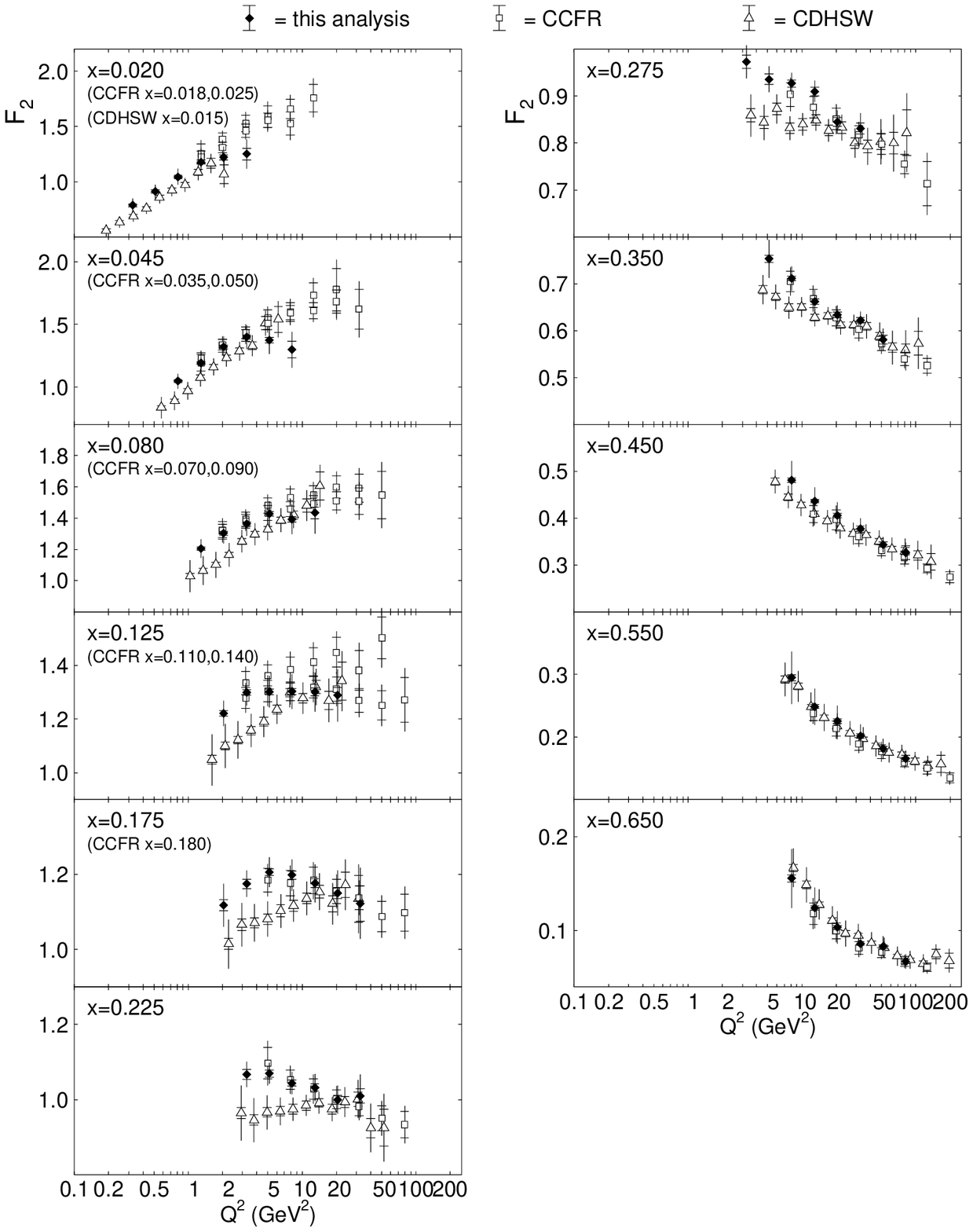,scale=0.48}
\epsfig{file=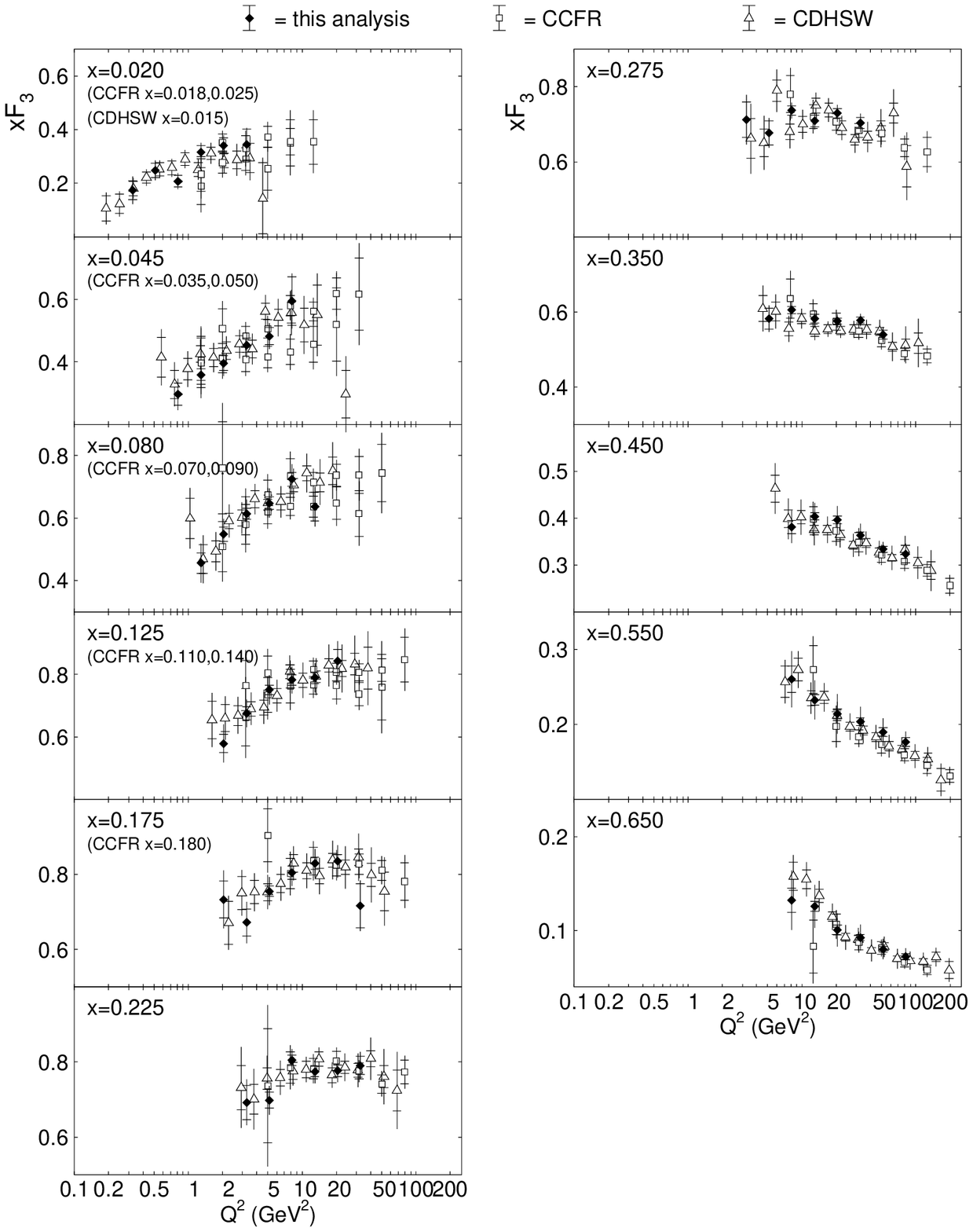,scale=0.48}
\end{minipage}
\begin{minipage}[t]{16.5 cm}
\caption{Large and smaller $x$ measurements of 
$F_2^{\nu N}$ (left) and $F_3^{\nu N}$ (right) in
neutrino scattering\cite{Onengut:2005kv}.   \label{CHORUS}}
\end{minipage}
\end{center}
\end{figure}

Complementary information can be obtained by fitting to charged-current 
(neutrino) DIS data. This is obtained by the CCFR\cite{Yang:2000ju}, 
NuTeV\cite{Tzanov:2005kr} and CHORUS\cite{Onengut:2005kv} collaborations. 
In this case the differential cross section is 
\be \frac{d^2 \sigma^{\nu,\bar\nu}}{dxdQ^2}\propto 
F^{\nu,\bar\nu}_2(x,Q^2)\biggl[(1-y) +\frac{y^2}{2(1+R(x,Q^2))}
\biggr]\pm xF^{\nu,\bar\nu}_3(x,Q^2)y(1-y/2),
\label{chargedcurrent}
\ee
where $R=F_L/(F_2-F_L)$ and $F_3$ appears due to 
parity violation. For the proton at LO
\begin{eqnarray} 
F_2^{\nu}&=& 2x[d+s+\bar u + \bar c] \nonumber \\
          F_2^{\bar \nu}&=& 2x[u+c+\bar d + \bar s] \nonumber \\
          xF_3^{\nu}&=& 2x[d+s-\bar u - \bar c] \nonumber \\
          xF_3^{\bar\nu}&=& 2x[u+c-\bar d - \bar s]. 
\label{protoncc}
\end{eqnarray}
Therefore
\begin{eqnarray} F_2^{\nu}+F_2^{\bar \nu} &=& 2x \sum_i (q + \bar q) 
= 2x\Sigma \nonumber \\             
F_3^{\nu}+F_3^{\bar \nu} &=& 2(u_V + d_V), \nonumber 
\label{ccsum}
\end{eqnarray}
assuming $\s=\bar s$ and $c = \bar c$ in the latter. 
In fact, in order to maximise the cross section the CCFR and
NuTeV measurements are made using an iron target and
the CHORUS measurements a lead target. Both be
corrected to an iso-scalar target, i.e. $F^N= \half(F^p+F^n)$,  so using the charge symmetry 
relationship in Eq. (\ref{isotransform}) we obtain. 
\begin{eqnarray} 
F_2^{N,\nu} &=&  F_2^{N,\bar \nu} = x \Sigma \nonumber \\
          xF_3^{N,\nu}&=& x(u_V +d_V) +2x[s-\bar c]\nonumber \\
          xF_3^{N, \bar\nu}&=& x(u_V +d_V) 
            -2x[s-\bar c] . 
\label{isoscalarcc}
 \end{eqnarray}
The results of the measurements by CCFR and CHORUS are shown in Fig. \ref{CHORUS}
for both $F^N_2(x,Q^2)$ and $F^N_3(x,Q^2)$. At high $x$ both structure functions 
are direct tests of the total valence quark distribution and are very similar 
to each other. At lower $x$ they begin to differ due to much more sea quark contribution to
 $F^N_2(x,Q^2)$.

\begin{figure}
\begin{center}
\begin{minipage}[t]{16.5 cm}
\epsfig{file=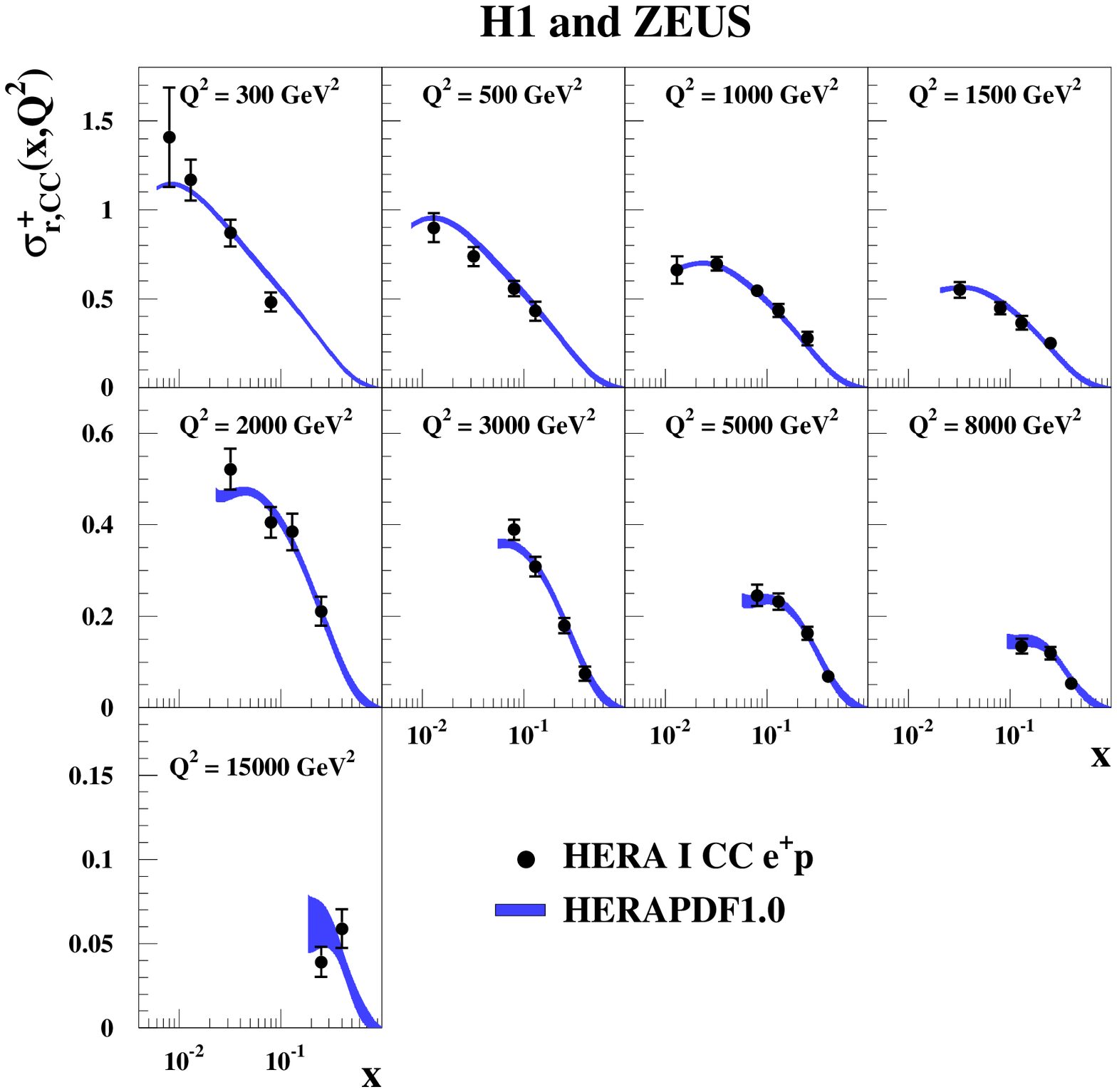,scale=0.42}
\epsfig{file=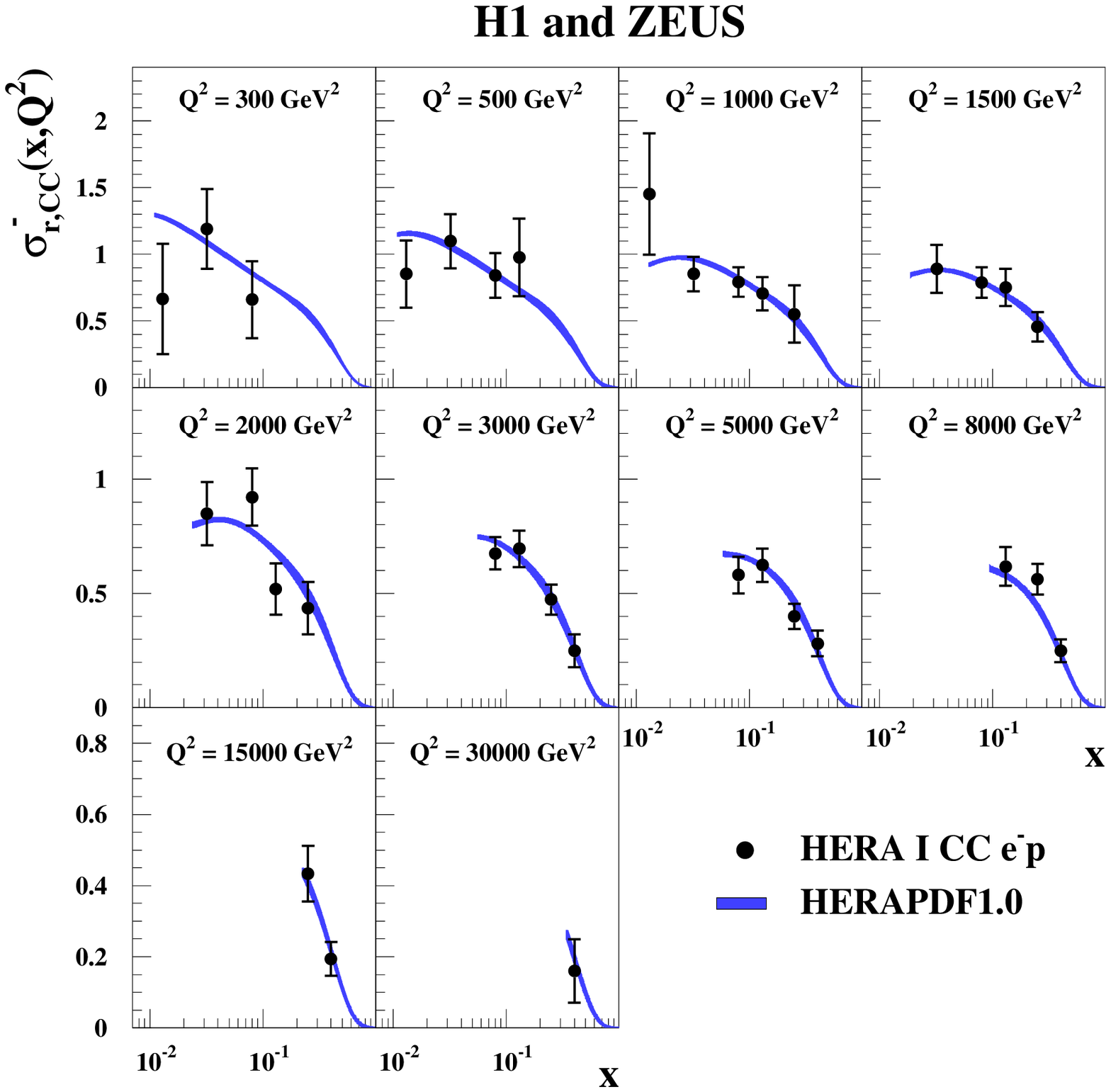,scale=0.42}
\end{minipage}
\begin{minipage}[t]{16.5 cm}
\caption{HERA charged current cross sections for $e^+$ (left) and $e^-$ (right)\cite{:2009wt}.   \label{HERACC}}
\end{minipage}
\end{center}
\end{figure}

As with the charged lepton scattering there are issues which
make the relationship of the structure functions to the 
quark distributions less direct. Again there are higher twist 
contributions, and these may be more significant
for $F_3(x,Q^2)$ than for $F_2(x,Q^2)$\cite{Dokshitzer:1995qm,Dasgupta:1996hh} 
(the latter being protected by the Adler sum rule,
i.e. $\int_0^1 F_2^{HT}(x,Q^2)\, dx = 0$). There is also the issue that 
parton distributions in nucleons in nuclei are 
not expected to be exactly the same as for free nucleons.
This means that global fits generally use use nuclear corrections determined by
fits to exclusively nuclear targets (this is not done
for the default set in\cite{Ball:2010de}). This is a subject worthy of review
in its own right. But examples of corrections can be found in \cite{deFlorian:2003qf,Eskola:2009uj,Hirai:2007sx}.
Comparison between theory and data is good, and leads additional 
information to help in the determination of the valence 
quarks at high $x$. (In principle nuclear corrections should also
be applied in fits to deuterium data. However, it is assumed
these are very small and global fits largely ignore them.)

There is also HERA charged current-data at high $Q^2$\cite{Adloff:2000qj,Adloff:2003uh,Chekanov:2002zs,
Chekanov:2003vw} which 
provides information on valence quarks and flavour decomposition. 
In principle it is superior to both the lower-energy fixed target
neutral current data and the charged current data from nuclear targets
since it is essentially free from higher twist corrections and 
is completely free from nuclear target corrections.  
However, the data analysed and published so far have low statistics,
even when combined\cite{:2009wt}. These data are shown compared to the fit in \cite{:2009wt} in Fig. \ref{HERACC}. They do not currently provide 
comparable constraint to the fixed target data even taking into account
the theoretical uncertainties inherent in fitting to the latter.
However, the precision of the HERA data is likely to improve significantly
once the full run II data has been fully analysed.

\subsection{\it Antiquarks at Large and Moderate $x$}

\begin{figure}
\begin{center}
\begin{minipage}[t]{13 cm}
\begin{picture}(360,290)(30,60)

\Text(62,343)[]{P}
\ArrowLine(70,335)(110,300)
\DashLine(110,300)(150,250){5}
\ArrowLine(150,250)(200,200)
\ArrowLine(110,300)(190,330)
\ArrowLine(110,300)(190,310)
\ArrowLine(110,300)(190,290)
\ArrowLine(110,300)(190,270)
\CCirc(110,300){10}{}{}
\Gluon(150,250)(230,250){5}{6}

\Photon(200,200)(275,200){5}{6}
\ArrowLine(275,200)(325,240)
\ArrowLine(325,160)(275,200)

\DashLine(110,100)(150,150){5}
\ArrowLine(200,200)(150,150)
\ArrowLine(70,65)(110,100)
\ArrowLine(110,100)(190,130)
\ArrowLine(110,100)(190,110)
\ArrowLine(110,100)(190,90)
\ArrowLine(110,100)(190,70)
\CCirc(110,100){10}{}{}
\Gluon(150,150)(230,150){5}{6}

\Text(62,67)[]{P}
\Text(238,218)[]{$\gamma^*(M^2)$}
\Text(202,170)[]{$\bar q_{i}(x_1)$}
\Text(202,230)[]{$q_{i}(x_2)$}
\Text(335,240)[]{$l^-$}
\Text(335,160)[]{$l^+$}
\end{picture}
\caption{The process of Drell-Yan annihilation to produce lepton pairs.   \label{DYfig}}
\end{minipage}
\end{center}
\end{figure}
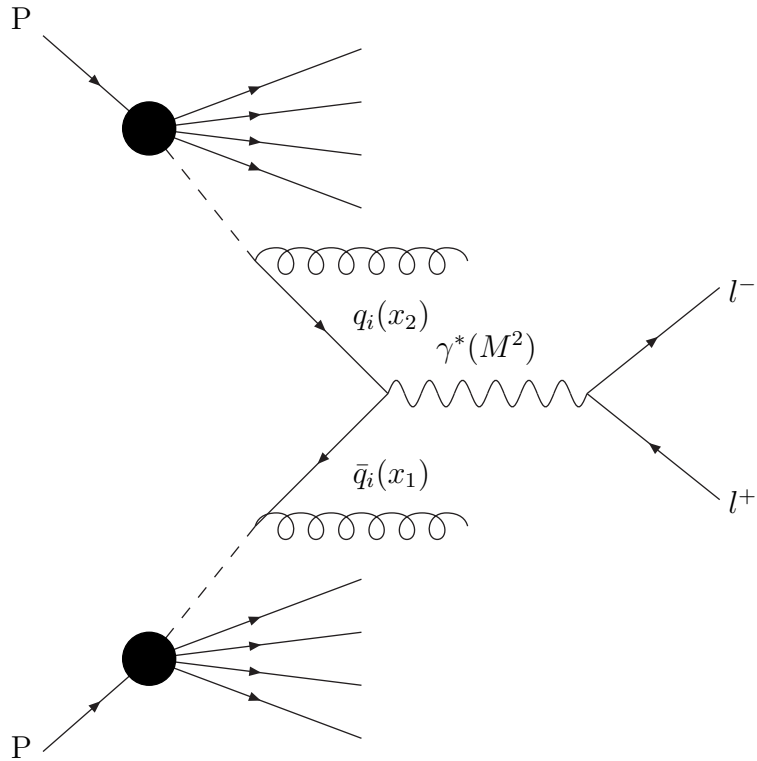

As $x$ decreases the sea quarks become more important. 
To a certain extent these are constrained by the difference between 
the valence quarks extracted from the fits to $F_3(x,Q^2)$ in 
charged current scattering and to $F_2(x,Q^2)$ in all
measurements. However, this is rather indirect, and the 
fit to $F_3(x,Q^2)$ tends to become more sensitive
to nuclear corrections at smaller $x$, i..e. this correction
is larger and more uncertain in this region. A more direct determination, 
which also probes the sea quarks in regions of $x\gsim 0.2$ where they are 
very small, comes from Drell-Yan scattering. 
The process is the production of lepton pairs from quark-antiquark 
annihilation in proton-proton scattering, shown in Fig. \ref{DYfig}.

\begin{figure}
\begin{center}
\begin{minipage}[t]{15 cm}
\epsfig{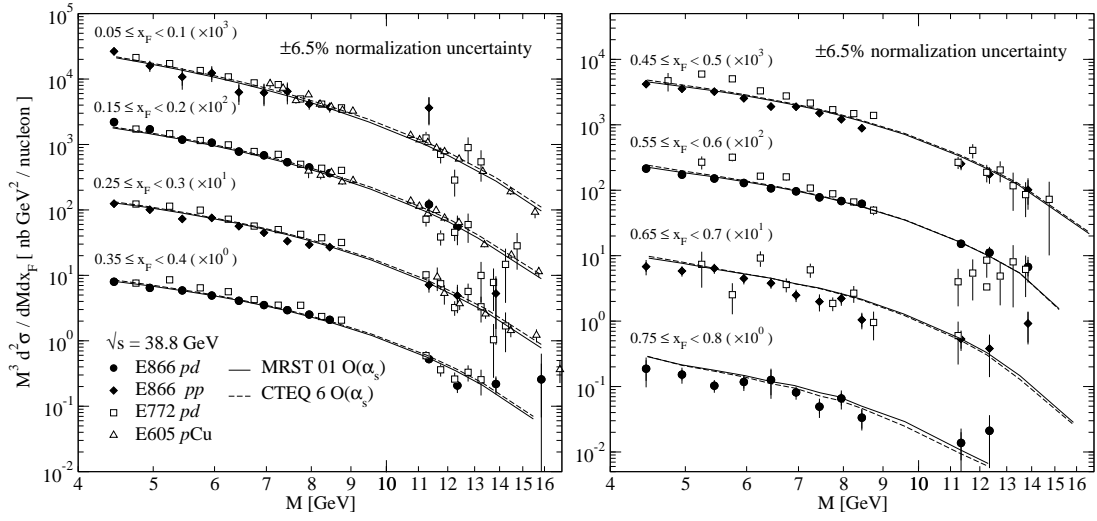}
\end{minipage}
\begin{minipage}[t]{16.5 cm}
\caption{The description of Drell-Yan scattering\cite{Webb:2003bj}.  \label{drellyan}}
\end{minipage}
\end{center}
\end{figure}

This is measured in fixed target experiments E605\cite{Moreno:1990sf},
E772\cite{McGaughey:1994dx} and E866\cite{Webb:2003bj}. In the first the 
target is copper, so nuclear target corrections are required, so this data 
is not always used. The second seems to have incompatibility issues with 
other data. The E866 data from $pp$ collisions is often used as seen 
in Fig. \ref{drellyan}. That from 
$pd$ collisions seems incompatible with structure function and other 
measurements, so is often neglected. 
For these fixed target experiments the kinematic variables are
Feynman $x$, i.e. $x_F$ and $\tau = M^2/s$, where $M^2$ is the invariant
mass of the dimuon pair. At LO these are related the the momentum fractions 
of the partons of the hadrons by   
$x_F = x_1 -x_2$ and $\tau = x_1x_2=M^2/s$. 
At LO the differential cross section is 
\be
\frac{d\sigma}{dM^2dx_F} \propto \sum e_q^2(q(x_1)\bar q(x_2) +
q(x_2)\bar q(x_1)).
\label{DrellYan}
\ee
The fixed target measurements cover $4.5 \GeV < M  < 14 \GeV$ and $0.02 < x_F <0.75$.

Assuming the total quark distributions are already well-known from structure 
function data this provides a 
probe of $\bar u$ and $\bar d$ in the proton for 
moderate $0.02<x_2<0.3$. For $x>0.1$ we find that very roughly 
$\bar q(x) \sim (1-x)^7$, much softer than the valence quarks, as expected.  
The NNLO correction for the total Drell Yan cross section has been 
known for many years\cite{Hamberg:1990np}.
More recently the fully differential Drell Yan cross-sections at 
NNLO were completed\cite{Anastasiou:2003ds,Melnikov:2006kv,Grazzini:2009nc,Catani:2010en},
so all data of this form can be included 
properly within an NNLO extraction of parton distributions.

\begin{figure}
\begin{center}
\begin{minipage}[t]{8 cm}
\epsfig{file=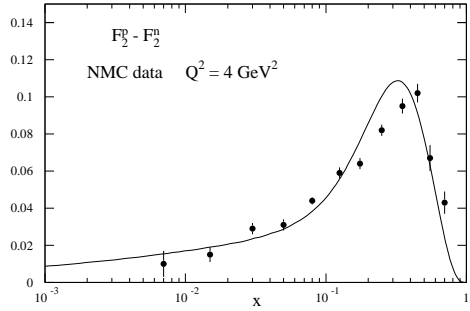,scale=0.4}
\end{minipage}
\begin{minipage}[t]{16.5 cm}
\caption{NMC $F_2^p-F_2^n$ data points compared to the 
a description using PDFS\cite{Martin:1998sq}.  \label{nmcn-p}}
\end{minipage}
\end{center}
\end{figure}

The Drell-Yan cross section gives information on the sum of the 
$\bar u$ and $\bar d$ distributions, with the former having the larger
charge weighting in the cross section, but we can also
consider the precise difference between 
$\bar u$ and $\bar d$. Some of this can be found from structure function measurements. 
The difference is related to the Gottfried sum rule, which at LO gives
\begin{eqnarray} 
I_{GS} = \int^1_0 \frac{dx}{x} (F^{\mu p}-F^{\mu n}) 
&=& \frac{1}{3} \int^1_0 dx (u_V-d_V+\bar u - \bar d)\nonumber \\
&=& \frac{1}{3} +\frac{1}{3}\int^1_0 dx (\bar u - \bar d).
\label{Gottfried}
\end{eqnarray}
The left-hand side of Eq. (\ref{Gottfried}) was measured by NMC
in the region $0.004-0.8$\cite{Arneodo:1994sh}  at $Q^2=4\GeV^2$
and was determined to be $0.258 \pm 0.017$
which implies $\int dx (\bar d -\bar u) \approx 0.2$.
This is shown in Fig. \ref{nmcn-p}, and relies on an 
extrapolation to high and particularly low $x$, which actually 
provides most of the uncertainty. Nevertheless, the evidence for 
$\int^1_0 dx (\bar u - \bar d)\not= 0$ is very strong. 

\begin{figure}
\begin{center}
\begin{minipage}[t]{15 cm}
\epsfig{file=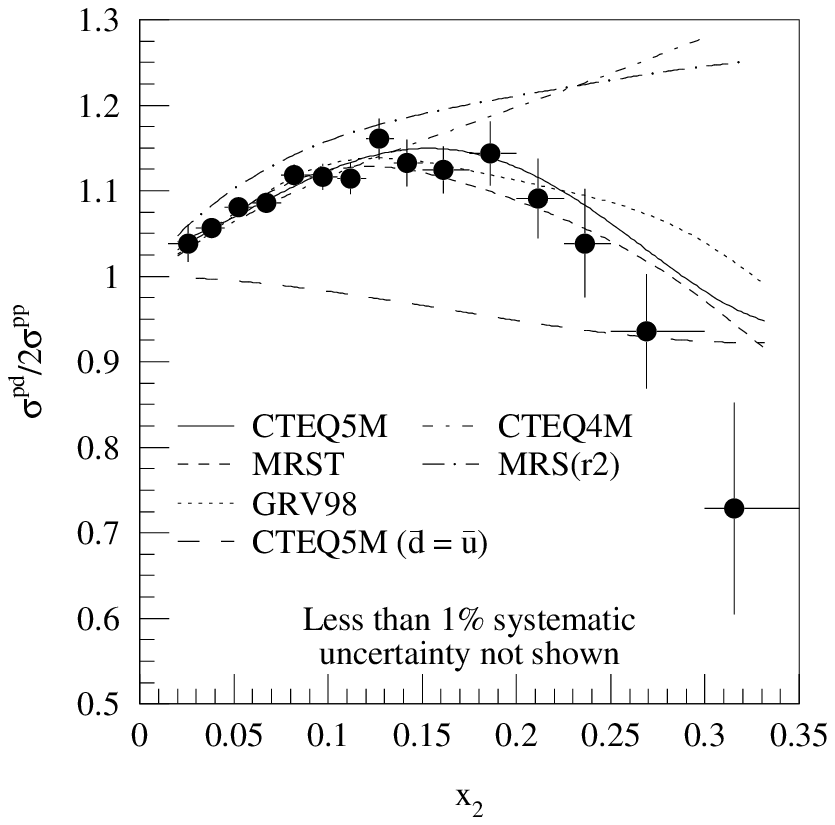,scale=0.75}
\hspace{1cm}
\epsfig{file=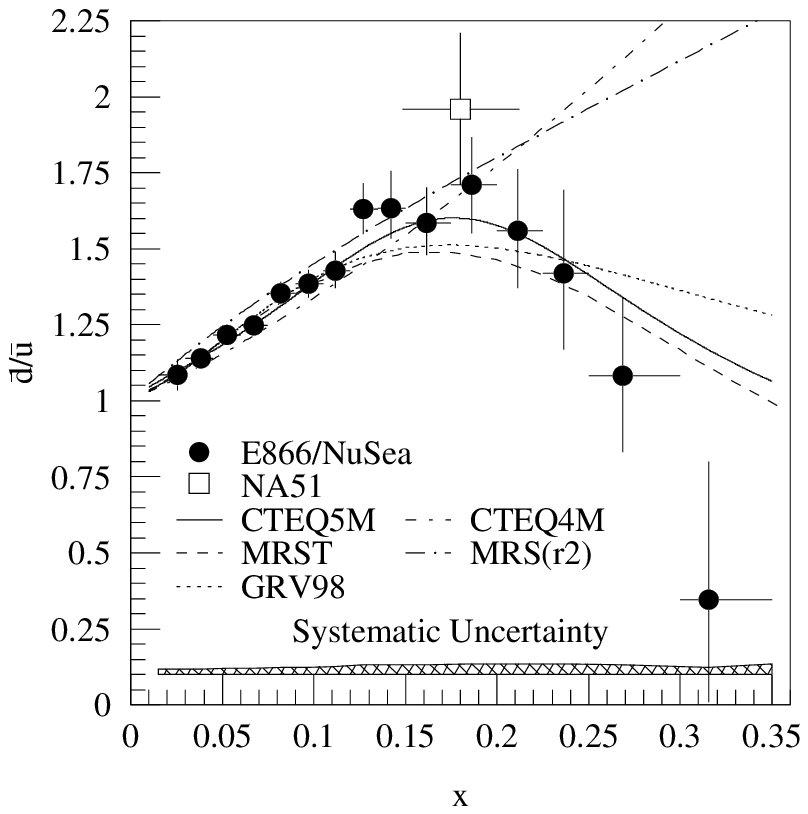,scale=0.75}
\end{minipage}
\begin{minipage}[t]{16.5 cm}
\caption{The Drell-Yan asymmetry predicted by various parton sets
compared to E866 data (left) and the $\bar d /\bar u$ ratio
compared to the data requirements\cite{Towell:2001nh}. \label{drellyanasymm}}
\end{minipage}
\end{center}
\end{figure}

Information on the $\bar d - \bar u$ difference is more directly 
available from Drell-Yan asymmetry
\be
A_{DY}= \frac{\sigma_{pp} -\sigma_{pn}}{\sigma_{pp} +\sigma_{pn}}=
\frac{1-r}{1+r},
\label{DYasymm} 
\ee
where
\be
r \approx \frac{4u_1\bar d_2+ d_1\bar u_2 +4\bar u_1 d_2+ \bar d_1
u_2}{4u_1\bar u_2+ d_1\bar d_2 +4\bar u_1 u_2+ \bar d_1
d_2},
\label{approxr}
\ee
and $1$ labels the proton and $2$ the neutron.
In fact it is the quantity 
\be
R_{dp}=\frac{\sigma_{pd}}{2\sigma_{pp}}= \frac{1}{2}(1+r),
\label{DYratio}
\ee
which is measured, which contains the same information.

\begin{figure}
\begin{center}
\begin{minipage}[t]{8 cm}
\epsfig{file=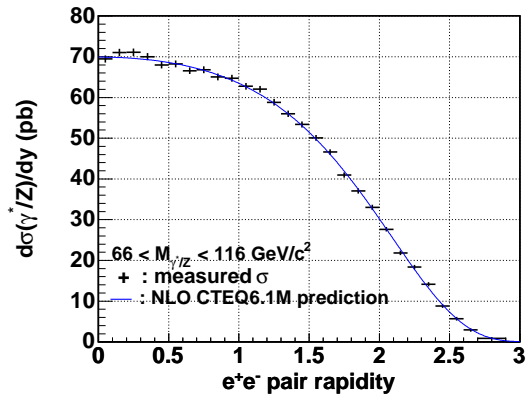,scale=0.4}
\end{minipage}
\begin{minipage}[t]{16.5 cm}
\caption{The rapidity distribution of the $Z$ boson\cite{Aaltonen:2010zza} produced a 
the CDF experiment compared with a prediction from using CTEQ PDFs. \label{d0rap}}
\end{minipage}
\end{center}
\end{figure}

There was originally one point for the Drell-Yan ratio measured by the NA51 experiment\cite{Baldit:1994jk} at $x_1=x_2=0.18$ which implied that 
$\bar d \approx 2 \bar u$ at this $x$ value. 
This was greatly improved by measurements from 
the E866/NuSea experiment\cite{Towell:2001nh} 
which made very accurate measurements from $0.04 < x <0.3$. 
This gives clear evidence of $\bar u-\bar d$ asymmetry, as seen in Fig. 
\ref{drellyanasymm} but not as much 
as suggested originally by the NA51 point. The asymmetry
seems to reach a maximum at $x\approx 0.2$. 
It is not currently clear what happens as $x \to 1$.
The asymmetry is becoming small at the smaller $x$ values, 
so the assumption that as a nonsinglet quantity $\bar d -\bar u$ 
will have the same general shape as valence distributions
implies it tends quickly to zero. This is what happens 
with the vast majority of parameterisations of this quantity in the 
parton distribution sets. However, there is no sum rule
constraint, so it is not impossible that the asymmetry becomes large,
in either direction at small $x$.

In the past couple of years supplementary information from an analysis
of the $Z$-boson rapidity data from from the D0\cite{Abazov:2007jy}
and CDF\cite{Aaltonen:2010zza} experiments
at the Tevatron collider has become available. 
This is dominated by a narrow range of invariant mass
near to $m_Z=91.1\GeV$ and is also a function of the
rapidity $y$ where 
\be
y = \ln((E+p_z)/(E-p_z)).
\label{defrapidity} 
\ee
At LO in the parton kinematics $x_{1,2}=x_0\exp(\pm y)$,
where $x_0=m_Z/\sqrt{s}$,
so since $\sqrt{s}=1.96\TeV$ $x_0= 0.05$ and corresponds to 
central rapidity. Over the full range of rapidity values of
$x$ from $0.003$ up to $0.7$ are probed, reaching smaller values of 
$x$ than the fixed target data. Since the Tevatron is a 
proton-antiproton collider at LO the differential cross section is given by
\be
\frac{d\sigma}{dM^2dy} \propto \sum_{q_i} (v_{q_i}^2 +
a_{q_i}^2) \bigl(q(x_1) q(x_2) +  \bar q(x_2)\bar q(x_1)\bigr),
\label{DrellYanZ}
\ee
where $v_{q_i}$ and $a_{q_i}$ are the vector and axial couplings 
respectively. As such it is largely sensitive to the larger quark 
distributions, rather than the antiquarks, particularly at high rapidity.
However, the fit to this data does rely on different flavour combinations 
from the previously discussed processes, so adds extra constraints, 
particularly for down type quarks and antiquarks due to the higher 
weighting in Eq. (\ref{DrellYanZ}) compared to the electric 
charge weighting for many other processes. A comparison of a prediction 
at NNLO to the D0 data is shown in Fig. \ref{d0rap}. Clearly the 
comparison is good, but the accurate data does add extra constraints.

\subsection{\it The Strange Quark Distribution}

\begin{figure}
\begin{center}
\begin{minipage}[t]{16 cm}
\epsfig{file=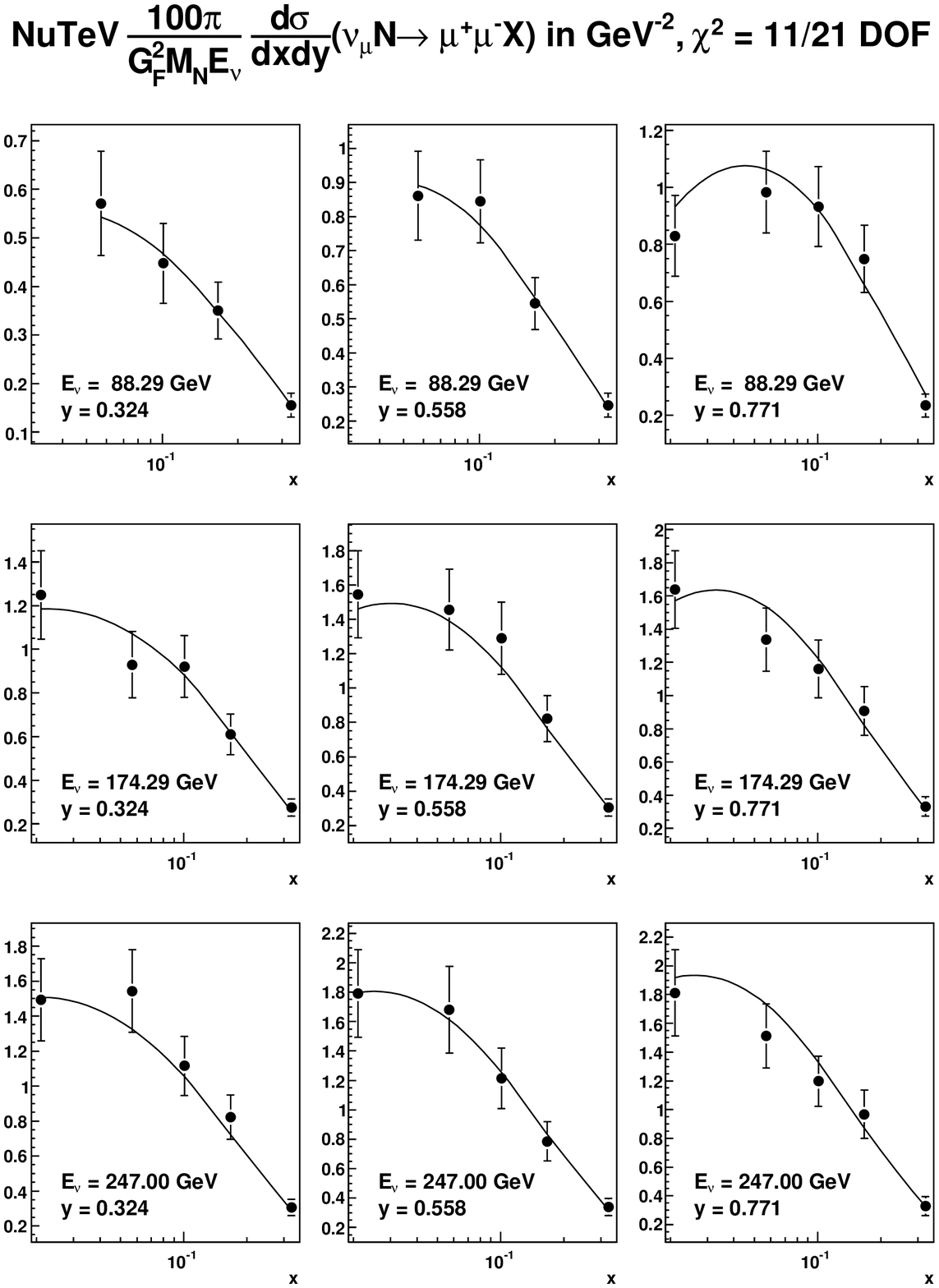,scale=0.42}
\epsfig{file=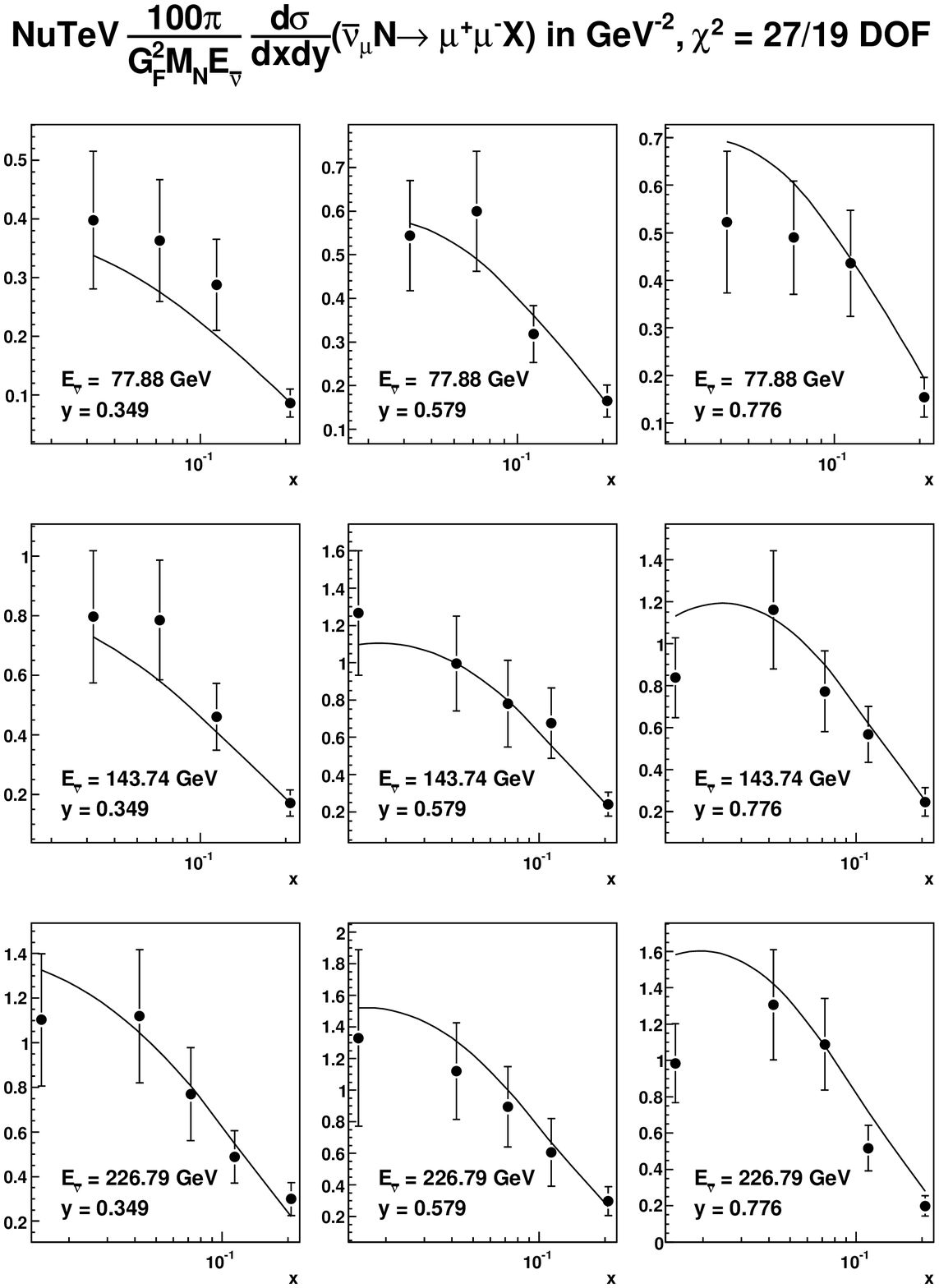,scale=0.42}
\end{minipage}
\begin{minipage}[t]{16.5 cm}
\caption{The MSTW08 fit\cite{Martin:2009iq} to NuTeV dimuon data\cite{Goncharov:2001qe} 
in neutrino and antineutrino scattering. 
\label{nutevdimuon}}
\end{minipage}
\end{center}
\end{figure}

\begin{figure}
\begin{center}
\begin{minipage}[t]{16.5 cm}
\epsfig{file=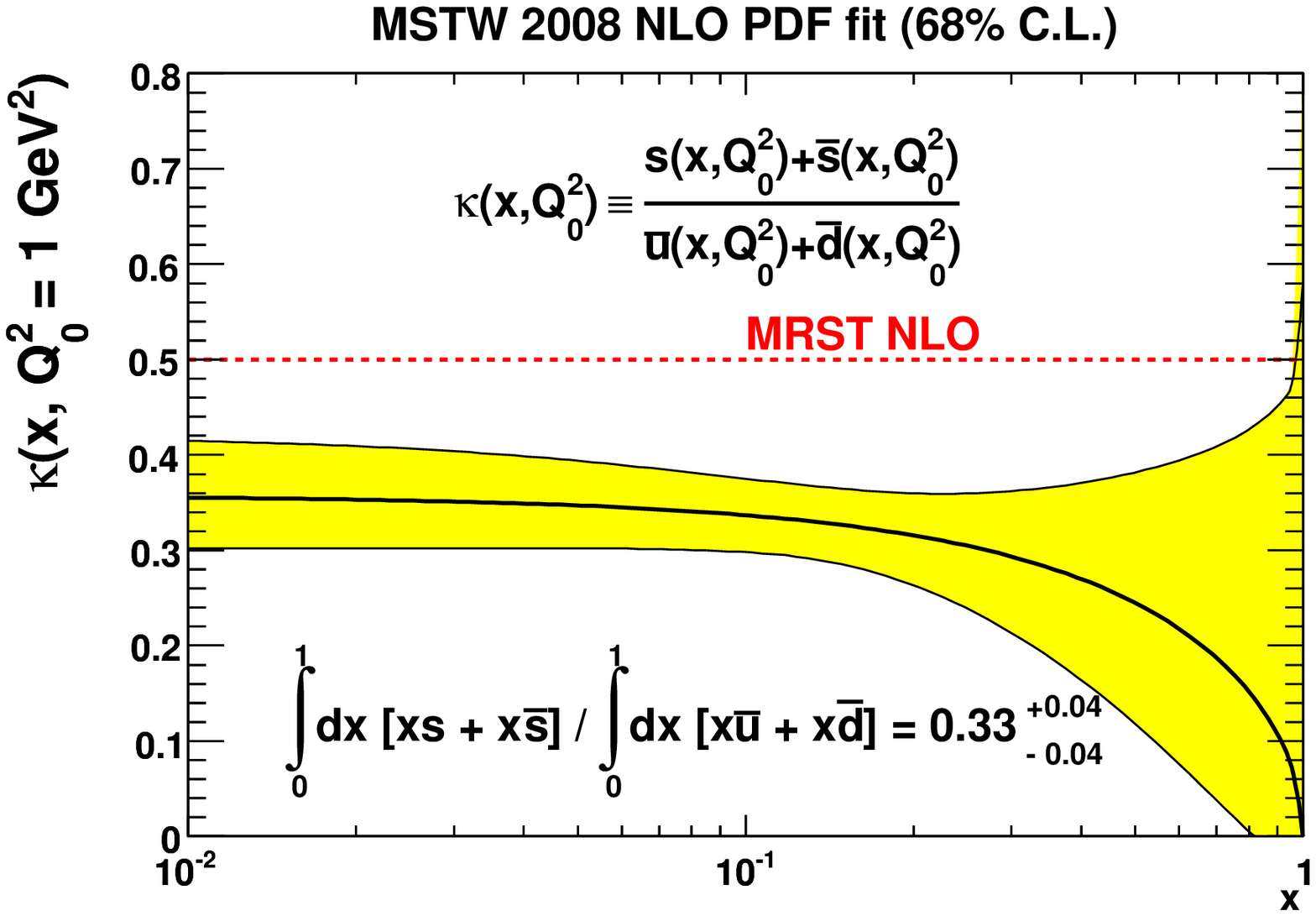,scale=0.45}
\epsfig{file=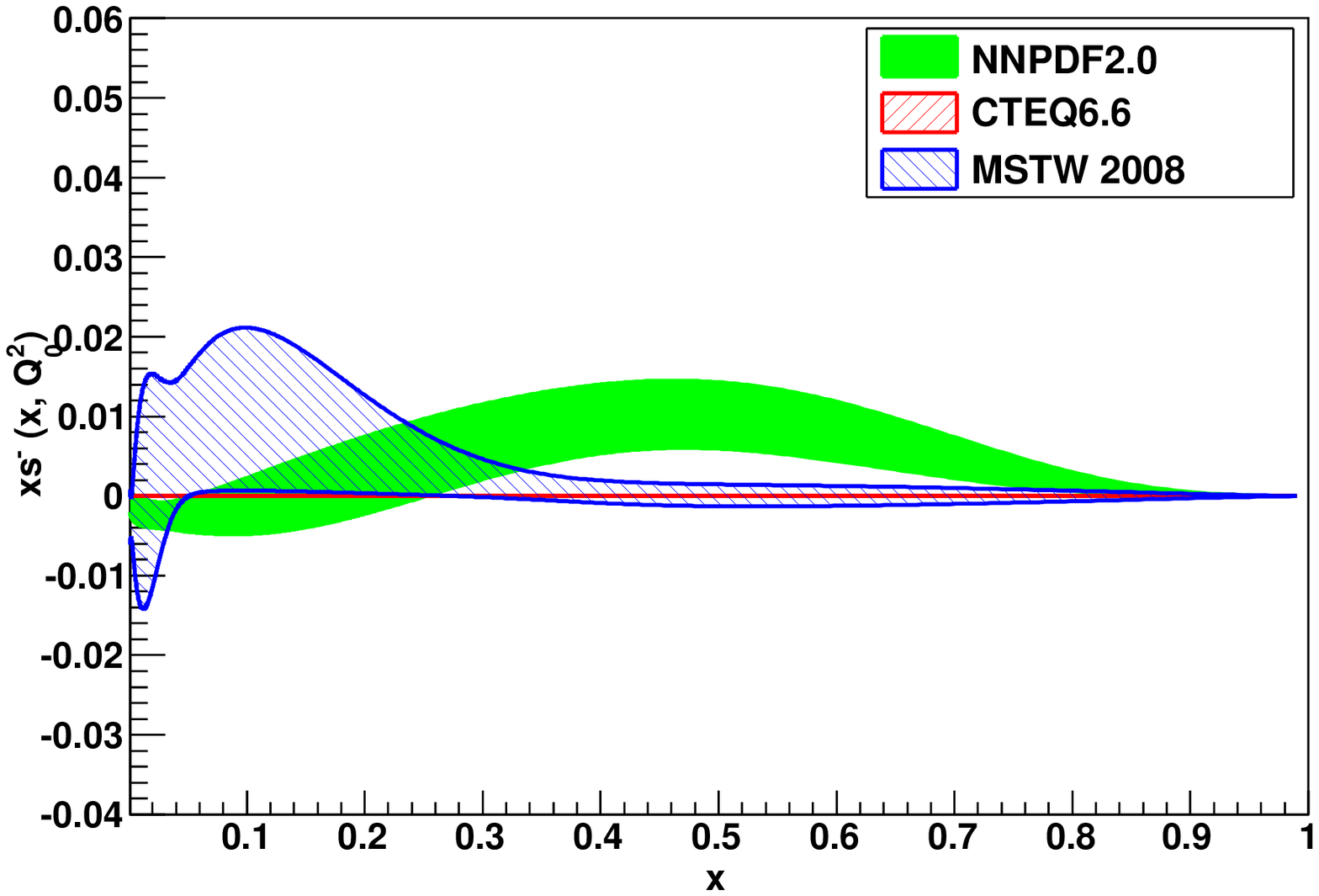,scale=0.43}
\end{minipage}
\begin{minipage}[t]{16.5 cm}
\caption{Ratio of strange quarks to light sea with uncertainty\cite{Martin:2009iq} (left)
and the strange-antistrange asymmetry\cite{Ball:2010de} from NNPDF2.0 and MSTW2008 (right). 
\label{splus}}
\end{minipage}
\end{center}
\end{figure}

It is now possible to find the strange quark 
distribution directly rather than simply making it an appropriate
fraction of the light sea.  
This is done by comparing to unlike sign dimuon production at CCFR and 
NuTeV, i.e.  
\be
\nu_{\mu} \to \mu^- + W^+
\label{dimoun1}
\ee
followed by
\be
W^+ + s \to c \to D^+ \to \mu^+.
\label{dimuon2}
\ee
or for antineutrinos
\be
\bar \nu_{\mu} \to \mu^+ + W^-
\label{dimuon3}
\ee
followed by
\be
W^- + \bar s \to \bar c \to D^- \to \mu^-.
\label{dimuon4}
\ee
In global fits it was previously 
assumed that at $Q_0^2$ we have 
$s(x)=\kappa 0.5(\bar u + \bar d)$. 
$\kappa$ was determined qualitatively by comparison to a model-dependent
extraction of the strange quark from CCFR dimuon data\cite{Bazarko:1994tt}. 
Using $Q_0^2= 1\GeV^2$ and $\kappa \approx 0.4$
worked well, i.e.
strange was about $18\%$ of the input sea at input.
Since all quarks evolve equally this fraction increases as  $Q^2$
increases. 
However, we can now do better and obtain a more precise
normalisation and also shape for the strange distribution by comparing
to the more detailed dimuon data obtained from NuTeV\cite{Goncharov:2001qe}, 
which also provides
a more thorough analysis of the CCFR data presented in \cite{Bazarko:1994tt}.
A comparison to the fit to both neutrino and
antineutrino data from NuTeV is shown in Fig. \ref{nutevdimuon}. 
generally one finds a reduced ratio of strange to non-strange sea compared to
the previous results.There is also some additional suppression at large $x$
i.e. lower $W^2$, as seen in the left of Fig. \ref{splus}. This is what one would expect 
of the suppression compared to  the light sea is the effect of non-zero strange 
quark mass $m_s$.

The data constrains the strange quark in the $x=$ range $0.01<x<0.2$.
From the figure one might possibly imply that the neutrino
data is sightly higher than the antineutrino data. This implies that
$s(x) \not= \bar s(x)$, though the lack of strangeness of the proton
requires that
\be
\int_0^1 (s(x) - \bar s(x))\, dx =0
\label{sasymmsum}
\ee
so an excess of $s(x)$ over $\bar s(x)$ in one $x$ range must be balanced by 
a deficit elsewhere. There have been numerous analyses to determine if 
this is indeed the case\cite{Olness:2003wz,Mason:2006qa,Lai:2007dq,Alekhin:2008mb}, 
with most finding evidence, if only at about $68\%$ confidence level,
for a positive momentum asymmetry
\be
\int_0^1 x(s(x) - \bar s(x))\, dx. 
\label{sasymmsummom}
\ee
Only NNPDF2.0\cite{Ball:2010de}
and MSTW2008\cite{Martin:2009iq} make the asymmetry part of their default
sets. These are shown in the right of Fig. \ref{splus}, and both give a positive 
momentum asymmetry, but are different at the highest $x$ (where the data 
constraint vanishes). This positive momentum asymmetry acts to reduce the 
anomaly in the NuTeV measurement of $\sin^2 \theta_W$ in neutrino
DIS, which is in 3 $\sigma$ disagreement with the world average\cite{Zeller:2001hh,Zeller:2002du}.

%\begin{figure}
%\begin{center}
%\begin{minipage}[t]{11 cm}
%\epsfig{file=nnpdfsminus.eps,scale=0.5}
%\end{minipage}
%\begin{minipage}[t]{16.5 cm}
%\caption{The strange-antistrange asymmetry\cite{Ball:2010de} from NNPDF2.0 and MSTW2008.  
%\label{strangeasymm}}
%\end{minipage}
%\end{center}
%\end{figure}

\subsection{\it More Quark Constraints}

The final independent experimental constraint on the light quarks at moderate 
to large 
$x$ comes from
$W$ asymmetry \cite{Aaltonen:2009ta}, or usually lepton asymmetry\cite{Acosta:2005ud,Abazov:2007pm}
at the Tevatron $p \bar p$ 
collider. At LO
\begin{eqnarray} 
A_W(y) &=& \frac{d\sigma(W^+)/dy-d\sigma(W^-)/dy}
{d\sigma(W^+)/dy +d\sigma(W^-)/dy}\nonumber \\
&\approx& \frac{u(x_1)d(x_2)-d(x_1)u(x_2)}{u(x_1)d(x_2)+d(x_1)u(x_2)},
\label{Wasymmetry}
\end{eqnarray}
where $x_{1,2}=x_0 \exp(\pm y), \quad x_0 
=\frac{m_W}{\sqrt{s}}$.
Since $u(x) > d(x)$ at large $x$, 
whereas they become roughly equal at 
smaller $x$, $A_W(y)$ is positive for $x_1 > x_0 = 0.05$ 
($y>1$) (and at a proton-antiproton collider the asymmetry is expected
to be exactly antisymmetric about $y=0$, so results in each half 
of the detector can be combined).
This helps pin down the $u$ and $d$ quarks in the region $x\sim 0.1$ 
as well as giving compatible information to NMC and CCFR/NuTeV at 
higher $x$, and 
thus contributes to the determination of the two valence quark distributions
without any complications due to higher twists or deuterium corrections.

\begin{figure}
\begin{center}
\begin{minipage}[t]{8 cm}
\epsfig{file=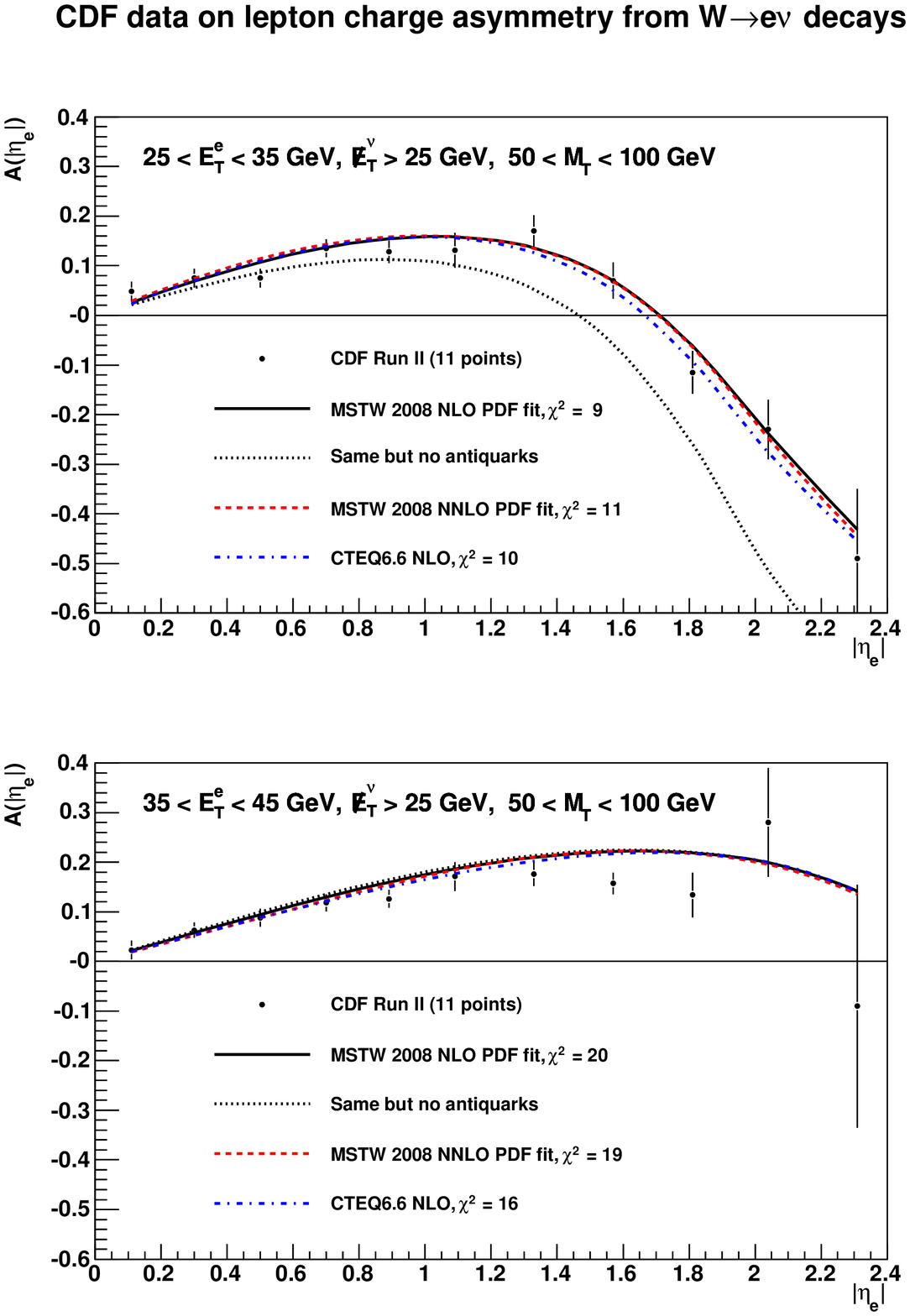,scale=0.4}
\end{minipage}
\begin{minipage}[t]{16.5 cm}
\caption{Comparison\cite{Martin:2009iq} of fits to CDF data\cite{Acosta:2005ud}
with various parton sets.  
\label{leptasy}}
\end{minipage}
\end{center}
\end{figure}

In practice it is usually the final state leptons that are detected, 
so it is really the lepton asymmetry 
\be
 A(y_l)= \frac{d\sigma(l^+)/dy_l-d\sigma(l^-)/dy_l}
{d\sigma(l^+)/dy_l +d\sigma(l^-)/dy_l}
\label{leptasymmetry}
\ee
which is measured
where $y_{l}$ is the rapidity of the charged lepton.  
Defining the angle of the lepton relative to the proton beam in the 
$W$ rest frame by $\cos^2\theta^* = 1 - 4E_T^2/m_W^2$ leads to
\begin{equation}
  y_{l}= y_{W} +  \frac{1}{2}
\ln\left(\frac{1+\cos\theta^*}{1-\cos\theta^*}\right).
\label{wtolrap}
\end{equation}
This means the valence-quark-only approximation to the lepton asymmetry, 
can be inaccurate particularly near the edges of phase space, i.e. 
$\cos\theta^* \approx \pm 1$, since sea-quark contributions can become 
significant. Neglecting overall factors, the numerator of 
Eq. (\ref{leptasymmetry}) can be approximated by ($s$, $c$ and $b$ quark 
contributions are very small at Tevatron energies)
\begin{equation}
  u(x_1)d(x_2)(1-\cos\theta^*)^2 + \bar{d}(x_1)\bar{u}(x_2)(1+\cos\theta^*)^2 - d(x_1)u(x_2)(1+\cos\theta^*)^2 
- \bar{u}(x_1)\bar{d}(x_2)(1-\cos\theta^*)^2,
\label{leptonasymmapprox}
\end{equation}
and for  $\cos\theta^* \approx 1$ the leading ($\bar d\bar u$) sea-sea  
contribution is enhanced relative to the valence-valence contribution by 
the large $(1+\cos\theta^*)^2$ term arising from the $V+A$ decay to 
leptons. The fit to the data in\cite{Acosta:2005ud} is shown in two $E_T$ 
bins in Fig. \ref{leptasy}, including the consequence of 
ignoring the antiquark contributions. In working back to the $W$-asymmetry 
in\cite{Aaltonen:2009ta} information on parton distributions must be used to 
infer the likelihood of a lepton having come from either quark-quark
or antiquark-antiquark annihilation. This implicitly loses some of the 
constraining power on the parton distributions, but makes the comparison 
between the data and the parton distributions more transparent. 
There is newer lepton asymmetry data from D0\cite{Abazov:2008qv}, but this 
has not yet been used in global fits 
(see however \cite{Lai:2010vv,:2010gb}), partially because there is a 
question-mark about how good a fit is possible when all other data is 
included.

\subsection{\it The Gluon Distribution}

\begin{figure}
\begin{center}
\begin{minipage}[t]{14.5 cm}
\begin{picture}(190,75)(0,-50)
\SetColor{Red}
\Text(4,75)[]{$q$}
\Text(4,5)[]{$g$}
\ArrowLine(8,75)(75,75)
\Gluon(8,5)(75,5){5}{6}
\ArrowLine(75,75)(75,5)
\ArrowLine(75,5)(142.5,5)
\Photon(75,75)(142.5,75){5}{6}
\Text(146,75)[]{$\gamma$}
\Text(146,5)[]{$q$}
\end{picture}
\epsfig{file=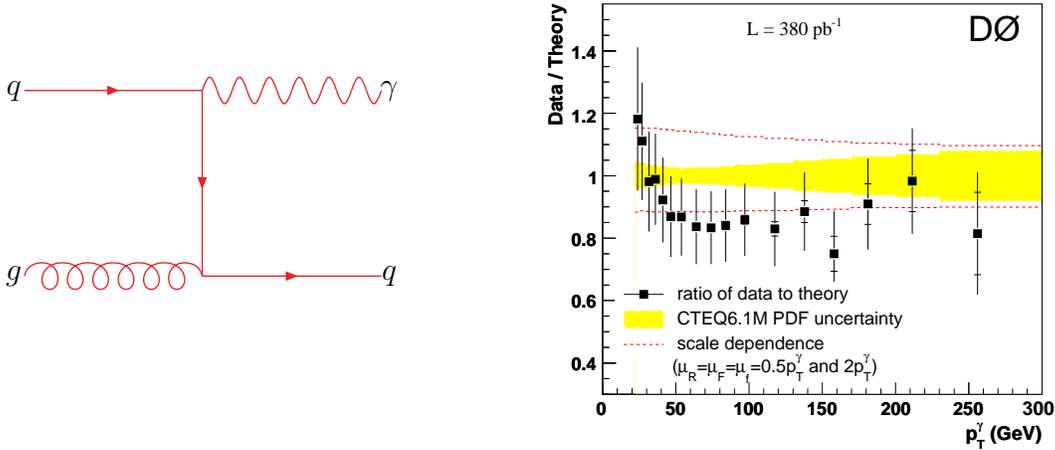,scale=0.4}
\end{minipage}
\begin{minipage}[t]{16.5 cm}
\caption{LO diagram for prompt photon production (left),
and comparison of the prediction to D0 data on direct photon
production\cite{Abazov:2005wc} (right).
\label{LOprompt}}
\end{minipage}
\end{center}
\end{figure}

The above measurements 
constrain the high and moderate $x$ quarks to a few
percent or better. It is far more difficult to obtain precise information 
on the form of the high $x$ gluon. In the early days of global
fits to parton distributions groups\cite{Martin:1998sq,Lai:1996mg} 
determined the 
gluon at high $x$ via prompt or direct photon production\cite{Bonesini:1987bv,Apanasevich:1997hm}, via the process 
shown in the left of Fig. \ref{LOprompt}.    
In principle this is a direct test of the large 
$x$ gluon - $x_T=2p_T/\sqrt{s}$.
However, $d^2\sigma/dE dp_T$ is sensitive to nonperturbative 
information 
about the intrinsic $k_T$ of the gluons in the proton, to resummation of
threshold logarithms, i.e. $\ln(1-x_T)$\cite{Laenen:2000de}, and to the 
interplay between the 
two. Also, some experiments probing similar regions of parameter space 
give results which are difficult to reconcile
without incorporating large changes in the corrections to the calculations
outlined above for only small changes in energy. Hence, due to 
intrinsic uncertainties this process gives only
a very rough indication of the gluon distribution. In order to be largely free 
from the theoretical ambiguities the data should be at $p_T$ such that 
an uncertainty in $p_T$ of about $1\GeV$ is not too important, i.e. at least a 
few tens of $\GeV$. There is more recent data 
from the Tevatron\cite{Abazov:2005wc}, which satisfies this. This is shown compared
to a prediction in the right of Fig. \ref{LOprompt}. There is good agreement, but the 
data is not precise enough to add any useful constraint. There is no 
obstacle to very precise direct photon production data from the Tevatron 
or LHC being used to constrain the gluon in the future\cite{Ichou:2010wc}.

\begin{figure}
\begin{center}
\begin{minipage}[t]{16.5 cm}
\begin{picture}(180,150)(0,20)
\SetColor{Red}
\Text(7,83)[]{$g$}
\Gluon(10,80)(50,50){5}{6}
\Gluon(50,50)(10,20){5}{6}
\Text(7,17)[]{$g$}
\Gluon(50,50)(100,50){5}{6}
\ArrowLine(100,50)(140,80)
\ArrowLine(140,20)(100,50)
\Text(143,83)[]{$q$}
\Text(143,17)[]{$\bar q$}
\Text(7,173)[]{$q$}
\ArrowLine(10,170)(50,140)
\Gluon(50,140)(10,110){5}{6}
\Text(7,107)[]{$g$}
\ArrowLine(50,140)(100,140)
\ArrowLine(100,140)(140,170)
\Gluon(140,110)(100,140){5}{6}
%\Gluon(100,240)(100,210){5}{2}
\Text(143,173)[]{$q$}
\Text(143,107)[]{$g$}
\end{picture}
\hspace{-0.2cm}
\epsfig{file=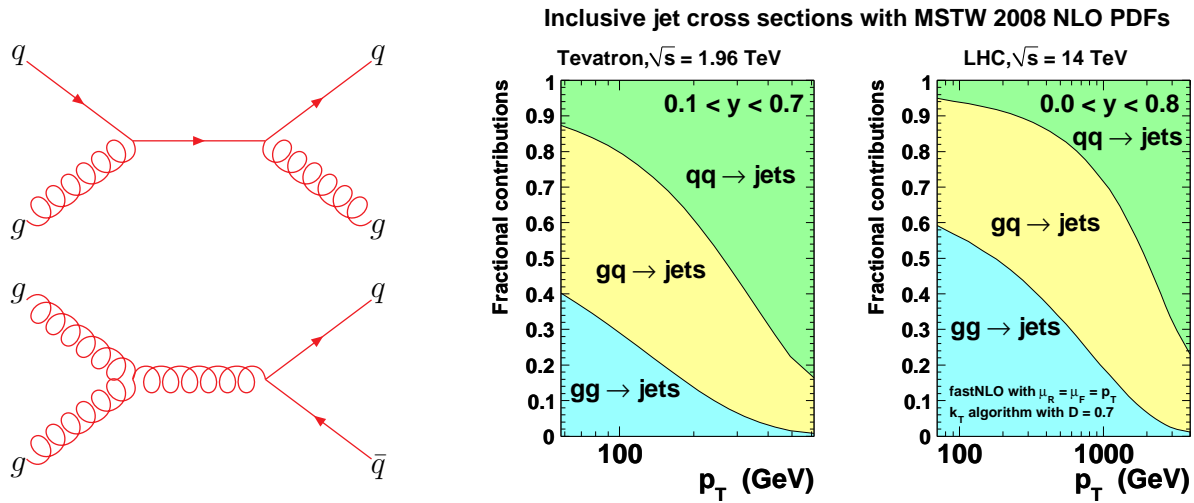,scale=0.5}
\end{minipage}
\begin{minipage}[t]{16.5 cm}
\caption{Some LO diagrams for jet production (left). 
The jet cross section fraction from different contributions 
as a function of $p_T$\cite{Martin:2009bu} (right).
\label{LOjet}}
\end{minipage}
\end{center}
\end{figure}

The current best direct determination of the high-$x$ 
gluon distribution is given by inclusive
jet measurements by D0 and CDF at the Tevatron,
where they measure $d \sigma/dp_Tdy$ and
$p_T$ is the transverse momentum of the jet.
For run I\cite{Abbott:2000ew,Affolder:2001fa} 
for D0 and for run II\cite{Abulencia:2007ez,Abazov:2008hu,Aaltonen:2008eq} 
for both experiments 
the measurements are in different bins of rapidity. 
This gives better coverage of $x$ since non-zero rapidity leads
to asymmetric $x$ values for the  incoming partons. 
At central rapidity $x_T= 2 p_T/\sqrt{s}$ the measurements extend up to
$p_T \sim 500 \GeV$, i.e. $x_T \sim 0.5$, and down to 
$p_T \sim 50 \GeV$, i.e. $x_T \sim 0.05$.

%\begin{figure}
%\begin{center}
%\begin{minipage}[t]{10 cm}
%\epsfig{file=jetspdffrac.eps,scale=0.5}
%\end{minipage}
%\begin{minipage}[t]{16.5 cm}
%\caption{The jet cross section fraction from different contributions 
%as a function of $p_T$\cite{Martin:2009bu}.  
%\label{jetcontribution}}
%\end{minipage}
%\end{center}
%\end{figure}

At matrix element level, where some LO diagrams are shown in the left of  
Fig. \ref{LOjet} gluon-gluon fusion dominates.
However, the gluon distribution
falls off more quickly as $x \to 1$ than quark distributions so  
there is 
a transition from gluon-gluon fusion at small $x_T$, to gluon-quark 
at 
intermediate $x_T$ to quark-quark at high $x_T$. 
However, even at the 
highest $x_T$ probed at the Tevatron, or
likely to be probed at the LHC, gluon-quark contributions are significant,
as shown in the right of Fig. \ref{LOjet}. This 
qualitative picture is not altered beyond LO, but the calculation of the 
cross section becomes much more complicated. It is aided immensely by the 
implementation of \textsc{fastnlo}\cite{Kluge:2006xs},
based on \textsc{nlojet++}~\cite{Nagy:2001fj,Nagy:2003tz}, which allows the 
inclusion of the exact NLO hard cross section corrections to jet data in 
the fitting routine. (We note that in \cite{Ball:2010de} a similar
numerical procedure is applied for Drell Yan-type processes, while the 
more general procedure is to use an $x$-dependent higher order $K$-factor.
So long as the latter is tuned to the parton distributions being used, as is 
usually the case, this should not lead to any noticeable inaccuracy.) NNLO
jet cross-sections are not known in full, but some threshold corrections are
calculated and can be applied\cite{Kidonakis:2000gi}, and are used in   
\textsc{fastnlo}. The LO $\to$ NLO cross section corrections are not very 
large, in general $\sim 20\%$, and the NNLO estimates are $5-10\%$ and in both
cases the corrections are smooth functions. 
This implies that missing NNLO corrections are similar in 
magnitude to correlated uncertainties on the data.

\begin{figure}
\begin{center}
\begin{minipage}[t]{16.5 cm}
\epsfig{file=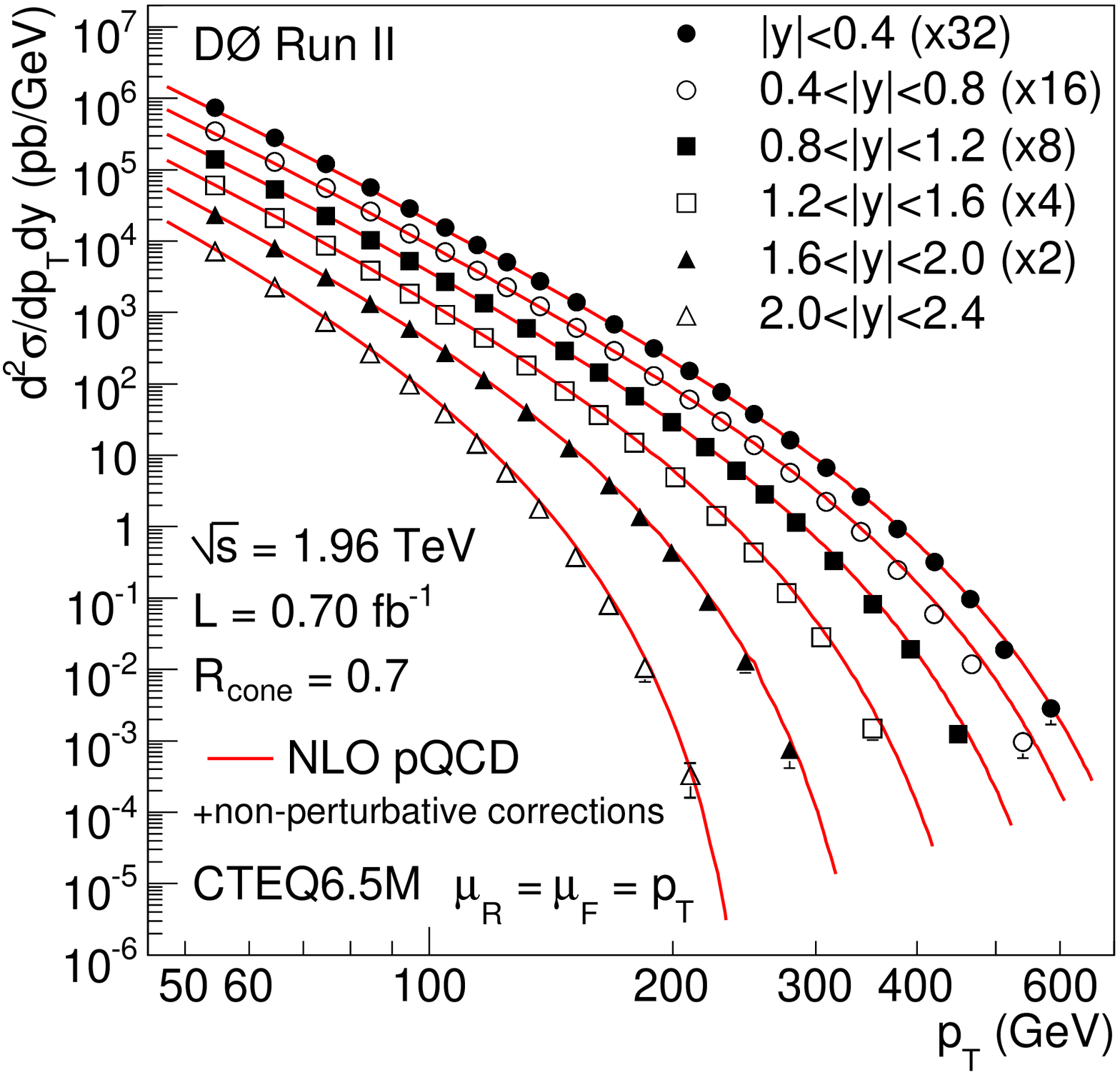,scale=0.36}
\epsfig{file=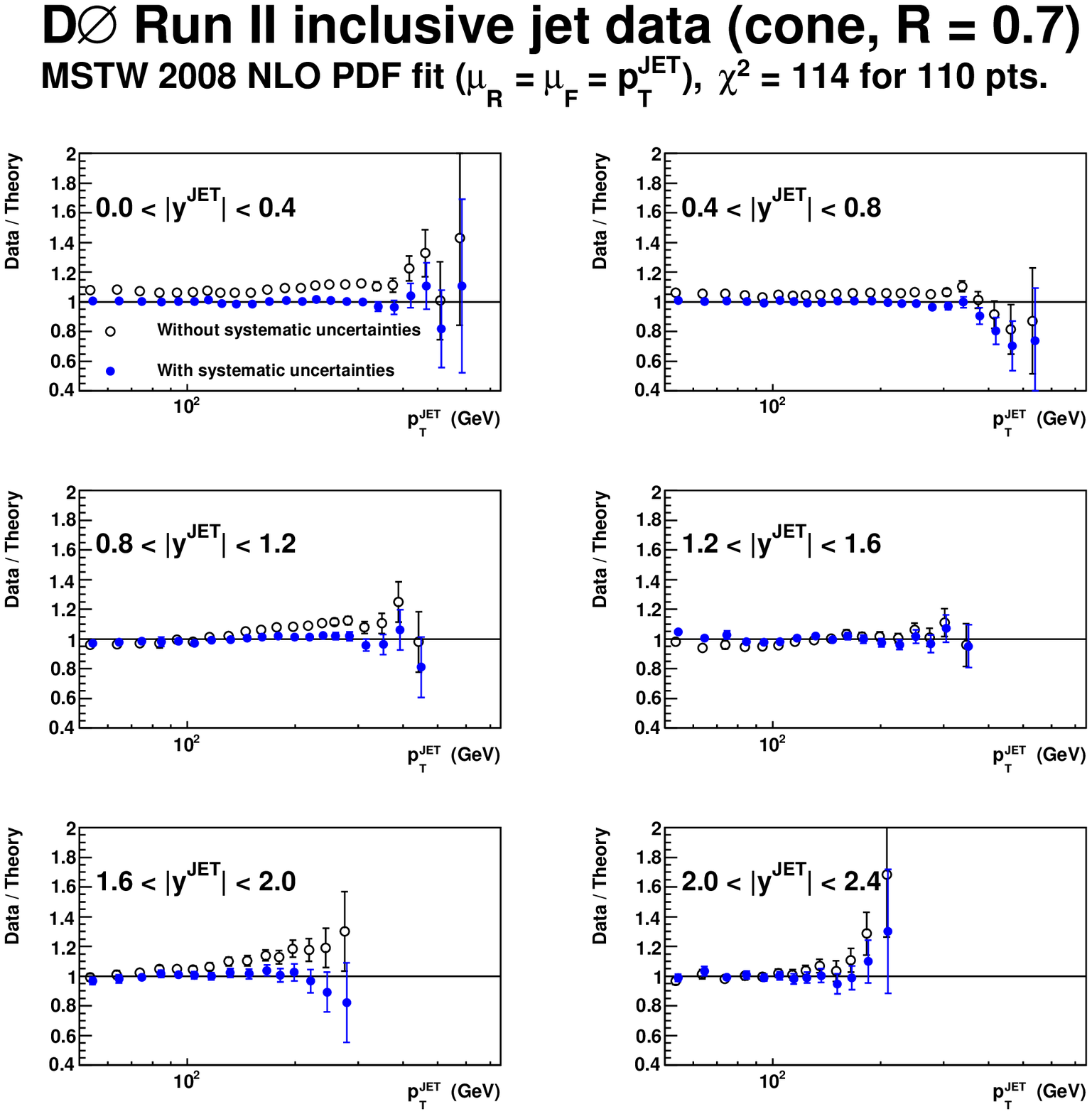,scale=0.4}
\end{minipage}
\begin{minipage}[t]{16.5 cm}
\caption{Comparison of theory to data for the most recent D0 jet data\cite{Abazov:2008hu},
as absolute values (left) and as a ratio of data to theory\cite{Martin:2009iq}(right). 
\label{D0jetfit}}
\end{minipage}
\end{center}
\end{figure}

For the run I data it was initially difficult to achieve a good fit at 
highest $E_T$, with theory undershooting data, 
due to indirect constraints on the gluon from the sum
rule or the limited sensitivity of the high-$x$ fixed target data to the 
gluon distribution. This was not a major problem and was solved by 
including additional flexibility in the gluon parameterisation and 
by a proper treatment of the large correlated systematic 
uncertainties on the jet data\cite{Lai:1999wy,Pumplin:2002vw,Martin:2001es,
Martin:2004ir}, though in many cases some tension between 
the fit to jet data and the best global fit was evident. The run II Tevatron
jet data is markedly softer at high $p_T$ than the run I data and 
consistency between the two is at best argued to be marginal\cite{Pumplin:2009nk}, and at worst poor\cite{Martin:2009iq}.   
This makes the run II data easier to fit in comparison to the other data.
An illustration of the fit quality is shown in Fig. \ref{D0jetfit}. 

%\end{document}

\begin{figure}
\begin{center}
\begin{minipage}[t]{7.5 cm}
\epsfig{file=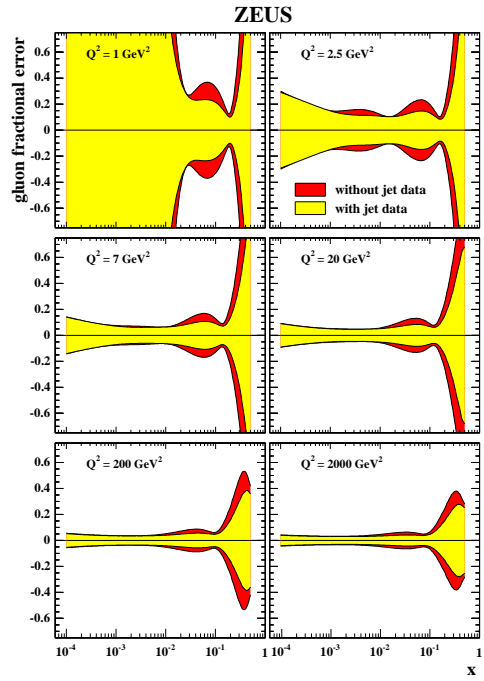,scale=0.35}
\end{minipage}
\begin{minipage}[t]{16.5 cm}
\caption{Reduction of the gluon uncertainty from addition of HERA jet
data to a fit with only HERA DIS data\cite{Chekanov:2005nn}. 
\label{herajet}}
\end{minipage}
\end{center}
\end{figure}

In principle there is also a direct constraint on the gluon distribution
from jet data in deep inelastic scattering at HERA\cite{Chekanov:2002be,Chekanov:2006xr,Adloff:2000tq,Wobisch:2000dk,
Aktas:2007pb}, 
primarily in the range $0.01<x<0.1$. However, 
within the context of a full global fit this does add much extra 
information\cite{Martin:2009iq}, actually being more important in the 
constraint on $\alpha_S$ within a fit to parton distributions\cite{Martin:2009bu}. When added to a fit containing only HERA structure
function data it does have a significant impact\cite{Chekanov:2005nn}, partially 
because it does constrain the strong coupling better than the structure 
function data alone. This impact is shown in Fig. \ref{herajet}, but the
maximum effect is obtained by including data from photo-production as well
as deep inelastic scattering, which includes the additional complication of 
having to use fits for the parton distributions of the photon. The NLO 
corrections are larger for HERA jet data than for Tevatron data, and there
is no approximation at NNLO. It is therefore difficult to include these data
in a NNLO fit.  
 
\subsection{\it Small-$x$ Parton Distributions}

\begin{figure}
\begin{center}
\begin{minipage}[t]{11 cm}
\epsfig{file=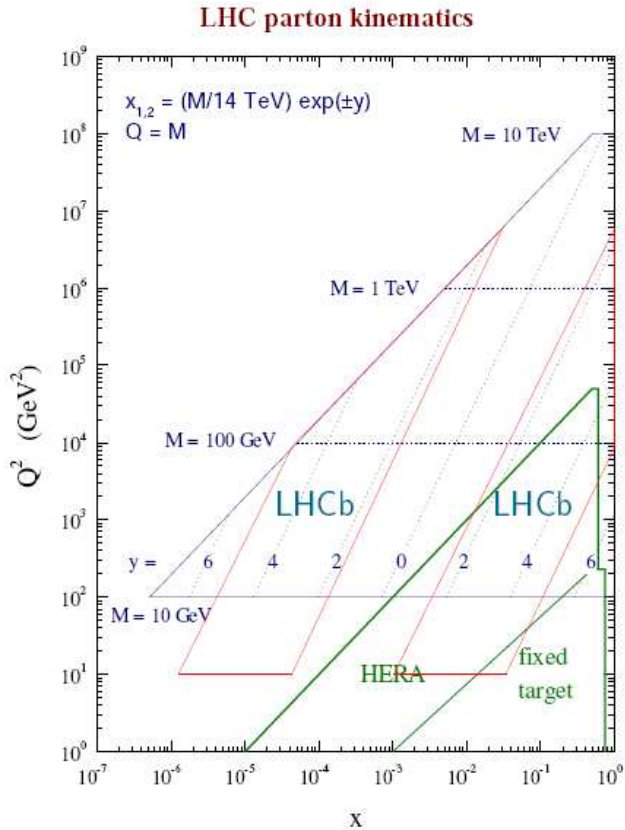,scale=0.55}
\end{minipage}
\begin{minipage}[t]{16.5 cm}
\caption{Kinematic region in $x$ and $Q^2$ probed by various experiments
for $14\TeV$ collisions at the LHC.  
\label{LHCkinematics}}
\end{minipage}
\end{center}
\end{figure}

All the above data constrain the partons mainly for 
$x >0.01$ (though Tevatron $Z$-rapidity data 
extend a little further, though with diminishing statistical precision).  
The extension to the
region of very low $x$ has been made in the past decade by HERA\cite{Adloff:2000qk,Breitweg:1998dz,Chekanov:2001qu,Chekanov:2003yv,Chekanov:2002ej}.
This region is very interesting for the study of QCD.
It is also vital for the LHC as seen from
the kinematic range as illustrated in Fig. \ref{LHCkinematics}. 
At the smallest $x$ values an extrapolation from the measurements at low 
scales to the LHC regions at higher scales is required. 

%\begin{figure}
%\begin{center}
%\begin{minipage}[t]{9 cm}
%\epsfig{file=combfig6.eps,scale=0.45}
%\end{minipage}
%\begin{minipage}[t]{16.5 cm}
%\caption{Comparison of HERAPDF1.0 at NLO to HERA data\cite{:2009wt}.  
%\label{HERAfit}}
%\end{minipage}
%\end{center}
%\end{figure}

In this region there is dramatic scaling violation of the partons from the
evolution equations and also a complex interplay between the quarks and 
gluons. The evolution equations for the singlet sector are coupled
\begin{eqnarray} \frac{d \Sigma}{d \ln Q^2} &=& P_{qq}\otimes \Sigma
+P_{qg} \otimes g \nonumber \\
\frac{d g}{d \ln Q^2} &=& P_{gq}\otimes \Sigma
+P_{gg} \otimes g 
\label{singlet evolution}
\end{eqnarray}
At very small $x$ and at LO the splitting functions tend to an effective limit
\be
P^0_{gg} \to \frac{3\alpha_S}{\pi}\frac{1}{x} \qquad P^0_{qg} \to
 2N_F\frac{\alpha_S}{6\pi}\delta(1-x),
\label{singeltsxlimitLO}
\ee
and so the gluon grows very quickly with increasing $Q^2$ 
while the quark distribution also grows quickly,
very largely driven by the gluon distribution.
This means there is a fairly direct correlation between $xg(x,Q^2)$ 
and $dF_2(x,Q^2)/d\ln Q^2$ at this order.
At NLO the splitting functions  as $x \to 0$ become
\be
P^1_{gg} \to -0.7 \alpha^2_S\frac{1}{x} \qquad P^1_{qg} \to
 2N_F\frac{\alpha^2_S}{6\pi}\frac{1.6}{x}.
\label{singeltsxlimitNLO} 
\ee
Hence, the gluon evolution is only slightly modified, 
whereas the quark 
evolution is greatly enhanced at NLO. At this order $dF_2(x,Q^2)/d\ln Q^2$ 
is no longer directly proportional to $xg(x,Q^2)$, but is sensitive
to the gluon at all higher values of $x$. The very precise data has already 
been 
illustrated in the left of Fig. \ref{fig:scal} compared to the HERAPDF1.0 fit. This data is 
a direct constraint on the charge weighted quark distributions down to
$x=5\times 10^{-5}$ and an indirect, but still very precise constraint
on the gluon. This is only for $x$ values of about an order of 
magnitude higher, however. This is partly because the accurate constraint 
on the gluon from evolution, i.e. from $dF_2(x,Q^2)/d\ln Q^2$, ceases when 
there is no longer a fairly large number of points at a given $x$ with different $Q^2$ 
values. It is also because the evolution involves a convolution of the gluon
so probes slightly higher $x$ values than the data, particularly at NLO
compared to LO.

\begin{figure}
\begin{center}
\begin{minipage}[t]{9 cm}
\epsfig{file=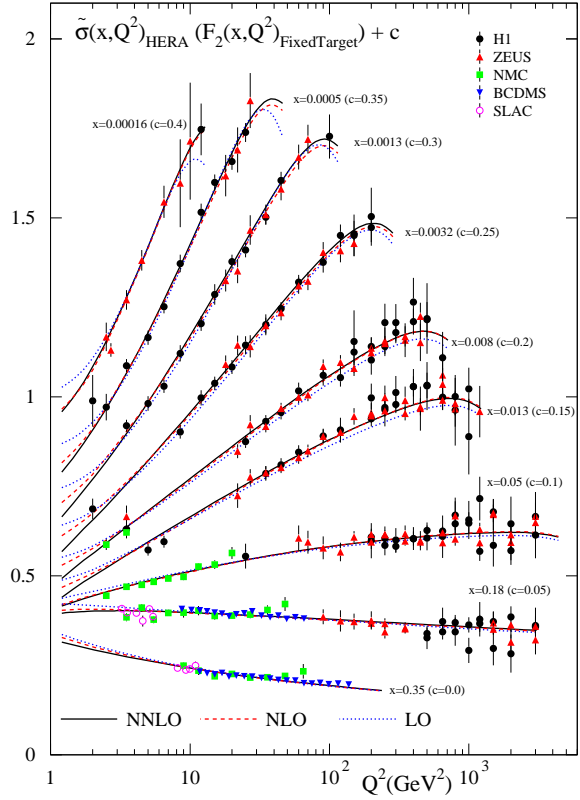,scale=0.5}
\end{minipage}
\begin{minipage}[t]{16.5 cm}
\caption{Comparison of MRST2008 at LO, NLO and NNLO and mainly HERA 
data\cite{Martin:2009iq}.  
\label{MSTWHERAfit}}
\end{minipage}
\end{center}
\end{figure}

At NNLO the small $x$ splitting functions become, as $x \to 0$\cite{Vogt:2004mw}
\be
P^1_{gg} \to -1.7 \alpha^3_S\frac{\ln(1/x)}{x} \qquad P^1_{qg} \to
 2N_F\frac{\alpha^3_S}{6\pi}\frac{1.4\ln(1/x)}{x}.
\label{singeltsxlimitNNLO} 
\ee
So at NNLO the quark evolution at the smallest $x$ values 
is enhanced yet again while the gluon evolution is suppressed. 
It is known that at each subsequent order in 
$\alpha_S$ each splitting 
function and coefficient function obtains an extra power of $\ln(1/x)$\cite{Lipatov:1976zz,Kuraev:1977fs,Balitsky:1978ic}
(there are some accidental zeros in $P_{gg}$), i.e. 
\be
P_{ij}(x,\alpha_S(Q^2)),\quad C^P_i(x,\alpha_S(Q^2)) \sim 
\alpha_S^m(Q^2)\ln^{m-1}(1/x).
\label{sxdivergences}
\ee
and hence the convergence at small $x$ is questionable.
The global fits usually assume that this turns out to be 
unimportant in practice, and proceed regardless. The fit is quite good, but 
could be improved as illustrated in Fig. \ref{MSTWHERAfit}. As seen in this 
figure the evolution becomes slightly steeper with $Q^2$ in general as the 
order increases, due to the extra contributions to the splitting functions 
discussed above, and the fit becomes slightly better. Further implications
for the small-$x$ behaviour will be discussed later.

\begin{figure}
\begin{center}
\begin{minipage}[t]{9.5 cm}
\epsfig{file=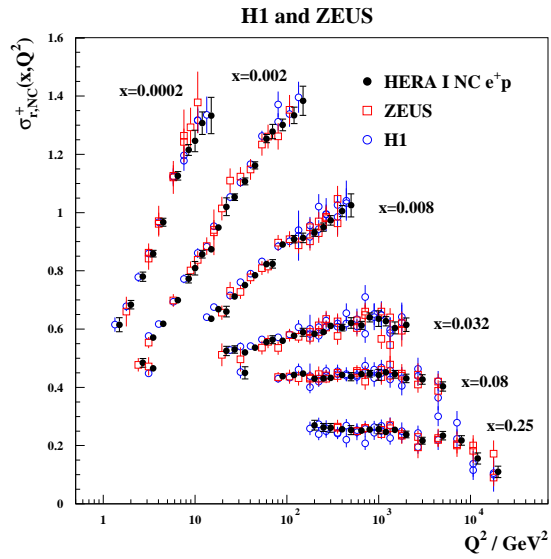,scale=0.4}
\end{minipage}
\begin{minipage}[t]{16.5 cm}
\caption{An illustration of the combination of HERA data\cite{:2009wt}.
The open circles and squares are the separate H1 and ZEUS data respectively 
and the closed points the combined data.   
\label{HERAcombination}}
\end{minipage}
\end{center}
\end{figure}

Very recently the H1 and ZEUS collaborations at HERA have combined their 
structure functions measurements into a single result\cite{:2009wt}. This 
is shown in Fig. \ref{HERAcombination}. This not only reduces the 
statistical uncertainty, but has a far more dramatic effect on the 
correlated systematic uncertainties, which are frequently far better 
understood for a given source by one collaboration, and can hence be 
reduced by considerably more than the statistical error. This change
in the correlated errors also means the average of two data points from 
the two collaborations is not the simply the weighted average of the central 
values of each. In particular, a better understanding of normalisations
has moved the data upwards slightly overall. This is reflected in slightly larger
quark distributions in most small-$x$ regions, most particularly near 
$x=0.01$. The data also leads to a reduction in uncertainty in parton 
distributions, 
but this depends on the other data used in the fit and the procedure used.

\begin{figure}
\begin{center}
\begin{minipage}[t]{11 cm}
\epsfig{file=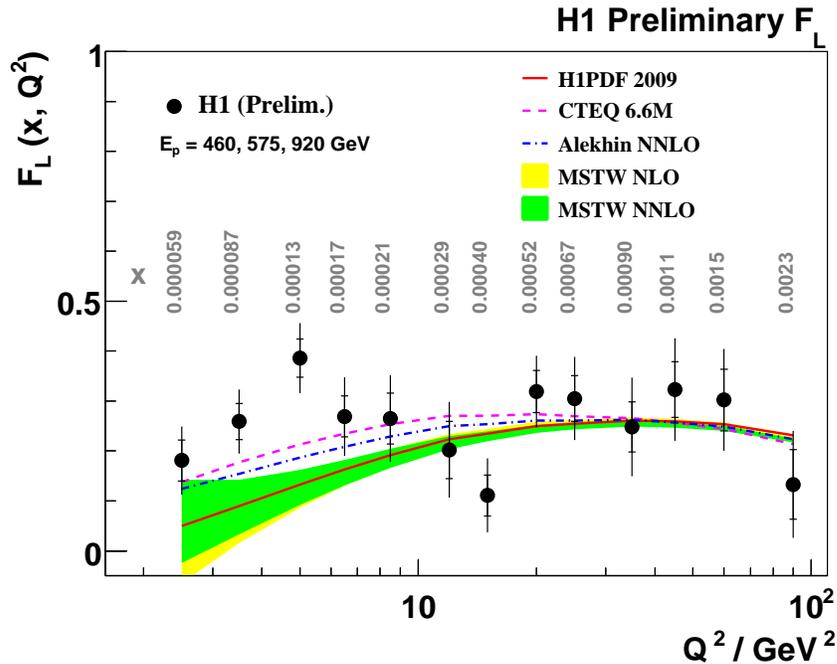,scale=0.6}
\end{minipage}
\begin{minipage}[t]{16.5 cm}
\caption{$F_L(x,Q^2)$ data from H1 compared to various predictions\cite{Raicevic:2010zz,Reisert:2009zz}.  
\label{H1FL}}
\end{minipage}
\end{center}
\end{figure}

There is also now a direct HERA measurement of $F_L(x,Q^2)$ 
which is a new and direct constraint on the 
small-$x$ gluon. The original published data\cite{Aaron:2008tx,Chekanov:2009na}
is not yet precise enough to add much real extra constraint, but is a vital
consistency check for the extracted gluon distribution. Existing parton 
distributions match this data well. However, the preliminary lower-$Q^2$
data from H1 (see e.g. \cite{Raicevic:2010zz,Reisert:2009zz}) 
shown in Fig. \ref{H1FL} 
seem to be in excess of the predictions from the majority of fixed-order parton
distribution sets, and imply some additional physics at small $x$ and/or
low $Q^2$.The data have very recently been 
finalised\cite{Collaboration:2010ry}, 
and change slightly, but the same general conclusion holds. 
This will be discussed briefly later. 

\begin{figure}
\begin{center}
\begin{minipage}[t]{15 cm}
\epsfig{file=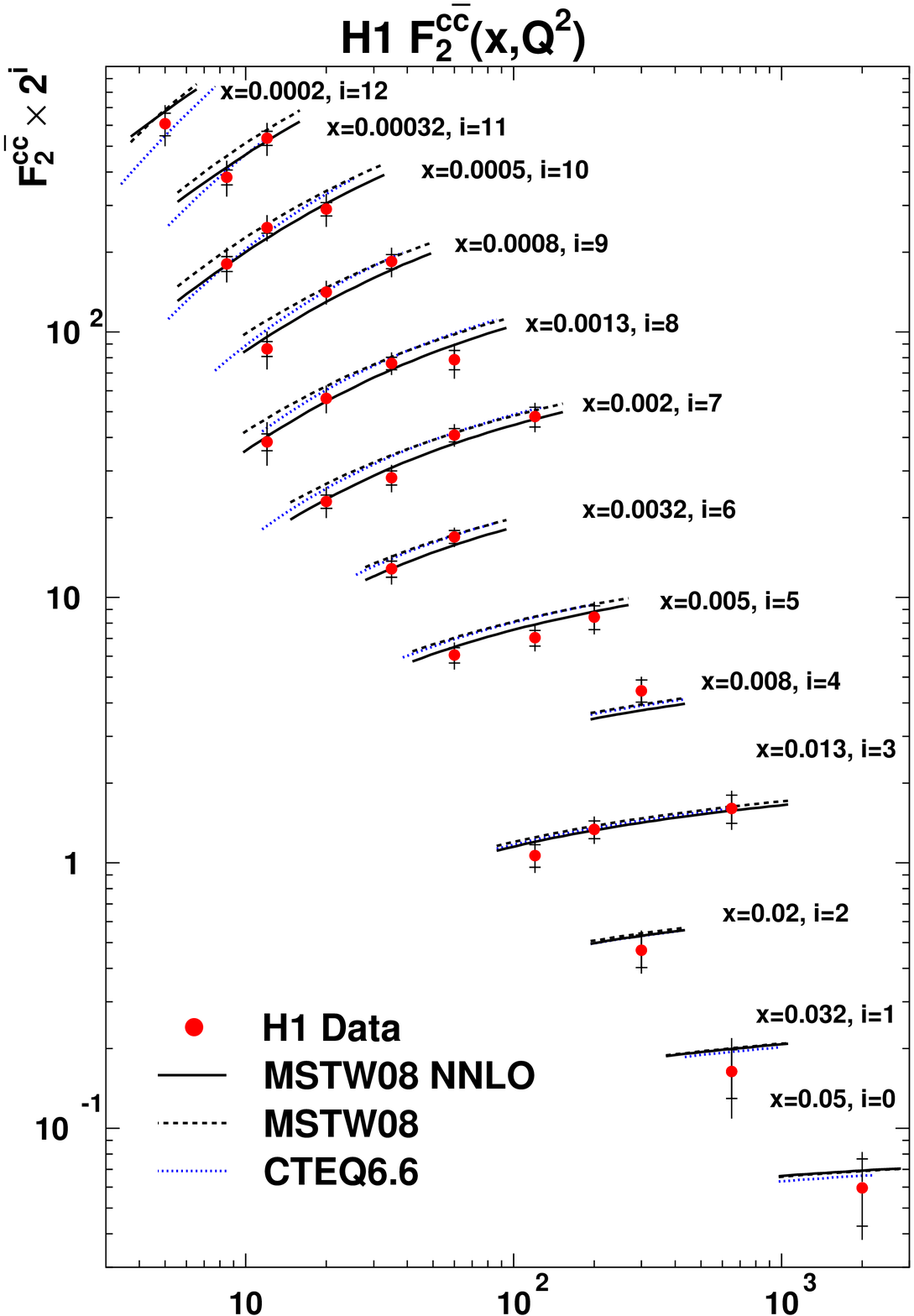,scale=0.36}
\epsfig{file=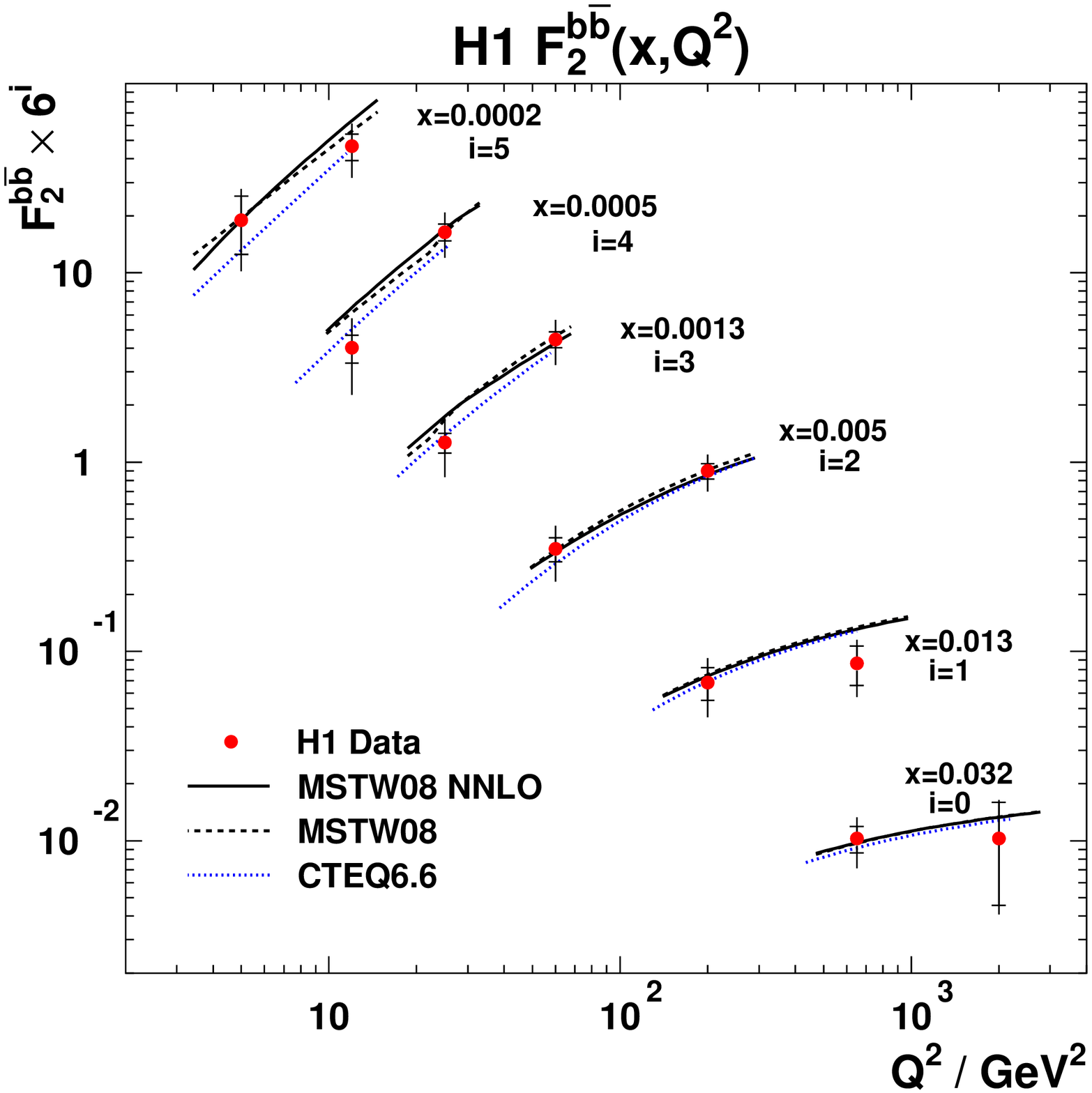,scale=0.36}
\end{minipage}
\begin{minipage}[t]{16.5 cm}
\caption{A comparison of the calculation of charm $F_2^c(x,Q^2)$
(left) and $F_2^b(x,Q^2)$ (right) using a variety of prescriptions and 
and recent H1 data\cite{Aaron:2009ut}. 
\label{HERAheavy}}
\end{minipage}
\end{center}
\end{figure}

\subsection{\it Heavy Flavours -- Quark Masses}

There is also data from HERA on the heavy flavour contribution to structure 
functions, i.e. on $F_2^c(x,Q^2)$\cite{Adloff:1996xq,Adloff:2001zj,Aktas:2005iw,Aktas:2004az,Breitweg:1999ad,Chekanov:2003rb,Chekanov:2007ch}, and also including
$F_2^b(x,Q^2)$\cite{Aaron:2009ut,Chekanov:2009kj}. One might consider 
this to be a constraint on the heavy flavour distributions. However, these 
are generated entirely from evolution from the light partons, mainly the 
gluon. The gluon is constrained by the data in the previous subsections, 
so the heavy flavour structure functions are very largely a prediction. They are,
however, a test of the quark masses, which set the boundary conditions for the 
heavy flavour evolution, and hence the size of the heavy quark PDFs, and of 
the theoretical procedure used to calculate 
heavy flavour structure functions. These will be discussed in more detail 
later. A comparison of theory to some of the most precise data is shown in  
Fig. \ref{HERAheavy}.

\section{The Variety of Parton Distributions}

\subsection{\it Different PDF Sets}

As outlined earlier there are a number of different groups which 
obtain full sets of parton distributions by fitting to structure function
and other data. Some of the differences were touched upon in the previous 
section when discussing the data sets which can be included in a fit. In this section
we will present the range of sets, and their similarities and differences 
rather more comprehensively. We choose not to dwell a great deal on the 
history of each, since in all cases the most up-to-date set is clearly the 
one which should be used, and in most cases there is a point before which 
there is a very good reason to no longer use the sets in a given series. 

The different sets and their most basic features are listed below.

\begin{itemize}

\item The MSTW group is on of two which has been producing 
parton distributions for many years from global fits to a wide variety of 
data. The group changed from MRST to MSTW in 2007, though the 
first set produced by the group\cite{Martin:2007bv} maintained the 
MRST nomenclature. The most recent set is MSTW2008\cite{Martin:2009iq}.
The group fits to essentially all the data sets listed in the previous 
section, including the up-to-date Tevatron jet data and $W$ and 
$Z$ data. The fit does not include the most recent HERA combination of 
structure function data (the effects have been investigated\cite{Thorne:2010kj}) or the HERA data on $F_L(x,Q^2)$, though it does 
include some fixed target data on $F_L(x,Q^2)$\cite{Benvenuti:1989rh,
Arneodo:1996qe,Whitlow:1991uw}. MSTW are the only group to include HERA jet 
data. The group produces PDFs at LO, NLO and NNLO. 

\item The CTEQ group is the other which has been performing global 
fits for may years, and is in many ways has an approach which is very 
similar to that of MSTW. Again the group fits to the vast majority of 
available data. The recent significant update in widest use is 
CTEQ6.6\cite{Nadolsky:2008zw}. This is slightly older than the MSTW2008 sets 
and is not quite as up-to-date on Tevatron data and likewise does not 
include the most recent HERA combination of structure function data, though
an updated set, CT10\cite{Lai:2010vv} has appeared recently and includes 
these data sets. PDFs are made available at NLO.

\item The NNPDF group uses a rather distinct 
procedure, as will be explained later in this section. It has continually
been developing for the past few years, but with NNPDF2.0\cite{Ball:2010de}
(and extremely recently NNPDF2.1\cite{Ball:2011mu}), 
which includes Tevatron data, they have reached 
the status of a global fit. Previous sets are based on rather smaller data 
sets, either mainly or entirely structure functions data.
The NNPDF2.0 fit includes all the data discussed above 
except HERA jet data and heavy flavour structure functions. 
It is sufficiently recent that it does include the HERA combined data,
and notices a moderate effect compared to the original 
individual data sets, most noticeably a smaller uncertainty in the 
gluon and singlet quarks below $x=5 \times 10^{-4}$. PDFs are made 
available at NLO.

\item There have been a variety of fits performed by the H1\cite{Adloff:2003uh} and ZEUS\cite{Chekanov:2002pv,Chekanov:2005nn} 
collaborations. These have sometimes included fixed target
structure function data\cite{Chekanov:2002pv,Adloff:2003uh}
or HERA jet data\cite{Chekanov:2005nn}. More recently, in order to analyse
the combined HERA structure function data the fitting groups have also 
converged. The HERAPDF1.0\cite{:2009wt} PDFs are based entirely on 
HERA inclusive structure function data, both neutral and charged current. 
PDFs are produced at NLO.  A preliminary update, including NNLO results, 
is in \cite{CooperSarkar:2010ik}.

\item The ABKM group provides a continuation of the fits performed in \cite{Alekhin:2002fv,Alekhin:2005gq,Alekhin:2006zm}. The first
set of PDFs obtained by the combined group ABKM09\cite{Alekhin:2009ni} comes from a fit to structure function, 
fixed target Drell-Yan, and dimuon data. No Tevatron data is included
(though preliminary results for some jet data sets can be seen in\cite{AlekhinPDF4LHC}). 
PDFs are produced with both NLO and NNLO evolution. 
There is a preliminary update in \cite{Alekhin:2010iu}. 
A parameterisation of ${\cal O}(1/Q^2)$ power corrections is employed at 
low $W^2$ rather than the more common kinematic cut.

\item The GJR\cite{Gluck:2007ck}, or dynamical parton distributions are
based on the idea, originally advocated in \cite{Gluck:1991ng}, that the 
PDFs are generated from a valence-like input form at some very low 
starting scale $Q_0^2 \lsim 0.5 \GeV^2$. They are obtained from a fit to 
structure function, fixed target Drell-Yan and 
Tevatron jet data. Sets from more conventional starting distributions are 
also obtained, though not advocated by the authors (despite providing much 
better fit quality and a more conventional value of the strong coupling). 
PDFs are made available with LO, NLO and NNLO\cite{JimenezDelgado:2008hf}
evolution. At NLO, PDFs are also made available in the DIS factorisation 
scheme as well as the $\overline{\rm MS}$ scheme used by all groups. 
This has been done in the past by some others, e.g.
\cite{Lai:1996mg,Martin:1998np}. In practice the transformation rule in 
\cite{Altarelli:1978id} (and defined up to NNLO in \cite{White:2005wm})
could be used to make the transformation between the two.

\end{itemize}

The evolution of the partons in each set should be exactly the same up to small
uncertainties. A cross-check of PDF evolution was first performed in
\cite{Blumlein:1996rp}, leading to the correction of bugs causing errors up to
a couple of percent. Now there is much better agreement. Benchmark tables 
have been constructed\cite{Jung:2009eq} 
using publicly available codes\cite{Vogt:2004ns,Salam:2008qg}, and checked against 
some sets (see e.g. \cite{Ball:2008by,Martin:2009iq}) and agreement to much better 
than $1\%$ is found.  
However, the PDF sets obtained differ by much more than this. As outlined above this 
can be due to  
the choices of data sets and kinematic cuts made, but can also occur for a large 
number of other reasons. This will be discussed in 
detail in the remainder of this section.  

\subsection{\it Parton Fitting and Uncertainties}

There are two main approaches to obtaining the most likely PDF sets and
the uncertainty. One, the original approach, is to find the PDFs by the best 
fit to the data, and then to perturb about this in some fashion. The other
is to obtain an ensemble of different PDF sets and to find the most likely
result by averaging and the uncertainty from the deviation of the predictions 
from the mean. This will be discussed in more detail below.  

In the former case the quality of the fit is traditionally 
determined by the $\chi^2$ of the fit to 
data, which may be calculated in various ways. 
The simplest is to add statistical and systematic errors in quadrature, 
which ignores correlations between data points, but is often
quite effective. Also, the information on the data 
means that sometimes only this method is available.
Being more complete one uses the full covariance matrix which 
is constructed as
\begin{equation}
C_{ij} = \delta_{ij} \sigma_{i,stat}^2 + \sum_{k=1}^n \rho^k_{ij}
\sigma_{k,i}\sigma_{k,j}, \qquad 
\chi^2 = \sum_{i=1}^N\sum_{j=1}^N (D_i-T_i(a))C^{-1}_{ij}
(D_j-T_j(a)),
\label{k}
\end{equation}
where $k$ runs over each source of correlated systematic error,
$\rho^k_{ij}$ are the correlation coefficients,
$N$ is the number of data points, $D_i$ is the 
measurement and $T_i(a)$ is the theoretical prediction depending on 
parton input parameters $a$. An alternative that produces identical results 
in the quadratic approximation is to 
incorporate the correlated errors into the theory prediction
\begin{equation}
f_i(a,s) = T_i(a) + \sum_{k=1}^n s_k \Delta_{ik}, \qquad \chi^2 = 
\sum_{i=1}^N \biggl(\frac{D_i-f_i(a,s)}{\sigma_{i,unc}}
\biggr)^2 + \sum_{k=1}^n s_k^2,
\label{m}
\end{equation}
where $\Delta_{ik}$ is the one-sigma correlated error for point 
$i$.
One can solve analytically for the $s_k$\cite{Stump:2001gu}.

\begin{figure}
\begin{center}
\begin{minipage}[t]{16.5 cm}
\epsfig{file=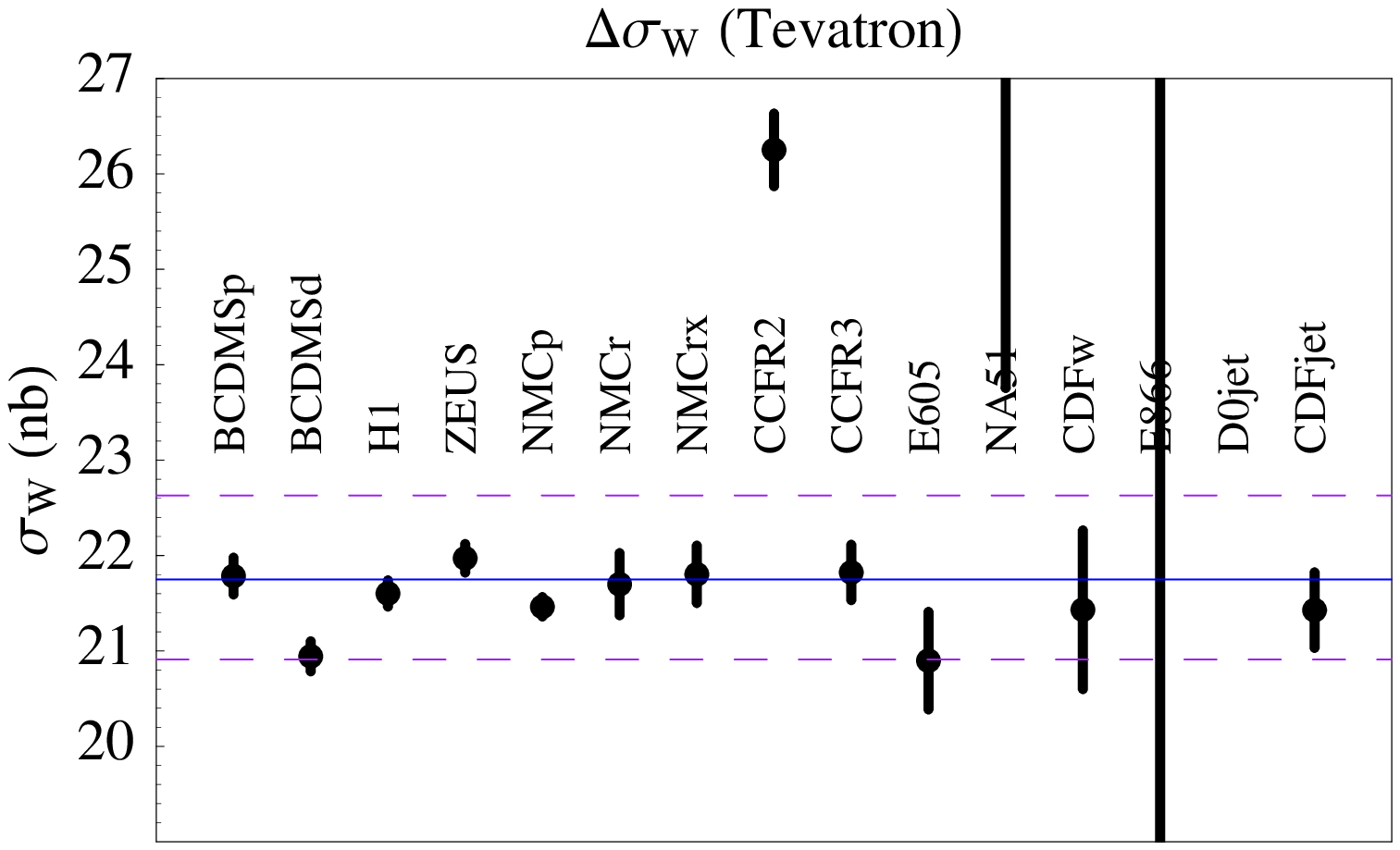,scale=0.5}
\epsfig{file=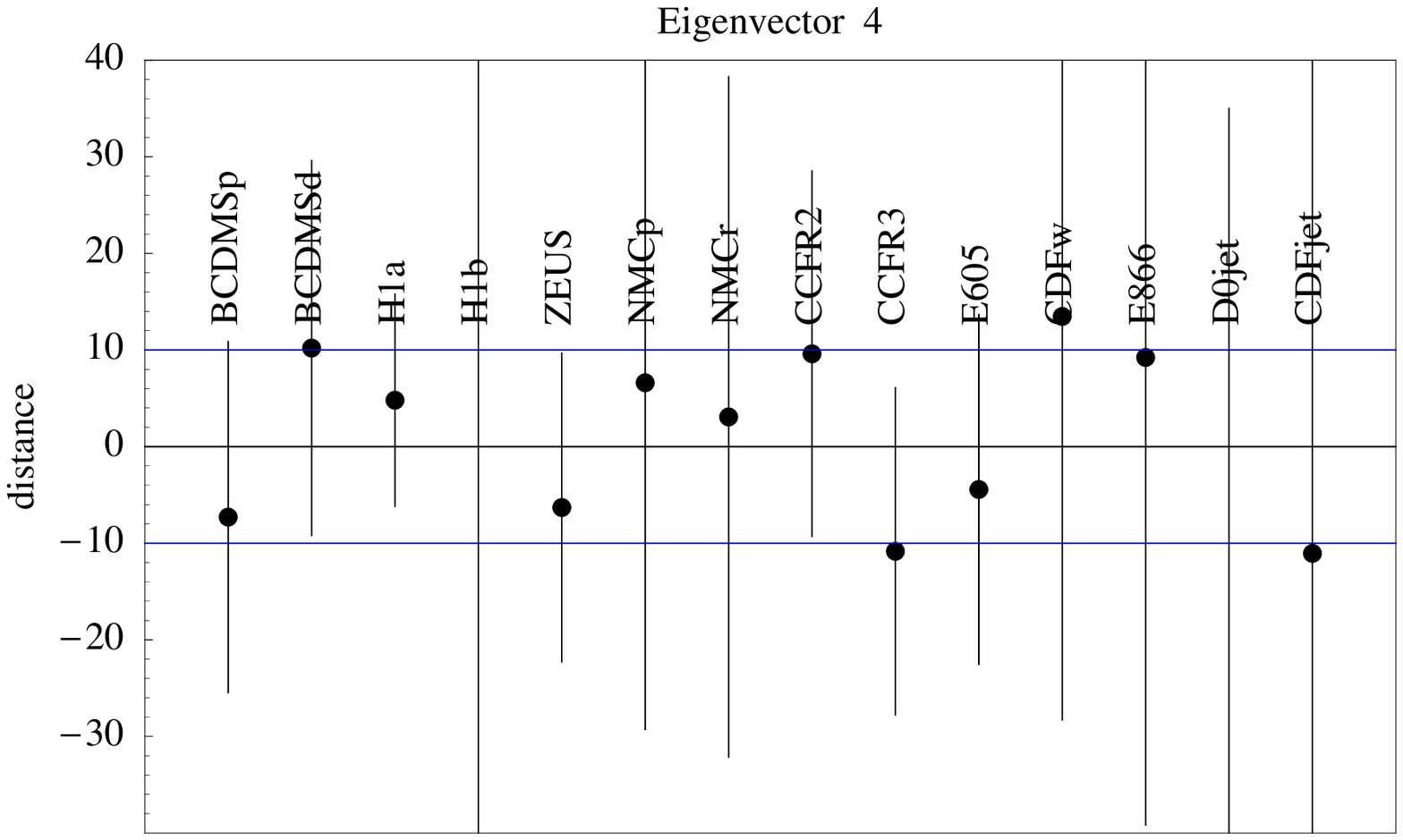,scale=0.43}
\end{minipage}
\begin{minipage}[t]{16.5 cm}
\caption{The best value of $\sigma_W$ and the uncertainty using a fit 
deterioration of 
$\Delta \chi^2=1$
for each data set in the CTEQ fit (left) and  the $90\%$ confidence limits
for each data set as a function of $\sqrt{\Delta \chi^2}$  
for one eigenvector (right)\cite{Stump:2001gu}. 
\label{cteq1sigma}}
\end{minipage}
\end{center}
\end{figure}

Having defined the fit quality
there are a number of different approaches for obtaining parton uncertainties. 
The most common is the Hessian (Error Matrix) approach. 
One defines the Hessian matrix by
\begin{equation}
\chi^2 -\chi_{min}^2 \equiv \Delta \chi^2 = \sum_{i,j} 
H_{ij}(a_i -a_i^{(0)})
(a_j -a_j^{(0)}).
\label{r}
\end{equation}
One can then use the standard formula for linear error propagation:  
\begin{equation}
(\Delta F)^2 = \Delta \chi^2 \sum_{i,j} \frac{\partial F}
{\partial a_i}(H)^{-1}_{ij}  
\frac{\partial F}{\partial a_j}.
\label{t}
\end{equation}
This was used to find partons with uncertainties in e.g.  
\cite{Adloff:2000qk,Alekhin:2002fv}. 
In practice this can be problematic due to extreme 
variations in $\Delta \chi^2$ in different directions in parameter space.
This can be improved by finding and rescaling the eigenvectors of $H$, 
a method developed by CTEQ\cite{Pumplin:2000vx,Pumplin:2001ct}, and now 
used by other groups. In terms of the rescaled eigenvectors $z_i$,
which are orthonormal combinations of the $a_i-a_i(0)$, the  
increase in $\chi^2$ is given simply by 
\begin{equation}
\chi^2 -\chi_{min}^2 \equiv \Delta \chi^2 = \sum_{i} z_i^2,
\label{rrescal}
\ee
i.e. constant $\Delta \chi^2$ is the surface of a hypersphere in the space 
of the parton parameter eigenvectors. 
The uncertainty on a physical quantity is then obtained using Pythagoras'
theorem,  
\begin{equation}
(\Delta F)^2 = \frac{1}{2}\sum_{i} \bigl(F(S_i^{(+)})-F(S_i^{(-)})\bigr)^2,
\label{v}
\end{equation}
where $S_i^{(+)}$ and $S_i^{(-)}$ are PDF sets 
displaced along eigenvector
directions by the given $\Delta \chi^2$. 

\begin{figure}
\begin{center}
\begin{minipage}[t]{12 cm}
\rotatebox{270}{\epsfig{file=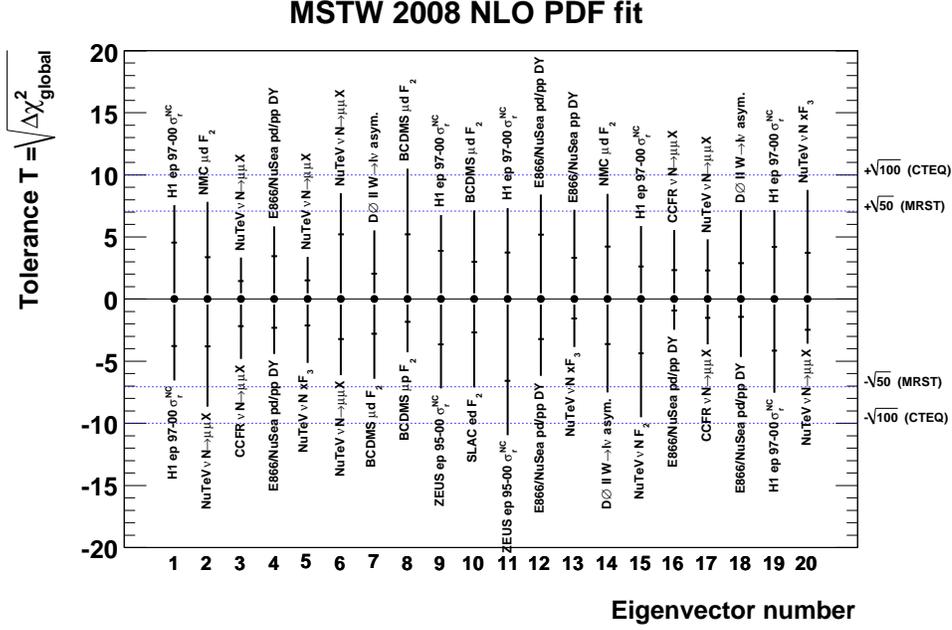,scale=0.5}}
\end{minipage}
\begin{minipage}[t]{16.5 cm}
\vspace{1.5cm}
\caption{The tolerance as a function of eigenvector number for the
MSTW2008 NLO PDFs\cite{Martin:2009iq}. 
The outer band is $90\%$ confidence level and the inner
band $68\%$. The label at each end of the bar is the data set 
providing the main constraint to that eigenvector in each direction.   
\label{tolerance}}
\end{minipage}
\end{center}
\end{figure}

One can also investigate 
the uncertainty on a given physical quantity using the 
Lagrange Multiplier method, first suggested by CTEQ\cite{Stump:2001gu}  
and also investigated in some detail by MRST\cite{Martin:2002aw}. 
One performs the global fit while constraining the value of some physical 
quantity, i.e.  minimise 
\begin{equation}
\Psi(\lambda,a) = \chi^2_{global}(a)  + \lambda F(a)
\label{ac}
\end{equation}
for various values of $\lambda$. This gives the set of best fits for 
particular values of the parameter $F(a)$ without relying on the quadratic 
approximation for $\Delta\chi^2$, but has to be done anew for each quantity.  

In each approach there is uncertainty in choosing 
the ``correct'' $\Delta \chi^2$. In principle this should be one unit, 
and some groups with smaller number of data sets use this. 
However, given the complications within a full global fit this gives 
unrealistically 
small uncertainties. This can be seen in the left of Fig.~\ref{cteq1sigma}.
where the variation in the predictions for $\sigma_W$ using $\Delta \chi^2 =1$
for each data set has an extremely wide scatter compared to the uncertainty. 
CTEQ choose $\Delta \chi^2 \sim 100$. The $90\%$ confidence 
limits for the fits to the larger individual data sets when 
$\sqrt{\Delta \chi^2}$ in the CTEQ fit is increased by a 
given amount are shown in 
the right of Fig.~\ref{cteq1sigma}. As one sees, a couple of sets may be some 
way beyond their $90\%$ confidence limit for $\Delta \chi^2 = 100$. The 
MRST group instead chose $\Delta \chi^2 = 50$
to represent the $90\%$ confidence limit for the fit\cite{Martin:2002aw}. However, 
the most recent fit\cite{Martin:2009iq} recognised that some eigenvectors are
constrained by many fewer data points than others, and modified the 
prescription to give a so-called ``dynamical tolerance'' where the 
$\Delta \chi^2$ depends on the eigenvector (and on the two orientations 
of the eigenvector). The values of $\sqrt{\Delta \chi^2}$ for the NLO MSTW
fit are shown in Fig. \ref{tolerance}. For $68\%$ confidence level they are 
usually of magnitude 2-4, suggesting, on average, a $\Delta \chi^2$ of 
about 10 for one-$\sigma$ uncertainties, somewhat smaller than previous 
MRST, and certainly CTEQ values. There has been recent work on how the 
increase in $\Delta \chi^2$ may be related to inconsistency of data sets\cite{Pumplin:2009sc} and limitations of a fixed number of parameters
for input parton distributions\cite{Pumplin:2009bb}, both in the context of 
the CTEQ global fit. The former implies that data set inconsistency
should give $\Delta \chi^2\approx 4$ for one $\sigma$, though this might be 
greater if more data sets or less conservative cuts are used than in the CTEQ
fit. The latter implies a relatively similar factor for the increase in 
$\chi^2$. In regions where there is little data constraint and PDFs are 
constrained by their limited parameterisation, e.g. very high $x$ or 
very small $x$ for valence quarks (and the gluon using some 
parameterisations) this can be due to quite large changes in PDFs making 
rather little difference to the fit quality. Where there is constraining data
it seems more likely to be the case only if a noticeably better fit can be found
by extending the parameterisation, an unsurprising result which then depends 
on how well the original parameterisation is performing.

\begin{figure}
\begin{center}
\begin{minipage}[t]{10 cm}
\epsfig{file=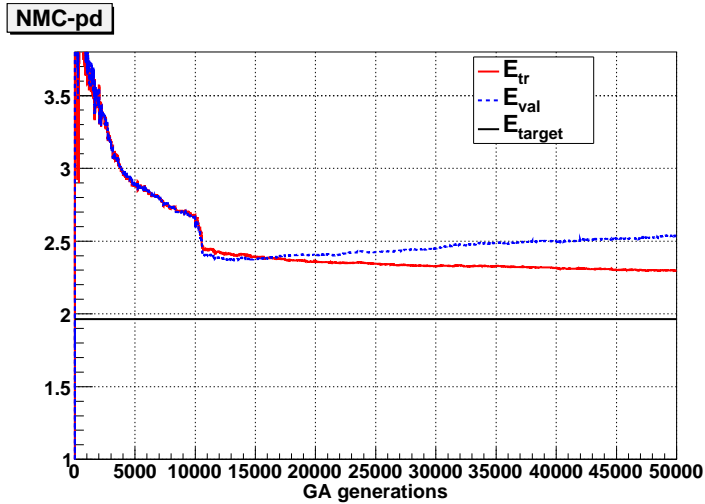,scale=0.5}
\end{minipage}
\begin{minipage}[t]{16.5 cm}
\caption{An illustration of the fit quality to the training (labelled $E_{\rm tr}$)
and validation set (labelled $E_{\rm val}$) for NMC data in the NNPDF fit\cite{Ball:2010de}.
The constant line labelled $E_{\rm target}$ determines the weight of this 
data set in the fit.    
\label{stopping}}
\end{minipage}
\end{center}
\end{figure}

There are other approaches to finding the uncertainties. In the offset method
the best fit is obtained by minimising the $\chi^2$ using
only uncorrelated errors. The systematic errors on the parton parameters 
$a_i$ are then determined by letting each $s_k = \pm 1$ and adding 
the deviations in quadrature. 
This method was used in some previous ZEUS fits\cite{Chekanov:2005nn}, 
and is used for the three ``procedural'' systematic uncertainties in the 
HERAPDF1.0 fit\cite{:2009wt}. 

There is also the 
statistical approach used by Neural Network group. Here one 
constructs a set of Monte Carlo replicas $\sigma^k(p_i)$ of the original 
data set  $\sigma^{data}(p_i)$ which gives a
representation of $P[\sigma(p_i)]$ at points $p_i$. 
Then one obtains a parton distribution function for each 
replica, obtaining a representation of the PDFs $q_i^{(k)}$.
The set of PDF replicas obtained is a representation of the probability 
density -- i.e. the mean $\mu_O$ and deviation 
$\sigma_O$ of an observable $O$ is given by
\begin{equation}
\mu_O = \frac {1}{N_{rep}}\sum_1^{N_{rep}}O[q_i^{(k)}], \quad
\sigma_O^2 =\frac {1}{N_{rep}} \sum_{1}^{N_{rep}}(O[q_i^{(k)}]-\mu_O)^2.
\end{equation}
One can incorporate full information about measurements and their error 
correlations in the distribution of $\sigma^{data}(p_i)$. 
This is does not rely on the approximation of linear propagation of errors 
(though the data replicas assume a Gaussian distribution) 
but is more time intensive. This basic idea was proposed in 
\cite{Giele:1998gw,Giele:2001mr}, but was performed using standard input
parameterisations for PDFs. 

The NNPDF group\cite{DelDebbio:2007ee,
Ball:2008by,Ball:2010de} has developed this 
philosophy and combined it with the standard input parameterisations being 
replaced by a neural network, or effectively a very much larger number 
of parameters than any other group (though there is some pre-processing which 
leads to the limits of $x\to 0$ and $x\to 1$ being related to the usual 
forms of $x^{\delta}$ and $(1-x)^{\eta}$ respectively). 
In principle this means that if the 
fit to data is left to converge for too long a time the input will start
to fit to fluctuations in the data. This is avoided by fitting
to both a training set and comparing to a validation set, each comprising
of half the data. The fit is then stopped when the quality of the fit to 
the training set may still be slowly improving, but that to the validation set
starts to deteriorate. This is one of the main sources of complication, and 
where there has been continual development. Ideally each data set within 
the global fit will reach its stopping point at the same time. In practice 
some will tend do so long before others and their validation sets can be 
progressively fit worse while the quality of the fit to 
the global validation set is still improving. Stopping while the 
global validation sets is still improving is probably somewhat 
analogous to the requirement for the other ``global'' fits to use
an inflated $\Delta \chi^2$ when perturbing about the best global fit. 
The NNPDF group have adopted a procedure called ``target weighted training''\cite{Ball:2010de} to minimise this problem, where the different data sets 
have different weights, which are determined iteratively, and aid the 
convergence of each set reaching the appropriate stopping point at the 
same time. The quality of the fit for the training and validation sets 
is shown for one data set in Fig. \ref{stopping}.
This new procedure leads to some data sets in \cite{Ball:2010de} to have a 
better fit quality than previous NNPDF fits, particular fixed target structure 
function data. It also leads to generally smaller uncertainties, perhaps being 
nearer in some sense to the criterion $\Delta \chi^2=1$ in the 
alternative procedure, 
than previous stopping criteria, as seen in Fig. 23 of \cite{Ball:2010de}. 
The most recent set has also included changes in the treatment of data set 
normalisations\cite{Ball:2009qv}, so it is difficult to appreciate the
change in PDFs between \cite{Ball:2010de}, and the previous set\cite{Ball:2009mk} due to the inclusion of new data by comparing the two sets. 
Helpfully, there are various illustrations in the effect of particular 
data sets in \cite{Ball:2010de}.      
 
To summarise, the procedure used to determine the uncertainty for each group
is:

\begin{itemize}

\item MSTW08 perturb around the best fit using 
$20$ orthonormal eigenvectors. Older data sets on structure functions use
data averaged over different energies and combine uncertainties for structure 
function data in quadrature, other than those on normalisation. 
It has been checked in \cite{Martin:2002aw,
Dittmar:2005ed,Fortepdf4lhc} that this has a small effect on the central 
values and uncertainties of the PDFs (in the last changes of up to $\sigma/4$
only were found), though it clearly affects the value of the $\chi^2$.  
Due to incompatibility of different sets, imperfect theory, 
and (to some extent) 
parameterisation inflexibility MSTW have an inflated $\Delta \chi^2$ 
of $\sim 5-20$ for one $\sigma$ uncertainty for the eigenvectors, the 
value being determined independently for each eigenvector and direction. 
Data set normalisation uncertainties are included in the 
determination of the best fit
and uncertainties, though a quartic penalty is applied to these to minimise
a drift to slightly low values, most notably in the LO fit where the theory is 
systematically low compared to data.   

\item CTEQ6.6 perturb around the best fit using $22$ orthonormal 
eigenvectors, using $\Delta \chi^2$ 
of $100$ for $90\%$ confidence level for the eigenvectors. There is some
unspecified weighting of data sets in the global fit.  
Data normalisation uncertainties are not included in the determination 
of the uncertainties, which might be an additional reason for 
the large tolerance. CT10 uses $26$ eigenvectors and an improved method of 
uncertainty determination, but still with a similar $\Delta \chi^2$.   

\item For NNPDF2.0 the uncertainty is determined by the deviation of 
either 100 or 1000 PDF replicas where all data uncertainty information has 
gone into generating the data and consequently PDF replicas. The ``best fit''
could be taken as the average of the replicas, but there is also one PDF 
replica which has been fit to the central value of all data points, as in 
the other procedures. The direct relationship to $\Delta \chi^2$ in 
alternative global fits is not trivial. 

\item HERAPDF1.0 perturb about the best fit using 
$9$ orthonormal eigenvectors. Most (110) systematic uncertainties are 
combined in quadrature with the statistical uncertainties. Since the data 
comes for one combined self-consistent set $\Delta \chi^2 =1$ is used to
determine the uncertainties from this source. The three procedural 
(and largest) systematic uncertainties are added using the offset method.
Hence, the uncertainty is ``slightly'' more conservative than a use 
of $\Delta \chi^2=1$ incorporating all uncertainties. Since the default
number of input parameters is small an additional parameterisation 
uncertainty is included by adding various other parameters one at a time,
and also by changing the starting scale for the evolution. Additional 
variations in  strange sea fraction, data cuts and quark masses are 
included. 

\item ABKM uses perhaps the most straightforward and conventional method of 
determining the PDF uncertainties. They perturb about the best fit using $21$ 
parton parameters and also include heavy quark 
masses and the strong coupling as free parameters. They publish the
correlation matrix of the fitted parameters.  
The strict criterion of $\Delta \chi^2=1$ is used for uncertainty 
determination.

\item The GJR08 set is also based on a perturbation about the best fit.
There are $20$ parton parameters and the strong coupling is also varied
when determining the uncertainty. They use $\Delta \chi^2=22$ in order to 
define a one $\sigma$ uncertainty, and seemingly add statistical and systematic 
uncertainties in quadrature for all data sets, including Tevatron data. 
The fact that they impose 
a strong theory constraint on the input form of PDFs results in a reduced 
uncertainty in the small-$x$ singlet distributions, particularly the gluon 
distribution. The error bands of their default ``dynamical'' PDFs do not 
always overlap with those in their ``standard'' determination, which 
uses a starting scale and parameterisation more similar to other groups.    

\end{itemize}

\begin{figure}
\begin{center}
\begin{minipage}[t]{16 cm}
\epsfig{file=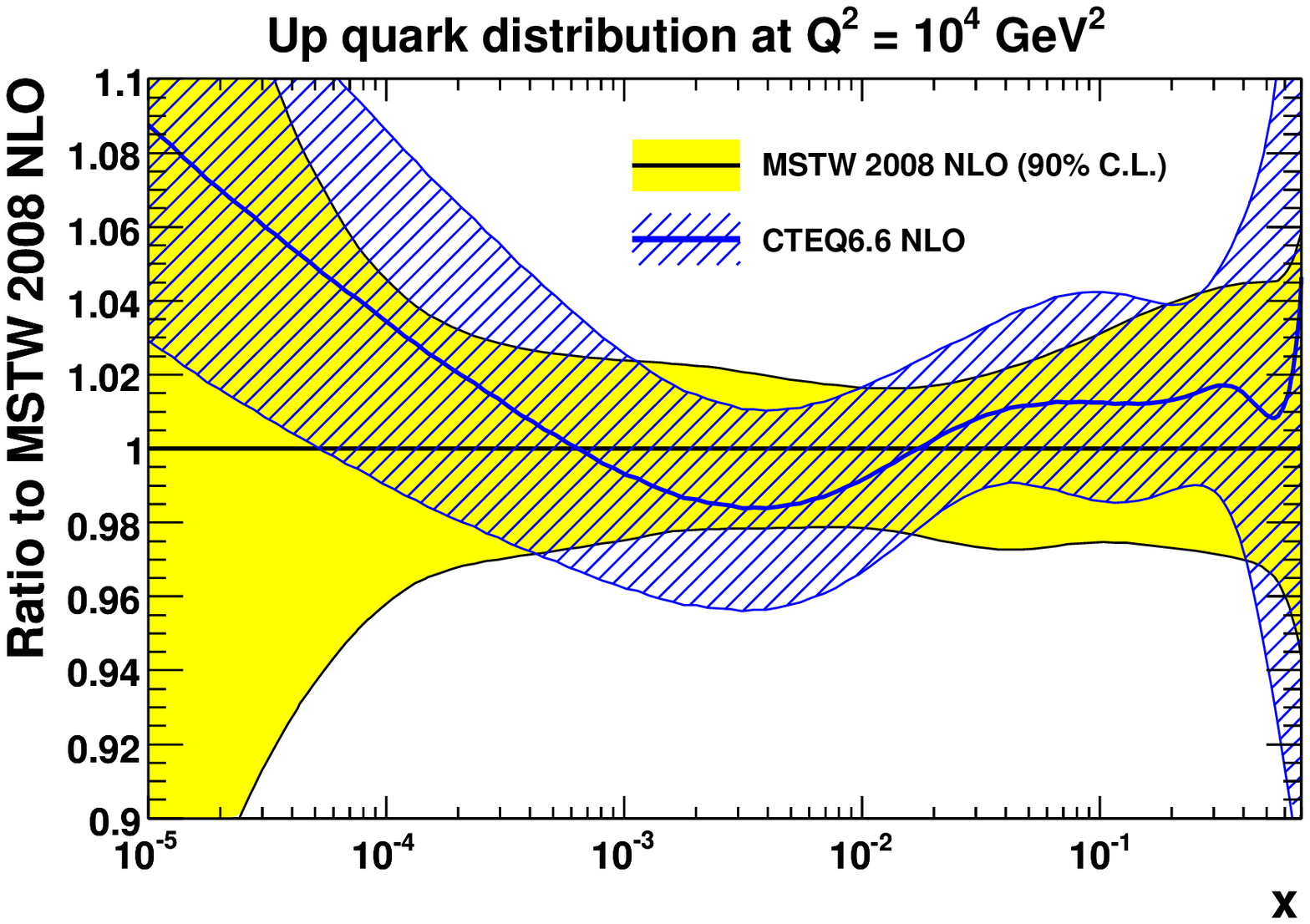,scale=0.42}
\epsfig{file=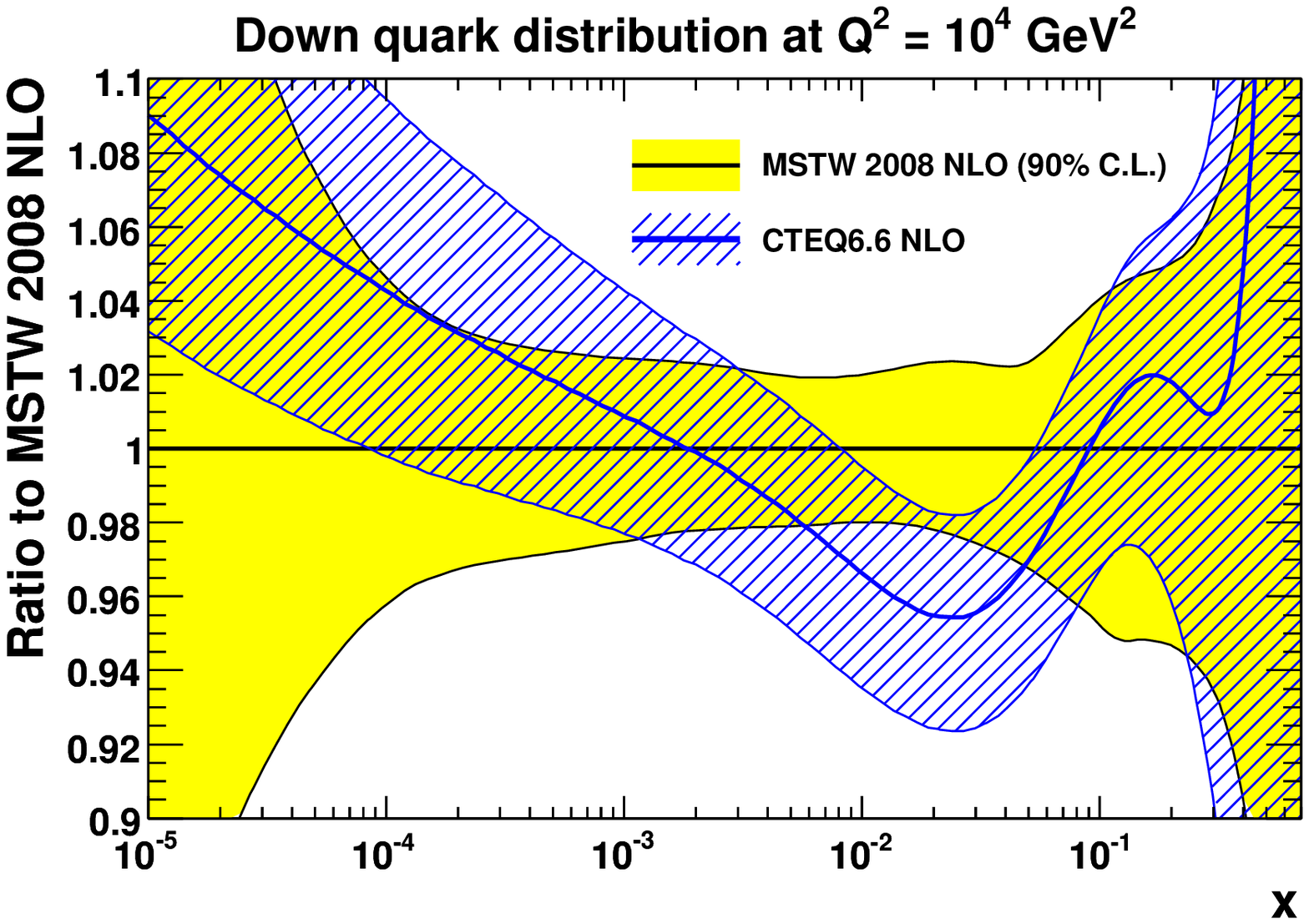,scale=0.42}
\end{minipage}
\begin{minipage}[t]{16.5 cm}
\caption{Uncertainty on the up quark (left) and down quark (right) 
from the CTEQ6.6 and MSTW08 PDF sets\cite{Martin:2009iq}.    
\label{upuncertainty}}
\end{minipage}
\end{center}
\end{figure}

Perhaps surprisingly, despite the very widely differing procedures for 
uncertainty determination described above, all PDF sets obtain rather 
similar uncertainties for the PDFs and predicted cross-sections. In fact the 
agreement in this respect is probably better than might be expected
given that some sets contain considerably  fewer data constrains than others.
This later impression is reinforced by the fact that the central values 
of the PDF sets do actually show some significant deviations, which are 
greater than the individual uncertainties, as seen in Figs. \ref{ABKMcomp} and
\ref{GJRcomp}.
As an example of some of the best agreement in Fig. \ref{upuncertainty}
we show the $90\%$ confidence level uncertainty on the MSTW08 NLO $u$ and $d$ 
distributions, along with CTEQ6.6, where the central line for the later 
represents the ratio of the CTEQ PDF to that of MSTW. 
There is clearly reasonable agreement between the two groups. 

\begin{figure}
\begin{center}
\begin{minipage}[t]{14.3 cm}
\epsfig{file=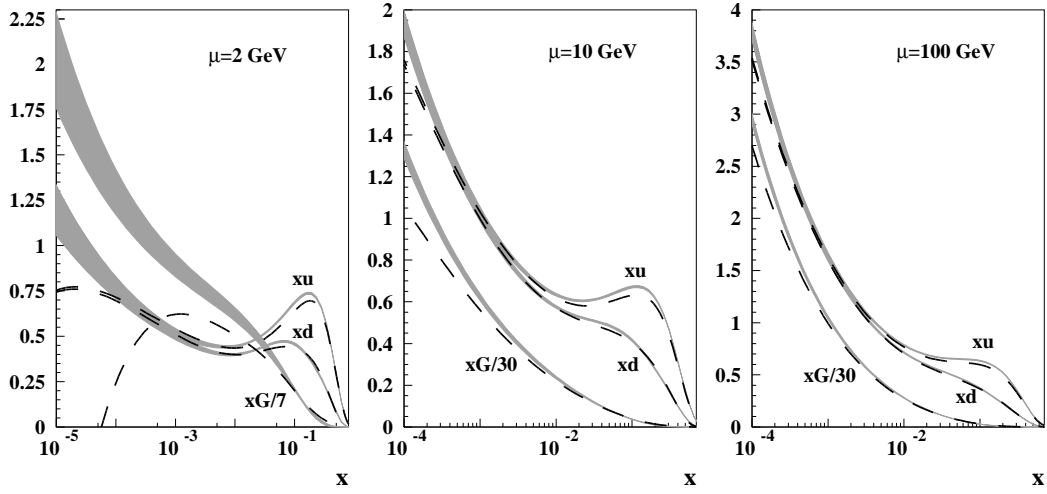,scale=0.37}
\end{minipage}
\begin{minipage}[t]{16.5 cm}
\caption{A comparison \cite{Alekhin:2009ni} of the ABKM09 PDFs with 
uncertainty bands to the MSTW08 central fit (dashed lines).    
\label{ABKMcomp}}
\end{minipage}
\end{center}
\end{figure}

\begin{figure}
\begin{center}
\begin{minipage}[t]{16.5 cm}
\epsfig{file=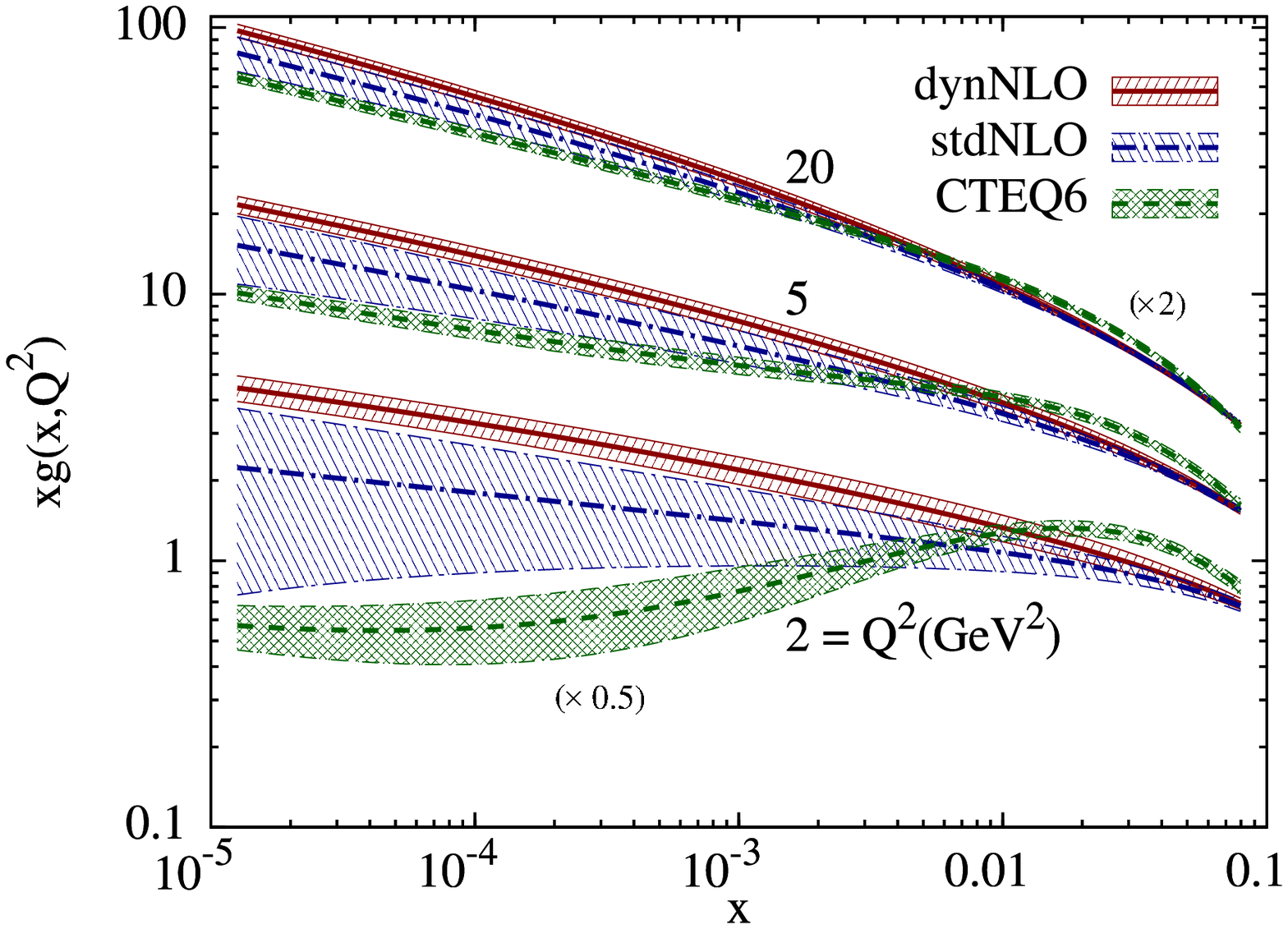,scale=0.41}
\epsfig{file=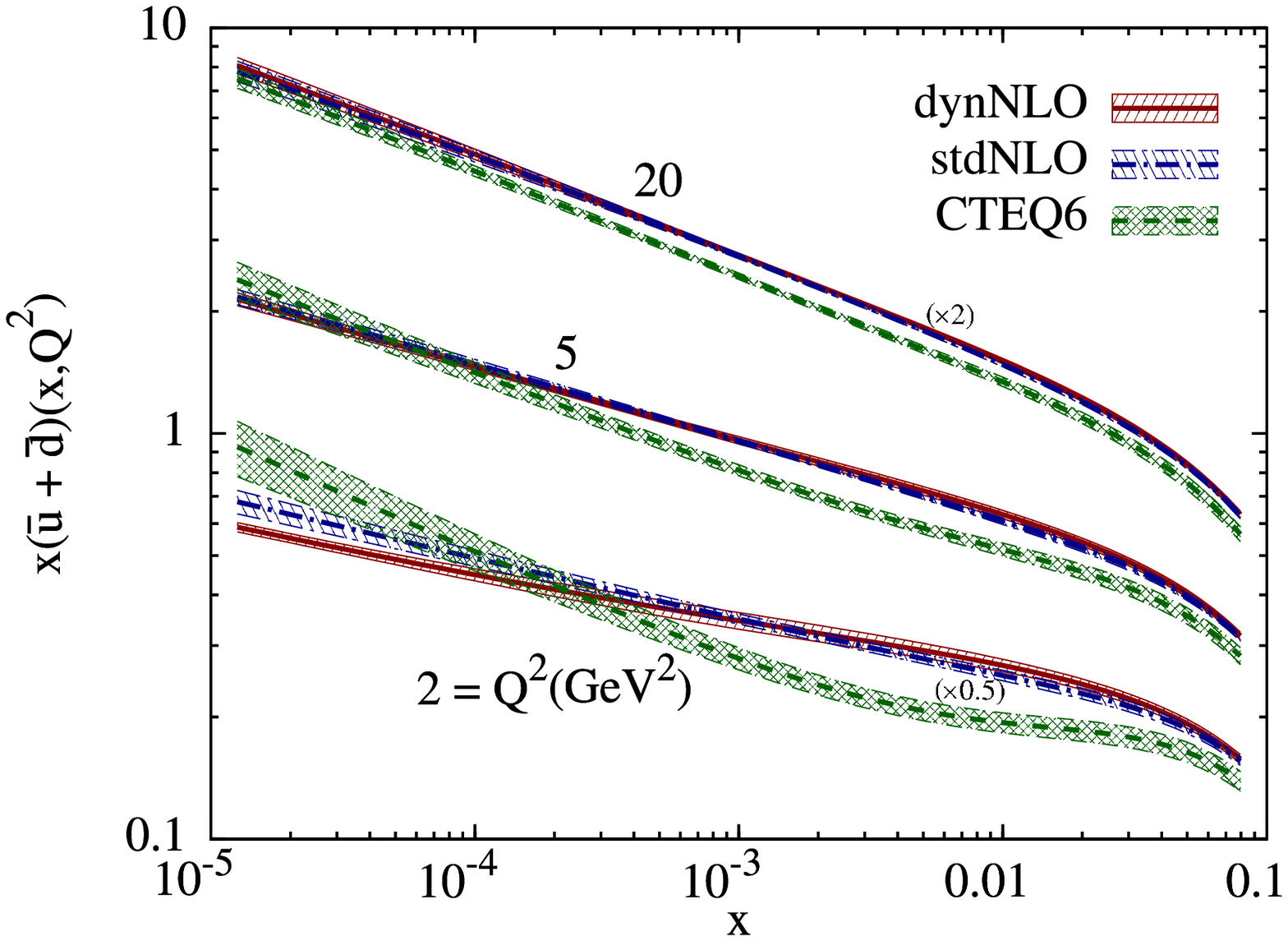,scale=0.41}
\end{minipage}
\begin{minipage}[t]{16.5 cm}
\caption{A comparison \cite{Gluck:2007ck} of the GJR08 PDFs, both the 
recommended ``dynamical set'' (dynNLO) and the more conventional ``standard 
set'' (stdNLO) to CTEQ6.    
\label{GJRcomp}}
\end{minipage}
\end{center}
\end{figure}

\begin{figure}
\begin{center}
\begin{minipage}[t]{16 cm}
\epsfig{file=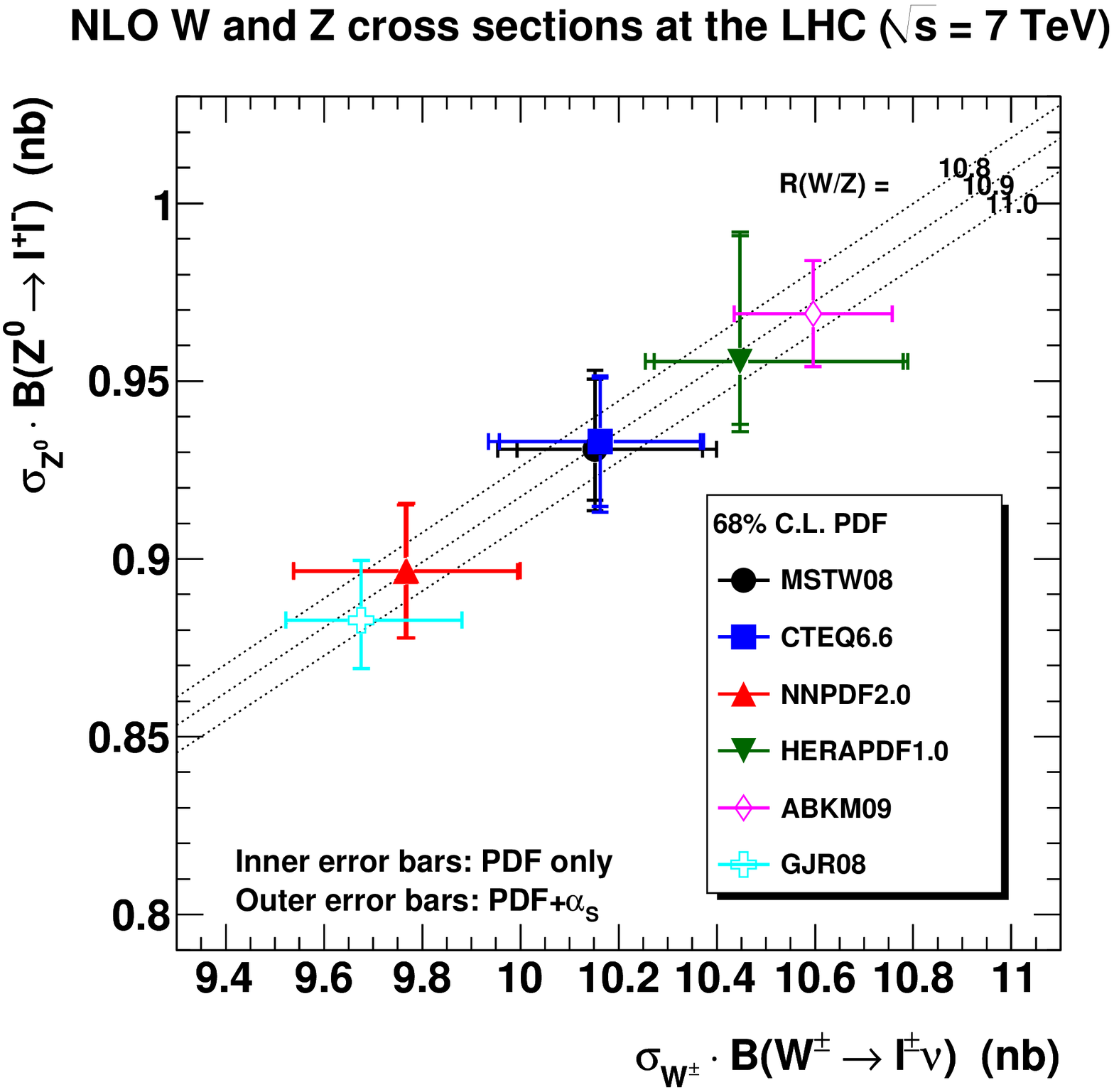,scale=0.4}
\epsfig{file=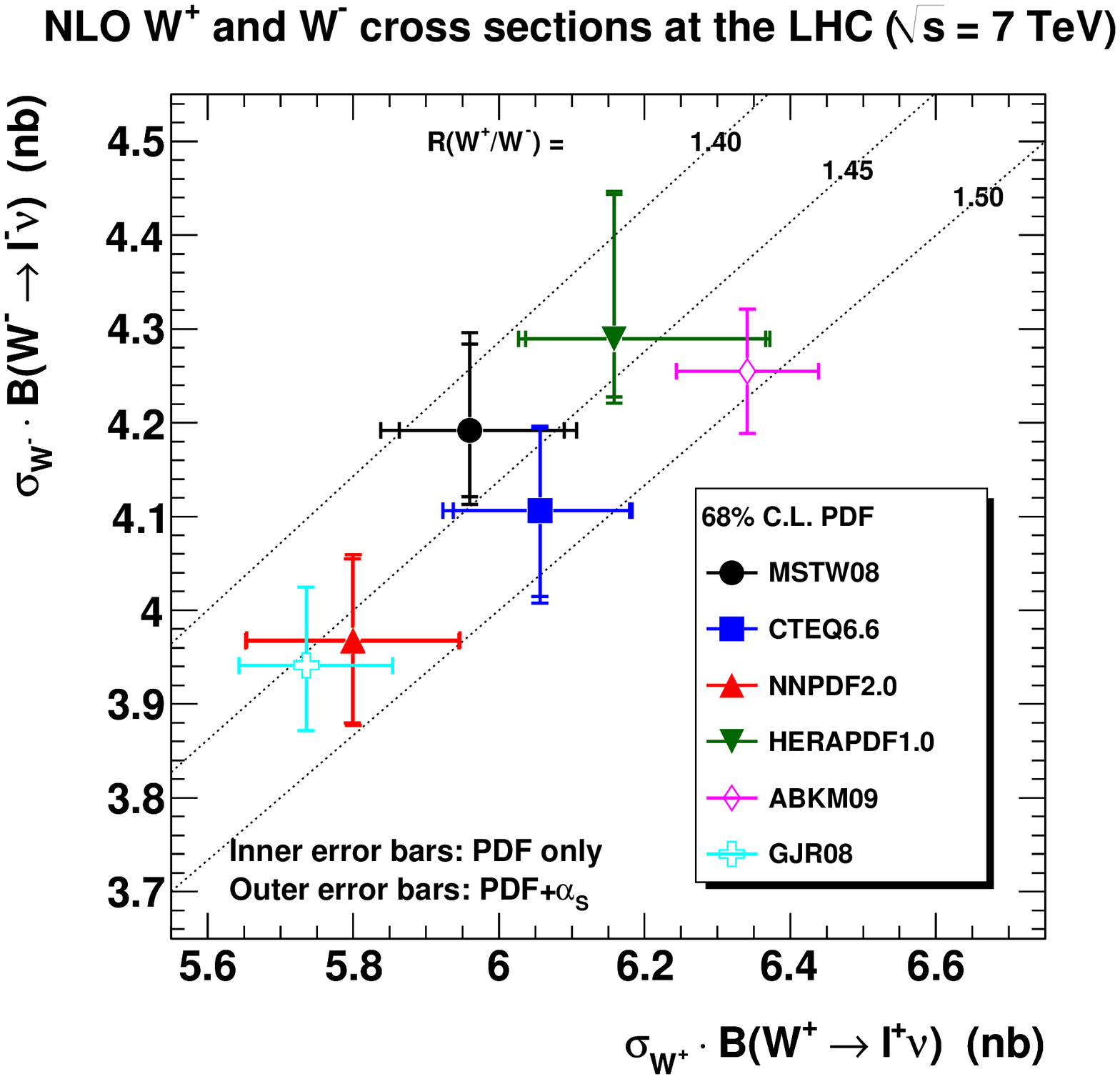,scale=0.4}
\end{minipage}
\begin{minipage}[t]{16.5 cm}
\caption{$W$ and $Z$ cross section predictions for the LHC (left)
and $W^+$ and $W^-$ cross section predictions for the LHC (right) using the PDFs from 
the different groups. Plots by G. Watt\cite{Wattplots}.   
\label{WZprediction}}
\end{minipage}
\end{center}
\end{figure}

The predictions at NLO for all the PDF sets for W and Z cross-sections
at the LHC at $7\TeV$ centre of mass energy, with 
common fixed order QCD and vector boson width effects, and common 
branching ratios are shown in the left of Fig. \ref{WZprediction},
(some similar results at NNLO can be found in \cite{Alekhin:2010dd}). 
There is fairly good agreement. However, there is a 3-4 $\sigma$ difference
between the extreme results. There is as much variation in the absolute
cross sections for $W^+$ and $W^-$ in the right of Fig. \ref{WZprediction},
but here there is also some significant variation in the ratio.  
These particular cross sections are 
primarily sensitive to the quark distributions in the region $x=0.01$, so
the inclusion or not of the combined HERA structure function data could have
some effect. However, the two PDFs which include these, HERAPDF1.0 and 
NNPDF2.0 are two of the furthest apart. Hence, there must be implicit 
differences in the PDFs obtained by groups, which for at least some can not be
fully reflected in the size of the uncertainties even with such measures as
inflated $\Delta \chi^2$ etc. One of the obvious examples is the theoretical 
constraint in the GJR ``dynamical PDFs''\cite{Gluck:2007ck}, which
have a hypothesis for the starting distributions 
not shared by other groups, but there are
other, less obvious examples.

\begin{figure}
\begin{center}
\begin{minipage}[t]{16.5 cm}
\epsfig{file=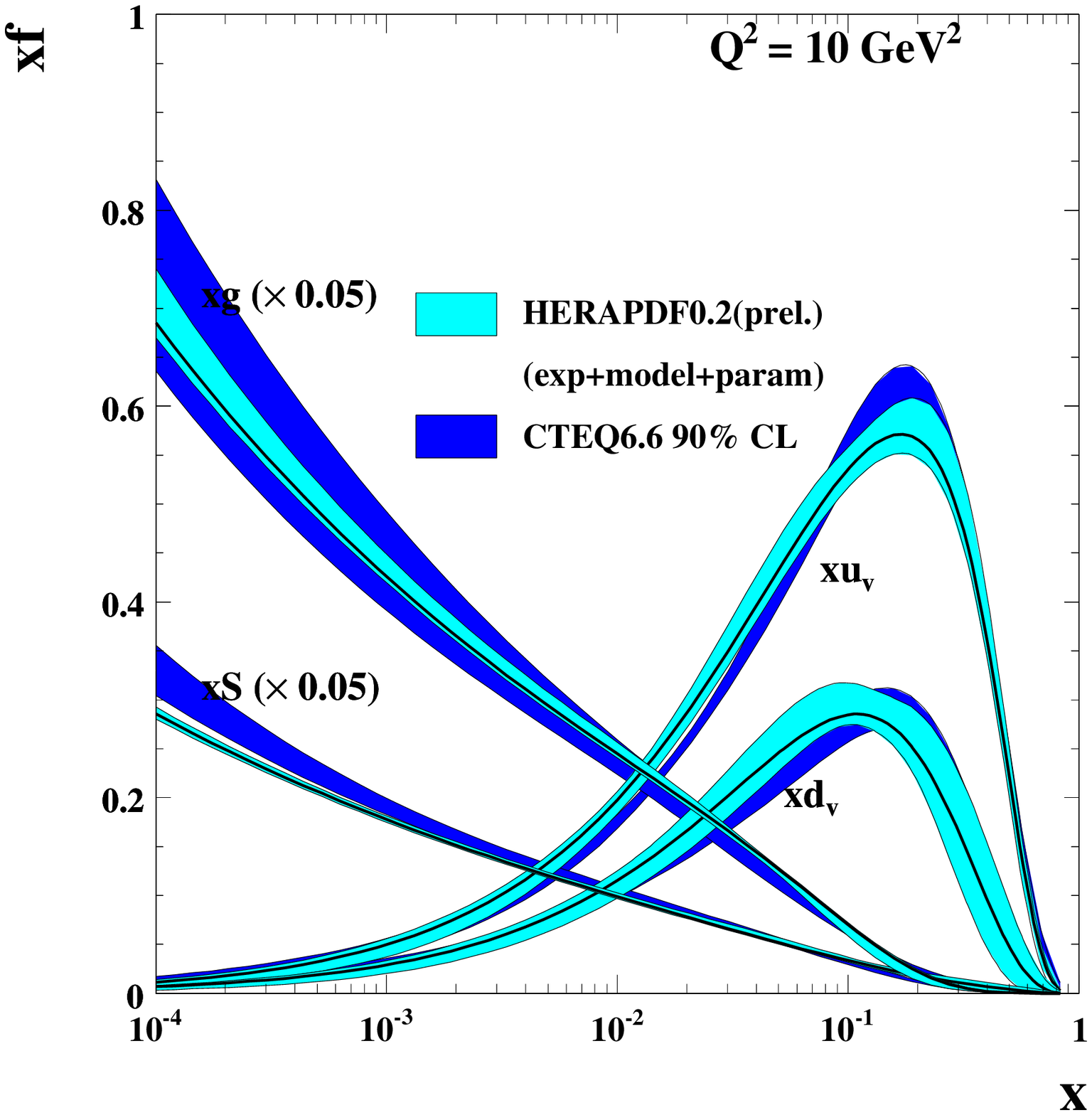, scale=0.4}
\epsfig{file=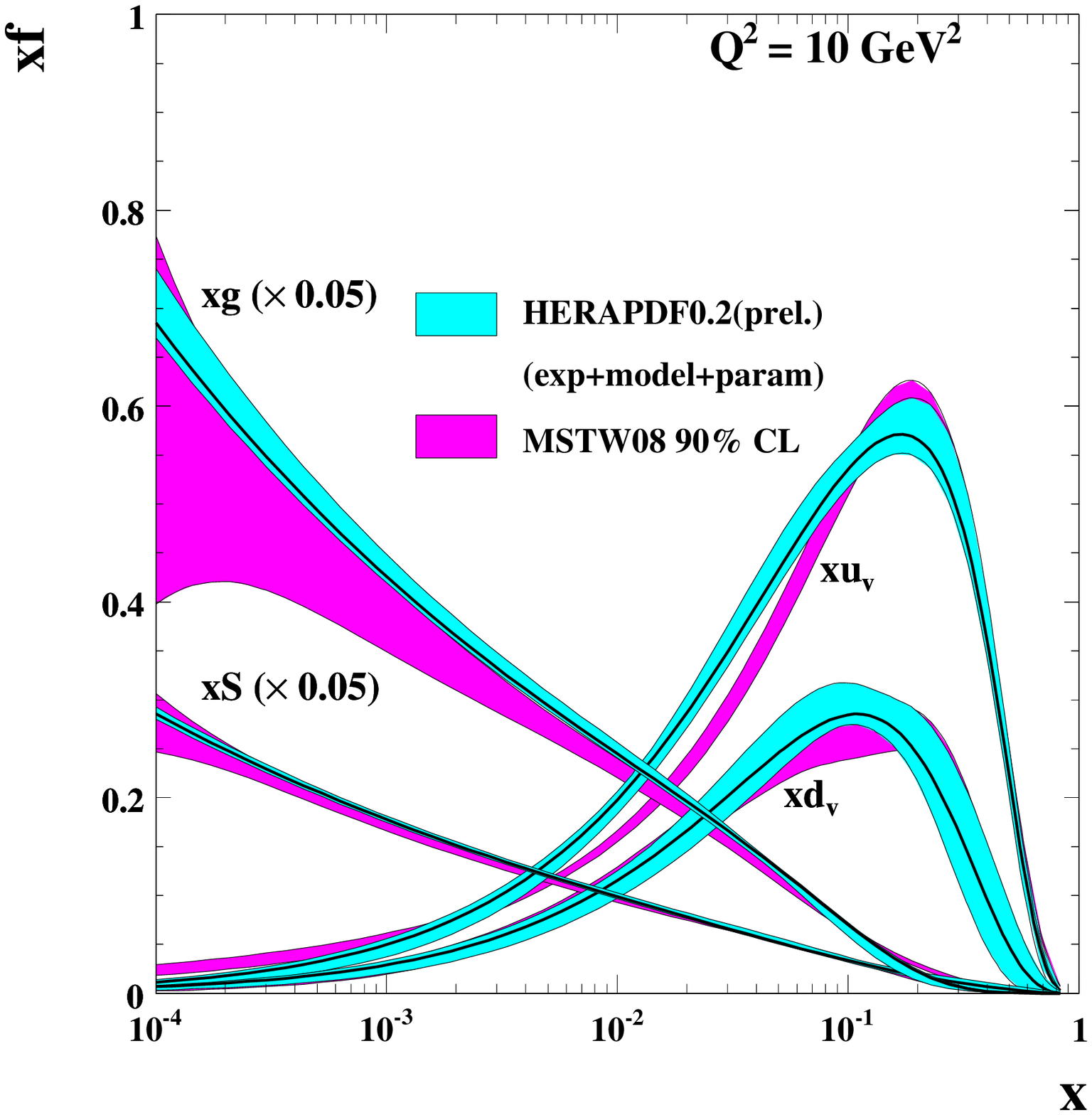, scale=0.4}
\end{minipage}
\begin{minipage}[t]{16.5 cm}
\caption{The HERA PDF compared to CTEQ (left) and MSTW (right)\cite{Raicevic:2010zz}.    
\label{HERAPDFcomp}}
\end{minipage}
\end{center}
\end{figure}

\subsection{\it Parameterisations}

One of the obvious sources of differences between the different PDF sets is
their parameterisations. Consider the example of the 
$\sigma(W^+)/\sigma(W^-)$ ratio already illustrated in the right
of \ref{WZprediction}. The 
MSTW08  prediction for this ratio has a very small quoted uncertainty of
$\approx 0.8\%$ for this\cite{Martin:2009iq}.
The prediction is sensitive to the $u$ and $d$ quarks --  
$\frac{\sigma(W^+)}{\sigma(W^-)}
\approx \frac{u(x)\bar d(x)} {d(x)\bar u(x)}
\approx \frac{u(x)} {d(x)},$ where in the last step we assume  
$\bar u(x) \to \bar d(x), x \to 0$, which data implies,
and most parameterisations assume. Hence, this ratio is sensitive to 
flavour in the proton, and on valence quarks at $x=0.01$, where
they are a small, but still significant contribution to the total 
quark distributions (this is more clear in the 
asymmetry of $\sigma(W^+)$ and $\sigma(W^-)$\cite{CooperSarkar:2007pj}).
The valence quarks of various groups can be seen, along with other
PDF comparisons, in Fig. \ref{HERAPDFcomp}, and there are 
appreciable differences at $x < 0.1$. It might be thought that since the 
small-$x$ valence quarks are only weakly constrained by data this 
variation in predictions is due to the valence quarks being overly 
constrained by a limited parameterisation. This was implied 
by the uncertainties in the earlier NNPDF sets, e.g.  \cite{DelDebbio:2007ee,
Ball:2008by}, but as we see in Fig. \ref{valenceuncertainty}, 
in the fully global 
fit, the NNPDF2.0 valence distribution, which has a much more flexible 
parameterisation, is no more uncertain that that of CTEQ and MSTW. This is
also clear from the uncertainties in the right of Fig. \ref{WZprediction}. Hence, 
PDF sets have differences in their valence quarks which are 
not entirely due to parameterisation inflexibility
(though this may still play a part, perhaps affecting even NNPDF2.0 to some 
small extent due to the preprocessing). It is true that they include 
different data sets, but the uncertainties should account for this.  
Differences are more likely to be due to 
generally unconsidered reasons such as implicit assumptions on 
nuclear corrections to neutrino DIS data, or deuterium corrections,
which it is difficult to account for. This not easily explained discrepancy
between predictions for $\sigma(W^+)/\sigma(W^-)$ makes early measurements of
this quantity at the LHC particularly interesting.  
This shows that although it is might be tempt to assign differences in predictions or 
PDFs which 
are larger than uncertainties to limited flexibility in parameterisations, 
there is not always much evidence that this is true. There are, 
however, a few explicit examples where the limitations on PDF 
parameterisations do affect the central values, but sometimes more 
particularly the size of the uncertainties of the PDFs

\begin{figure}
\begin{center}
\begin{minipage}[t]{9 cm}
\epsfig{file=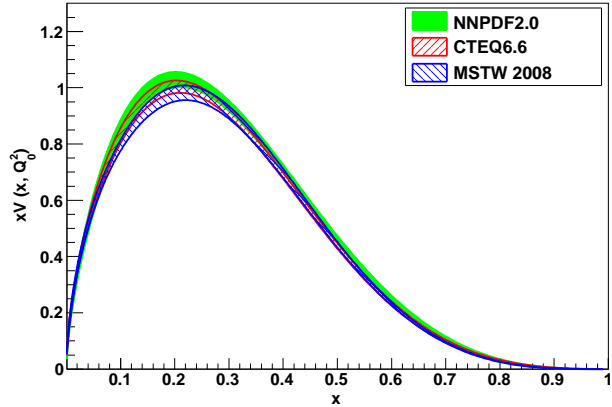,scale=0.45}
\end{minipage}
\begin{minipage}[t]{16.5 cm}
\caption{The valence quark combination from various groups 
and its uncertainty\cite{Ball:2010de}.      
\label{valenceuncertainty}}
\end{minipage}
\end{center}
\end{figure}

\begin{figure}
\begin{center}
\begin{minipage}[t]{16 cm}
\epsfig{file=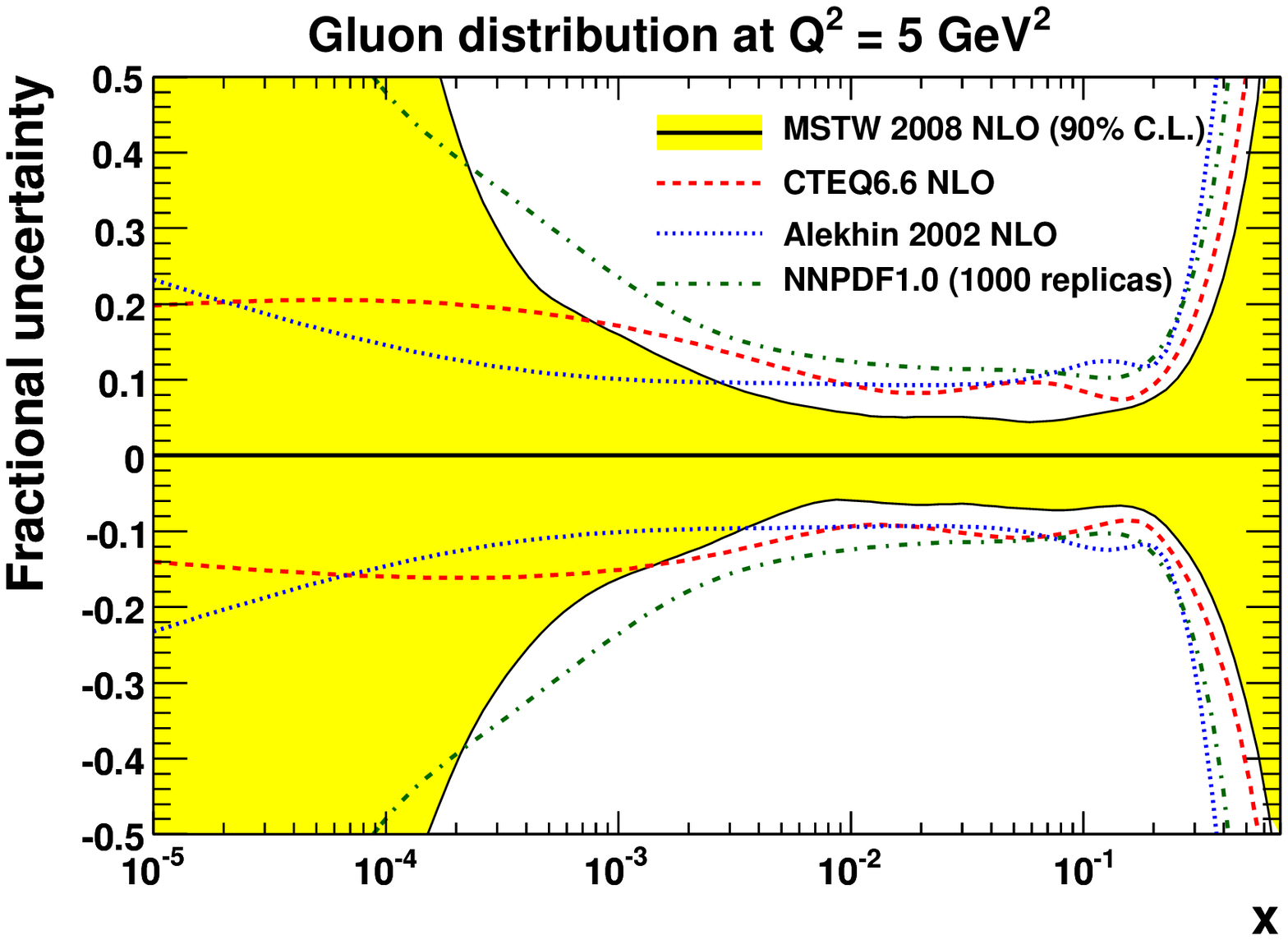,scale=0.5}
\epsfig{file=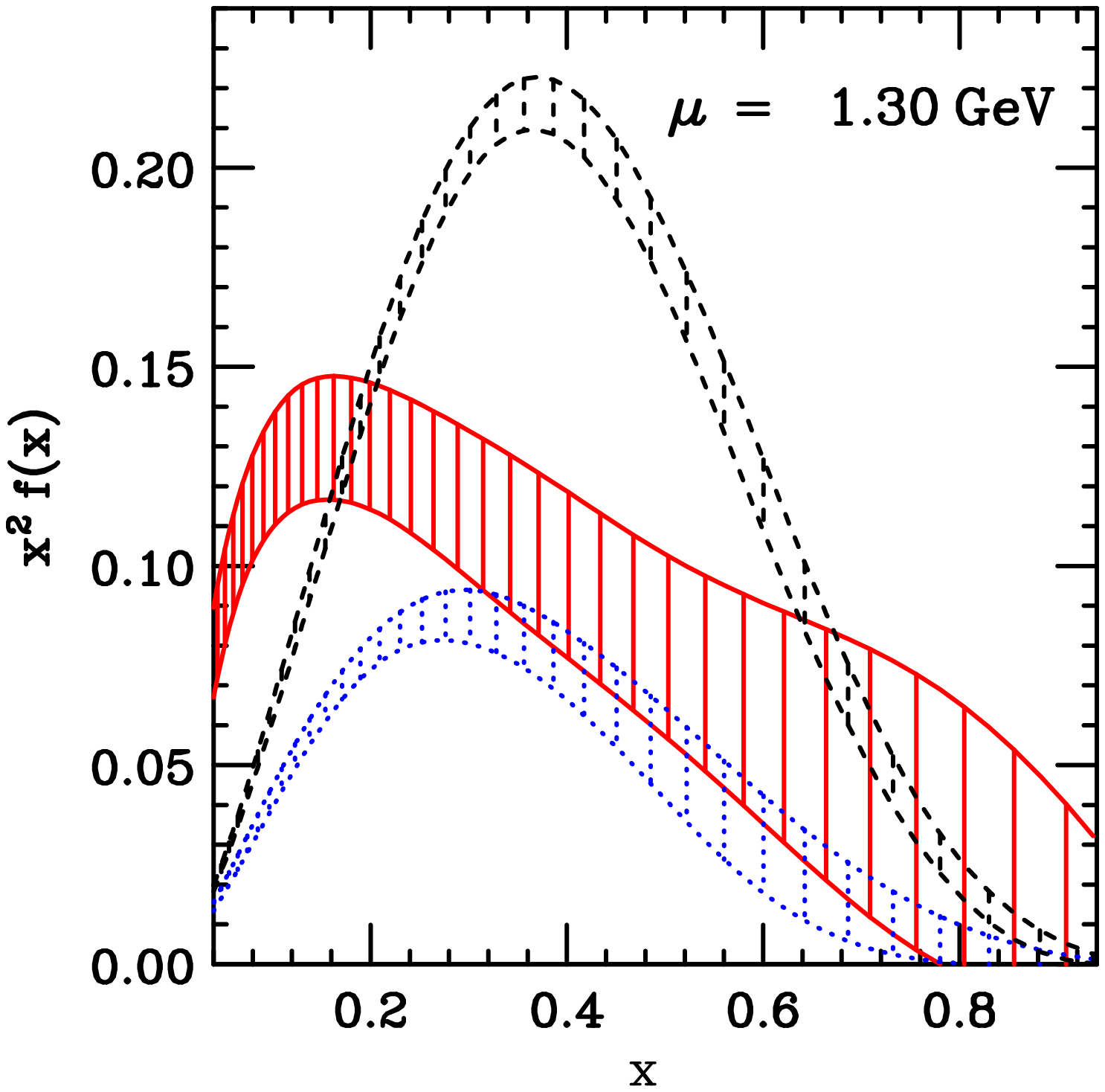,scale=0.38}
\end{minipage}
\begin{minipage}[t]{16.5 cm}
\caption{The gluon fractional uncertainty from a variety of PDF sets 
at small $x$\cite{Martin:2009iq} (left). The gluon and its uncertainty 
at large $x$ (solid) compared the the 
up quark (dashed) and down quark (dotted)\cite{Pumplin:2009nk}(right).   
\label{fracgluonsx}}
\end{minipage}
\end{center}
\end{figure}

One clear example of this is the gluon parameterisation at small $x$. 
In this case different parameterisations can lead to different central value
and also a very different 
uncertainty for the small $x$ gluon distribution. 
This can be seen in Fig. \ref{HERAPDFcomp} where the MSTW gluon distribution at 
low $x$ is a little smaller but where the  
gluon uncertainties have
a very different shape (though the HERAPDF1.0 gluon has a clearer smaller
uncertainty in magnitude due to inclusion of combined data and 
more stringent tolerance). 
Most parameterisations assume the gluon behaves like a  single power 
$x^{\lambda}$ at input. If $g(x) \propto x^{\lambda \pm \Delta \lambda}$ then 
$\Delta g(x) = \Delta \lambda \ln(1/x)  g(x)$. So this form of 
parameterisation by definition leads to a
limited fractional uncertainty, growing fairly slowly as $x$ becomes very 
small. This is represented by the ``Alekhin'' curve in Fig. \ref{fracgluonsx}.
The HERAPDF1.0 and ABKM gluon uncertainties would have similar shape.
If the input for the gluon at low $Q^2$ actually has $\lambda$ positive then 
the small-$x$ input gluon is effectively fine-tuned to be $\sim 0$. 
In this case the very small-$x$ gluon at higher scales is entirely generated 
by evolution from the more precisely determined gluon at higher $x$ values 
and the uncertainty is even smaller, as in the ``CTEQ'' curve in  the left of 
Fig. \ref{fracgluonsx}. The GJR uncertainty would be a similar form.
The MSTW and NNPDF parameterisations are more flexible (the gluon can be 
negative) and this leads to a smaller distribution at the lowest $x$ and 
a rapid expansion of the 
uncertainty where the data constraint runs out. Indeed, the two powers in 
the MSTW parameterisation allow more flexibility at the smallest $x$ than 
NNPDF, perhaps because of the pre-processing power for the latter. 

%\begin{figure}
%\begin{center}
%\begin{minipage}[t]{8 cm}
%\epsfig{file=CTEQhighxgud.ps,scale=0.45}
%\end{minipage}
%\begin{minipage}[t]{16.5 cm}
%\caption{The gluon and its uncertainty at large $x$ (solid) compared the the 
%up quark (dashed) and down quark (dotted)\cite{Pumplin:2009nk}.      
%\label{gluonlx}}
%\end{minipage}
%\end{center}
%\end{figure}

There are also parameterisation variations in the high-$x$ gluon. 
Generally high-$x$ PDFs are parameterised so they will behave like
$(1-x)^{\eta}$ as $x \to 1$. Even though the parameterisation does contain 
a term of this form there is more flexibility in the CTEQ parameterisation
(again seemingly even more than for NNPDF).
This allows a very hard high-$x$ gluon distribution, as in 
the right of Fig. \ref{fracgluonsx},  which is still
consistent with published Tevatron jet data\cite{Pumplin:2009nk}. 
However, one might ask whether the gluon, which is usually though of
as being radiated from quarks, should be allowed to be harder than the up 
valence distribution for $x \to 1$. This excess of the gluons does actually
disappear by $Q^2=100\GeV^2$\cite{Pumplin:2009nk} due to fast 
radiation of very high-$x$ gluons to smaller-$x$ gluons.

\begin{figure}
\begin{center}
\begin{minipage}[t]{10 cm}
\epsfig{file=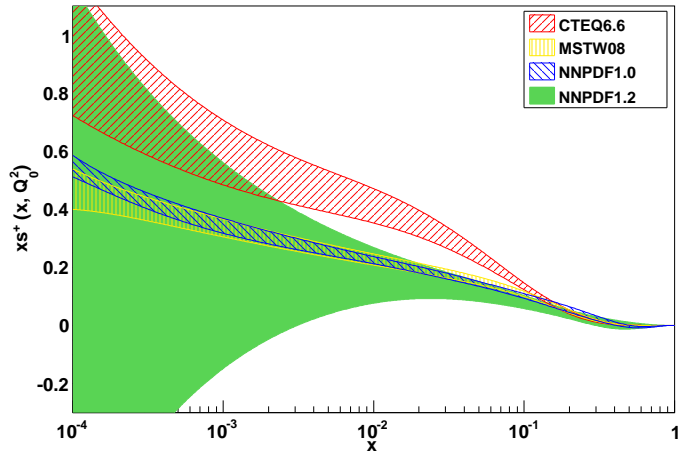,scale=0.5}
\end{minipage}
\begin{minipage}[t]{16.5 cm}
\caption{The strange quark and its uncertainty\cite{Ball:2010de} from various PDF sets.     
\label{strangeuncertainty}}
\end{minipage}
\end{center}
\end{figure}

The other parton distribution  with significant dependence on
choice of parameterisations is the 
strange quark distribution. 
In fact the direct fit to $s, \bar s$ from dimuon data 
has tended to lead to a significant 
uncertainty increase compared to previous assumptions of a 
fixed fraction of strange in the well-constrained total light sea. 
This direct constraint is only for for $x \geq 0.01$ and there are a wide
variety of assumptions about what happens below this.  
In the MSTW08 PDFs it is assumed that the shape of the strange 
distribution can be largely inferred from the theory assumption that 
the suppression of this distribution is of the same form as 
the distributions of the massive quarks charm and bottom. This implies that
below $x=0.01$ the strange distribution is a fixed fraction of the total sea. 
As seen in Fig. \ref{strangeuncertainty} this results in a shape which is 
significantly different to that in CTEQ6.6 despite a fit to the same data. In 
the CTEQ6.6 distributions the assumption is weaker, i.e. only that 
there is the same small-$x$ power for strange as light quarks.
However, there is even a significant difference in the region of the 
data, which must be due to the effect of nuclear corrections 
and/or the heavy quark treatment.
NNPDF2.0, which also includes dimuon data, 
impose no theoretical constraint (other than positivity of the dimuon 
cross section) on the strange quark distribution at small $x$. 
This results in a very large uncertainty which then  impacts on 
the other small-$x$ light quarks since it is only the charge weighted 
sum which is constrained by HERA structure function data at small $x$. 
Due to their simple choice of parameterisation of the strange distribution 
and the fact that at small $x$ it is all generated by evolution the strange 
distribution in the HERAPDF1.0 and GJR08 PDF sets will have an uncertainty at 
small $x$ similar to that of MSTW08. Due to a lack of any theory constraint 
that of ABKM09 is similar to NNPDF2.0, though since the small-$x$ 
behaviour is a single power the variation is not quite as large.

\subsection{\it Heavy Quarks}

The treatment of heavy quarks is something that nearly every group does
slightly, or sometimes significantly differently, and it can lead to 
perhaps surprisingly different results for the parton distributions extracted. 
In treating heavy quarks in parton scattering there are two distinct regimes:
Near threshold for the quark production, i.e. $Q^2\sim m_h^2$ 
massive quarks are not treated as not partons. 
They are entirely created in the final state and are described using 
the so-called Fixed Flavour Number Scheme
(FFNS), e.g. for structure functions
\be
F^h(x,Q^2)=C^{FF}_k(Q^2/m_h^2)\otimes f^{n_f}_k(Q^2),
\label{FFNS}
\ee
where $f^{n_k}$ represents the light partons only. 
This is exact, but at each perturbative order there are $\ln^n (Q^2/m_h^2)$ 
terms which are not resummed. There is argument about the importance 
of these, but it is unlikely that resummation is universally unimportant. 
For structure functions the coefficient functions have been calculated to 
NLO\cite{Laenen:1992zk,Harris:1995tu}, and there is some progress at 
NNLO\cite{Bierenbaum:2009mv}, but the coefficient functions are 
not calculated yet for many processes beyond LO. 
Alternatively, at very high scales $Q^2 \gg m_h^2$ heavy quarks 
can be assumed to behave like the massless quarks. In this case
we have heavy quark parton distributions and sum the $\ln(Q^2/m_h^2)$ 
terms via evolution. The simplest form of this is known as 
the Zero Mass Variable Flavour Number Scheme (ZM-VFNS), though mass 
dependence does come into the boundary conditions for evolution 
(calculated up to ${\cal O}(\alpha_S^2)$ in \cite{Buza:1996wv} and to 
${\cal O}(\alpha_S^3)$ in \cite{Bierenbaum:2009mv}). 
This scheme is the normal assumption in calculations at high scales. 
It is not exact since it ignores ${\cal O}(m_h^2/Q^2)$ corrections, e.g.
for structure functions 
\be
F(x,Q^2) = C^{ZMVF}_j\otimes f^{n_f+1}_j(Q^2).
\label{ZMVFNS}
\ee
This approximation does not matter if the scale of physics is 
$\gg m_h^2$. However, in fitting structure function data in global fits one
goes from the region $Q^2\sim m_h^2$ to  $Q^2 \gg m_h^2$ via the less
clear region in between. Hence, for maximum precision one needs 
a General Mass Variable Flavour Number Scheme 
(GM-VFNS) interpolating between the two well-defined limits.

\begin{figure}
\begin{center}
\begin{minipage}[t]{16.5 cm}
\epsfig{file=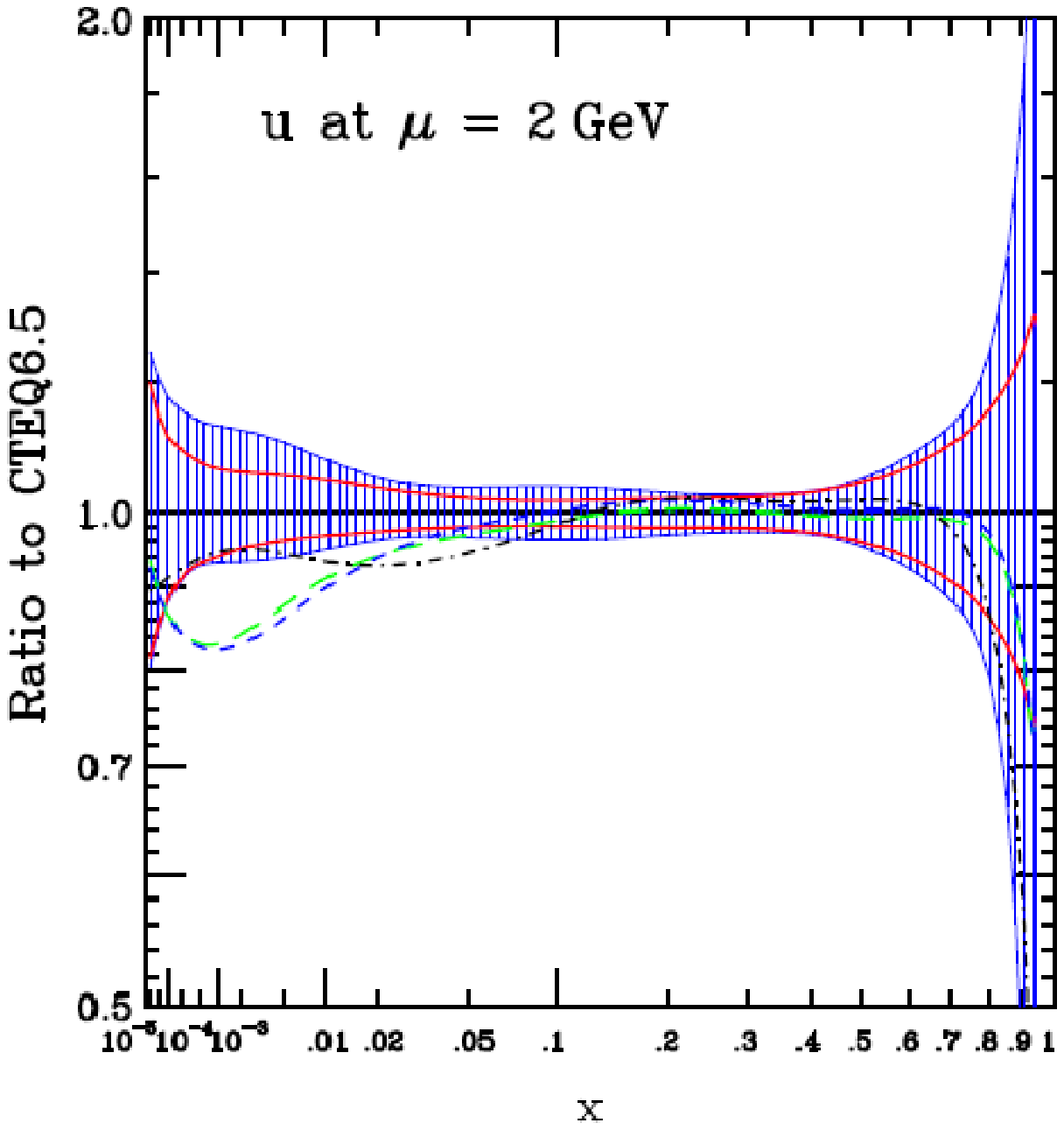,scale=0.45}
\epsfig{file=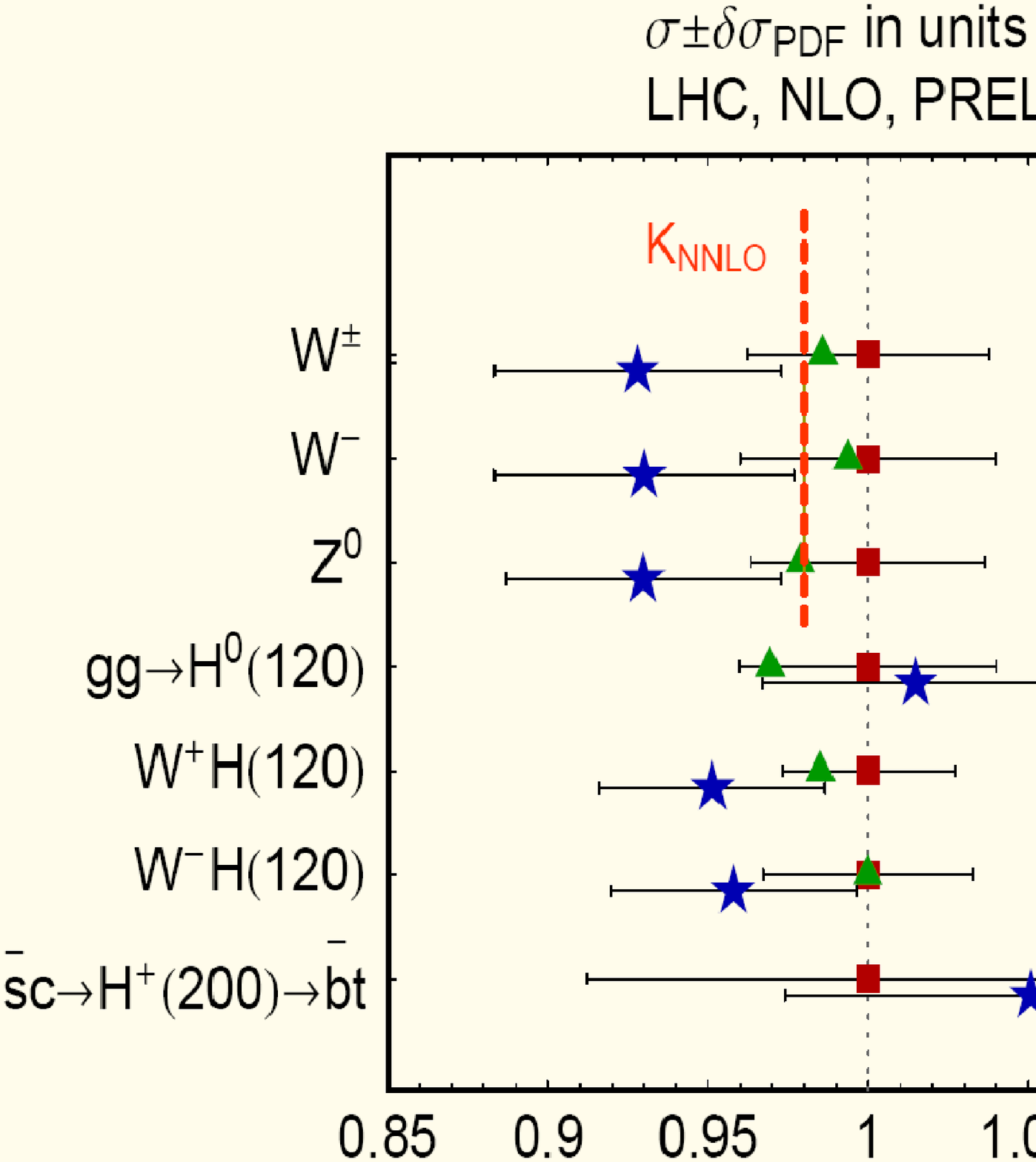,scale=0.25}
\end{minipage}
\begin{minipage}[t]{16.5 cm}
\caption{The comparison of the CTEQ up quark in a fit using the ZM-VFNS 
(dashed) and a GM-VFNS (left) (MRST is dashed-dotted) and a comparison 
of predictions at the LHC from the fit 
using the ZM-VFNS (labelled CTEQ6.1) and and a GM-VFNS (labelled CTEQ 6.5) along
with the prediction when a large intrinsic charm distribution is included (labelled
IC-sea) (right)\cite{Tung:2006tb}. 
\label{ZMVFNScomp}}
\end{minipage}
\end{center}
\end{figure}

There are various definitions possible\cite{Aivazis:1993pi,Buza:1996wv,Thorne:1997ga,Chuvakin:1999nx,Kramer:2000hn,Tung:2001mv,Thorne:2006qt,Alekhin:2009ni,Forte:2010ta}, 
and there is a review in 
\cite{Thorne:2008xf} and a numerical comparison of alternatives in 
\cite{Binoth:2010ra}. (A theoretical underpinning is provided in
\cite{Collins:1998rz}.) The versions used by MSTW
(TR/TR')\cite{Thorne:1997ga,Thorne:2006qt} and CTEQ (ACOT)\cite{Aivazis:1993pi,Tung:2001mv} 
have converged somewhat in recent years. Initially
the ACOT prescriptions did not incorporate the correct kinematics, term by
term in the expansions (though violations were limited
by cancellations between terms). This was rectified, but in a complicated 
fashion, in \cite{Thorne:1997ga}. The simplest choice in  
the heavy flavour coefficient functions is now commonly 
based on the ACOT($\chi$) prescription\cite{Tung:2001mv}, i.e. 
the scaling variable $x$ is replaced by $\chi \equiv x(1+4m_h^2/Q^2)$,
which automatically incorporates the correct kinematic limit.
However, there are still choices of $m_h^2/Q^2$-dependent factors, 
ordering of the perturbation series, and even subtle changes in scaling variable
(see e.g. \cite{Nadolsky:2009ge}).
Various significant differences still exist as illustrated
by comparison to the recent H1 data on heavy flavour production in Fig. 
\ref{HERAheavy}.

The importance of using a GM-VFNS instead of massless
approach was illustrated by CTEQ\cite{Tung:2006tb} (it had been assumed by 
MRST/MSTW since \cite{Martin:1998sq} that the GM-VFNS was preferable, but 
a detailed study of the difference not presented). This can be seen in 
the left of Fig. \ref{ZMVFNScomp} where the 
up quark with uncertainties compared with previous versions, e.g. 
CTEQ6.1 (dashed). Clearly the use of the ZM-VFNS
can lead to a considerable error in the PDFs. This consequently  
can lead to a large change in predictions using CTEQ partons at
LHC of $5-10 \%$ as seen in the right of Fig. \ref{ZMVFNScomp}, where
we also note the possible effects of {\it intrinsic charm}.

\begin{figure}
\begin{center}
\begin{minipage}[t]{13 cm}
\epsfig{file=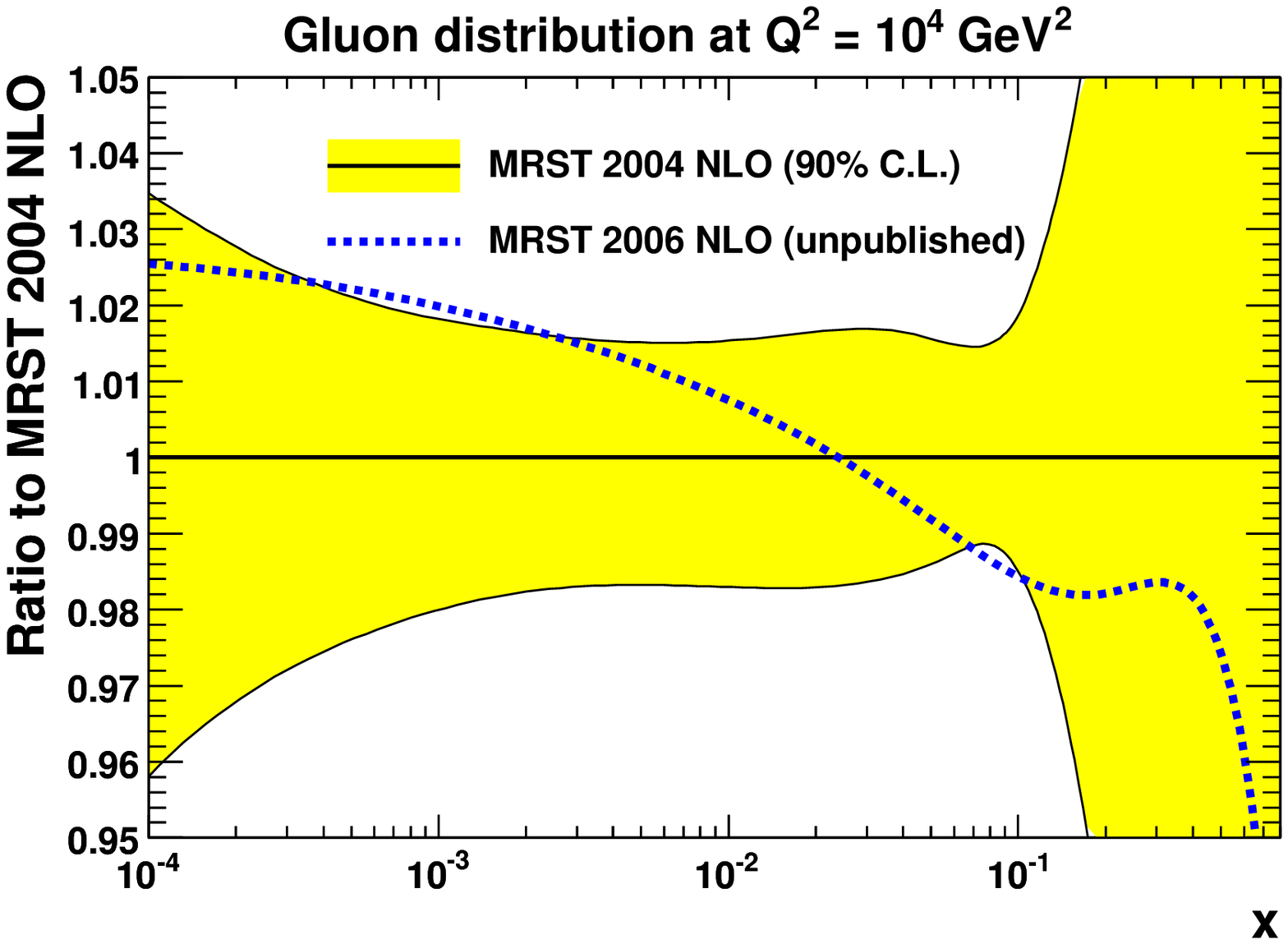,scale=0.32}
\epsfig{file=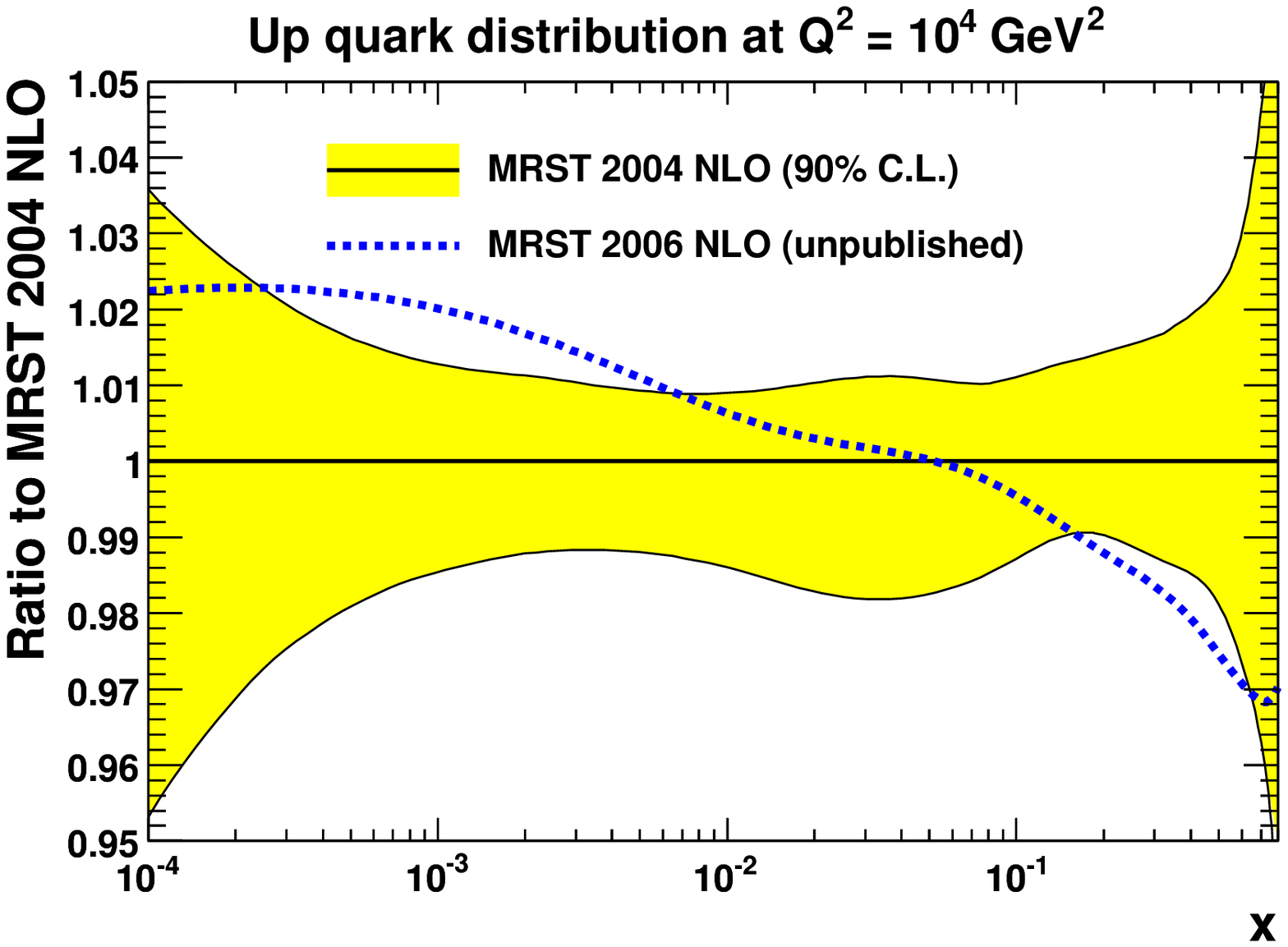,scale=0.32}
\end{minipage}
\begin{minipage}[t]{16.5 cm}
\caption{A comparison of the gluon (left) 
and up quark (right) in two versions of a GM-VFNS\cite{Martin:2007bv}, where the 
band represents the uncertainty on the MRST2004 PDF due to experimental errors. 
\label{GMVFNScomp}}
\end{minipage}
\end{center}
\end{figure}

Although this large change in improving from the ZM-VFNS to a GM-VFNS
can be viewed as a correction due to the missing physics in the GM-VFNS 
the freedom in defining a GM-VFNS at finite perturbative order, means there
is still an associated theoretical uncertainty. This was studied briefly in 
\cite{Martin:2007bv} where the differences in PDFs obtained using the NLO
prescriptions in \cite{Thorne:1997ga} and \cite{Thorne:2006qt}, but much the 
same data sets, was investigated. The change of scheme could lead to  
changes of up to $2\%$ in PDFs, as seen in Fig. \ref{GMVFNScomp}, and this can 
lead to up to a $3\%$ change in $\sigma_W$ and $\sigma_Z$
at the LHC. This is a genuine 
theory uncertainty due to competing but equally valid choices of a definition
of a GM-VFNS, and is analogous to the freedom in making a choice of 
factorisation and renormalisation scale. The variation PDFs obtained from fits 
using different GM-VFNS choices has recently been investigated in more detail 
in \cite{Thorne:2010pa}, and $2-3\%$ seems a reasonable estaimate at NLO
(though changes from ZM-VFNS to GM-VFNS were found to be typically $\sim
5-6\%$ in this case, slightly smaller than for CTEQ). Moreover, at NNLO the
variation was reduced to $\sim 1\%$, so the expected reduction in ambiguity
at higher orders in perturbation theory is verified.      

There could also be some nonperturbative (intrinsic) heavy flavour,
as well as that generated by perturbative evolution. This is 
suppressed by $\Lambda_{QCD}^2/Q^2$ or possibly
$\Lambda_{QCD}^2/W^2 \sim \Lambda_{QCD}^2/(Q^2(1-x))$, and hence 
likely to be enhanced at high $x$\cite{Brodsky:1980pb}. 
CTEQ have investigated the possibility\cite{Pumplin:2007wg} 
by constraining the intrinsic charm in a normal global fit (and 
considering an effect which can be large at all $x$). This suggests a 
maximum $1\%$ integrated momentum density contribution of intrinsic
charm (and the same for anticharm). The possible effects of this are shown in 
the right of Fig. \ref{ZMVFNScomp}. However, MSTW\cite{Martin:2009iq}
have checked against against old EMC 
data\cite{Aubert:1982tt}, finding at most 
$(1/10)th$ this value. 

%\begin{figure}
%\begin{center}
%\begin{minipage}[t]{10 cm}
%\epsfig{file=charmemc.ps,scale=0.5}
%\end{minipage}
%\begin{minipage}[t]{16.5 cm}
%\caption{Comparison of fits to EMC data with and without intrinsic charm.  
%\label{intrinsiccharm}}
%\end{minipage}
%\end{center}
%\end{figure}

To summarise, the different fitting groups have 
different ways of dealing with heavy flavours. 

\begin{itemize}

\item MSTW08 use the definition of a GM-VFNS in \cite{Thorne:2006qt} at 
LO, NLO and NNLO, and precise details are described in \cite{Martin:2009iq}. 
The group have used a GM-VFNS for all partons 
since MRST98\cite{Martin:1998sq}, but the details changed
in 2006. Before 2006 the NNLO GM-VFNS prescription was approximate, i.e. 
the first NNLO distributions correct in this sense are in 
\cite{Martin:2007bv}, and the correction led to a few percent change
compared to \cite{Martin:2004ir}. Even now the NNLO GM-VFNS requires
some modelling at low $Q^2$ due to the absence of the full ${\cal O}(\alpha_S^3)$
FFNS coefficient functions (though some GM-VFNS definitions would not 
require these at NNLO). The information on the small-$x$\cite{Catani:1990eg}
and threshold limits\cite{Laenen:1998kp} at NNLO
are used. Since the massless splitting and coefficient functions are known at 
NNLO the GM-VFNS becomes exact at this order well above $Q^2=m_h^2$. 
PDFs are also made available for 3 and 4 flavours\cite{Martin:2006qz,Martin:2010db}. 

\item CTEQ6.6 (and CT10) use the definition of a GM-VFNS in \cite{Tung:2006tb} at NLO 
as default. The GM-VFNS version was only used as a special case 
(e.g. \cite{Kretzer:2003it}) in 
the pre-CTEQ6.5 sets, where the ZM-VFNS was always used as default.  

\item NNPDF2.0 uses the ZM-VFNS. The group has versions of 
of a GM-VFNS\cite{Forte:2010ta} at NLO and one at NNLO bench-marked\cite{Binoth:2010ra} along with MSTW and CTEQ, and there is a very new set using these \cite{Ball:2011mu}.

\item HERAPDF1.0 uses the same GM-VFNS as MSTW, i.e. that in 
\cite{Thorne:2006qt}. Previous fits have used the older TR
prescription\cite{Thorne:1997ga}, but usually compared to ZM-VFNS and
FFNS.     

\item ABKM09 perform their fit using FFNS. They compare to the GM-VFNS
defined in \cite{Alekhin:2009ni} and claim insensitivity to 
using GM-VFNS. However, in this definition, although charm and bottom 
quark distributions are defined this is only to fixed order, i.e. unlike other
variable-scheme deviations resummation of the $\ln(Q^2/m_h^2)$ terms in the
parton distributions is not performed in the fit comparisons (it is ultimately
when generating 4- and 5-flavour PDF sets from the inputs obtained from 
the FFNS fit). While the PDFs are obtained at NLO and at NNLO, the 
heavy quark treatment is identical at both orders (see \cite{Alekhin:2010iu}
for developments). 

\item GJR08 use the FFNS exclusively, and as ABKM using the same definition, i.e
up to ${\cal O}(\alpha_S^2)$, for the heavy flavour coefficient functions for
both NLO and NNLO. PDFs are converted into variable flavour scheme evolution 
at NLO\cite{Gluck:2008gs} and NNLO\cite{JimenezDelgado:2009tv}.

\end{itemize}

The different groups also use values of the charm quark mass varying from
$1.3\GeV \geq m_c \geq 1.65 \GeV$, and the bottom quark mass varying from 
$4.3 \GeV \geq m_b \geq 5\GeV$. In \cite{:2009wt} and \cite{Alekhin:2009ni}
variation is allowed, and varying the value of $m_c$ by $0.2\GeV$ can change 
PDFs and predictions by up to a couple of percent. MSTW have recently 
completed a detailed study of mass dependence in the PDFs
and predictions, agreeing with the HERA fit results\cite{Martin:2010db}, and also make PDFs available for 3 and 4 flavours for 
these different masses. There are also NNPDF results 
\cite{Ball:2011mu} on mass dependence.

\subsection{\it PDFs and the Strong Coupling}

\begin{figure}
\begin{center}
\begin{minipage}[t]{7 cm}
\epsfig{file=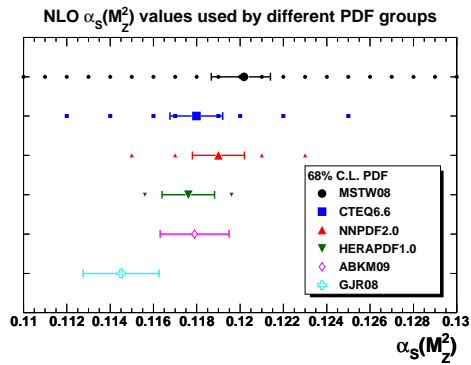,scale=0.36}
\end{minipage}
\begin{minipage}[t]{16.5 cm}
\caption{The values of $\alpha_S(m_Z^2)$ for PDFs of different groups. 
The error bar represents the central value and uncertainty for each set, 
and the points the values of $\alpha_S(m_Z^2)$ at which extra sets
are made available. Plots by G. Watt 
\cite{Wattplots}.    
\label{alphas}}
\end{minipage}
\end{center}
\end{figure}

Each group deals with the strong coupling in a slightly different manner. 
For MSTW08, ABKM09 and GJR08 the $\alpha_S(m_Z^2)$ values and uncertainty are 
determined by the fit both at NLO and at NNLO 
(in each case the NNLO value is about $0.002-0.004$ lower than the NLO value -- see
later). However, the values are rather different, i.e. $\alpha_S(m_Z^2) = 0.1202, 
0.1179$  and $0.1145$ respectively at NLO ($0.1171$, $0.1135$ and $0.112$ at NNLO).      
The other groups pick standard values and uncertainties, i.e. $0.118$ for CTEQ6.6,
0.119 for NNPDF2.0 and 0.1176 for HERAPDF1.0. In addition some groups provide additional
sets at a variety of $\alpha_S$ values\cite{Martin:2009bu,Ball:2010de,Lai:2010nw}.
The respective NLO values of $\alpha_S(m_Z^2)$, the uncertainties, and other values
available are summarised in Fig. \ref{alphas}\cite{Wattplots}.

\begin{figure}
\begin{center}
\begin{minipage}[t]{16.5 cm}
\epsfig{file=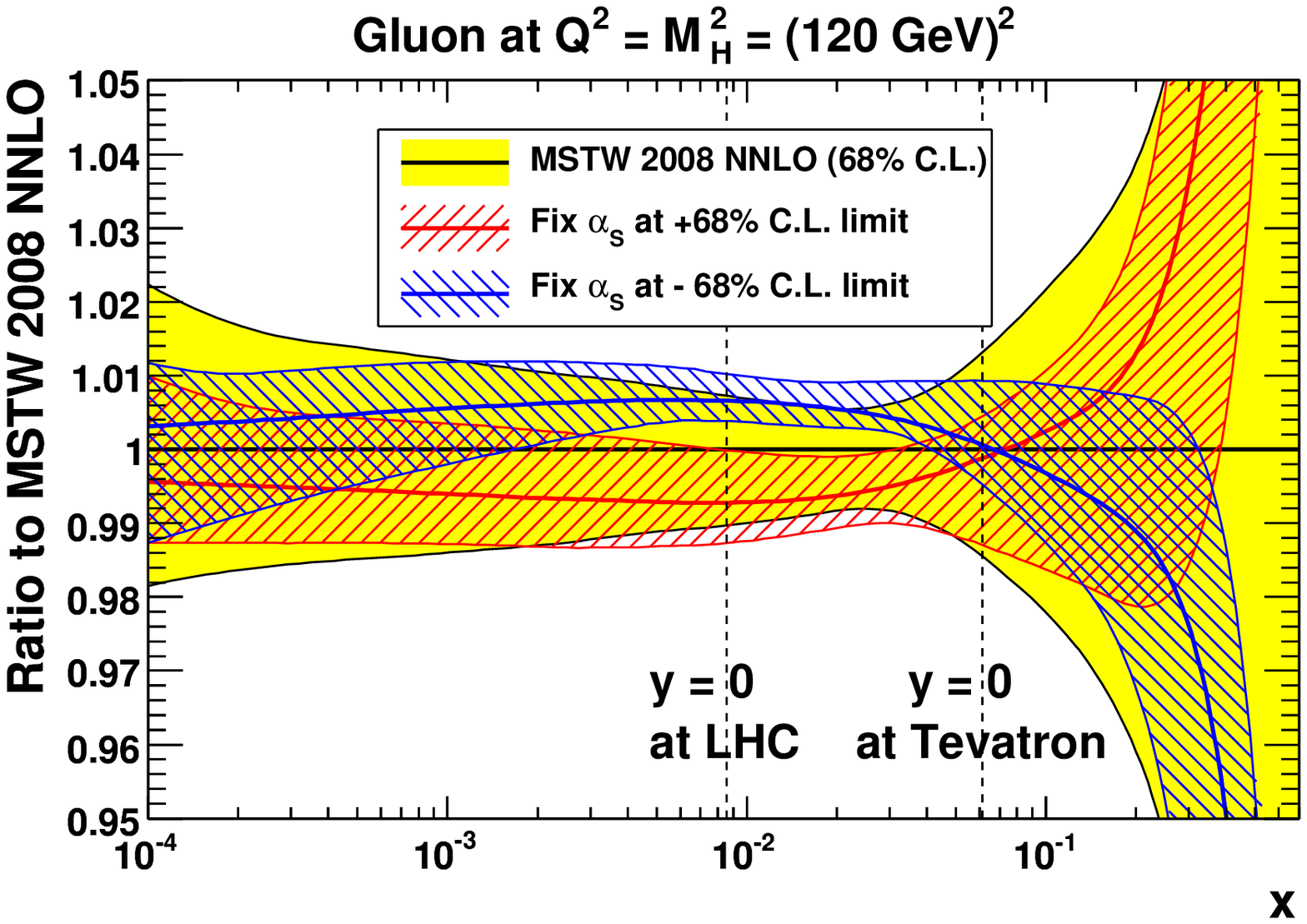,scale=0.42}
\epsfig{file=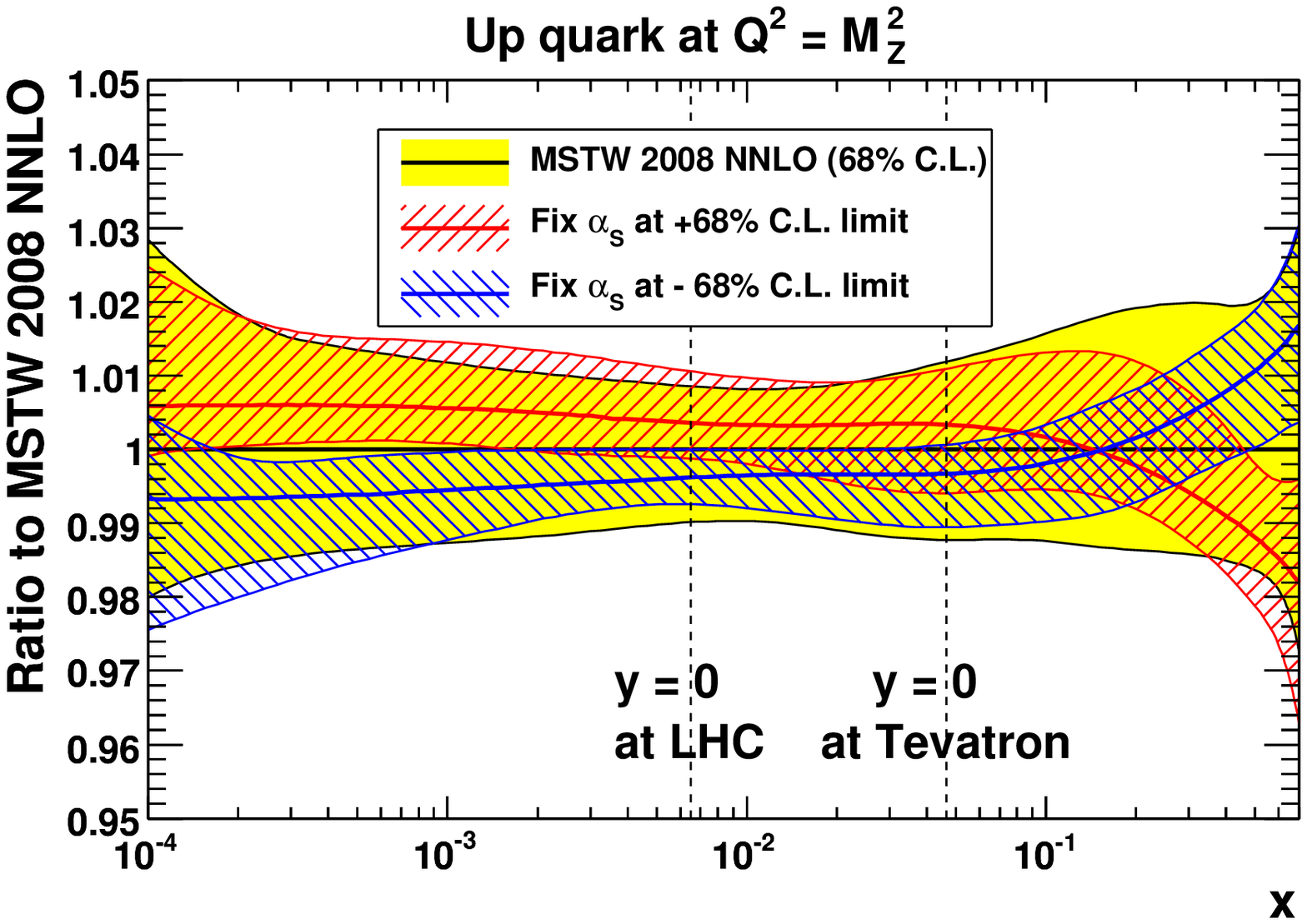,scale=0.42}
\end{minipage}
\begin{minipage}[t]{16.5 cm}
\caption{The correlation of $\alpha_S$ and the gluon distribution
(left) and the up quark distribution (right) for the MSTW2008 PDFs\cite{Martin:2009bu}.
In each case the solid line represents the central PDF at given $\alpha_S(m_Z^2)$ and 
the band the uncertainty due to experimental errors.     
\label{alphasgluon}}
\end{minipage}
\end{center}
\end{figure}

\begin{figure}
\begin{center}
\begin{minipage}[t]{9 cm}
\epsfig{file=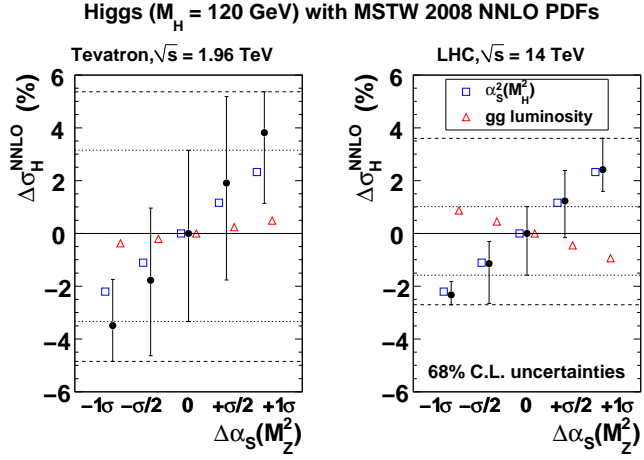,scale=0.45}
\end{minipage}
\begin{minipage}[t]{16.5 cm}
\caption{The correlation of $\alpha_S$ and the Higgs cross section at the Tevatron 
and LHC \cite{Martin:2009bu}. The closed point and bands represent the best prediction and 
uncertainty at each $\alpha_S(m_Z^2)$ value, the inner dotted lines the uncertainty at 
the fixed best value of $\alpha_S(m_Z^2)$ and the outer dotted line the uncertainty including
that on the coupling which is the envelope of the predictions. The triangles represent the
variation due to changes in the gluon only and the open squares the variation in the factor
of $\alpha_S^2$ alone.     
\label{alphasHiggs}}
\end{minipage}
\end{center}
\end{figure}

One can also look at PDF changes and correlations in uncertainties for different 
$\alpha_S(m_Z^2)$. ABKM09 and GJR08 simply include $\alpha_S$ as an additional 
parameter in their error determinations, and uncertainties on physical quantities are
obtained by summing over all free parameters in the error matrix. 
Due to the more complicated dynamical 
tolerance procedure this is not so straightforward for MSTW08. As stated the coupling 
uncertainty is determined from fit, i.e. 
$\alpha_S(m_Z^2) = 0.1202^{+0.0012}_{-0.0015}$ at NLO 
($\alpha_S(m_Z^2) = 0.1171^{+0.0014}_{-0.0014}$ at NNLO)
and the PDFs are presented for the $\pm \half\sigma$ and $\pm\sigma$ uncertainty 
$\alpha_S(m_Z^2)$ values, and similarly for $90\%$ confidence level values. As 
$\alpha_S(m_Z^2)$ departs from its best value the 
PDF uncertainties reduce since the quality of the fit is already 
worse than the best global fit. The PDFs and their uncertainties for different 
$\alpha_S(m_Z^2)$ values are shown in Fig. \ref{alphasgluon}. The expected 
gluon--$\alpha_S(m_Z^2)$ 
small--$x$ anti-correlation is seen and this also leads to a high-$x$ 
gluon--$\alpha_S(m_Z^2)$ correlation from the momentum sum rule. 
The up quark at high $Q^2$ is also shown.
The gluon feeds into the evolution of the quarks, but change in 
$\alpha_S(m_Z^2)$ just outweighs gluon change,
i.e. a larger $\alpha_S(m_Z^2)$ $\to$ slightly more 
evolution. There is a strong anti-correlation at high-$x$ due to evolution 
and positive coefficient function.
The MSTW08 NNLO predictions for Higgs ($120 \GeV$) production for different
allowed $\alpha_S(m_Z^2)$ values and their uncertainties are shown for both the 
Tevatron and LHC in Fig. \ref{alphasHiggs}.   
The uncertainty increases by a factor of $2-3$ (up more than down) at 
the LHC. The direct $\alpha_S(m_Z^2)$ dependence is mitigated somewhat by 
the anti-correlated small-$x$ gluon. At the Tevatron 
the two effects add due to the correlation at high $x$ but in this case the 
intrinsic gluon uncertainty dominates.

\begin{figure}
\begin{center}
\begin{minipage}[t]{16 cm}
\epsfig{file=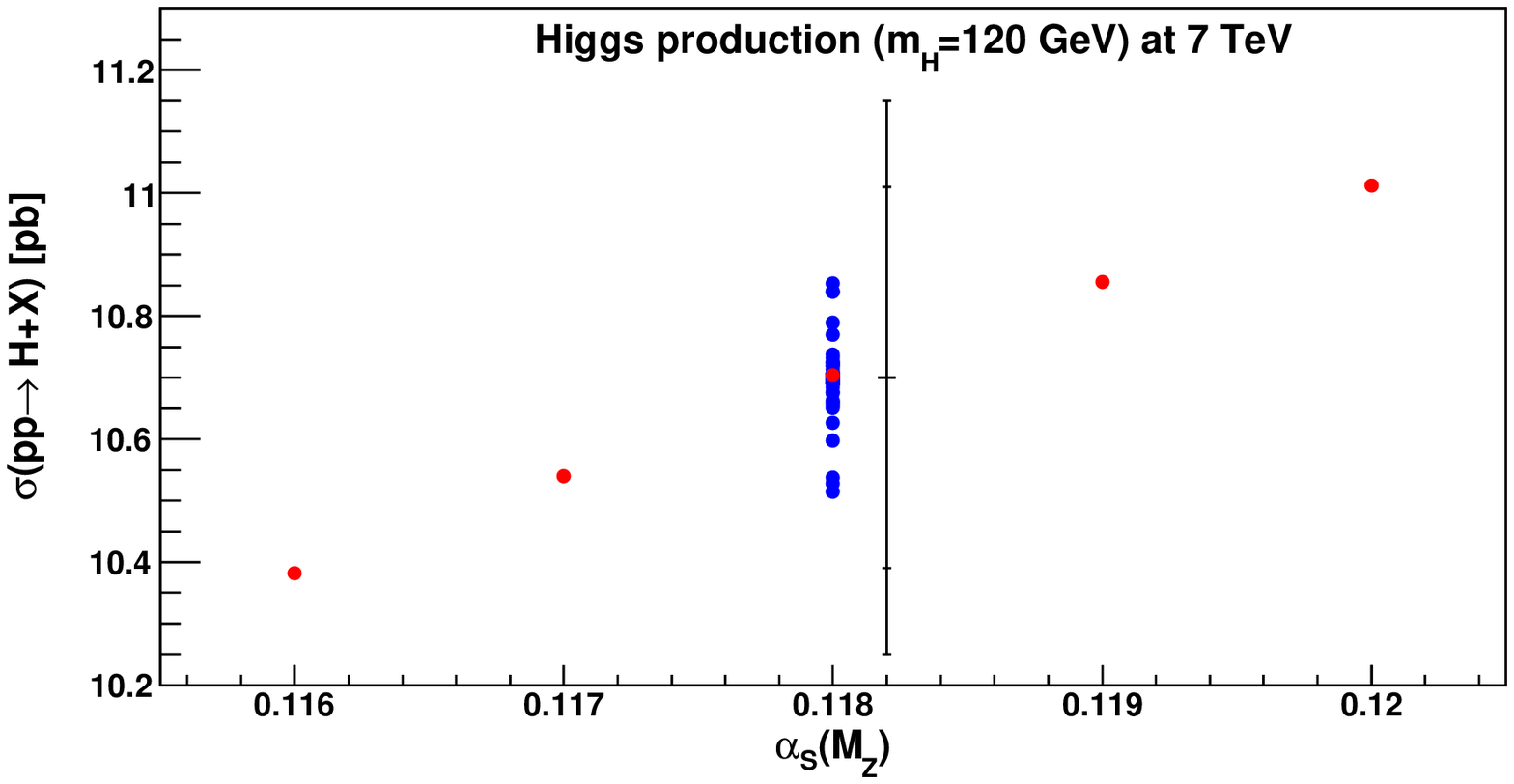,scale=0.4}
\epsfig{file=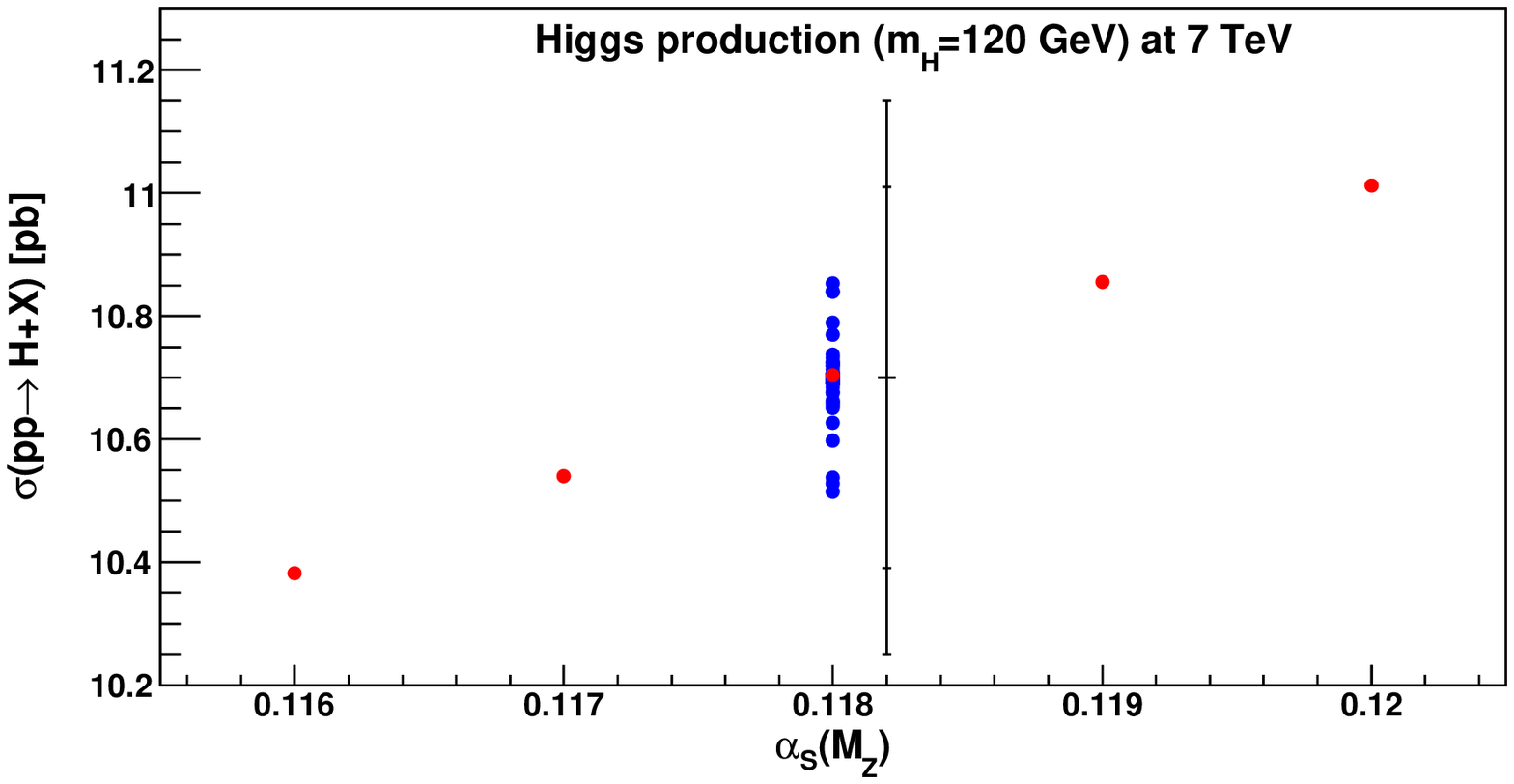,scale=0.4}
\end{minipage}
\begin{minipage}[t]{16.5 cm}
\caption{The uncertainty in the CTEQ prediction for the $120\GeV$ Higgs cross section at 
the LHC from each of the 22 PDF eigenvectors (blue) and variations in $\alpha_S$ (red)
at $7\TeV$ (left) and $14\TeV$ (right)\cite{Lai:2010nw}.    
\label{CTEQashiggs}}
\end{minipage}
\end{center}
\end{figure}

The HERAPDF1.0 fit considers $\alpha_S=0.1176\pm 0.002$, basing their central value on 
the world average\cite{Amsler:2008zz}, and add the results from 
the two extreme values in quadrature with other uncertainties. 
CTEQ also, in principle, base their value on the world average, choosing the similar
value of $0.118\pm 0.002$\cite{Lai:2010nw}
where the uncertainty is the $90\%$ confidence level, though also point out the central
value is the same as their best fit. They also prove analytically that adding the 
uncertainties from the fits at their limits of $\alpha_S(m_Z^2)$ in quadrature with
those from the other orthogonal parameter eigenvectors is, in the quadratic approximation
for $\Delta \chi^2$, exactly equivalent to fitting with $\alpha_S(m_Z^2)$ free 
and constructing
the orthonormal eigenvectors from scratch using the extra variable. The one caveat to this
is that for the latter case the $90\%$ confidence level on $\alpha_S(m_Z^2)$ must 
correspond to a 
deterioration in the fit quality which is exactly the same as that for the $90\%$ confidence
level for the parton parameters, i.e. $\Delta \chi^2=100$. This means the value of 
$\alpha_S(m_Z^2)$ must be included as a data point in the fit with the appropriate
(presumably rather large) weighting factor. CTEQ examine the uncertainty in 
Higgs and $t \bar t$ cross sections using both approaches, CTEQ6.6AS and CTEQ6.6FAS
respectively (F stands for floating $\alpha_S$ ), 
finding that the uncertainty is the same up to 
at most $10\%$. The uncertainty on the cross section for a 120$\GeV$ Higgs boson at the 
LHC as a function of the parton eigenvectors and $\alpha_S$ is shown in Fig. 
\ref{CTEQashiggs}.

\begin{figure}
\begin{center}
\begin{minipage}[t]{16 cm}
\epsfig{file=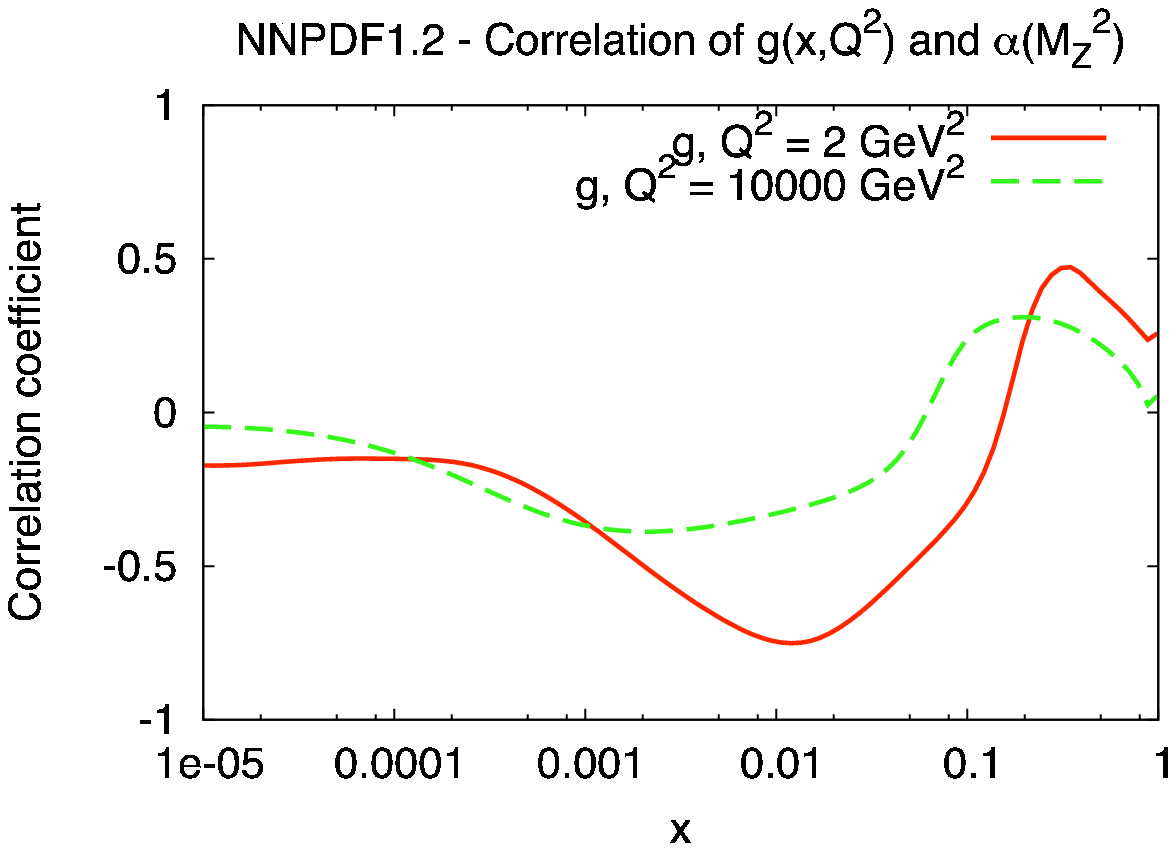,scale=0.6}
\epsfig{file=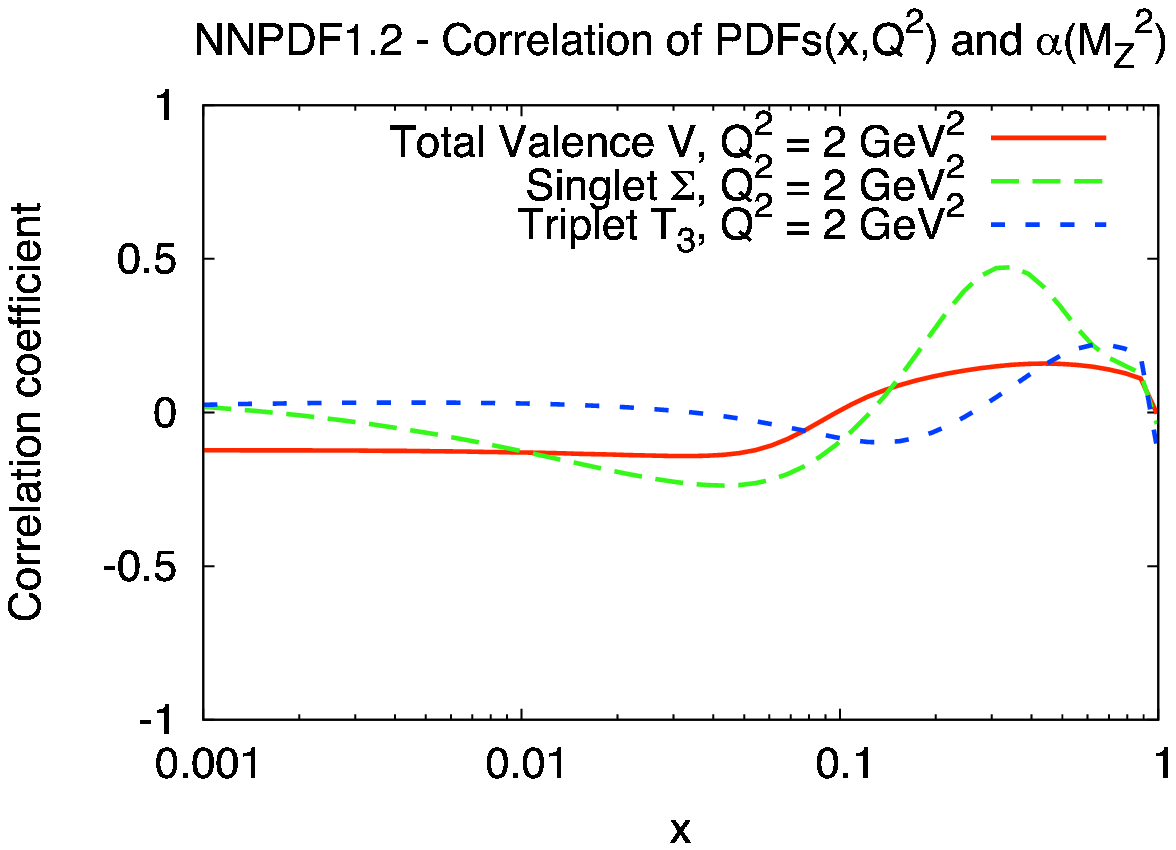,scale=0.6}
\end{minipage}
\begin{minipage}[t]{16.5 cm}
\caption{The correlation coefficient for $\alpha_S$ and the gluon (left) and various quark
distributions (right) from NNPDF\cite{Demartin:2010er}.    
\label{NNPDFcorr}}
\end{minipage}
\end{center}
\end{figure}

NNPDF2.0 also choose a particular external value of $\alpha_S(m_Z^2)$, in their 
case 0.119, with a one $\sigma$ uncertainty of $0.0012$ or $0.002$ for $90\%$
confidence level. Due to their method of determining uncertainties via replicas 
they have an alternative manner of dealing with the $\alpha_S$ 
uncertainty\cite{Demartin:2010er}. 
In order to calculate a quantity they use the PDF sets obtained at different
$\alpha_S(m_Z^2)$ value with the number at a particular value of $\alpha_S(m_Z^2)$ 
determined by the probability of  $\alpha_S(m_Z^2)$ taking that value, i.e.
\be 
N_{\rm rep}^{\alpha_S} \propto \exp \biggl(- \frac{(\alpha_S  - \alpha_S^{(0)})^2}
{2(\delta \alpha_S^{(68)})^2}\biggr),
\ee
where $\alpha_S^{(0)}$ is the central value and $\delta \alpha_S^{(68)}$ the $68\%$
confidence level uncertainty. They verify that there is an anti-correlation
between the small-$x$ gluon and $\alpha_S$, and the opposite at high-$x$, Fig. 
\ref{NNPDFcorr}. They also compare cross-sections for the Higgs boson from different
groups and using different prescriptions for the uncertainty due to the coupling
(there are also results in \cite{Binoth:2010ra}). It is found that adding the deviations 
at the extreme values of $\alpha_S$ in quadrature with other uncertainties is generally
a good approximation, as is now better understood from the CTEQ result above. 
Some of the worse discrepancies between MSTW, CTEQ and NNPDF, e.g. 
the predicted NLO Higgs cross section from CTEQ can be $7\%$ lower than MSTW, are seen
to be about halved by comparing results at the same $\alpha_S$ value. These results are 
using NNPDF1.2, but more recent results can be found in \cite{Ubiali:2010xc}, and are
shown in Fig. \ref{NNPDFCTEQMSTW}. There is clearly some discrepancy, e.g. the 
lower CTEQ value of $\sigma_H$ (due to lower $\alpha_S$ and probably related to
gluon parameterisation and default charm mass) and the lower NNPDF value of $\sigma_Z$
(similar for $W^{\pm}$, probably related to use of the ZM-VFNS).

\begin{figure}
\begin{center}
\begin{minipage}[t]{14 cm}
\epsfig{file=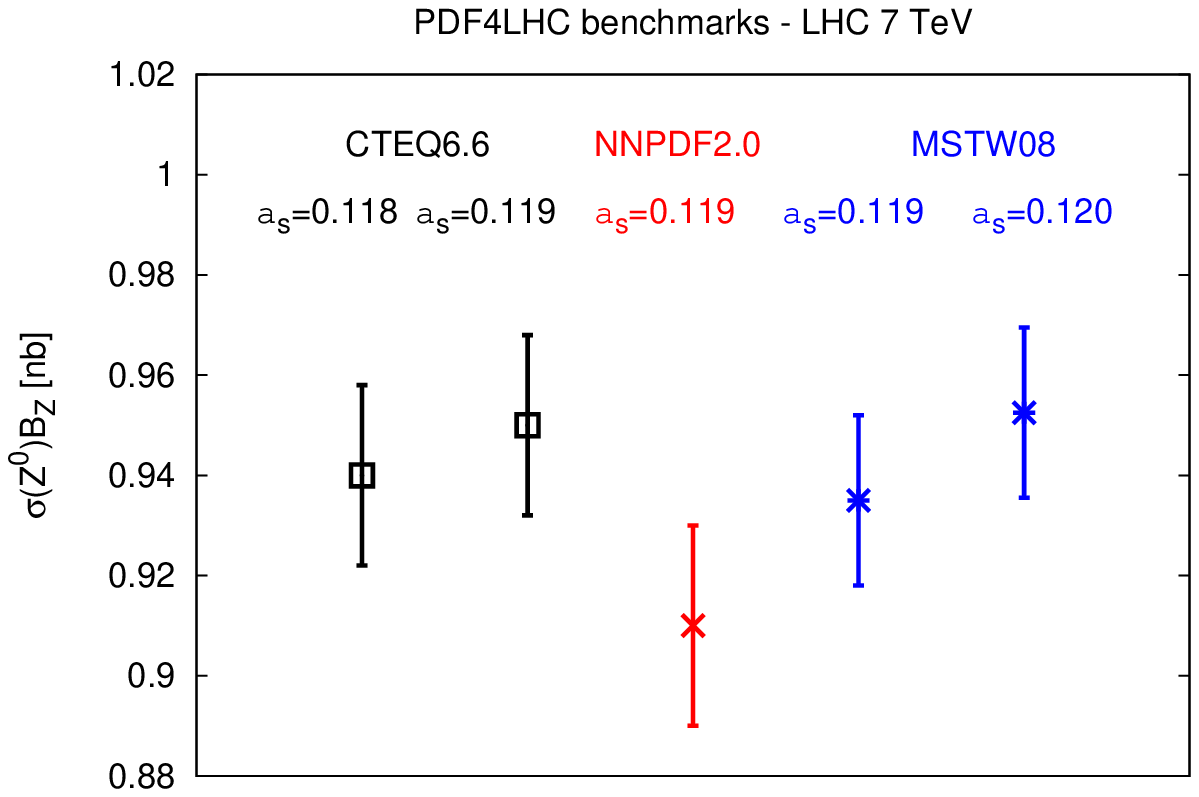,scale=0.5}
\epsfig{file=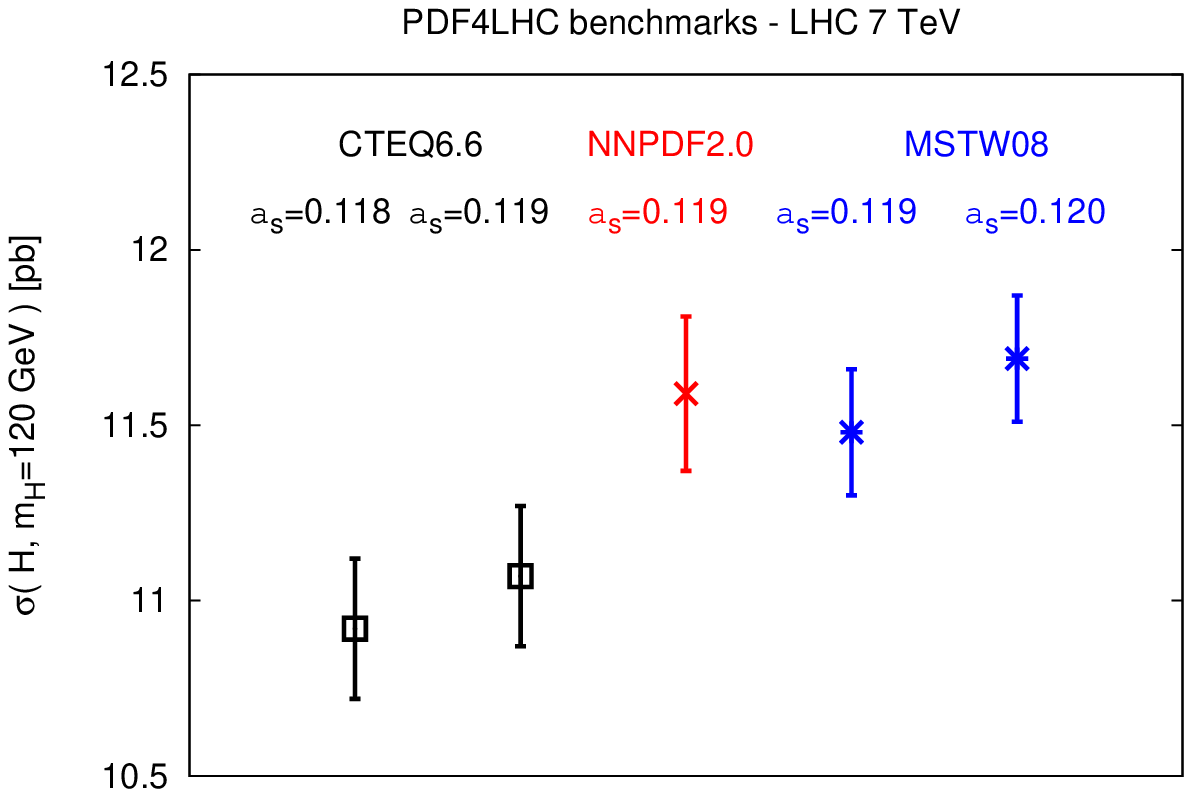,scale=0.5}
\end{minipage}
\begin{minipage}[t]{16.5 cm}
\caption{The variety in predictions from different groups and central values of
$\alpha_S$ for the $Z$ cross section (left) and the $120\GeV$ Higgs cross section
(right) at the LHC ($7\TeV$)\cite{Ubiali:2010xc}.      
\label{NNPDFCTEQMSTW}}
\end{minipage}
\end{center}
\end{figure}

\section{Other sources of Uncertainty}

In the previous section we have discussed various factors of 
what might be deemed theoretical uncertainty, such as parameterisations, 
heavy quark treatments, choices in strong coupling, which are unavoidable, and
lead to differences between the PDFs obtained by groups and the resulting 
predictions. As well as these there are additional theoretical corrections which
can lead to further changes or corrections. These are systematic 
corrections which will lead to the PDFs and their predictions 
from all groups being modified.
Some are already investigated, perhaps partially by some groups or by others
working in the field of perturbative QCD and parton distributions.    
Some of the most important are:

\begin{itemize}

\item Standard higher orders, i.e  NNLO in perturbation theory and beyond. 

\item QED and weak corrections, which are nominally small, but where there might 
sometimes be enhancements.

\item Resummations, e.g. small $x$ ($\alpha_S^n \ln^{n-1}(1/x)$), 
or large $x$ ($\alpha_S^n \ln^{2n-1}(1-x)$)

\item  Corrections at low $Q^2$, e.g.  higher twist and possible saturation effects.

\end{itemize}

We will now discuss each of these briefly.

\begin{figure}
\begin{center}
\begin{minipage}[t]{14.8 cm}
\epsfig{file=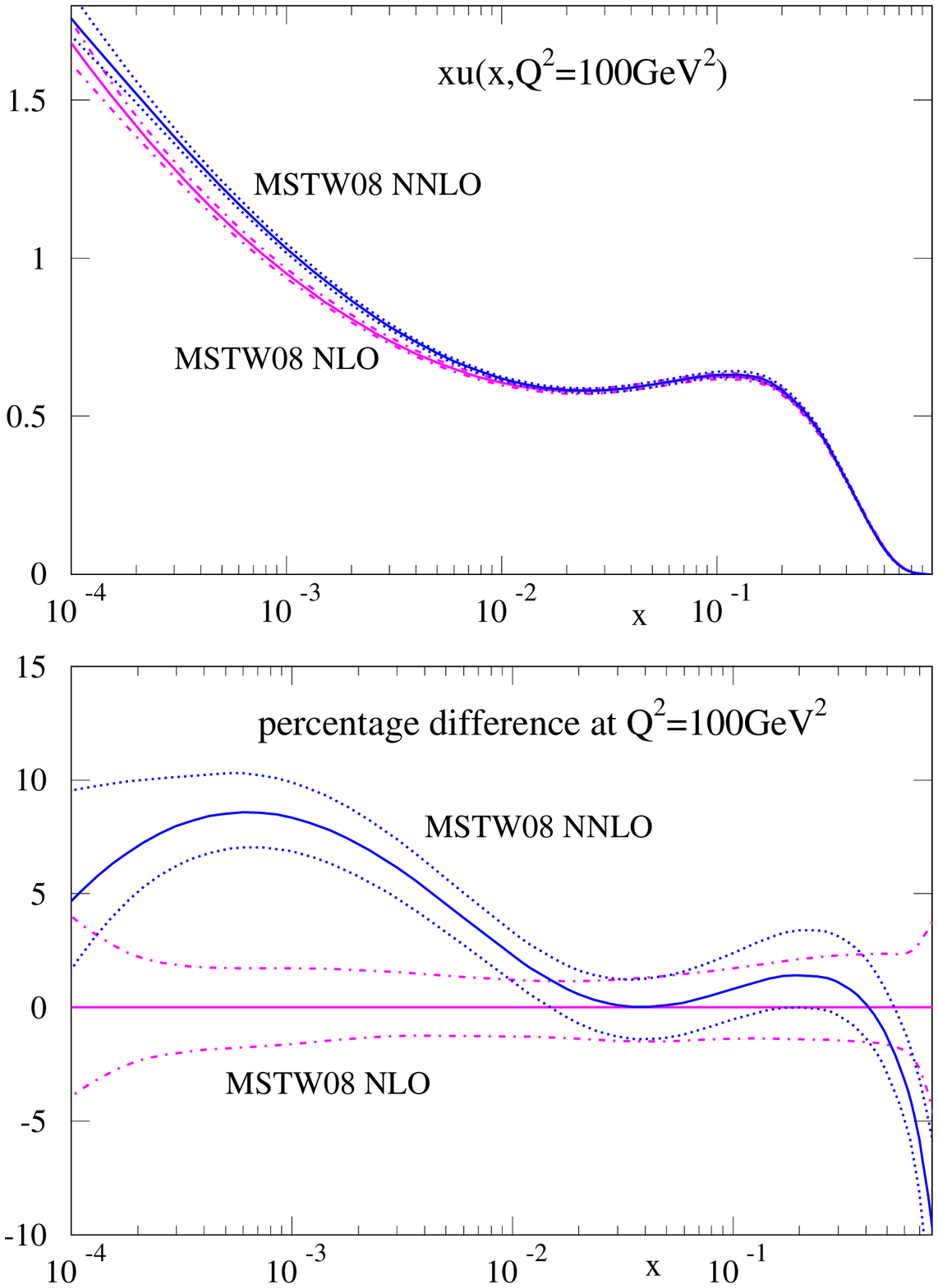,scale=0.42}
\hspace{0.8cm}
\epsfig{file=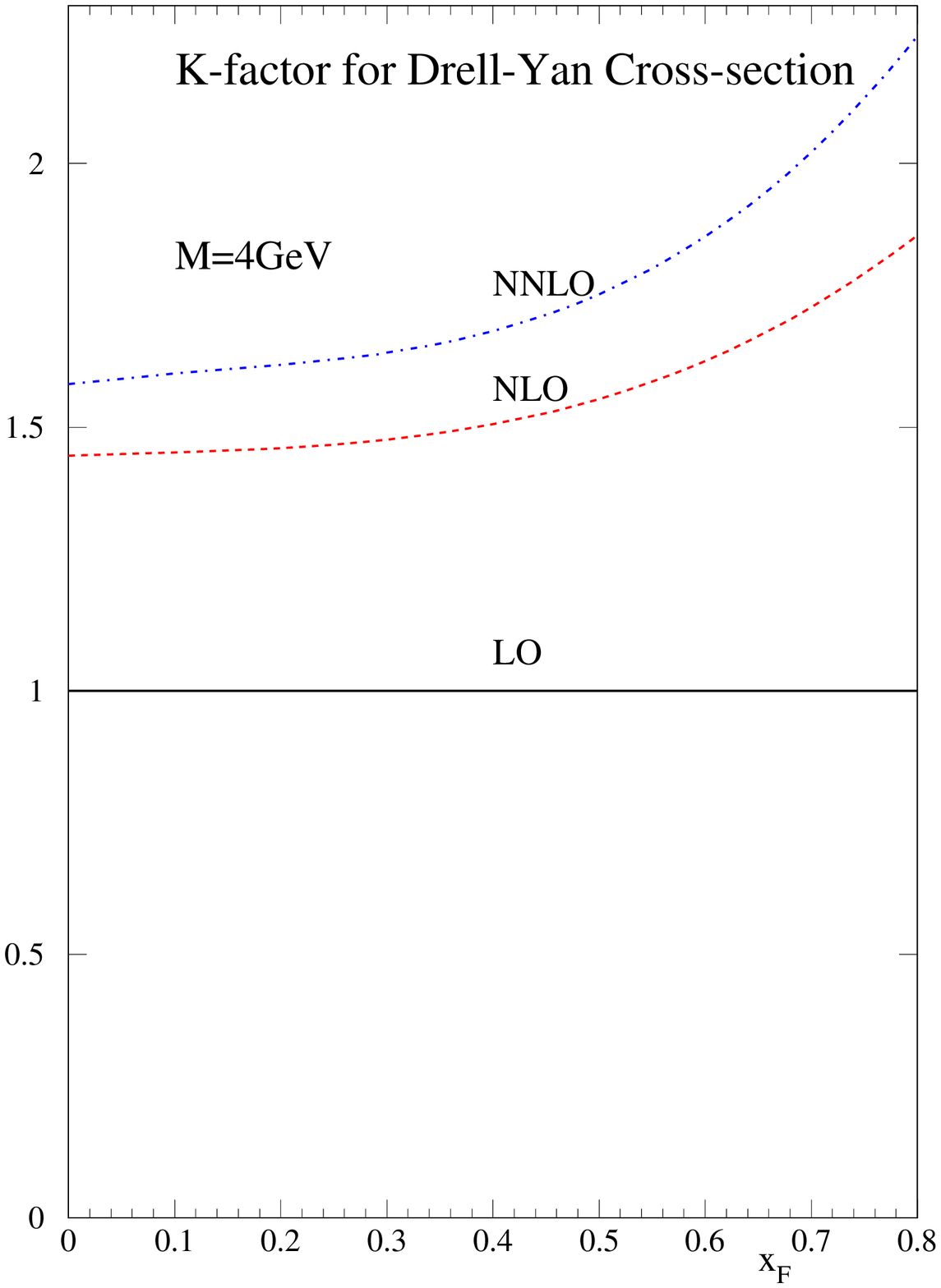,scale=0.4}
\end{minipage}
\begin{minipage}[t]{16.5 cm}
\caption{The MSTW2008 up quark distributions at NLO and NNLO (left). The absolute values, with 
$1\sigma$ uncertainties are shown in the upper plot, and the ratio compared 
to the central value at NLO are shown in the lower plot. 
The $K$-factors for Drell-Yan production at NLO and NNLO\cite{Martin:2007bv}
for a muon pair of invariant mass $4\GeV$ (right).
\label{NNLOup}}
\end{minipage}
\end{center}
\end{figure}

\subsection{\it NNLO corrections}

We have already pointed out that some groups  produce
NNLO PDFS, though the results vary quite a lot. 
As noted the extraction can be from a nearly complete NNLO definition within
the global fit procedure, i.e. NNLO evolution, massless coefficient functions for
structure functions and Drell-Yan vector boson production are known exactly, 
while some approximation and or modelling is required for massive quark 
coefficient functions or jet production. So the degree of theoretical
approximation required in an NNLO fit, particularly if a GM-VFNS is used, is 
not large and the PDFs can certainly be taken seriously. However, we have so 
far not considered the change in the PDFs as one goes from NLO to NNLO
and the consequences.  

%\begin{figure}
%\begin{center}
%\begin{minipage}[t]{5.2 cm}
%\epsfig{file=dynnlo.ps,scale=0.3}
%\end{minipage}
%\begin{minipage}[t]{16.5 cm}
%\caption{The $K$-factors for Drell-Yan production at NLO and NNLO\cite{Martin:2007bv}
%for a muon pair of invariant mass $4\GeV$.
%\label{NNLODY}}
%\end{minipage}
%\end{center}
%\end{figure}

It is important to note that because the PDFs are not 
physical quantities the NNLO PDFS are not simply 
a more accurate version of the 
NLO PDFs. There can be systematic differences. This is illustrated in the left of Fig. 
\ref{NNLOup}, where we compare the up quark, the most accurately determined 
parton distribution, at NLO and NNLO. One can see that the shape is different 
as a function of $x$ and the central value of the NLO and NNLO PDFs can 
differ by 3-4 times the uncertainty. This does not necessarily mean a large
change in any physical quantities however, as the change in PDFs can be  
compensated by a change in the coefficient functions. Indeed, both the PDFs 
shown fit the same structure function data with a similar quality. As well as
coefficient functions and PDFs the physical quantities implicitly depend on 
the strong coupling constant $\alpha_S$. The NNLO corrections are largely 
positive, i.e. evolution of PDFs increases in speed at both large and small
$x$ and many cross section corrections are positive, e.g the $K$-factor
for NLO and NNLO fixed target Drell-Yan production is shown in the right of Fig. 
\ref{NNLOup}, the latter being $\sim 10\%$ bigger. Hence, it seems very likely 
that the coupling will have to become smaller at NNLO in order to compensate. 
Indeed, this is seen in all cases -- MSTW\cite{Martin:2009iq}, 
ABKM\cite{Alekhin:2009ni} and GJR\cite{Gluck:2007ck} all see a fall in the
value of $\alpha_S(m_Z^2)$ of 0.002--0.004 at NNLO compared to NLO.
Hence, precise predictions using NNLO PDFs require the simultaneous use of 
NNLO cross-sections (or {\it vice versa}) and the appropriate coupling at NNLO.

\begin{figure}
\begin{center}
\begin{minipage}[t]{8.5 cm}
\epsfig{file=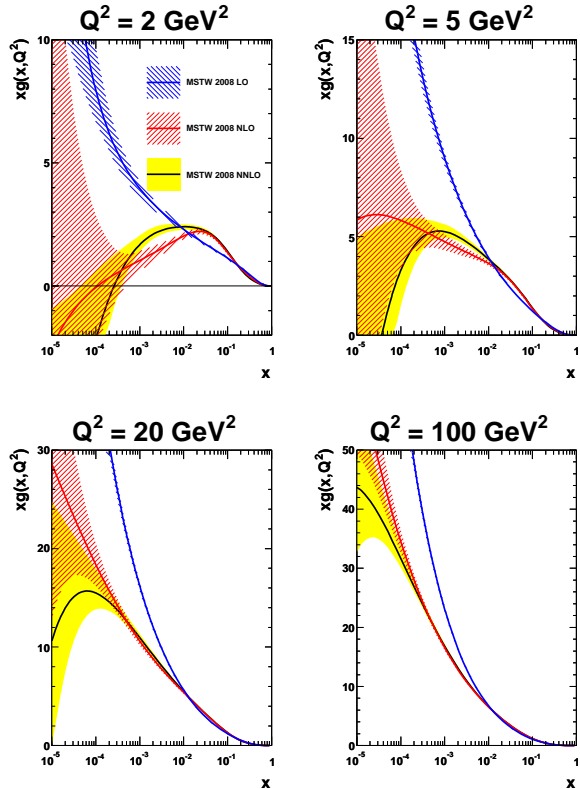,scale=0.42}
\end{minipage}
\begin{minipage}[t]{16.5 cm}
\caption{Gluon distributions at LO, NLO and NNLO along with their 
uncertainties at each order due to experimental errors.
\label{NNLOgluon}}
\end{minipage}
\end{center}
\end{figure}

Since, we do have these PDF sets determined at NNLO 
surely it is best, i.e. most accurate, to make use of these.
In principle this is correct, however, we 
only know some hard cross-sections at NNLO.
Processes with two strongly interacting particles 
are largely completed -- DIS coefficient functions and sum rules,
$pp(\bar p) \to$ $\gamma^{\star}, W, Z$
(including rapidity dist.), $H, A^0, WH, ZH$\cite{Catani:2001ic,Harlander:2002wh,Anastasiou:2002yz,Ravindran:2003um,Anastasiou:2004xq,Anastasiou:2005qj,Brein:2003wg}. 
For other final states the NNLO coefficient 
functions are not known and so NLO PDFs are still more appropriate.
There are even some processes where only LO is known, particularly those with 
large number of final state particles or very exclusive final states.  

However, as well as providing us with maximum precision, if it is available   
NNLO also tells us about the convergence of perturbation 
theory. For most structure functions convergence is guaranteed, because the
PDFs are obtained by comparison to the very accurate data. Hence, it is 
predictions (including the normalisation of the total $W$ and $Z$ 
cross-sections at the Tevatron since the normalisation uncertainty of the data
is large), which are more illuminating. For $W$ and $Z$ cross-sections at
both the Tevatron and the LHC the perturbation series is reasonably 
convergent. In both cases the prediction is 3-4$\%$ higher at NNLO than 
at NLO\cite{Martin:2009iq}, but this change is about twice the 
uncertainty quoted at each order. The NNLO prediction is in better agreement 
with the measurements at the Tevatron\cite{Abulencia:2005ix,D0wztot}.  
The change is dominated by the change in the PDFs as one goes from NLO to NNLO
rather than the change in the cross section. The NNLO Higgs cross section is 25$\%$ bigger than at NLO. In 
this case the change is completely dominated by the NNLO contribution to 
the cross section. In both cases better stability is achieved by allowing the 
NNLO strong coupling value $\alpha_S(m_Z^2)$ to be 0.003 lower than at NLO
than by using the same value in both cases.

\begin{figure}
\begin{center}
\begin{minipage}[t]{8 cm}
\epsfig{file=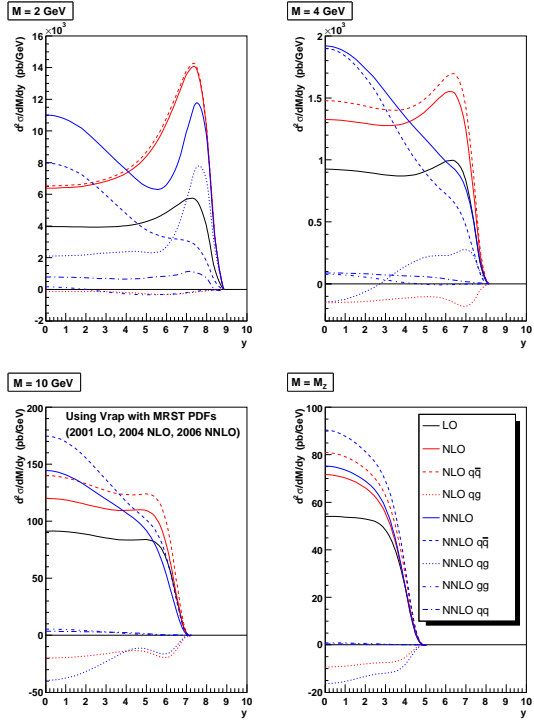,scale=0.4}
\end{minipage}
\begin{minipage}[t]{16.5 cm}
\caption{The Drell Yan cross section at LO, NLO and NNLO at the LHC\cite{Thorne:2008am}.
The dotted lines show the contributions from various subprocess at each order, 
e.g. NNLO qg is the contribution from a quark and gluon initial state at NNLO.
\label{NNLODYcross}}
\end{minipage}
\end{center}
\end{figure}

Hence, there is some question about the stability of 
cross-sections as one progresses to higher orders in the perturbative series. 
This may sometimes be related to the issue of resummations at large or 
small $x$ and whether these are important. In Section 2.7 we gave a 
preliminary discussion of the convergence at small $x$. In Fig. 
\ref{NNLOgluon} we see a comparison of the    
gluon distribution extracted from the global fit at LO, NLO and 
NNLO\cite{Martin:2009iq}. The additional positive small-$x$ contributions 
in the splitting function $P_{qg}$ at each order lead to a smaller very 
small-$x$ gluon at each successive order. Hence, in this regime there is 
clearly fairly poor stability. This is similar for $F_L(x,Q^2)$, though there
is some compensation due to the NNLO coefficient function. 
A more dramatic consequence for this lack of convergence can be seen 
for the LHC. In Fig. \ref{NNLODYcross}\cite{Thorne:2008am} 
we show the LO, NLO and 
NNLO predictions for for $Z$ and $\gamma^{\star}$ production at the LHC for
$14\TeV$ centre-of-mass energy, as a function of rapidity and invariant
mass. There is good stability in the predictions for very high final state masses, 
but it becomes worse at lower scales where $\alpha_S$ is larger and 
large $\ln (s/M^2)$ terms appear in the perturbative expansion, which are 
equivalent to $\ln (1/x)$. This suggests resummation  may be necessary in this
regime.  

We note that at large $x$ there is enough information available to perform 
reliable approximations to fits at NNNLO, see e.g 
\cite{Blumlein:2006be,Khorramian:2009xz}, in the nonsinglet sector .
This seems to lead to no significant difference to the results at NNLO, 
with a stabilisation in the values of $\alpha_S$, implying good 
theoretical convergence 
in the the kinematic region where these fits are performed.

\subsection{\it Electroweak corrections}

In principle the smallness of the electromagnetic and weak couplings should 
mean the corrections from this source are small. However, at high scales
$\alpha \sim \alpha_S^3$ so they may be comparable to NNLO QCD 
corrections. Additionally there can be some enhancements, or violations of
symmetry not present in QCD which can be important. 
The simplest thing one can consider is the QED--improved DGLAP equations which 
are, 

\bea 
 \partial q_i(x,\mu^2) 
\over \partial \log \mu^2 &=& {\alpha_S\over 2\pi}
\int_x^1 \frac{dy}{y} \Big\{
    P_{q q}(y)\; q_i(\frac{x}{y},\mu^2)
     +  P_{q g}(y,\alpha_S)\; g(\frac{x}{y},\mu^2)
    \Big\} \nonumber \\
& + &
   {\alpha\over 2\pi} \int_x^1 \frac{dy}{y} \Big\{
    \tilde{P}_{q q}(y)\; e_i^2 q_i(\frac{x}{y},\mu^2)  +  P_{q \gamma}(y)\;
e_i^2 \gamma(\frac{x}{y},\mu^2)         \Big\}  \nonumber \\
{\partial g(x,\mu^2) \over \partial \log \mu^2} &=& 
{\alpha_S\over 2
\pi} \int_x^1 \frac{dy}{y} \Big\{
    P_{g q}(y)\; \sum_j q_j(\frac{x}{y},\mu^2) + 
    P_{g g}(y)\; g(\frac{x}{y},\mu^2)
    \Big\} \nonumber \\
 {\partial \gamma(x,\mu^2) \over \partial \log \mu^2} 
&=   & {\alpha
\over 2\pi} \int_x^1 \frac{dy}{y} 
   \Big\{ P_{\gamma q}(y)\; \sum_j e_j^2\; q_j(\frac{x}{y},\mu^2)
+  P_{\gamma \gamma}(y)\; \gamma(\frac{x}{y},\mu^2) \Big\} 
\label{QEDDGLAP}
\eea
at leading order in $\alpha_S$ and $\alpha$, where
\be
\tilde{P}_{qq} = C_F^{-1} P_{qq}, \quad   P_{\gamma q} = C_F^{-1} P_{g q},
\quad P_{q\gamma} = T_R^{-1} P_{q g} , \quad  P_{\gamma \gamma} = - 
\frac{2}{3}\; \sum_i e_i^2\; \delta(1-x),
\label{QEDsplit}
\ee
$\gamma(x,\mu^2)$ is the photon distribution and momentum is conserved:
\be
\int_0^1 dx\;  x\; \Big\{\sum_i q_i(x,\mu^2) + g(x,\mu^2) 
+ \gamma(x,\mu^2)
\Big\}  = 1.
\label{QEDmomentum}
\ee

\begin{figure}
\begin{center}
\begin{minipage}[t]{8 cm}f
\epsfig{file=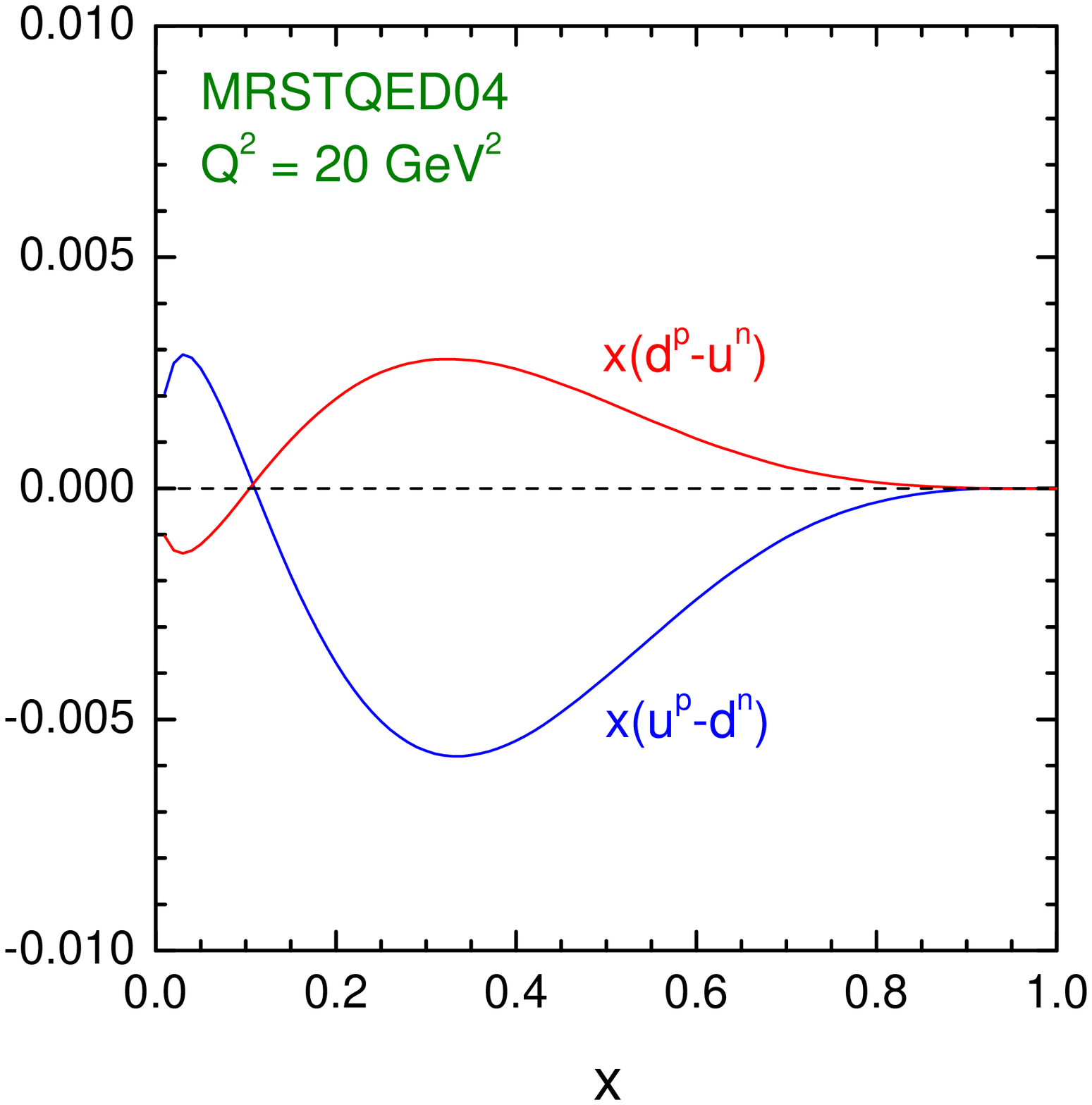,scale=0.35}
\end{minipage}
\begin{minipage}[t]{16.5 cm}
\caption{QED induced different in valence quarks in the proton compared 
to those in the neutron\cite{Martin:2004dh}.  
\label{isoviolation}}
\end{minipage}
\end{center}
\end{figure}

There has been one detailed study of the effects of the QED evolution
on the PDFs obtained via a global fit in \cite{Martin:2004dh}.
Because there are not any enhancements in Eq. (\ref{QEDDGLAP})
the effect on the quark distributions of the QED corrections is 
negligible at small $x$ where the gluon contribution
dominates the evolution. The main effect is that the gluon loses a 
little momentum to the photon in order for the momentum sum rule to
be satisfied. At large $x$, photon radiation from quarks leads to faster 
evolution, roughly equivalent to a slight shift  in $\alpha_S$, i.e.
$\Delta \alpha_S(m_Z^2)  \simeq + 0.0003 $.
Overall, the QED effects are much smaller than many sources of 
uncertainty. However, the up quarks at high $x$ radiate more
photons than down quarks due to the higher charge weighting. 
This leads to an automatic violation in the charge symmetry 
assumed in Eq. (\ref{isotransform}), as seen in Fig. \ref{isoviolation},
and this reduces the 
NuTeV anomaly in the measurement of $\sin^2 \theta_W$\cite{Zeller:2001hh,Zeller:2002du}. 
The other place where QED corrected PDFs are important is where 
an initial state photon plays a role. Consider the electroweak corrections 
to lepton pair production\cite{CarloniCalame:2007cd}.
In the hard cross-sections the QED effects are typically a few percent and
negative, becoming larger in magnitude at high transverse momentum.  
However, one also needs to consider photon-induced 
processes driven by the photon distribution of the proton, as shown in 
Fig. \ref{QEDDYinitial}. Can be a significant fraction of the  
other electroweak corrections, and in the opposite direction, i.e. positive.

\begin{figure}
\begin{center}
\begin{minipage}[t]{13 cm}
\epsfig{file=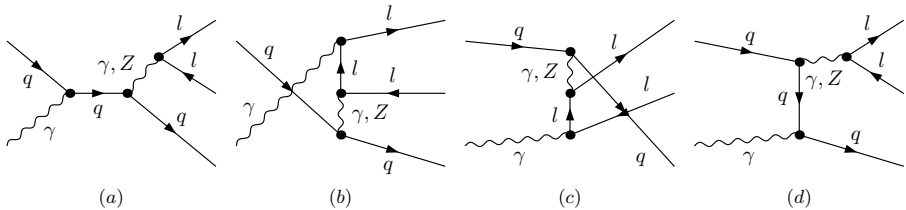,scale=0.8}
\end{minipage}
\begin{minipage}[t]{16.5 cm}
\caption{Drell Yan cross-sections with an initial state photon\cite{CarloniCalame:2007cd}. 
\label{QEDDYinitial}}
\end{minipage}
\end{center}
\end{figure}

Large Electroweak corrections are potentially possible due to enhanced
logarithms of the form $\alpha^n_W\log^{2n}(E_T^2/M_W^2))$
in the perturbative series\cite{Ciafaloni:2000df}. 
Jet cross-sections are an example\cite{Moretti:2006ea} 
where there is potentially a big effect at LHC energies where 
$\log^2(E_T^2/M_W^2)$ is a very large number, as seen in Fig. 
\ref{EWjets}.  Similar results exist for corrections to other processes 
with a hard scale, e.g. di-boson production\cite{Accomando:2004de}
and large-$p_T$ vector bosons in 
conjunction with jets\cite{Kuhn:2007cv} (though very sensitive to jet vetoes). These could potentially 
affect the extraction of PDFs at the LHC if they are not taken account
of properly. In order to do this though, 
one must not only have virtual corrections 
for $W,Z$, but must have contributions of the form where the bosons
are emitted as extra final state particles, which will certainly 
cancel the loop virtual corrections to some significant 
extent\cite{Baur:2006sn}. 
Whether the consideration of parton distributions with weak bosons as well as 
the photon will aid maximum precision since it is known that the evolution
will produce electroweak double logarithms\cite{Ciafaloni:2005fm}.

\begin{figure}
\begin{center}
\begin{minipage}[t]{8 cm}
\rotatebox{90}{\epsfig{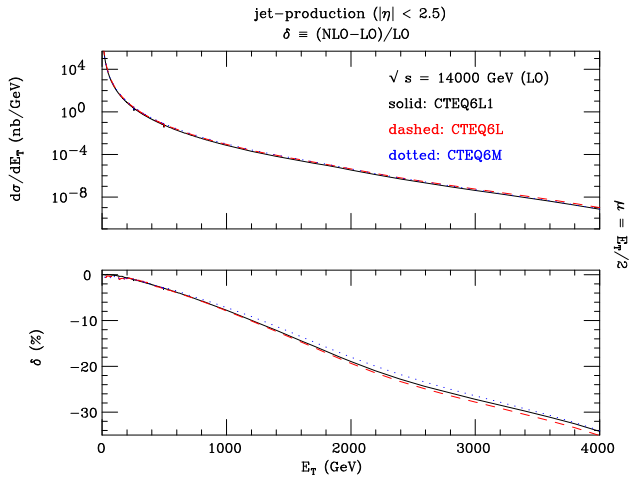}}
\end{minipage}
\begin{minipage}[t]{16.5 cm}
\caption{The fractional size of the NLO electroweak effect on high-$E_T$ 
jet cross-sections\cite{Moretti:2006ea}. 
\label{EWjets}}
\end{minipage}
\end{center}
\end{figure}

\subsection{\it Small-$x$ Theory}

%\begin{figure}
%\begin{center}
%\begin{minipage}[t]{5 cm}
%\epsfig{file=bfkllad.ps,scale=0.6}
%\end{minipage}
%\begin{minipage}[t]{16.5 cm}
%\caption{The gluon ladder represented by the unintegrated gluon $f$ 
%linked to the bare gluon distribution and the hard cross 
%section impact factor 
%in scattering from a proton.
%\label{BFKLladder}}
%\end{minipage}
%\end{center}
%\end{figure}

As seen in Section 4.1 there are fairly strong hints of some instability, or 
lack of convergence in perturbation theory, when small-$x$ PDFs are probed. 
The reason for this instability was outlined in Section 2.7 -- 
as known since\cite{Lipatov:1976zz,Kuraev:1977fs,Balitsky:1978ic}
at each order in $\alpha_S$ each splitting 
function and coefficient function generally
obtains an extra power of $\ln(1/x)$. For the 
parton distributions these leading logarithms can be 
obtained from the BFKL equation for the high-energy limit of the unintegrated
(in transverse momentum $k$) distribution 
\be
f(k^2,x) =
f_I(Q_0^2) +\! \int_x^1\frac{dx'}{x'}\bar\alpha_S\!\! \int_0^{\infty} 
\frac{dq^2}{q^2}K(q^2,k^2)f(q^2,x)
\label{BFKL}
\ee
where $f(k^2,x)$ is the unintegrated gluon distribution 
$g(x,Q^2) = \int^{Q^2}_0 \, (dk^2/k^2) f(x,k^2)$, and 
$K(q^2,k^2)$ is a calculated kernel now known to NLO\cite{Fadin:1998py,Ciafaloni:1998gs}. 
The physical structure functions are then obtained from 
\be
\sigma(Q^2,x) = \int (d k^2/k^2) \,h(k^2/Q^2)  f(k^2,x)
\label{BFKLconvolution}
\ee
$h(k^2/Q^2)$ is a calculable impact factor, known for 
structure functions\cite{Catani:1990eg,Catani:1994sq} and
some other processes, e.g. \cite{Ball:2001pq}.  
As mentioned, the global fits usually assume that this is not significant
in the region of interest, though a purely phenomenological investigation\cite{Martin:2003sk} did find that resummed terms were preferred by data.

\begin{figure}
\begin{center}
\begin{minipage}[t]{5.8 cm}
\epsfig{file=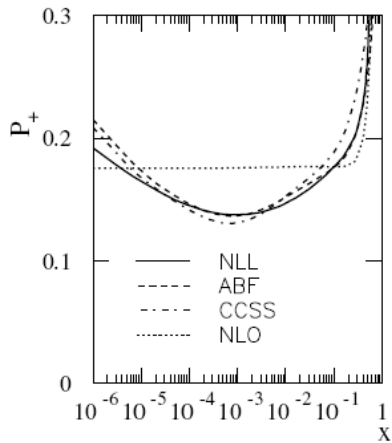,scale=0.5}
\end{minipage}
\begin{minipage}[t]{16.5 cm}
\caption{Comparison of the leading splitting function $P_+\approx P_{gg}+4/9P_{qg}$) from
different groups\cite{White:2006yh}. NLO is the standard fixed order NLO result, NLL
is from \cite{White:2006yh}, CCSS from \cite{Ciafaloni:2007gf} and ABF 
from \cite{Altarelli:2008aj}.
\label{BFKLsplit}}
\end{minipage}
\end{center}
\end{figure}

The inclusion of the NLO corrections to the BFKL equation and the 
consequent scale breaking made the solution much more difficult to obtain
due to the difficulty of avoiding nonperturbative contamination from 
the infrared region. This led to a concentration of effort more closely
related to the collinear factorisation of the usual perturbative ordering, 
and in particular the assumption that input PDFs should be fit and 
splitting functions and coefficient functions calculated. On this basis there
has been good progress in incorporating $\ln(1/x)$ resummation
from essentially three groups \cite{White:2006yh,Ciafaloni:2007gf,Altarelli:2008aj} 
with results roughly in agreement,
 despite some differences in technique. In order to achieve stable results
additional effects such as running coupling\cite{Thorne:2001nr} effects 
and (depending on group) other corrections such as resummation of dominant 
collinear logarithms\cite{Salam:1998tj} are included.
A comparison of the leading (mainly
gluon) splitting function compared to the standard NLO result is shown in 
Fig. \ref{BFKLsplit}. It is a common result that the small-$x$ resummation 
leads to a dip for $x \sim 10^{-3}$ before the expected rise at very low $x$
in splitting functions and coefficient functions (though a full set 
of coefficient functions is still to come in some cases). A recent review of this
work can be found in \cite{Jung:2009eq}. There are also approaches 
which attempt to predict the full structure functions, rather than just 
coefficient functions and splitting functions, e.g. 
\cite{GolecBiernat:2009be,Kowalski:2010ue}, though this must necessarily 
introduce some assumption about or modelling of the infrared physics. 
Results are encouraging, but it is more difficult to directly relate the
PDFs and structure functions obtained to the standard fixed order ones using 
this type of approach. 

%\begin{figure}
%\begin{center}
%\begin{minipage}[t]{8 cm}
%\epsfig{file=data2nllb.ps,scale=0.45}
%\end{minipage}
%\begin{minipage}[t]{16.5 cm}
%\caption{Comparison of a fit with $\ln(1/x)$ resummation (NLL+)
%and at fixed order (NLO+) to HERA data\cite{White:2006yh}.
%\label{smallxfit}}
%\end{minipage}
%\end{center}
%\end{figure}

\begin{figure}
\begin{center}
\begin{minipage}[t]{13 cm}
\epsfig{file=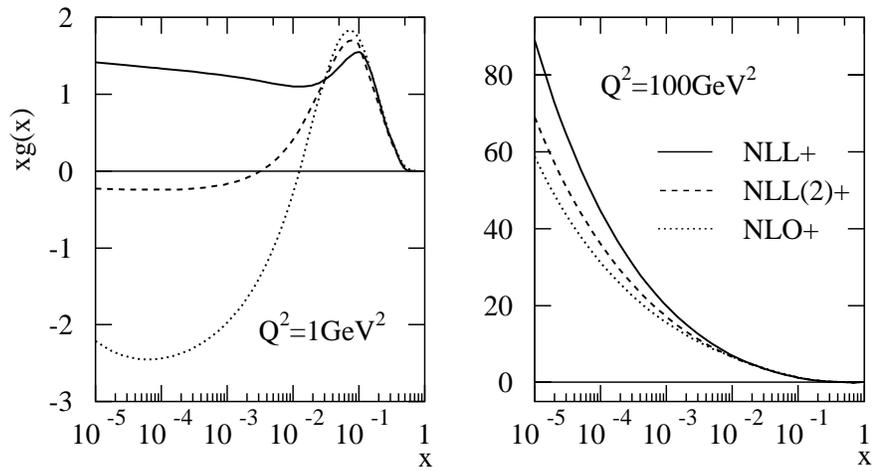,scale=0.73}
\end{minipage}
\begin{minipage}[t]{16.5 cm}
\caption{Comparison of the gluons from a fit with $\ln(1/x)$ 
resummation (NLL+) and at fixed order (NLO+)\cite{White:2006yh}.
\label{smallxgluons}}
\end{minipage}
\end{center}
\end{figure}

A fit to data at NLO plus NLO resummation (in DIS scheme to NLO and a 
DIS-type resummed scheme beyond) with full resummation for
heavy quarks included\cite{White:2006yh} has been performed. 
It leads to significant improvement in the fit to HERA data within
a global fit and a change in the extracted 
gluon (Fig. \ref{smallxgluons}), making it steeper at low $Q^2$ and 
consequently slightly larger than the fixed-order gluon below $x=0.005$
at higher $Q^2$. Together with indications from Drell Yan resummation 
calculations\cite{Marzani:2008uh} this suggests at least a few percent 
effect due to small-$x$ resummation at the LHC is quite possible, even for
$W$ and $Z$ bosons at higher rapidity. The resummed fit also produced a 
prediction for the HERA data on $F_L(x,Q^2)$ at 
low $Q^2$\cite{Raicevic:2010zz,Reisert:2009zz,Collaboration:2010ry}. 
The results are
seen in Fig. \ref{smallxFL}. The prediction is clearly successful, and
gives additional evidence that resummation may be important, though there
are other possible ways of explaining the excess over the NLO and NNLO
perturbative predictions at small $x$, which also means 
low $Q^2$, which leads us to the next topic.

\begin{figure}
\begin{center}
\begin{minipage}[t]{9.5 cm}
\epsfig{file=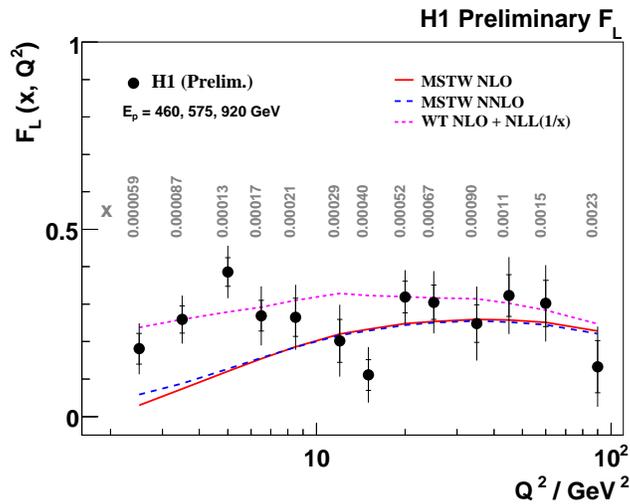,scale=0.45}
\end{minipage}
\begin{minipage}[t]{16.5 cm}
\caption{Comparison of the predictions for $F_L(x,Q^2)$ from a fit 
with $\ln(1/x)$ 
resummation (WT NLO+NLL($1/x$)) and at fixed order, either NLO or NNLO, 
to HERA data\cite{Raicevic:2010zz}.
\label{smallxFL}}
\end{minipage}
\end{center}
\end{figure}

\subsection{\it Low $Q^2$, Higher Twist, Saturation}

As noted in Section 1.2, the factorisation into coefficient functions and 
parton distributions is formally broken by corrections of 
${\cal O}(\Lambda_{\rm QCD}^2/Q^2)$. This effect is expected to be enhanced
at high values of $x$, and is related to the resummation of the
$\alpha_S^2\ln^{2n-1}(1-x)$ perturbative corrections\cite{Sterman:1986aj,
Appell:1988ie,Catani:1989ne} which is formally 
divergent, leading to an ambiguity in the perturbation series which can be
interpreted as power corrections or infrared renormalons\cite{Dokshitzer:1995qm,Dasgupta:1996hh}. There have been numerous studies of 
higher twist contributions at high $x$, e.g. \cite{Kataev:1999bp,Alekhin:2002fv,Martin:2003sk,Blumlein:2008kz,Accardi:2009br,Khorramian:2009xz}, 
and there is general 
agreement in the results. All studies find that the higher twist effects 
for $F_2$ and $F_3$ appear to be as expected from renormalon 
calculations, seen in Fig. \ref{renormalon}, i.e. 
most important for high $x$. They diminish
with the perturbative order, i.e. at lower order they are mimicking the missing
higher order effects, and appear to be stabilising in size by NNLO. Indeed, 
in those which have approximations to NNNLO\cite{Kataev:1999bp,Blumlein:2008kz,Khorramian:2009xz} there is little 
difference, within uncertainties, to the higher twist extracted at NNLO. 
Hence, the series presumably reaches maximal convergence near NNNLO at 
high $x$. For $F_3$ higher twist is a slightly larger effect at moderate 
$x\sim 0.01-0.1$, and this certainly seems to be the case in $F_L$\cite{Alekhin:2002fv,Martin:2006qv}, as predicted by the renormalon 
calculation\cite{Stein:1996wk} which is has no protection from any sum 
rule at small $x$, as the Adler sum rule provides for nonsinglet $F_2$. 
This nonsinglet higher twist correction for $F_L$ (which is unrelated to
the gluon distribution) is another possible explanation (at least in part) of the 
apparent low-$Q^2$ results in \cite{Raicevic:2010zz,Reisert:2009zz,Collaboration:2010ry}.

\begin{figure}
\begin{center}
\begin{minipage}[t]{9 cm}
\epsfig{file=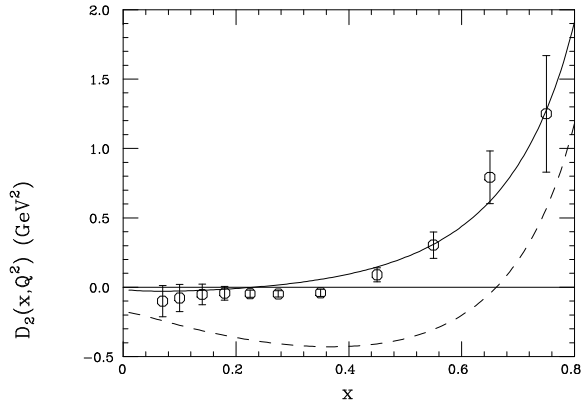,scale=0.45}
\end{minipage}
\begin{minipage}[t]{16.5 cm}
\caption{Predictions for the power corrections to $F_2(x,Q^2)$
(solid) and $F_1(x,Q^2)$ and  $F_3(x,Q^2)$ (dashed)\cite{Dasgupta:1996hh}
compared to an extraction from data. 
\label{renormalon}}
\end{minipage}
\end{center}
\end{figure}

At low $x$ there has long been an expectation that higher twist effects
related to the gluon should be important, mainly due to the fact that 
the gluon distribution is expected to be very large and hence
recombination\cite{Mueller:1985wy} and ultimately saturation effects 
should set in. Empirical investigations have suggested\cite{Martin:2003sk}
that this is not the case, with higher twist effects again diminishing
with order, but not being very significant beyond the LO estimate. 
Additionally, a study of absorptive corrections\cite{Watt:2005iu} does not
imply a very big effect. It has been suggested that this is due to 
an accidental cancellation of large terms in $F_2(x,Q^2)$, e.g. 
\cite{Bartels:2009tu}, and that large gluon induced higher twist will persist
in other structure functions. However, it may also be related to the fact
that the small-$x$ low-$Q^2$ gluon extracted from full NLO and NNLO fits
is actually not nearly as large as once expected, or often assumed in 
attempts to calculate higher twist, which are often built upon LO
perturbative expansions where the gluon is much larger.

The subject of the gluon at small-$x$, and the degree of saturation has 
become a very large topic of study, inspired by the discovery that within the 
dipole model for DIS scattering a simple model incorporating saturation could 
give a good fit to both inclusive and diffractive structure function data\cite{GolecBiernat:1998js}. This has branched out into the colour glass 
condensate\cite{Iancu:2000hn} approach to the small-$x$ gluon 
and is far too extensive a topic to summarise here.  A brief discussion and 
summary of recent results and fits using the dipole model, saturation effects
and the colour glass condensate can be found in \cite{Motyka:2008jk}, and
a more recent review of the colour glass condensate in particular can be 
found in \cite{Gelis:2010nm}. Here we simply note a couple of points. 
More recent and sophisticated treatments, including certainly
heavy quarks (sometimes missed in early studies) and impact 
parameter-dependence, seem to find saturation being associated with rather 
lower $x$ and $Q^2$. The saturation scale scale for various treatments
is illustrated in Fig. \ref{saturation} 
is seen to be at very low $x$ even for $b=0$, falling to even lower $x$ as $b$ 
rises (the average for inclusive processes is $b \sim 2-3 \GeV^{-1}$). 
For truly quantitative results it is also necessary to match the calculations
in these approaches to the PDFs obtained at higher $x$ and $Q^2$
from the reliable results using the collinear factorisation theory, 
which is by no means trivial and certainly not automatic\cite{Thorne:2005kj}. 

\begin{figure}
\begin{center}
\begin{minipage}[t]{9 cm}
\epsfig{file=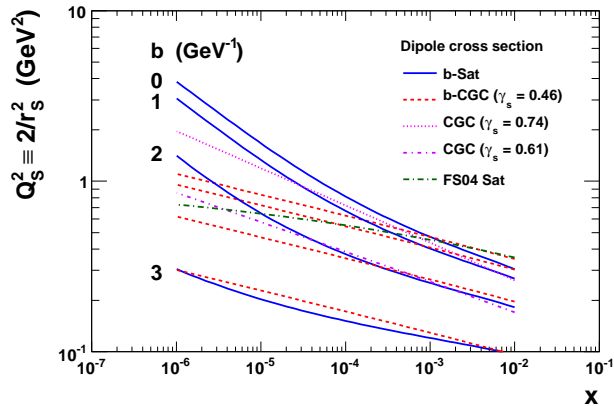,scale=0.45}
\end{minipage}
\begin{minipage}[t]{16.5 cm}
\caption{The line denoting the saturation scale as a function of $x$ and $Q^2$ for various 
approaches and different values of impact parameter. See \cite{Watt:2007nr}
for details.
\label{saturation}}
\end{minipage}
\end{center}
\end{figure}

\section{PDFs for LO Monte Carlo generators}

\begin{figure}
\begin{center}
\begin{minipage}[t]{12 cm}
\epsfig{file=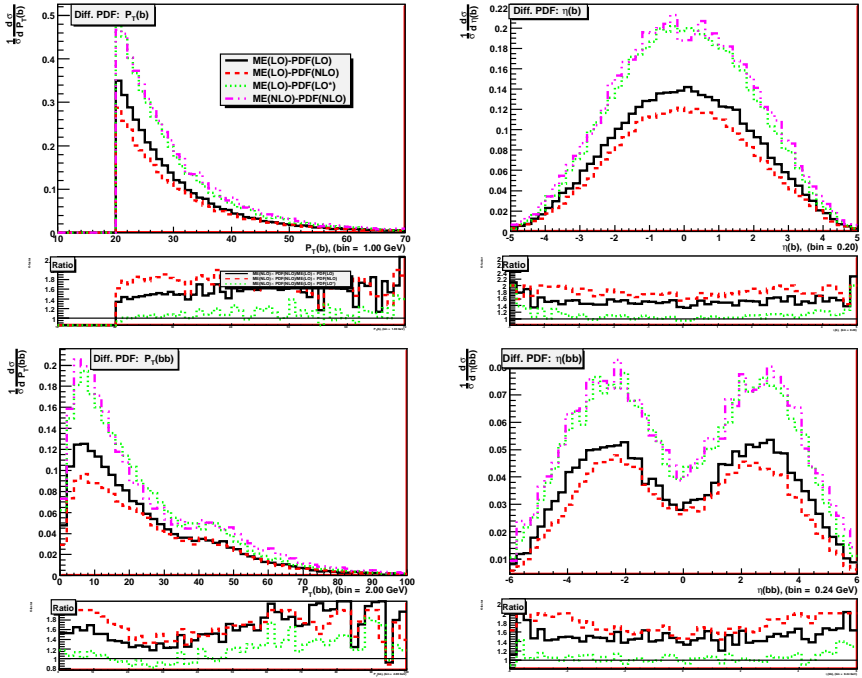,scale=0.6}
\end{minipage}
\begin{minipage}[t]{16.5 cm}
\caption{Comparison of the predictions for $b \bar b$ production using a LO generator 
and various PDFs and a NLO generator with NLO PDFs\cite{Sherstnev:2007nd}. The upper plots 
are absolute cross-sections with the variety of combinations of order of matrix element
(either ME(LO) or ME(NLO)) and type of PDF (LO, LO* and NLO). The lower plots are the
ration of the full NLO prediction to each prediction using LO matrix elements and different
PDFs.  
\label{LO*PDF}}
\end{minipage}
\end{center}
\end{figure}

A recent development in the study of PDFs 
has been the introduction of a different definition
of parton distributions generally known as modified LO PDFs. 
These have arisen due to the frequent need 
to use generators for events at particle colliders which perform 
the cross section calculation only at LO in
QCD. LO cross sections combined with 
LO PDFs is often a very inaccurate approximation, usually being rather too small
in normalisation and sometimes also with the wrong shape.
This can easily be understood if one considers that NLO matrix elements (and beyond)
often give 
large positive corrections: at small $x$ due to $1/x$ divergent terms in the matrix elements;
near threshold due to large corrections from soft-gluon emissions near the edge of 
phase space; and  there can be numerically large corrections from analytic continuation from 
the space-like to time-like region, e.g. a $(1+\alpha_S \pi C_F/2)$ factor in Drell-Yan 
production. 
Cross sections for hadro-production of $W$, $Z$, Higgs bosons, $t \bar t$, 
$b \bar b$-production and
jet production (including $W/Z+$ jets) all have NLO enhancements from
at least one of these sources. $t$-channel processes do not have these type of large 
corrections, and for e.g. single top or Higgs via vector boson fusion
the NLO matrix-element correction is small. Such processes probe partons usually in the 
range of $x=0.1$, i.e. neither very large or small.

\begin{figure}
\begin{center}
\begin{minipage}[t]{14 cm}
\epsfig{file=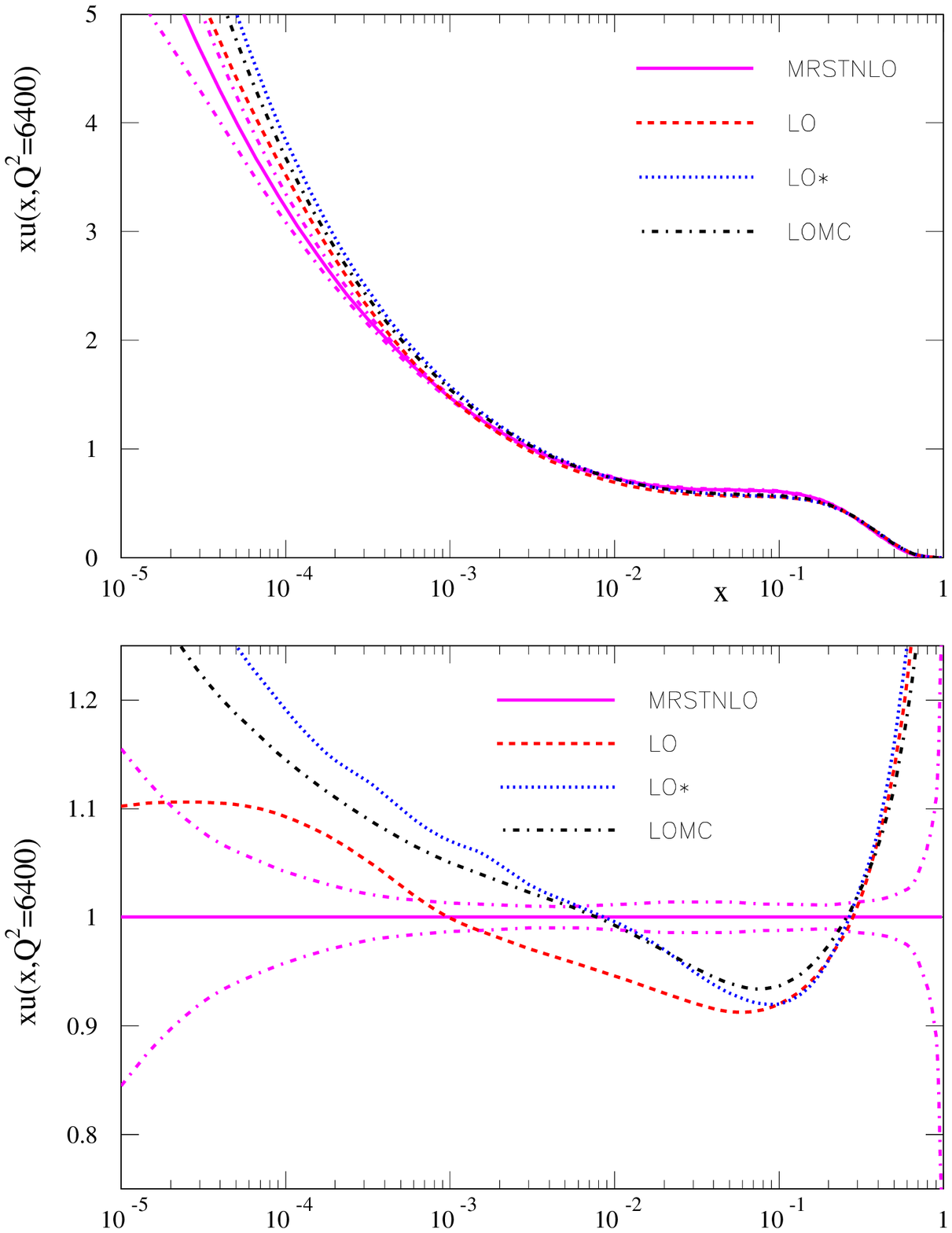,scale=0.37}
\hspace{1cm}
\epsfig{file=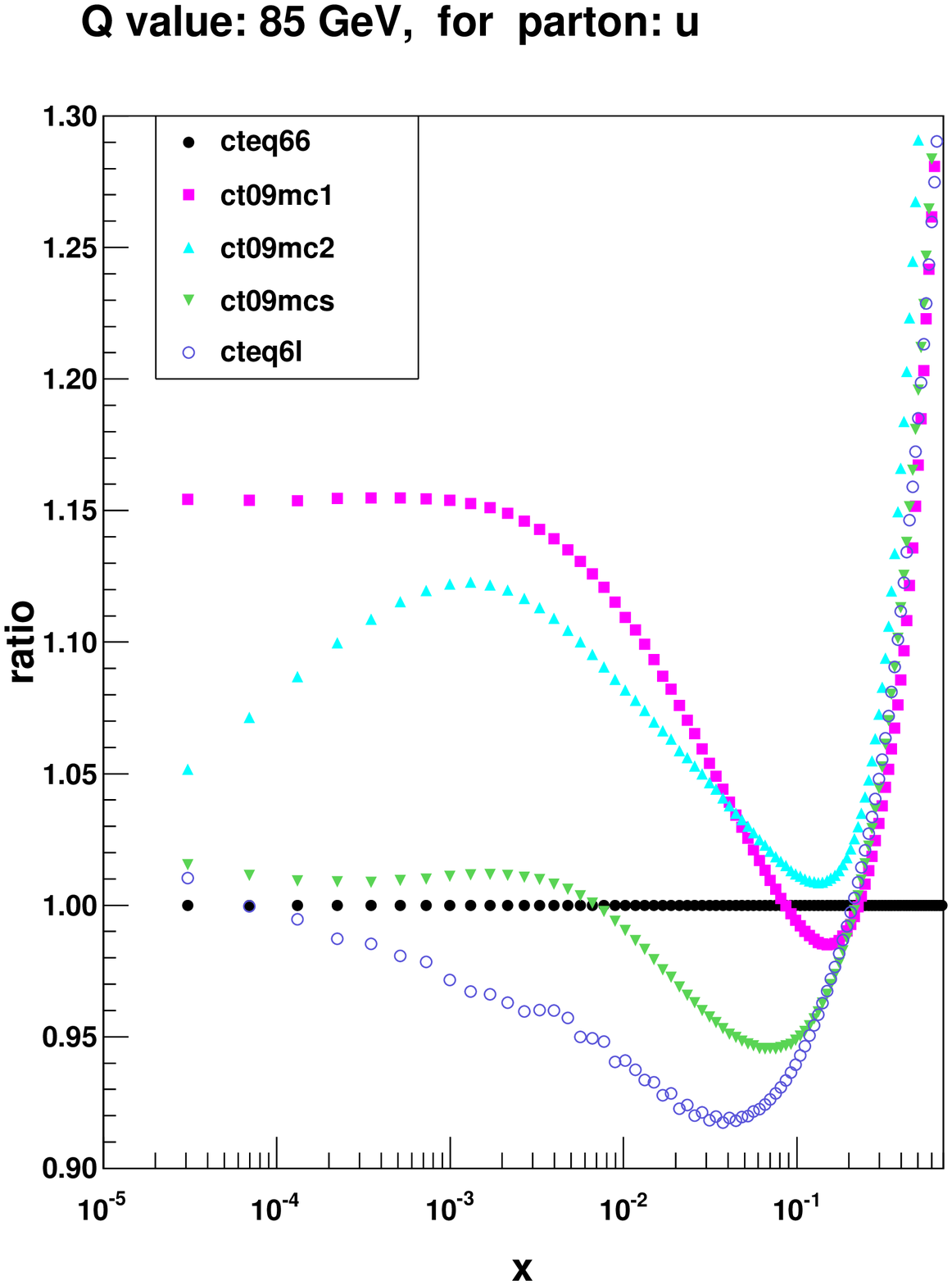,scale=0.35}
\end{minipage}
\begin{minipage}[t]{16.5 cm}
\caption{Comparison of the up quark in conventional LO and NLO sets
to the MRST LO* set\cite{Sherstnev:2007nd}, either absolute value (upper) or
ratio to NLO (lower), each with the NLO uncertainty included (left) and the CTEQ modified LO 
sets\cite{Lai:2009ne} as a ratio to NLO (right). In the latter case pink gives the MC@NLO result,
green, blue, red and black the result using a LO generator and LO*, LO**, NLO and LO PDFs
respectively. 
\label{LOmodcteq}}
\end{minipage}
\end{center}
\end{figure}

The use of NLO PDFS in LO Monte Carlo generators has been suggested to counter this,
since NLO PDFs are larger in some regions than at LO. 
Sometimes it does lead to better results, but sometimes even 
worse, particularly at small $x$ where NLO PDFs, especially the gluon distribution,
can be much smaller than at NLO because they have been extracted with large 
positive contributions to quark evolution included at NLO.
An alternative argument is that rather than the normal fixed order definitions
this situation requires the introduction of new type of modified LO (LO*) PDF\cite{Sherstnev:2007nd}.
These allow the LO PDFs to be generally bigger by allowing momentum 
violation in global fits performed at LO, and can also use the NLO 
definition of $\alpha_S$, which is larger at low scales where much of 
the DIS data is fit than a LO $\alpha_S$ with the same value at $m_Z^2$. 
As a further development one
can also make the evolution more ``Monte Carlo like'', by changing the 
renormalisation scale in the coupling from $Q^2$ to something more like $k_T^2$ 
resulting in the LO**
distributions\cite{Sherstnev:2008dm}. 
In both cases the PDFs are obtained entirely from a fit to existing data, the quality
of the LO* fit being much better than the rather poor LO result, but not quite as 
good as NLO. 
It was hoped that in the modified PDFs the enhancement in the partons
compared to standard LO should compensate to some extent the missing NLO
enhancements in the matrix elements. 
In\cite{Sherstnev:2007nd} there was an extensive investigation of whether
this idea works in practice. 
Comparison was made between predictions for a wide variety
of processes made using the MC@NLO generator\cite{Frixione:2002ik}, which combines NLO matrix elements with parton showering, 
using NLO PDFs, and predictions using LO generators and LO, NLO and LO* PDFs. 
Taking the MC@NLO results to be the most accurate representation it was found that 
for the vast majority of cases the LO* gave the best results for the 
LO generators, particularly for gluon initiated processes, while sometimes 
standard LO and sometimes NLO gave the worst results. 
As an example we see in Fig. \ref{LO*PDF} the final state 
distributions for single $b$  and $b \bar b$ pairs\cite{Sherstnev:2007nd},
where the results using the LO* PDFs are almost identical to NLO.

The LO* MRST PDFs have been followed by parton distributions for 
event generators from the CTEQ collaboration\cite{Lai:2009ne}. The reasoning 
for the need for
these PDFs is the same as for the LO* sets. However, the manner of obtaining 
them follows some of the same principles, but also has some different ones. 
Various sets have been produced, CT09MCS, CT09MC1 and CT09MC2. 
Some (CT09MC1 and CT09MC2) do allow the violation of the momentum sum
rule and in CT09MC2 the NLO definition of the coupling is used. 
A major difference is that the PDFs are obtained by fits also including LHC ``pseudodata''
generated using full NLO calculations. This is different in philosophy to the LO*
sets. It is noted that there is significant tension between the best fits to
pseudodata and to the existing, largely structure function data.  
There is no modification 
of the scale of the coupling 
to make it more Monte Carlo like, but scales are varied to obtain 
the best quality fit to the pseudodata. An example of the comparison of the modified 
PDFs to standard LO and NLO PDFs is shown for the up quark in the right of 
Fig. \ref{LOmodcteq}. Clearly when momentum is violated 
the CT09MC quark is much larger than the fixed order over most
$x$ values. This is unlike the MRST LO*/LO** quark in the left of    
Fig. \ref{LOmodcteq}, which is constrained at $x=0.01$ to give a good fit to 
high-$Q^2$ HERA data. Inclusion of existing Tevatron data on $Z$ rapidity,
which postdates the LO* set would add some tension and should 
raise the quark distribution in this region a small amount. The gluon distributions 
in the two approaches are more similar, both being much bigger than fixed order at 
small $x$. This is not surprising since the lack of direct constraint on 
the gluon distribution renders the differences in the approaches less important. 

\begin{figure}
\begin{center}
\begin{minipage}[t]{15 cm}
\epsfig{file=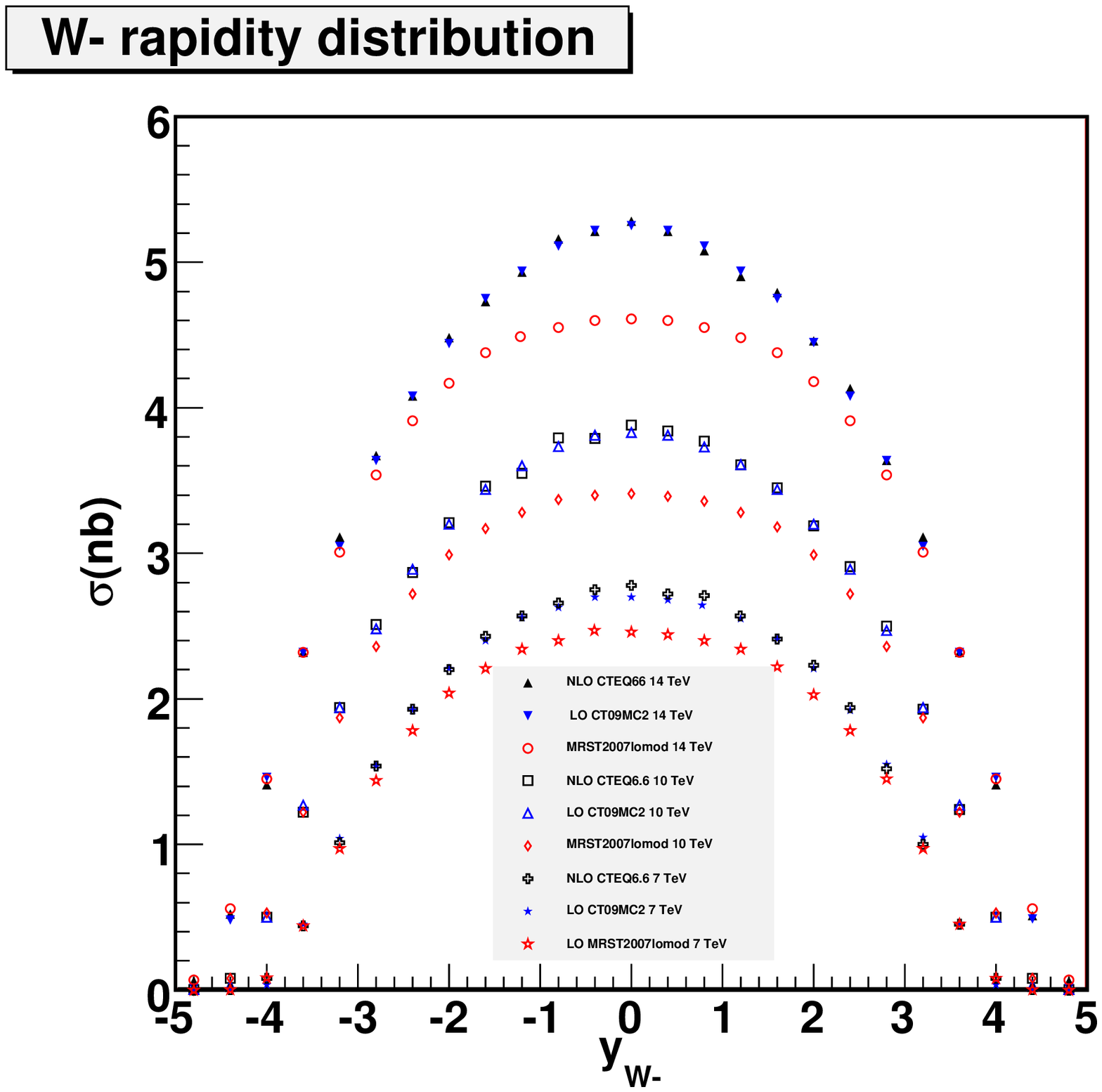,scale=0.38}
\epsfig{file=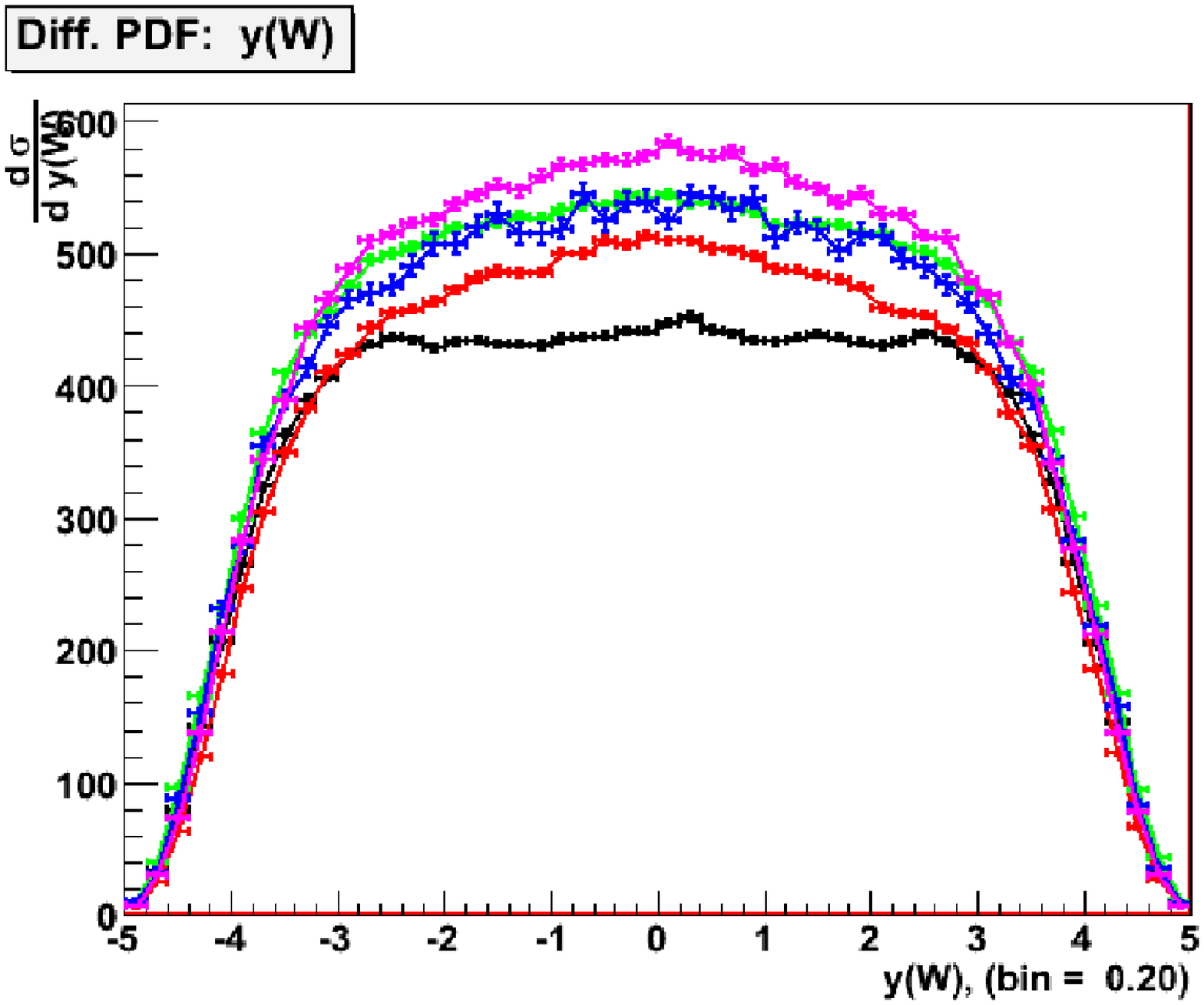,scale=0.38}
\end{minipage}
\begin{minipage}[t]{16.5 cm}
\caption{Comparison of the W rapidity using conventional LO and NLO sets
that using the modified LO sets from CTEQ\cite{Lai:2009ne} 
(left) and MRST\cite{Sherstnev:2008dm} (right).
\label{LOmodWcomp}}
\end{minipage}
\end{center}
\end{figure}

The comparison of the two types of PDF for gluon initiated processes is very similar. 
There is more difference in quark driven processes, such as $W$ and $Z$ production. 
An example is shown in Fig. \ref{LOmodWcomp} which shows the $W$ rapidity distribution. 
The left-hand figure is from \cite{Lai:2009ne}, and shows excellent agreement using 
CT09MC2, which is not surprising since the PDF as been obtained by fitting data of this type 
very well. The MRST LO* looks less successful in this plot. 
However, these results are from inclusive calculation with no parton showering as applied in 
generators. This can have a few percent effect, as seen in Section 3 of \cite{Lai:2009ne},
and automatically includes some higher orders in the LO calculation.
The right-hand plot from  \cite{Sherstnev:2008dm} does include parton showering 
in both LO and NLO calculations. The LO* results are somewhat nearer to NLO in this case
since the LO calculation is a little nearer to NLO in this framework. Applying parton showering 
to the left hand plot must, on this evidence, improve the LO* comparison, and will affect the 
CT09MC2 comparison to some extent as well. There is no comparison to $t$-channel processes in 
\cite{Lai:2009ne}. In \cite{Sherstnev:2007nd} the enhancement of the PDFs led to these 
being marginally worse than LO PDFs. Even further enhancement is unlikely to be helpful. 

An entirely different alternative is to obtain PDFs from fits using Monte Carlo generators
directly. In detail this will then produce a slightly different PDF set for each 
generator, the details of parton showering differing between each. Work on this 
approach has been ongoing. However, it is rather more time intensive than 
normal fits, though there have been helpful developments\cite{Bacchetta:2010hh}, and results
are so far limited. As noted near the beginning of Section 2 of this article, a very wide variety of 
data needs to be fit to provide true constraints on any PDFs, so it is not clear if this approach
will lead to useful results in the immediate future.

\section{Outlook}
This review demonstrates the vast amount of 
progress that has taken place
in the last years on pinning down the PDFs of the proton, as well as
the dramatic increase in 
awareness of the impact of PDFs on  the physics program
of  LHC experiments. LHC will need the best PDFs,
 especially for precision measurements, setting of limits in searches, 
and even for discoveries.
Ideally the ATLAS and CMS (and LHCb and ALICE) analyses should follow a common
procedure for using PDFs and their uncertainties in their key analyses. 
Also, changing frequently the PDFs in the software of the experiments, 
e.g. for cross--checks or the determination  of
error bands, is often non-trivial (e.g. due to the inter-connection 
with parameter choices for underlying event modelling, showering
parameters and so on) and sometimes impractical if CPU intensive 
detector simulations are involved. 
LHC studies therefore will need both good central values for the PDFs 
to start with, and a
good estimate of the associated uncertainties. 

This has triggered the so called PDF4LHC initiative. PDF4LHC offers a 
discussion forum for PDF studies and information exchange between all 
stake-holders in the field. More details and links to the meetings so far
can be found on the PDF4LHC web site~\cite{pdf4lhcweb}.
Apart from getting the best PDFs, including the PDF uncertainties, based on the 
present data, another important deliverable is to devise strategies to use future LHC data to improve the PDFs.
All this needs a close collaboration between
theorists and those that are preparing to make the measurements.
Such measurements include $W$ and $Z$ production and asymmetries, di-jet 
production, hard prompt photons, Drell-Yan production, 
bottom  and top quark production, Z-shape fits and Z+jets measurements.
One expects that some of these channels can 
already be studied  with first data at the LHC.

The final HERA data of run II (2004-2007) are still being analysed and will be very instrumental for future 
PDF fits, particularly for the high $Q^2$ region. These data will become available in the next few years.
Meanwhile interest is growing worldwide for a novel electron-ion collider. Design concepts exist at CERN, with ideas to intersect an electron accelerator with the Large Hadron Collider (LHeC)\cite{Newman:2009mb}, and in the U.S. to either add an electron accelerator to the Relativistic Heavy Ion Collider  at BNL, or an ion accelerator to the upgraded 12-GeV Continuous Electron Beam Accelerator Facility at JLab. The US project is generically called the EIC\cite{Deshpande:2008zz}.

\begin{figure}[t]
\centerline{\includegraphics[width=0.35\columnwidth]{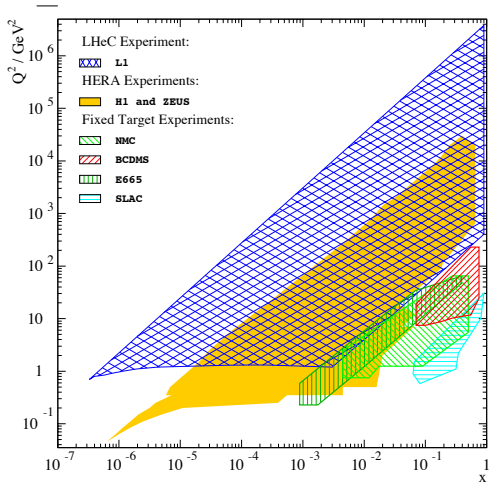}}
\caption{The kinematic reach for LHeC}\label{Fig:LHeC_kinem}
\end{figure}

The LHeC has two alternative scenarios: a ring-ring (RR) scenario and a linac-ring (LR) scenario. For 
the RR scenario one typically  has a 50 GeV electron beam on a 7 TeV proton beam (or 2.75 TeV heavy ion 
beam), and a peak luminosity of around $5 . 10^{33}$ cm$^{-2}$s$^{-1}$ for 50 MW power.
The LR scenario has the  potential to reach larger electron energies, perhaps up to 150 GeV, but in general
the total integrated luminosity will by a factor 5 to 10 lower compared to the RR option.

The EIC projects discussed by BNL and JLab describe an electron beam of 4 to 20 GeV on a proton beam of
50 to 250 GeV. The peak luminosity aimed for is similar to the RR LHeC option. Polarization is an integral part of the proposal, aiming for 70\% of polarization for each beam.

The kinematic reach covered by the LHeC is shown in Fig.~\ref{Fig:LHeC_kinem} and for the 
EIC in Fig.~\ref{Fig:EIC_kinem}. 
The high energy of the LHeC will 
allow to explore a new kinematic area for $ep$ collisions, with $Q^2$ values and $x$ values down to 
a few times $10^{-7}$. The high luminosity anticipated for the EIC and the possibility for polarized beams will allow for number a precision and novel measurements.
It will however take us into the next decade before any future DIS data with much higher precision or
larger kinematic domain will be available.
 
Further possible future constraints may come from the JLab experiments for the high-$x$ range with the new 
high intensity 12 GeV electron beam upgrade.
The MINERvA experiment will use neutrino beams on nuclear targets at FNAL and is set to make 
 precise measurements of neutrino cross sections with neutrino beams of energies up to roughly 30 GeV.
 MINERvA will collect 6M events on a carbon target in the transition (not so deep DIS) and DIS region plus an additional  6.5 M events in four nuclear targets and
 will significantly increase the existing neutrino data set available to the community.
E906 is a new experiment at FNAL and is set to measure Drell-Yan production via muons.  E906 will measure 
 the ratio of the $\overline{d}$  to $\overline{u}$ distributions in the proton and the modifications to the quark sea in a nucleus. The expected statistics that will be collected is a factor 50 larger than that of E866/NuSea. 
 The incoming proton beam energy will be only 120 GeV, reducing in the energy squared $s$ by a factor of 7 with respect to  E866/NuSea. Thus
 the measurements of E906 will cover a different kinematic range, namely
 up to  high $x$  values of 0.5. The run of E906 is scheduled to start in 2010 and to last for 2 years.

All these new data. together with ongoing theory developments  will ensure that further improvements on the 
understanding and precision of the protons structure will continue in the next 1-2 decades.

\begin{figure}[t]
\centerline{\includegraphics[width=0.48\columnwidth]{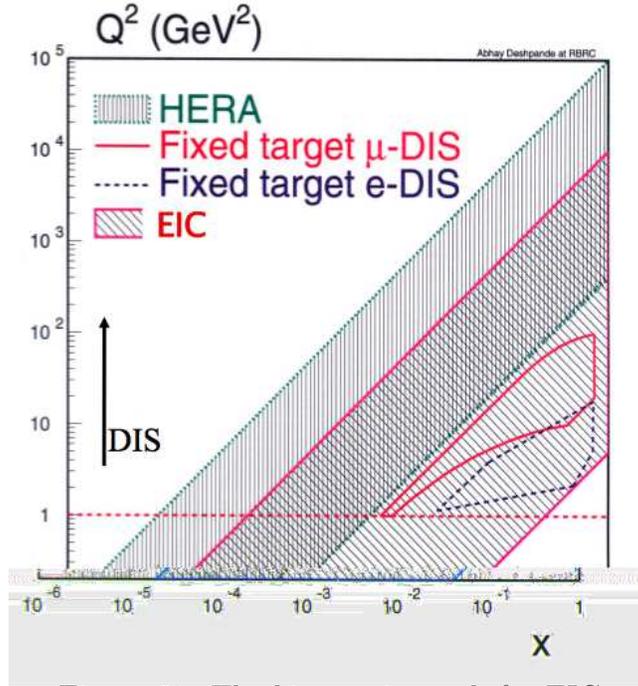}}
\caption{The kinematic reach for  EIC}\label{Fig:EIC_kinem}
\end{figure}

\section{Conclusions}

Structure function measurements are a measure of the partonic structure of the proton and are 
instrumental in parton distribution function fits. The PDFs allow us to predict cross sections at particle colliders and
a good knowledge of PDFs and their 
uncertainties is of prime importance for the success of the physics program of  e.g. proton-proton collisions 
at the newly commissioned collider LHC.

Several versions of fits to data relevant for the hadron structure exist and all show that overall good
quality fits  using NLO or NNLO QCD can be obtained.  Apart from the central values of the PDFs, 
it is essential to have also a good understanding of the uncertainties. 
Various ways of looking at uncertainties have been discussed in this paper, as used by the different
fitting groups. The uncertainties are naively rather small   
-- {$\sim 1-5 \%$} for a large number of PDFs and the predicted LHC quantities. Measurements of ratios, e.g.  
$W^+/W^-$ ratios can give extremely tight constraint on partons distributions. 
However, there are many effects in the fitting  procedure that
can contribute to the uncertainties, e.g.  
effects from input assumptions, in particular the selection of data fitted, cuts applied to the data  
and input parameterisation choices
can shift the central values of predictions significantly and affect the size of uncertainties. 
During the last years it has also become clear that a complete heavy flavour 
treatment is essential in the extraction and use of the  PDFs. 
Furthermore  PDFs and $\alpha_S$ are correlated and the uncertainties must be considered in tandem.    
These are all part of the Fixed Order QCD analysis, but there are additional effects.
Electroweak corrections are often neglected but can potentially be large at very high energies
following $\ln^2(E^2/M_W^2)$. Equally care must be taken for 
errors from higher orders/resummation, and power corrections/higher twist, 
which can be potentially large in certain phase space regions. 
Direct measurement of $F_L(x,Q^2)$ at HERA now give us some scope to test these predictions.

Since there is a spread in the PDFs obtained by the ``global'' fits which is indeed 
actually somewhat larger than the quoted uncertainties of each some procedure is required to estimate
the ``true'' uncertainty associated with PDFs. If one is comparing to a measured cross-section
then, of course, the comparison can and should be made to the prediction using any PDF set. 
In fact it is ideal to check as widely as possible to help determine which PDFs are most 
accurate in their predictions. However, if one is trying to determine the best value and 
uncertainty on a prediction in order to help set limits, determine the significance of 
a signal, or estimate uncertainty from extrapolation into certain regions of phase space, there
is the need for some recommendation for a {\it best} prediction and a conservative, but 
representative uncertainty.  This was requested, for example, for making benchmark 
predictions for Higgs boson cross-sections\cite{LHCHiggsCrossSectionWorkingGroup:2011ti},
and has been a major focus of the efforts of
the afore-mentioned PDF4LHC group. If very high or very small $x$ PDFs are probed the lack of
significant data constraints in some PDFs can lead to big variations -- the minimum variation seems 
to be for $x=0.01-0.001$, since all fits include HERA data, but also sum rules impose crossing points.
However, the spread of the predictions\cite{Wattplots,Alekhin:2011sk,Alekhin:2010dd} can be  
significant, even when the 
PDFs probed are in this $x$ range, as seen in
Fig \ref{NNPDFCTEQMSTW}. Hence, it is the interim PDF4LHC 
recommendation\cite{pdf4lhcweb,Botje:2011sn} that at NLO
a conservative uncertainty should be the envelope of the predictions using NNPDF2.0, CTEQ6.6 and 
MSTW08 PDFs and their uncertainty, including that due to variations in $\alpha_S(m_Z^2)$. The centre 
of the envelope can be taken as the best prediction.  
An example of predictions using this is seen in Table~\ref{tab:LHCcs}.
Since of these three sets MSTW is the one available at NNLO at present the central value should be taken from this, but the
fractional uncertainty should be the same as at NLO in order to be conservative, also seen in 
Table~\ref{tab:LHCcs}. 
NNLO sets are available from ABKM, JR and most recently HERA, as described earlier in this article, and so ideally checks should also be made against these, though some deviations can be large. 
As seen the NNLO corrections themselves can be large for some processes.
This recommendation is to be viewed as temporary and updates are expected.   
Another aim of the PDF4LHC group is to investigate, and hopefully minimise the spread between different sets, 
or at least to understand it as fully as possible. This will be aided by additional data as well as 
theoretical improvements, and will also inform future recommendations.   

\begin{table}
\begin{center}
\begin{minipage}[t]{16.5 cm}
\caption{Cross-sections predictions and uncertainties at the LHC at NLO 
and NNLO using
the PDF4LHC prescription for the central value and uncertainties.
(Note the NNLO result for the $t \bar t$ cross-section in the table uses the NLO matrix elements. 
An approximation including a variety of NNLO contributions is available in \cite{Aliev:2010zk},
but if NNLO PDFs are used the correction to NLO at the LHC is only a couple of percent, small compared to the uncertainty, though is larger at the Tevatron.)}
\label{tab:LHCcs}
\vspace{0.4cm}
\end{minipage}
\begin{tabular}{|r|r|r|r|r|r|r|}
\hline
%\hline
$\sigma(W^+)$ nb & $\sigma(W^-)$ nb & $\sigma(Z)$ nb& $\sigma(t\bar t)$ pb 
& $\sigma(H_{120})$ pb&
$\sigma(H_{180})$ pb&  $\sigma(H_{240})$ pb\\
\hline
\multicolumn{7}{c} {7 TeV centre of mass energy  (NLO)}\\
\hline
$54.7\pm 2.5$ & $37.8 \pm 2.0$ & $27.6 \pm 1.1$ & $162 \pm 14$ & $12.06 
\pm 0.75 $ & $5.04 \pm 0.32$ & $ 2.75 \pm 0.20$ \\
\hline
\multicolumn{7}{c} {7 TeV centre of mass energy (NNLO)}\\
\hline
$56.9\pm 2.6$ & $40.0 \pm 2.1$ & $28.9 \pm 1.2$ & $169^{*} \pm 15$   & $15.7  
\pm 0.98$ & $6.53 \pm 0.41$ & $3.52 \pm 0.26$  \\
\hline
%\multicolumn{7}{c} {14TeV centre of mass energy}\\
%\hline
%$111.5\pm 6.6$ & $82.0 \pm 4.7$ & $59.2 \pm 2.8$ & $915 \pm 60$ & $37.4 
%\pm 2.1 $ & $16.6 \pm 0.90$ & $ 8.87 \pm 0.49$ \\
%\hline
%\hline
\end{tabular}
\end{center}
%\caption{W cross sections using the PDF4LHC prescription calculating the uncertainties}
%\label{cross-sections}
\end{table}

Hence, as well as testing for Beyond the Standard Model Physics the LHC can add 
significantly to our knowledge of the proton structure especially at 
low $x$ through measurement at high rapidities of hard probes, e.g. $W, Z$ 
and Drell-Yan events. Indeed, the former is largely a precursor for the latter. 
The extraction of PDFs from existing data and use for LHC  is still a far from 
straightforward procedure. We currently have the bare minimum of constraints required for 
constraining all PDFs over the full range required, and the a full understanding
of PDF uncertainties related to experimental errors is still being developed. 
In addition there are many theoretical issues to consider to obtain real  
precision in some cases. At the LHC there will be relatively few cases where 
Standard Model discrepancies will not require some significant, and decidedly nontrivial
input from PDF physics to determine their true significance. 

\newpage

\section*{Acknowledgements}

RST would like to thank Alan Martin, Dick Roberts, James Stirling and Graeme 
Watt for many discussions on this subject during collaboration. Both authors 
would like to thank  the members of the PDF4LHC steering committee for 
fruitful interaction over the past few years, as well as numerous other
colleagues. RST would like to thank the IPPP, Durham for the award of a 
Research Associateship to aid travel funding.


\begin{thebibliography}{99}
\itemsep -2pt 

\bibitem{Breidenbach:1969kd} M. Breidenbach, et al., 
{\it Phys. Rev. Lett.} 23 (1969) 935


\bibitem{Eichten:1984eu}
E. Eichten, I. Hinchliffe, K. D. Lane,  C. Quigg,
{\it Rev. Mod. Phys.} 56 (1984) 579

\bibitem{Morfin:1990ck}
J. G. Morfin, W. Tung
{\it Z. Phys.} C 52 (1991) 13

\bibitem{Martin:1988nk}
A. D. Martin, R.G. Roberts,  W.J. Stirling,
{\it Phys. Lett.} B206 (1988) 327


%\cite{Bjorken:1968dy}
\bibitem{Bjorken:1968dy}
  J.~D.~Bjorken,
  %``Asymptotic Sum Rules At Infinite Momentum,''
  {\it Phys.\ Rev.\ } 179 (1969) 1547
  %%CITATION = PHRVA,179,1547;%%

%\cite{Feynman}
\bibitem{Feynman}
    R.~P.~Feynman, {\it Photon hadron Interactions}, W.A. Benjamin, New York 
(1972)  

%\cite{Miller:1971qb}
\bibitem{Miller:1971qb}
  G.~Miller {\it et al.},
  %``Inelastic Electron-Proton Scattering At Large Momentum Transfers,''
  {\it Phys.\ Rev.\ }  D 5 (1972) 528
  %%CITATION = PHRVA,D5,528;%%

%\cite{Callan:1969uq}
\bibitem{Callan:1969uq}
  C.~G.~.~Callan and D.~J.~Gross,
  %``High-energy electroproduction and the constitution of the electric
  %current,''
{\it Phys.\ Rev.\ Lett.\ }  22 (1969) 156
  %%CITATION = PRLTA,22,156;%%

%\cite{Ellis:1982cd}
\bibitem{Ellis:1982cd}
  R.~K.~Ellis, W.~Furmanski and R.~Petronzio,
  %``Unraveling Higher Twists,''
 {\it Nucl.\ Phys.\ }  B { 212} (1983) 29
  %%CITATION = NUPHA,B212,29;%%


%\cite{Altarelli:1977zs}
\bibitem{Altarelli:1977zs}
  G.~Altarelli and G.~Parisi,
  %``Asymptotic Freedom In Parton Language,''
  {\it Nucl.\ Phys.\ }  B { 126} (1977) 298
  %%CITATION = NUPHA,B126,298;%%


%\cite{Lipatov:1974qm}
\bibitem{Lipatov:1974qm}
  L.~N.~Lipatov,
  %``The parton model and perturbation theory,''
  {\it Sov.\ J.\ Nucl.\ Phys.\ }  { 20} (1975) 94
  [{\it Yad.\ Fiz.\ }  {20} (1974) 181]
  %%CITATION = YAFIA,20,181;%%

%\cite{Gribov:1972ri}
\bibitem{Gribov:1972ri}
  V.~N.~Gribov and L.~N.~Lipatov,
  %``Deep Inelastic E P Scattering In Perturbation Theory,''
  {\it Sov.\ J.\ Nucl.\ Phys.\ }  { 15} (1972) 438
  [{\it Yad.\ Fiz.\ }  { 15} (1972) 781]
  %%CITATION = YAFIA,15,781;%%

%\cite{Dokshitzer:1977sg}
\bibitem{Dokshitzer:1977sg}
  Y.~L.~Dokshitzer,
  %``Calculation Of The Structure Functions For Deep Inelastic Scattering And E+
  %E- Annihilation By Perturbation Theory In Quantum Chromodynamics,''
  {\it Sov.\ Phys.\ JETP}  46 (1977) 641
  [{\it Zh.\ Eksp.\ Teor.\ Fiz.\ }  {73} (1977) 1216]
  %%CITATION = ZETFA,73,1216;%%

%\cite{Gross:1973ju}
\bibitem{Gross:1973ju}
  D.~J.~Gross and F.~Wilczek,
  %``Asymptotically Free Gauge Theories. 1,''
 {\it  Phys.\ Rev.\ }  D { 8} (1973) 3633
  %%CITATION = PHRVA,D8,3633;%%

%\cite{Georgi:1951sr}
\bibitem{Georgi:1951sr}
  H.~Georgi and H.~D.~Politzer,
  %``Electroproduction scaling in an asymptotically free theory of strong
  %interactions,''
  {\it Phys.\ Rev.\ }  D {9} (1974) 416
  %%CITATION = PHRVA,D9,416;%%

%\cite{Bardeen:1978yd}
\bibitem{Bardeen:1978yd}
  W.~A.~Bardeen, A.~J.~Buras, D.~W.~Duke and T.~Muta,
  %``Deep Inelastic Scattering Beyond The Leading Order In Asymptotically Free
  %Gauge Theories,''
 {\it  Phys.\ Rev.\ } D {18} (1978) 3998
  %%CITATION = PHRVA,D18,3998;%%

%\cite{Floratos:1977au}
\bibitem{Floratos:1977au}
  E.~G.~Floratos, D.~A.~Ross and C.~T.~Sachrajda,
  %``Higher Order Effects In Asymptotically Free Gauge Theories: The Anomalous
  %Dimensions Of Wilson Operators,''
 {\it  Nucl.\ Phys.\ }  B {129} (1977) 66
  [{\it Erratum-ibid.\ }  B {139} (1978) 545]
  %%CITATION = NUPHA,B129,66;%%

%\cite{Floratos:1978ny}
\bibitem{Floratos:1978ny}
  E.~G.~Floratos, D.~A.~Ross and C.~T.~Sachrajda,
  %``Higher Order Effects In Asymptotically Free Gauge Theories. 2. Flavor
  %Singlet Wilson Operators And Coefficient Functions,''
  {\it Nucl.\ Phys.\ }  B {152} (1979) 493
  %%CITATION = NUPHA,B152,493;%%

%\cite{GonzalezArroyo:1979df}
\bibitem{GonzalezArroyo:1979df}
  A.~Gonzalez-Arroyo, C.~Lopez and F.~J.~Yndurain,
  %``Second Order Contributions To The Structure Functions In Deep Inelastic
  %Scattering. 1. Theoretical Calculations,''
  {\it Nucl.\ Phys.\ }  B {153} (1979) 161
  %%CITATION = NUPHA,B153,161;%%

%\cite{Curci:1980uw}
\bibitem{Curci:1980uw}
  G.~Curci, W.~Furmanski and R.~Petronzio,
  %``Evolution Of Parton Densities Beyond Leading Order: The Nonsinglet Case,''
  {\it Nucl.\ Phys.\ }  B { 175} (1980) 27
  %%CITATION = NUPHA,B175,27;%%

%\cite{GonzalezArroyo:1979he}
\bibitem{GonzalezArroyo:1979he}
  A.~Gonzalez-Arroyo and C.~Lopez,
  %``Second Order Contributions To The Structure Functions In Deep Inelastic
  %Scattering. 3. The Singlet Case,''
  {\it Nucl.\ Phys.\ }  B {166} (1980) 429
  %%CITATION = NUPHA,B166,429;%%

%\cite{Furmanski:1980cm}
\bibitem{Furmanski:1980cm}
  W.~Furmanski and R.~Petronzio,
  %``Singlet Parton Densities Beyond Leading Order,''
  {\it Phys.\ Lett.\ }  B {97} (1980) 437
  %%CITATION = PHLTA,B97,437;%%

%\cite{Floratos:1981hs}
\bibitem{Floratos:1981hs}
  E.~G.~Floratos, C.~Kounnas and R.~Lacaze,
  %``Higher Order QCD Effects In Inclusive Annihilation And Deep Inelastic
  %Scattering,''
 {\it  Nucl.\ Phys.\ }  B {192} (1981) 417
  %%CITATION = NUPHA,B192,417;%%

%\cite{Collins:1989gx}
\bibitem{Collins:1989gx}
  J.~C.~Collins, D.~E.~Soper and G.~Sterman,
  %``Factorisation of Hard Processes in QCD,''
  {\it Adv.\ Ser.\ Direct.\ High Energy Phys.\ } {5} (1988) 1
  [arXiv:hep-ph/0409313]
  %%CITATION = 00319,5,1;%%

%\cite{Altarelli:1978id}
\bibitem{Altarelli:1978id}
  G.~Altarelli, R.~K.~Ellis and G.~Martinelli,
  %``Leptoproduction And Drell-Yan Processes Beyond The Leading Approximation In
  %Chromodynamics,''
  {\it Nucl.\ Phys.\ }  B {143} (1978) 521
  [{\it Erratum-ibid.\ }  B {146} (1978) 544]
  %%CITATION = NUPHA,B143,521;%% 


%\cite{Martin:2009iq}
\bibitem{Martin:2009iq}
  A.~D.~Martin, W.~J.~Stirling, R.~S.~Thorne and G.~Watt,
  %``Parton distributions for the LHC,''
  {\it Eur.\ Phys.\ J.\ } C {63} (2009) 189
  [arXiv:0901.0002 [hep-ph]]
  %%CITATION = EPHJA,C63,189;%%


%\cite{Nadolsky:2008zw}
\bibitem{Nadolsky:2008zw}
  P.~M.~Nadolsky {\it et al.},
  %``Implications of CTEQ global analysis for collider observables,''
 {\it  Phys.\ Rev.\ } D {78}, (2008) 013004 
  [arXiv:0802.0007 [hep-ph]]
  %%CITATION = PHRVA,D78,013004;%%

%\cite{Ball:2010de}
\bibitem{Ball:2010de}
  R.~D.~Ball, L.~Del Debbio, S.~Forte, A.~Guffanti, J.~I.~Latorre, J.~Rojo and M.~Ubiali,
  %``A first unbiased global NLO determination of parton distributions and their
  %uncertainties,''
  {\it Nucl.\ Phys.\ }  B {838} (2010) 136
  [arXiv:1002.4407 [hep-ph]]
  %%CITATION = NUPHA,B838,136;%%

%\cite{:2009wt}
\bibitem{:2009wt}
  F.~D.~Aaron {\it et al.}  [H1 Collaboration and ZEUS Collaboration],
  %``Combined Measurement and QCD Analysis of the Inclusive ep Scattering Cross
  %Sections at HERA,''
 {\it  JHEP} {1001} (2010) 109
  [arXiv:0911.0884]
  %%CITATION = JHEPA,1001,109;%%



%\cite{Alekhin:2009ni}
\bibitem{Alekhin:2009ni}
  S.~Alekhin, J.~Blumlein, S.~Klein and S.~Moch,
  %``The 3-, 4-, and 5-flavor NNLO Parton from Deep-Inelastic-Scattering Data
  %and at Hadron Colliders,''
  {\it Phys.\ Rev.\ } D {81} (2010) 014032
  [arXiv:0908.2766 [hep-ph]]
  %%CITATION = PHRVA,D81,014032;%%



%\cite{Gluck:2007ck}
\bibitem{Gluck:2007ck}
  M.~Gluck, P.~Jimenez-Delgado and E.~Reya,
  %``Dynamical parton distributions of the nucleon and very small-x physics,''
  {\it Eur.\ Phys.\ J.\ } C {53} (2008) 355
  [arXiv:0709.0614 [hep-ph]]
  %%CITATION = EPHJA,C53,355;%%

%\cite{vanNeerven:1991nn}
\bibitem{vanNeerven:1991nn}
  W.~L.~van Neerven and E.~B.~Zijlstra,
  %``Order Alpha-S**2 Contributions To The Deep Inelastic Wilson Coefficient,''
  {\it Phys.\ Lett.\ } B {272} (1991) 127
  %%CITATION = PHLTA,B272,127;%%

%\cite{Zijlstra:1991qc}
\bibitem{Zijlstra:1991qc}
  E.~B.~Zijlstra and W.~L.~van Neerven,
  %``Contribution Of The Second Order Gluonic Wilson Coefficient To The Deep
  %Inelastic Structure Function,''
{\it   Phys.\ Lett.\ } B {273} (1991) 476
  %%CITATION = PHLTA,B273,476;%%

%\cite{Zijlstra:1992kj}
\bibitem{Zijlstra:1992kj}
  E.~B.~Zijlstra and W.~L.~van Neerven,
  %``Order alpha-s**2 correction to the structure function F3 (x, Q**2) in deep
  %inelastic neutrino - hadron scattering,''
 {\it  Phys.\ Lett.\ } B {297} (1992) 377
  %%CITATION = PHLTA,B297,377;%%

%\cite{Zijlstra:1992qd}
\bibitem{Zijlstra:1992qd}
  E.~B.~Zijlstra and W.~L.~van Neerven,
  %``Order Alpha-S**2 QCD Corrections To The Deep Inelastic Proton Structure
  %Functions F2 And F(L),''
 {\it  Nucl.\ Phys.\ } B {383} (1992) 525
  %%CITATION = NUPHA,B383,525;%%



%\cite{Moch:2004xu}
\bibitem{Moch:2004xu}
  S.~Moch, J.~A.~M.~Vermaseren and A.~Vogt,
  %``The longitudinal structure function at the third order,''
 {\it  Phys.\ Lett.\ } B {606} (2005) 123
  [arXiv:hep-ph/0411112]
  %%CITATION = PHLTA,B606,123;%%

%\cite{Vermaseren:2005qc}
\bibitem{Vermaseren:2005qc}
  J.~A.~M.~Vermaseren, A.~Vogt and S.~Moch,
  %``The third-order QCD corrections to deep-inelastic scattering by photon
  %exchange,''
 {\it  Nucl.\ Phys.\ } B {724} (2005) 3
  [arXiv:hep-ph/0504242]
  %%CITATION = NUPHA,B724,3;%%

%\cite{Moch:2004pa}
\bibitem{Moch:2004pa}
  S.~Moch, J.~A.~M.~Vermaseren and A.~Vogt,
  %``The three-loop splitting functions in QCD: The non-singlet case,''
  {\it Nucl.\ Phys.\ } B {688} (2004) 101
  [arXiv:hep-ph/0403192]
  %%CITATION = NUPHA,B688,101;%%

%\cite{Vogt:2004mw}
\bibitem{Vogt:2004mw}
  A.~Vogt, S.~Moch and J.~A.~M.~Vermaseren,
  %``The three-loop splitting functions in QCD: The singlet case,''
 {\it  Nucl.\ Phys.\ } B {691} (2004) 129
  [arXiv:hep-ph/0404111]
  %%CITATION = NUPHA,B691,129;%%

%\cite{Brodsky:1980pb}
\bibitem{Brodsky:1980pb}
  S.~J.~Brodsky, P.~Hoyer, C.~Peterson and N.~Sakai,
  %``The Intrinsic Charm Of The Proton,''
 {\it  Phys.\ Lett.\ } B {93} (1980) 451
  %%CITATION = PHLTA,B93,451;%%


%\cite{Catani:2004nc}
\bibitem{Catani:2004nc}
  S.~Catani, D.~de Florian, G.~Rodrigo and W.~Vogelsang,
  %``Perturbative generation of a strange-quark asymmetry in the nucleon,''
  {\it Phys.\ Rev.\ Lett.\ } {93} (2004) 152003
  [arXiv:hep-ph/0404240]
  %%CITATION = PRLTA,93,152003;%%


%\cite{Martin:1998sq}
\bibitem{Martin:1998sq}
  A.~D.~Martin, R.~G.~Roberts, W.~J.~Stirling and R.~S.~Thorne,
  %``Parton distributions: A new global analysis,''
  {\it Eur.\ Phys.\ J.\ } C {4} (1998) 463
  [arXiv:hep-ph/9803445]
  %%CITATION = EPHJA,C4,463;%%


%\cite{Whitlow:1991uw}
\bibitem{Whitlow:1991uw}
  L.~W.~Whitlow, E.~M.~Riordan, S.~Dasu, S.~Rock and A.~Bodek,
  %``Precise measurements of the proton and deuteron structure functions from a
  %global analysis of the SLAC deep inelastic electron scattering
  %cross-sections,''
  {\it Phys.\ Lett.\ } B {282} (1992) 475
  %%CITATION = PHLTA,B282,475;%%


%\cite{Benvenuti:1989rh}
\bibitem{Benvenuti:1989rh}
  A.~C.~Benvenuti {\it et al.}  [BCDMS Collaboration],
  %``A High Statistics Measurement of the Proton Structure Functions F(2) (x,
  %Q**2) and R from Deep Inelastic Muon Scattering at High Q**2,''
  {\it Phys.\ Lett.\ } B {223} (1989) 485
  %%CITATION = PHLTA,B223,485;%%

%\cite{Arneodo:1996qe}
\bibitem{Arneodo:1996qe}
  M.~Arneodo {\it et al.}  [New Muon Collaboration],
  %``Measurement of the proton and deuteron structure functions, F2(p) and
  %F2(d), and of the ratio sigma(L)/sigma(T),''
  {\it Nucl.\ Phys.\ } B {483} (1997) 3
  [arXiv:hep-ph/9610231]
  %%CITATION = NUPHA,B483,3;%%

%\cite{Adams:1996gu}
\bibitem{Adams:1996gu}
  M.~R.~Adams {\it et al.}  [E665 Collaboration],
  %``Proton and deuteron structure functions in muon scattering at 470-GeV,''
 {\it  Phys.\ Rev.\ }  D {54} (1996) 3006
  %%CITATION = PHRVA,D54,3006;%%


%\cite{Whitlow:1990dr}
\bibitem{Whitlow:1990dr}
  L.~W.~Whitlow, {\it Ph.D.~thesis, Stanford University,} 1990, SLAC-0357
  %``DEEP INELASTIC STRUCTURE FUNCTIONS FROM ELECTRON SCATTERING ON HYDROGEN,
  %DEUTERIUM, AND IRON AT 0.6-GeV**2 <= Q**2 <= 30-GeV**2,''
  %%CITATION = SLAC-R-357;%%


%\cite{Benvenuti:1989fm}
\bibitem{Benvenuti:1989fm}
  A.~C.~Benvenuti {\it et al.}  [BCDMS Collaboration],
  %``A HIGH STATISTICS MEASUREMENT OF THE DEUTERON STRUCTURE FUNCTIONS F2 (x,
  %Q**2) AND R FROM DEEP INELASTIC MUON SCATTERING AT HIGH Q**2,''
  {\it Phys.\ Lett.\ } B {237} (1990) 592
  %%CITATION = PHLTA,B237,592;%%

%\cite{Arneodo:1996kd}
\bibitem{Arneodo:1996kd}
  M.~Arneodo {\it et al.}  [New Muon Collaboration],
  %``Accurate measurement of F2(d)/F2(p) and R(d)-R(p),''
  {\it Nucl.\ Phys.\ } B {487} (1997) 3
  [arXiv:hep-ex/9611022]
  %%CITATION = NUPHA,B487,3;%%

%\cite{Yang:2000ju}
\bibitem{Yang:2000ju}
  U.~K.~Yang {\it et al.}  [CCFR/NuTeV Collaboration],
  %``Measurements of F2 and xF3(nu) - xF3(anti-nu) from CCFR nu/mu Fe and
  %anti-nu/mu Fe data in a physics model independent way,''
  {\it Phys.\ Rev.\ Lett.\ } {86} (2001) 2742
  [arXiv:hep-ex/0009041]
  %%CITATION = PRLTA,86,2742;%%

%\cite{Tzanov:2005kr}
\bibitem{Tzanov:2005kr}
  M.~Tzanov {\it et al.}  [NuTeV Collaboration],
  %``Precise measurement of neutrino and anti-neutrino differential cross
  %sections,''
  {\it Phys.\ Rev.\ } D {74} (2006) 012008
  [arXiv:hep-ex/0509010]
  %%CITATION = PHRVA,D74,012008;%%



%\cite{Onengut:2005kv}
\bibitem{Onengut:2005kv}
  G.~Onengut {\it et al.}  [CHORUS Collaboration],
  %``Measurement of nucleon structure functions in neutrino scattering,''
  {\it Phys.\ Lett.\ } B {632} (2006) 65
  %%CITATION = PHLTA,B632,65;%%


%\cite{Dokshitzer:1995qm}
\bibitem{Dokshitzer:1995qm}
  Y.~L.~Dokshitzer, G.~Marchesini and B.~R.~Webber,
  %``Dispersive Approach to Power-Behaved Contributions in QCD Hard Processes,''  
  {\it Nucl.\ Phys.\ } B {469} (1996) 93
  [arXiv:hep-ph/9512336]
  %%CITATION = NUPHA,B469,93;%%

%\cite{Dasgupta:1996hh}
\bibitem{Dasgupta:1996hh}
  M.~Dasgupta and B.~R.~Webber,
  %``Power Corrections and Renormalons in Deep Inelastic Structure Functions,''
  {\it Phys.\ Lett.\ } B {382} (1996) 273
  [arXiv:hep-ph/9604388]
  %%CITATION = PHLTA,B382,273;%%

%\cite{deFlorian:2003qf}
\bibitem{deFlorian:2003qf}
  D.~de Florian and R.~Sassot,
  %``Nuclear parton distributions at next to leading order,''
  {\it Phys.\ Rev.\ } D {69} (2004) 074028
  [arXiv:hep-ph/0311227]
  %%CITATION = PHRVA,D69,074028;%%

%\cite{Eskola:2009uj}
\bibitem{Eskola:2009uj}
  K.~J.~Eskola, H.~Paukkunen and C.~A.~Salgado,
  %``EPS09 - a New Generation of NLO and LO Nuclear Parton Distribution
  %Functions,''
 {\it  JHEP} {0904} (2009) 065
  [arXiv:0902.4154 [hep-ph]]
  %%CITATION = JHEPA,0904,065;%%


%\cite{Hirai:2007sx}
\bibitem{Hirai:2007sx}
  M.~Hirai, S.~Kumano and T.~H.~Nagai,
  %``Determination of nuclear parton distribution functions and their
  %uncertainties at next-to-leading order,''
  {\it Phys.\ Rev.\ } C {76} (2007) 065207
  [arXiv:0709.3038 [hep-ph]]
  %%CITATION = PHRVA,C76,065207;%%


%\cite{Adloff:2000qj}
\bibitem{Adloff:2000qj}
  C.~Adloff {\it et al.}  [H1 Collaboration],
  %``Measurement of neutral and charged current cross sections in electron
  %proton collisions at high Q**2,''
  {\it Eur.\ Phys.\ J.\ } C {19} (2001) 269
  [arXiv:hep-ex/0012052]
  %%CITATION = EPHJA,C19,269;%%

%\cite{Adloff:2003uh}
\bibitem{Adloff:2003uh}
  C.~Adloff {\it et al.}  [H1 Collaboration],
  %``Measurement and QCD analysis of neutral and charged current cross sections
  %at HERA,''
 {\it  Eur.\ Phys.\ J.\ } C {30} (2003) 1
  [arXiv:hep-ex/0304003]
  %%CITATION = EPHJA,C30,1;%%

%\cite{Chekanov:2002zs}
\bibitem{Chekanov:2002zs}
  S.~Chekanov {\it et al.}  [ZEUS Collaboration],
  %``Measurement of high-Q**2 charged current cross sections in e- p deep
  %inelastic scattering at HERA,''
  {\it Phys.\ Lett.\ } B {539} (2002) 197
  [{\it Erratum-ibid.\ } B {552} (2003) 308]
  [arXiv:hep-ex/0205091]
  %%CITATION = PHLTA,B539,197;%%

%\cite{Chekanov:2003vw}
\bibitem{Chekanov:2003vw}
  S.~Chekanov {\it et al.}  [ZEUS Collaboration],
  %``Measurement of high-Q**2 charged current cross sections in e+ p deep
  %inelastic scattering at HERA,''
  {\it Eur.\ Phys.\ J.\ } C {32} (2003) 1
  [arXiv:hep-ex/0307043]
  %%CITATION = EPHJA,C32,1;%%

%\cite{Moreno:1990sf}
\bibitem{Moreno:1990sf}
  G.~Moreno {\it et al.},
  %``Dimuon production in proton - copper collisions at $\sqrt{s}$ = 38.8-GeV,''
  {\it Phys.\ Rev.\ } D {43} (1991) 2815
  %%CITATION = PHRVA,D43,2815;%%



%\cite{McGaughey:1994dx}
\bibitem{McGaughey:1994dx}
  P.~L.~McGaughey {\it et al.}  [E772 Collaboration],
  %``Cross-Sections For The Production Of High Mass Muon Pairs From 800-Gev
  %Proton Bombardment Of H-2,''
 {\it  Phys.\ Rev.\ } D {50} (1994) 3038
  [{\it Erratum-ibid.\ } D {60} (1999) 119903]
  %%CITATION = PHRVA,D50,3038;%%

%\cite{Webb:2003bj}
\bibitem{Webb:2003bj}
  J.~C.~Webb,
  %``Measurement of continuum dimuon production in 800-GeV/c proton nucleon
  %collisions,''
  {\it Ph.D.~thesis, New Mexico State University,} 2002,
  arXiv:hep-ex/0301031; Paul E.~Reimer, private communication (for the radiative corrections)
  %%CITATION = HEP-EX/0301031;%%

%\cite{Hamberg:1990np}
\bibitem{Hamberg:1990np}
  R.~Hamberg, W.~L.~van Neerven and T.~Matsuura,
  %``A Complete calculation of the order alpha-s**2 correction to the Drell-Yan
  %K factor,''
 {\it  Nucl.\ Phys.\ } B {359} (1991) 343
  [{\it Erratum-ibid.\ } B {644} (2002) 403];
  \\{}original code from \href{http://www.lorentz.leidenuniv.nl/~neerven/}{\tt http://www.lorentz.leidenuniv.nl/$\sim$neerven/}
  %%CITATION = NUPHA,B359,343;%%


%\cite{Anastasiou:2003ds}
\bibitem{Anastasiou:2003ds}
  C.~Anastasiou, L.~J.~Dixon, K.~Melnikov and F.~Petriello,
  %``High-precision QCD at hadron colliders: Electroweak gauge boson  rapidity
  %distributions at NNLO,''
 {\it  Phys.\ Rev.\ } D {69} (2004) 094008
  [arXiv:hep-ph/0312266];
  %%CITATION = PHRVA,D69,094008;%%
  \\{}\textsc{Vrap} code from \href{http://www.slac.stanford.edu/~lance/Vrap/}{\tt http://www.slac.stanford.edu/$\sim$lance/Vrap/}

%\cite{Melnikov:2006kv}
\bibitem{Melnikov:2006kv}
  K.~Melnikov and F.~Petriello,
  %``Electroweak gauge boson production at hadron colliders through
  %O(alpha(s)**2),''
  {\it Phys.\ Rev.\ } D {74} (2006) 114017
  [arXiv:hep-ph/0609070];
  %%CITATION = PHRVA,D74,114017;%%
  \\{}\textsc{Fewz} code from \href{http://www.phys.hawaii.edu/~kirill/FEHiP.htm}{\tt http://www.phys.hawaii.edu/$\sim$kirill/FEHiP.htm}

%\cite{Grazzini:2009nc}
\bibitem{Grazzini:2009nc}
  M.~Grazzini,
  %``The Drell-Yan process in NNLO QCD,''
  arXiv:0908.1336 [hep-ph]
  %%CITATION = ARXIV:0908.1336;%%


%\cite{Catani:2010en}
\bibitem{Catani:2010en}
  S.~Catani, G.~Ferrera and M.~Grazzini,
  %``W boson production at hadron colliders: the lepton charge asymmetry in NNLO
  %QCD,''
 {\it  JHEP} {1005} (2010) 006
  [arXiv:1002.3115]
  %%CITATION = JHEPA,1005,006;%%


%\cite{Arneodo:1994sh}
\bibitem{Arneodo:1994sh}
  M.~Arneodo {\it et al.}  [New Muon Collaboration],
  %``A Reevaluation of the Gottfried sum,''
 {\it  Phys.\ Rev.\ } D {50} (1994) 1
  %%CITATION = PHRVA,D50,1;%%


%\cite{Baldit:1994jk}
\bibitem{Baldit:1994jk}
  A.~Baldit {\it et al.}  [NA51 Collaboration],
  %``Study of the isospin symmetry breaking the in the light quark sea of the
  %nucleon from the Drell-Yan process,''
 {\it  Phys.\ Lett.\ }  B {332} (1994) 244
  %%CITATION = PHLTA,B332,244;%%

%\cite{Towell:2001nh}
\bibitem{Towell:2001nh}
  R.~S.~Towell {\it et al.}  [FNAL E866/NuSea Collaboration],
  %``Improved measurement of the anti-d/anti-u asymmetry in the nucleon sea,''
  {\it Phys.\ Rev.\ } D {64} (2001) 052002
  [arXiv:hep-ex/0103030]
  %%CITATION = PHRVA,D64,052002;%%

%\cite{Abazov:2007jy}
\bibitem{Abazov:2007jy}
  V.~M.~Abazov {\it et al.}  [D0 Collaboration],
  %``Measurement of the shape of the boson rapidity distribution for $p \bar{p}
  %\to Z/gamma^* \to e^{+} e^{-}$ + $X$ events produced at $\sqrt{s}$ of
  %1.96-TeV,''
 {\it  Phys.\ Rev.\ } D {76} (2007) 012003
  [arXiv:hep-ex/0702025]
  %%CITATION = PHRVA,D76,012003;%%


%\cite{Aaltonen:2010zza}
\bibitem{Aaltonen:2010zza}
  T.~A.~Aaltonen {\it et al.}  [CDF Collaboration],
  %``Measurement of $d\sigma/dy$ of Drell-Yan $e^+e^-$ pairs in the $Z$ Mass
  %Region from $p\bar{p}$ Collisions at $\sqrt{s}=1.96$ TeV,''
 {\it Phys.\ Lett.\ }  B {692} (2010) 232
  [arXiv:0908.3914 [hep-ex]]
  %%CITATION = PHLTA,B692,232;%%



%\cite{Bazarko:1994tt}
\bibitem{Bazarko:1994tt}
  A.~O.~Bazarko {\it et al.}  [CCFR Collaboration],
  %``Determination Of The Strange Quark Content Of The Nucleon From A
  %Next-To-Leading Order QCD Analysis Of Neutrino Charm Production,''
  {\it Z.\ Phys.\ } C {65} (1995) 189
  [arXiv:hep-ex/9406007]
  %%CITATION = ZEPYA,C65,189;%%

\cite{Goncharov:2001qe}
%\cite{Goncharov:2001qe}
\bibitem{Goncharov:2001qe}
  M.~Goncharov {\it et al.}  [NuTeV Collaboration],
  %``Precise measurement of dimuon production cross-sections in nu/mu Fe and
  %anti-nu/mu Fe deep inelastic scattering at the Tevatron,''
 {\it  Phys.\ Rev.\ } D {64} (2001) 112006
  [arXiv:hep-ex/0102049]
  %%CITATION = PHRVA,D64,112006;%%

%\cite{Olness:2003wz}
\bibitem{Olness:2003wz}
  F.~Olness {\it et al.},
  %``Neutrino dimuon production and the strangeness asymmetry of the nucleon,''
  {\it Eur.\ Phys.\ J.\ } C {40} (2005) 145
  [arXiv:hep-ph/0312323]
  %%CITATION = EPHJA,C40,145;%%

%\cite{Mason:2006qa}
\bibitem{Mason:2006qa}
  D.~A.~Mason,
 {\it  ``Measurement of the strange--antistrange asymmetry at NLO in QCD from
  NuTeV dimuon data,''}
  FERMILAB-THESIS-2006-01;
  %%CITATION = UMI-32-11223;%%
  %
  {\it ``Final strange asymmetry results from NuTeV,''
  Proceedings of 14th International Workshop on Deep Inelastic Scattering (DIS 2006)}, Tsukuba, Japan, 20-24 Apr 2006;\\
  %
  D.~A.~Mason {\it et al.},
  %``Measurement of the strange - antistrange asymmetry at NLO in QCD from
  %NuTeV dimuon data,''
  {\it Phys.\ Rev.\ Lett.\ } {99} (2007) 192001
  %%CITATION = PRLTA,99,192001;%%

%\cite{Lai:2007dq}
\bibitem{Lai:2007dq}
  H.~L.~Lai, P.~Nadolsky, J.~Pumplin, D.~Stump, W.~K.~Tung and C.~P.~Yuan,
  %``The Strange Parton Distribution of the Nucleon: Global Analysis and
  %Applications,''
  {\it JHEP} {0704} (2007) 089
  [arXiv:hep-ph/0702268]
  %%CITATION = JHEPA,0704,089;%%

%\cite{Alekhin:2008mb}
\bibitem{Alekhin:2008mb}
  S.~Alekhin, S.~Kulagin and R.~Petti,
  %``Determination of Strange Sea Distributions from Neutrino-Nucleon Deep
  %Inelastic Scattering,''
  {\it Phys.\ Lett.\ } B {675} (2009) 433
  [arXiv:0812.4448 [hep-ph]]
  %%CITATION = PHLTA,B675,433;%%


%\cite{Zeller:2001hh}
\bibitem{Zeller:2001hh}
  G.~P.~Zeller {\it et al.}  [NuTeV Collaboration],
  %``A precise determination of electroweak parameters in neutrino nucleon
  %scattering,''
 {\it  Phys.\ Rev.\ Lett.\ } {88} (2002) 091802
  [{\it Erratum-ibid.\ } {90} (2003) 239902]
  [arXiv:hep-ex/0110059]
  %%CITATION = PRLTA,88,091802;%%

%\cite{Zeller:2002du}
\bibitem{Zeller:2002du}
  G.~P.~Zeller {\it et al.}  [NuTeV Collaboration],
  %``On the effect of asymmetric strange seas and isospin-violating parton
  %distribution functions on sin**2(Theta(W)) measured in the NuTeV
  %experiment,''
{\it Phys.\ Rev.\ } D {65} (2002) 111103
  [{\it Erratum-ibid.\ } D {67} (2003) 119902]
  [arXiv:hep-ex/0203004]
  %%CITATION = PHRVA,D65,111103;%%



%\cite{Aaltonen:2009ta}
\bibitem{Aaltonen:2009ta}
  T.~Aaltonen {\it et al.}  [CDF Collaboration],
  %``Direct Measurement of the $W$ Production Charge Asymmetry in $p\bar{p}$
  %Collisions at $\sqrt{s} = 1.96$ TeV,''
 {\it  Phys.\ Rev.\ Lett.\ } {102} (2009) 181801
  [arXiv:0901.2169 [hep-ex]]
  %%CITATION = PRLTA,102,181801;%%

%\cite{Acosta:2005ud}
\bibitem{Acosta:2005ud}
  D.~Acosta {\it et al.}  [CDF Collaboration],
  %``Measurement of the forward-backward charge asymmetry from $W \to e \nu$
  %production in $p\bar{p}$ collisions at $\sqrt{s} = 1.96$ TeV,''
  {\it Phys.\ Rev.\ } D {71} (2005) 051104
  [arXiv:hep-ex/0501023]
  %%CITATION = PHRVA,D71,051104;%%


%\cite{Abazov:2007pm}
\bibitem{Abazov:2007pm}
  V.~M.~Abazov {\it et al.}  [D0 Collaboration],
  %``Measurement of the muon charge asymmetry from $W$ boson decays,''
 {\it  Phys.\ Rev.\ } D {77} (2008) 011106
  [arXiv:0709.4254 [hep-ex]]
  %%CITATION = PHRVA,D77,011106;%%


%\cite{Abazov:2008qv}
\bibitem{Abazov:2008qv}
  V.~M.~Abazov {\it et al.}  [D0 Collaboration],
  %``Measurement of the electron charge asymmetry in $p \bar{p} \to W + X \to e
  %\nu + X$ events at $\sqrt{s}$ = 1.96-TeV,''
  {\it Phys.\ Rev.\ Lett.\ }  {101} (2008) 211801
  [arXiv:0807.3367 [hep-ex]]
  %%CITATION = PRLTA,101,211801;%%

%\cite{Lai:2010vv}
\bibitem{Lai:2010vv}
  H.~L.~Lai, M.~Guzzi, J.~Huston, Z.~Li, P.~M.~Nadolsky, J.~Pumplin and C.~P.~Yuan,
  %``New parton distributions for collider physics,''
 {\it  Phys.\ Rev.\ }  D {82} (2010) 074024
  [arXiv:1007.2241 [hep-ph]]
  %%CITATION = PHRVA,D82,074024;%%


%\cite{:2010gb}
\bibitem{:2010gb} R.~D.~Ball
   {\it et al.}  [The NNPDF Collaboration],
  %``Reweighting NNPDFs: the W lepton asymmetry,''
  arXiv:1012.0836 [hep-ph]
  %%CITATION = ARXIV:1012.0836;%%


%\cite{Lai:1996mg}
\bibitem{Lai:1996mg}
  H.~L.~Lai {\it et al.},
  %``Improved parton distributions from global analysis of recent deep inelastic
  %scattering and inclusive jet data,''
  {\it Phys.\ Rev.\ } D {55}, 1280 (1997)
  [arXiv:hep-ph/9606399]
  %%CITATION = PHRVA,D55,1280;%%

%\cite{Bonesini:1987bv}
\bibitem{Bonesini:1987bv}
  M.~Bonesini {\it et al.}  [WA70 Collaboration],
  %``Production of High Transverse Momentum Prompt Photons and Neutral Pions in
  %Proton Proton Interactions at 280-GeV/c,''
  {\it Z.\ Phys.\ } C {38} (1988) 371
  %%CITATION = ZEPYA,C38,371;%%

%\cite{Apanasevich:1997hm}
\bibitem{Apanasevich:1997hm}
  L.~Apanasevich {\it et al.}  [Fermilab E706 Collaboration],
  %``Evidence for parton $k_{T}$ effects in high $p_{T}$ particle production,''
  {\it Phys.\ Rev.\ Lett.\ } {81} (1998) 2642
  [arXiv:hep-ex/9711017]
  %%CITATION = PRLTA,81,2642;%%

%\cite{Laenen:2000de}
\bibitem{Laenen:2000de}
  E.~Laenen, G.~Sterman and W.~Vogelsang,
  %``Higher-order QCD corrections in prompt photon production,''
  {\it Phys.\ Rev.\ Lett.\ } {84} (2000) 4296
  [arXiv:hep-ph/0002078]
  %%CITATION = PRLTA,84,4296;%%


%\cite{Abazov:2005wc}
\bibitem{Abazov:2005wc}
  V.~M.~Abazov {\it et al.}  [D0 Collaboration],
  %``Measurement of the isolated photon cross section in $p \bar{p}$ collisions
  %at $\sqrt{s}$ = 1.96-TeV,''
  {\it Phys.\ Lett.\ } B {639} (2006) 151
  [{\it Erratum-ibid.\ } B {658} (2008) 285]
  [arXiv:hep-ex/0511054]
  %%CITATION = PHLTA,B639,151;%%

%\cite{Ichou:2010wc}
\bibitem{Ichou:2010wc}
  R.~Ichou and D.~d'Enterria,
  %``Sensitivity of isolated photon production at TeV hadron colliders to the
  %gluon distribution in the proton,''
  {\it Phys.\ Rev.\ }  D {82} (2010) 014015
  [arXiv:1005.4529 [hep-ph]]
  %%CITATION = PHRVA,D82,014015;%%




%\cite{Abbott:2000ew}
\bibitem{Abbott:2000ew}
  B.~Abbott {\it et al.}  [D0 Collaboration],
  %``Inclusive jet production in $p\bar{p}$ collisions,''
  {\it Phys.\ Rev.\ Lett.\ } {86} (2001) 1707
  [arXiv:hep-ex/0011036]
  %%CITATION = PRLTA,86,1707;%%

%\cite{Affolder:2001fa}
\bibitem{Affolder:2001fa}
  A.~A.~Affolder {\it et al.}  [CDF Collaboration],
  %``Measurement of the inclusive jet cross section in $\bar{p}p$ collisions at
  %$\sqrt{s} = 1.8$ TeV,''
  {\it Phys.\ Rev.\ } D {64} (2001) 032001
  [{\it Erratum-ibid.}\  D {65} (2002) 039903]
  [arXiv:hep-ph/0102074]
  %%CITATION = PHRVA,D64,032001;%%

%\cite{Abulencia:2007ez}
\bibitem{Abulencia:2007ez}
  A.~Abulencia {\it et al.}  [CDF - Run II Collaboration],
  %``Measurement of the Inclusive Jet Cross Section using the {\boldmath $k_{\rm
  %T}$} algorithmin{\boldmath $p\overline{p}$} Collisions at{\boldmath
  %$\sqrt{s}$} = 1.96 TeV with the CDF II Detector,''
  {\it Phys.\ Rev.\ } D {75} (2007) 092006
  [{\it Erratum-ibid.\ } D {75} (2007) 119901]
  [arXiv:hep-ex/0701051]
  %%CITATION = PHRVA,D75,092006;%%

%\cite{Abazov:2008hu}
\bibitem{Abazov:2008hu}
  V.~M.~Abazov {\it et al.}  [D0 Collaboration],
  %``Measurement of the inclusive jet cross section in $p \bar{p}$ collisions at
  %$\sqrt{s}=1.96 {\rm TeV}$,''
  {\it Phys.\ Rev.\ Lett.\ } {101} (2008) 062001
  [arXiv:0802.2400 [hep-ex]]
  %%CITATION = PRLTA,101,062001;%%

%\cite{Aaltonen:2008eq}
\bibitem{Aaltonen:2008eq}
  T.~Aaltonen {\it et al.}  [CDF Collaboration],
  %``Measurement of the Inclusive Jet Cross Section at the Fermilab Tevatron $p
  %\bar{p}$ Collider Using a Cone-Based Jet Algorithm,''
  {\it Phys.\ Rev.\ } D {78} (2008) 052006
  [arXiv:0807.2204 [hep-ex]]
  %%CITATION = PHRVA,D78,052006;%%



%\cite{Kluge:2006xs}
\bibitem{Kluge:2006xs}
  T.~Kluge, K.~Rabbertz and M.~Wobisch,
  %``fastNLO: Fast pQCD calculations for PDF fits,''
  % Presented at 14th International Workshop on Deep Inelastic Scattering (DIS 2006), Tsukuba, Japan, 20-24 Apr 2006.
  arXiv:hep-ph/0609285;\\ \href{http://projects.hepforge.org/fastnlo/}{\tt http://projects.hepforge.org/fastnlo/}.
  %%CITATION = HEP-PH/0609285;%%

%\cite{Nagy:2003tz}
\bibitem{Nagy:2003tz}
  Z.~Nagy,
  %``Next-to-leading order calculation of three-jet observables in hadron-hadron  %collision,''
  {\it Phys.\ Rev.\ } D {68} (2003) 094002
  [arXiv:hep-ph/0307268]
  %%CITATION = PHRVA,D68,094002;%%

%\cite{Nagy:2001fj}
\bibitem{Nagy:2001fj}
  Z.~Nagy,
  %``Three-jet cross sections in hadron hadron collisions at next-to-leading
  %order,''
 {\it  Phys.\ Rev.\ Lett.\ } {88} (2002) 122003
  [arXiv:hep-ph/0110315]
  %%CITATION = PRLTA,88,122003;%%

%\cite{Kidonakis:2000gi}
\bibitem{Kidonakis:2000gi}
  N.~Kidonakis and J.~F.~Owens,
  %``Effects of higher-order threshold corrections in high-E(T) jet
  %production,''
 {\it  Phys.\ Rev.\ } D {63} (2001) 054019
  [arXiv:hep-ph/0007268]
  %%CITATION = PHRVA,D63,054019;%%


%\cite{Lai:1999wy}
\bibitem{Lai:1999wy}
  H.~L.~Lai {\it et al.}  [CTEQ Collaboration],
  %``Global QCD analysis of parton structure of the nucleon: CTEQ5 parton
  %distributions,''
 {\it  Eur.\ Phys.\ J.\ } C {12} (2000) 375
  [arXiv:hep-ph/9903282]
  %%CITATION = EPHJA,C12,375;%%

%\cite{Pumplin:2002vw}
\bibitem{Pumplin:2002vw}
  J.~Pumplin, D.~R.~Stump, J.~Huston, H.~L.~Lai, P.~M.~Nadolsky and W.~K.~Tung,
  %``New generation of parton distributions with uncertainties from global  QCD
  %analysis,''
 {\it  JHEP} {0207} (2002) 012
  [arXiv:hep-ph/0201195]
  %%CITATION = JHEPA,0207,012;%%

%\cite{Martin:2001es}
\bibitem{Martin:2001es}
  A.~D.~Martin, R.~G.~Roberts, W.~J.~Stirling and R.~S.~Thorne,
  %``MRST2001: Partons and alpha(s) from precise deep inelastic scattering  and
  %Tevatron jet data,''
 {\it  Eur.\ Phys.\ J.\ } C {23} (2002) 73
  [arXiv:hep-ph/0110215]
  %%CITATION = EPHJA,C23,73;%%

%\cite{Martin:2004ir}
\bibitem{Martin:2004ir}
  A.~D.~Martin, R.~G.~Roberts, W.~J.~Stirling and R.~S.~Thorne,
  %``Physical gluons and high-E(T) jets,''
  {\it Phys.\ Lett.\ } B {604}, 61 (2004)
  [arXiv:hep-ph/0410230]
  %%CITATION = PHLTA,B604,61;%%

%\cite{Pumplin:2009nk}
\bibitem{Pumplin:2009nk}
  J.~Pumplin, J.~Huston, H.~L.~Lai, P.~M.~Nadolsky, W.~K.~Tung and C.~P.~Yuan,
  %``Collider Inclusive Jet Data and the Gluon Distribution,''
  {\it Phys.\ Rev.\ } D {80} (2009) 014019
  [arXiv:0904.2424 [hep-ph]]
  %%CITATION = PHRVA,D80,014019;%%

%\cite{Chekanov:2002be}
\bibitem{Chekanov:2002be}
  S.~Chekanov {\it et al.}  [ZEUS Collaboration],
  %``Inclusive jet cross sections in the Breit frame in neutral current deep
  %inelastic scattering at HERA and determination of alpha(s),''
  {\it Phys.\ Lett.\ } B {547} (2002) 164
  [arXiv:hep-ex/0208037]
  %%CITATION = PHLTA,B547,164;%%

%\cite{Chekanov:2006xr}
\bibitem{Chekanov:2006xr}
  S.~Chekanov {\it et al.}  [ZEUS Collaboration],
  %``Inclusive-jet and dijet cross sections in deep inelastic scattering at
  %HERA,''
 {\it   Nucl.\ Phys.\ } B {765} (2007) 1
  [arXiv:hep-ex/0608048]
  %%CITATION = NUPHA,B765,1;%%

%\cite{Adloff:2000tq}
\bibitem{Adloff:2000tq}
  C.~Adloff {\it et al.}  [H1 Collaboration],
  %``Measurement and QCD analysis of jet cross sections in deep-inelastic
  %positron proton collisions at s**(1/2) of 300-GeV,''
  {\it Eur.\ Phys.\ J.\ } C {19} (2001) 289
  [arXiv:hep-ex/0010054]
  %%CITATION = EPHJA,C19,289;%%



%\cite{Wobisch:2000dk}
\bibitem{Wobisch:2000dk}
  M.~Wobisch, {\it Ph.D.~thesis}, RWTH Aachen, 2000, DESY-THESIS-2000-049
  %``Measurement and QCD analysis of jet cross sections in deep-inelastic
  %positron proton collisions at s**(1/2) = 300-GeV,''
  %%CITATION = DESY-THESIS-2000-049;%%

%\cite{Aktas:2007pb}
\bibitem{Aktas:2007pb}
  A.~Aktas {\it et al.}  [H1 Collaboration],
  %``Measurement of Inclusive Jet Production in Deep-Inelastic Scattering at
  %High Q^2 and Determination of the Strong Coupling,''
 {\it  Phys.\ Lett.\ }  B {653} (2007) 134
  [arXiv:0706.3722 [hep-ex]]
  %%CITATION = PHLTA,B653,134;%%

%%%%%%%%%%%%%%%%%%%%%%%%%%%%%%%%



%\cite{Chekanov:2005nn}
\bibitem{Chekanov:2005nn}
  S.~Chekanov {\it et al.}  [ZEUS Collaboration],
  %``An NLO QCD analysis of inclusive cross section and jet-production data
  %from the ZEUS experiment,''
 {\it  Eur.\ Phys.\ J.\ } C {42} (2005) 1
  [arXiv:hep-ph/0503274]
  %%CITATION = EPHJA,C42,1;%%



%\cite{Adloff:2000qk}
\bibitem{Adloff:2000qk}
  C.~Adloff {\it et al.}  [H1 Collaboration],
  %``Deep-inelastic inclusive e p scattering at low x and a determination of
  %alpha(s),''
 {\it  Eur.\ Phys.\ J.\ } C {21} (2001) 33
  [arXiv:hep-ex/0012053]
  %%CITATION = EPHJA,C21,33;%%



%\cite{Breitweg:1998dz}
\bibitem{Breitweg:1998dz}
  J.~Breitweg {\it et al.}  [ZEUS Collaboration],
  %``ZEUS results on the measurement and phenomenology of F2 at low x and  low
  %Q**2,''
 {\it  Eur.\ Phys.\ J.\ } C {7} (1999) 609
  [arXiv:hep-ex/9809005]
  %%CITATION = EPHJA,C7,609;%%

%\cite{Chekanov:2001qu}
\bibitem{Chekanov:2001qu}
  S.~Chekanov {\it et al.}  [ZEUS Collaboration],
  %``Measurement of the neutral current cross section and F2 structure function
  %for deep inelastic e+ p scattering at HERA,''
 {\it  Eur.\ Phys.\ J.\ } C {21} (2001) 443
  [arXiv:hep-ex/0105090]
  %%CITATION = EPHJA,C21,443;%%

%\cite{Chekanov:2003yv}
\bibitem{Chekanov:2003yv}
  S.~Chekanov {\it et al.}  [ZEUS Collaboration],
  %``High-Q**2 neutral current cross sections in e+ p deep inelastic scattering
  %at s**(1/2) = 318-GeV,''
  {\it Phys.\ Rev.\ } D {70} (2004) 052001
  [arXiv:hep-ex/0401003]
  %%CITATION = PHRVA,D70,052001;%%

%\cite{Chekanov:2002ej}
\bibitem{Chekanov:2002ej}
  S.~Chekanov {\it et al.}  [ZEUS Collaboration],
  %``Measurement of high-Q**2 e- p neutral current cross sections at HERA and
  %the extraction of xF3,''
 {\it  Eur.\ Phys.\ J.\ } C {28} (2003) 175
  [arXiv:hep-ex/0208040]
  %%CITATION = EPHJA,C28,175;%%

%\cite{Martin:2009bu}
\bibitem{Martin:2009bu}
  A.~D.~Martin, W.~J.~Stirling, R.~S.~Thorne and G.~Watt,
  %``Uncertainties on alpha_S in global PDF analyses and implications for
  %predicted hadronic cross sections,''
  {\it Eur.\ Phys.\ J.\ } C {64} (2009) 653
  [arXiv:0905.3531 [hep-ph]]
  %%CITATION = EPHJA,C64,653;%%

%\cite{Lipatov:1976zz}
\bibitem{Lipatov:1976zz}
  L.~N.~Lipatov,
  %``Reggeisation Of The Vector Meson And The Vacuum Singularity In Nonabelian
  %Gauge Theories,''
  {\it Sov.\ J.\ Nucl.\ Phys.\ } {23} (1976) 338
  [{\it Yad.\ Fiz.\ } {23} (1976) 642]
  %%CITATION = YAFIA,23,642;%%

%\cite{Kuraev:1977fs}
\bibitem{Kuraev:1977fs}
  E.~A.~Kuraev, L.~N.~Lipatov and V.~S.~Fadin,
  %``The Pomeranchuk Singularity In Nonabelian Gauge Theories,''
  {\it Sov.\ Phys.\ JETP} {45} (1977) 199
  [{\it Zh.\ Eksp.\ Teor.\ Fiz.\ } {72} (1977) 377]
  %%CITATION = ZETFA,72,377;%%

%\cite{Balitsky:1978ic}
\bibitem{Balitsky:1978ic}
  I.~I.~Balitsky and L.~N.~Lipatov,
  %``The Pomeranchuk Singularity In Quantum Chromodynamics,''
 {\it  Sov.\ J.\ Nucl.\ Phys.\ } {28} (1978) 822
  [{\it Yad.\ Fiz.\ } {28} (1978) 1597]
  %%CITATION = YAFIA,28,1597;%%




%\cite{Aaron:2008tx}
\bibitem{Aaron:2008tx}
  F.~D.~Aaron {\it et al.}  [H1 Collaboration],
  %``Measurement of the Proton Structure Function F_L at Low x,''
  arXiv:0805.2809 [hep-ex]
  %%CITATION = ARXIV:0805.2809;%%

%\cite{Chekanov:2009na}
\bibitem{Chekanov:2009na}
  S.~Chekanov {\it et al.}  [ZEUS Collaboration],
  %``Measurement of the Longitudinal Proton Structure Function at HERA,''
 {\it  Phys.\ Lett.\ } B {682} (2009) 8
  [arXiv:0904.1092 [hep-ex]]
  %%CITATION = PHLTA,B682,8;%%

%\cite{Raicevic:2010zz}
\bibitem{Raicevic:2010zz}
  N.~Raicevic  [H1 and ZEUS Collaborations],
  %``HERA results and their impact for LHC,''
  {\it Nucl.\ Phys.\ Proc.\ Suppl.\ } {198} (2010) 75
  %%CITATION = NUPHZ,198,75;%%

%\cite{Reisert:2009zz}
\bibitem{Reisert:2009zz}
  B.~Reisert  [H1 and ZEUS Collaborations],
  %``Measurements of the longitudinal proton structure function F(L) at HERA,''
  {\it PoS E} {PS-HEP2009} (2009) 309
  %%CITATION = POSCI,EPS-HEP2009,309;%%
      


%\cite{Collaboration:2010ry}
\bibitem{Collaboration:2010ry}
  H1.~Collaboration,
  %``Measurement of the Inclusive e{\pm}p Scattering Cross Section at High
  %Inelasticity y and of the Structure Function FL,''
  arXiv:1012.4355 [hep-ex]
  %%CITATION = ARXIV:1012.4355;%%


% H1 F2charm

%\cite{Adloff:1996xq}
\bibitem{Adloff:1996xq}
  C.~Adloff {\it et al.}  [H1 Collaboration],
  %``Inclusive D0 and D*+- production in neutral current deep inelastic  e p
  %scattering at HERA,''
  {\it Z.\ Phys.\ }  C {72} (1996) 593
  [arXiv:hep-ex/9607012]
  %%CITATION = ZEPYA,C72,593;%%

%\cite{Adloff:2001zj}
\bibitem{Adloff:2001zj}
  C.~Adloff {\it et al.}  [H1 Collaboration],
  %``Measurement of D*+- meson production and F2(c) in deep inelastic
  %scattering at HERA,''
  {\it Phys.\ Lett.\ } B {528} (2002) 199
  [arXiv:hep-ex/0108039]
  %%CITATION = PHLTA,B528,199;%%

%\cite{Aktas:2005iw}
\bibitem{Aktas:2005iw}
  A.~Aktas {\it et al.}  [H1 Collaboration],
  %``Measurement of F2(c anti-c) and F2(b anti-b) at low Q**2 and x using  the
  %H1 vertex detector at HERA,''
  {\it Eur.\ Phys.\ J.\ } C {45} (2006) 23
  [arXiv:hep-ex/0507081]
  %%CITATION = EPHJA,C45,23;%%

%\cite{Aktas:2004az}
\bibitem{Aktas:2004az}
  A.~Aktas {\it et al.}  [H1 Collaboration],
  %``Measurement of F2(c anti-c) and F2(b anti-b) at high Q**2 using the H1
  %vertex detector at HERA,''
  {\it Eur.\ Phys.\ J.\ } C {40} (2005) 349
  [arXiv:hep-ex/0411046]
  %%CITATION = EPHJA,C40,349;%%

% ZEUS F2charm

%\cite{Breitweg:1999ad}
\bibitem{Breitweg:1999ad}
  J.~Breitweg {\it et al.}  [ZEUS Collaboration],
  %``Measurement of D*+- production and the charm contribution to F2 in  deep
  %inelastic scattering at HERA,''
  {\it Eur.\ Phys.\ J.\ } C {12} (2000) 35
  [arXiv:hep-ex/9908012]
  %%CITATION = EPHJA,C12,35;%%

%\cite{Chekanov:2003rb}
\bibitem{Chekanov:2003rb}
  S.~Chekanov {\it et al.}  [ZEUS Collaboration],
  %``Measurement of D*+- production in deep inelastic e+- p scattering at
  %HERA,''
  {\it Phys.\ Rev.\ }  D {69} (2004) 012004
  [arXiv:hep-ex/0308068]
  %%CITATION = PHRVA,D69,012004;%%

%\cite{Chekanov:2007ch}
\bibitem{Chekanov:2007ch}
  S.~Chekanov {\it et al.}  [ZEUS Collaboration],
  %``Measurement of D mesons production in deep inelastic scattering at HERA,''
  {\it JHEP} {0707} (2007) 074
  [arXiv:0704.3562 [hep-ex]]
  %%CITATION = JHEPA,0707,074;%%

%\cite{Aaron:2009ut}
\bibitem{Aaron:2009ut}
  F.~D.~Aaron {\it et al.}  [H1 Collaboration],
  %``Measurement of the Charm and Beauty Structure Functions using the H1 Vertex
  %Detector at HERA,''
  {\it Eur.\ Phys.\ J.\ } C {65} (2010) 89
  [arXiv:0907.2643 [hep-ex]]
  %%CITATION = EPHJA,C65,89;%%

%\cite{Chekanov:2009kj}
\bibitem{Chekanov:2009kj}
  S.~Chekanov {\it et al.}  [ZEUS Collaboration],
  %``Measurement of charm and beauty production in deep inelastic ep scattering
  %from decays into muons at HERA,''
  {\it Eur.\ Phys.\ J.\ } C {65} (2010) 65
  [arXiv:0904.3487 [hep-ex]]
  %%CITATION = EPHJA,C65,65;%%



%\cite{Martin:2007bv}
\bibitem{Martin:2007bv}
  A.~D.~Martin, W.~J.~Stirling, R.~S.~Thorne and G.~Watt,
  %``Update of Parton Distributions at NNLO,''
  {\it Phys.\ Lett.\ } B {652} (2007) 292
  [arXiv:0706.0459 [hep-ph]]
  %%CITATION = PHLTA,B652,292;%%


%\cite{Thorne:2010kj}
\bibitem{Thorne:2010kj}
  R.~S.~Thorne, A.~D.~Martin, W.~J.~Stirling and G.~Watt,
  %``The effects of combined HERA and recent Tevatron W - > lepton neutrino
  %charge asymmetry data on the MSTW PDFs,''
  {\it PoS} D {IS2010}, 052 (2010)
  [arXiv:1006.2753 [hep-ph]]
  %%CITATION = POSCI,DIS2010,052;%%




%\cite{Ball:2011mu}
\bibitem{Ball:2011mu}
  R.~D.~Ball {\it et al.},
  %``Impact of Heavy Quark Masses on Parton Distributions and LHC
  %Phenomenology,''
  arXiv:1101.1300 [hep-ph]
  %%CITATION = ARXIV:1101.1300;%%



%\cite{Chekanov:2002pv}
\bibitem{Chekanov:2002pv}
  S.~Chekanov {\it et al.}  [ZEUS Collaboration],
  %``A ZEUS next-to-leading-order QCD analysis of data on deep inelastic
  %scattering,''
  {\it Phys.\ Rev.\ }  D {67} (2003) 012007
  [arXiv:hep-ex/0208023]
  %%CITATION = PHRVA,D67,012007;%%

%\cite{CooperSarkar:2010ik}
\bibitem{CooperSarkar:2010ik}
  A.~M.~Cooper-Sarkar,
  %``HERAPDF fits including $F_2$(charm) data,''
  {\it PoS} D {IS2010} (2010) 023
  [arXiv:1006.4471 [hep-ph]]
  %%CITATION = POSCI,DIS2010,023;%%



%\cite{Alekhin:2002fv}
\bibitem{Alekhin:2002fv}
  S.~Alekhin,
  %``Parton distributions from deep-inelastic scattering data,''
 {\it  Phys.\ Rev.\ } D {68} (2003) 014002
  [arXiv:hep-ph/0211096]
  %%CITATION = PHRVA,D68,014002;%%

%\cite{Alekhin:2005gq}
\bibitem{Alekhin:2005gq}
  S.~Alekhin,
  %``Parton distribution functions from the precise NNLO QCD fit,''
  {\it JETP Lett.\ } {82} (2005) 628
  [{\it Pisma Zh.\ Eksp.\ Teor.\ Fiz.\ }  {82} (2005) 710]
  [arXiv:hep-ph/0508248]
  %%CITATION = ZFPRA,82,710;%%

%\cite{Alekhin:2006zm}
\bibitem{Alekhin:2006zm}
  S.~Alekhin, K.~Melnikov and F.~Petriello,
  %``Fixed target Drell-Yan data and NNLO QCD fits of parton distribution
  %functions,''
 {\it  Phys.\ Rev.\ } D {74} (2006) 054033
  [arXiv:hep-ph/0606237]
  %%CITATION = PHRVA,D74,054033;%%

%\cite{AlekinPDF4LHC}
\bibitem{AlekhinPDF4LHC}
S.~Alekhin, presented at PDF4LHC meeting, DESY, Hamburg, Nov. 2010, 
{\href http://indico.cern.ch/conferenceDisplay.py?confId=103872}   


%\cite{Alekhin:2010iu}
\bibitem{Alekhin:2010iu}
  S.~Alekhin, J.~Blumlein and S.~O.~Moch,
  %``Update of the NNLO PDFs in the 3-, 4-, and 5-flavour scheme,''
  {\it PoS} D {IS2010} (2010) 021
  [arXiv:1007.3657 [hep-ph]]
  %%CITATION = POSCI,DIS2010,021;%%





%\cite{Gluck:1991ng}
\bibitem{Gluck:1991ng}
  M.~Gluck, E.~Reya and A.~Vogt,
  %``Parton distributions for high-energy collisions,''
  {\it Z.\ Phys.\ } C {53} (1992) 127
  %%CITATION = ZEPYA,C53,127;%%


%\cite{JimenezDelgado:2008hf}
\bibitem{JimenezDelgado:2008hf}
  P.~Jimenez-Delgado and E.~Reya,
  %``Dynamical NNLO parton distributions,''
  {\it Phys.\ Rev.\ } D {79} (2009) 074023
  [arXiv:0810.4274 [hep-ph]]
  %%CITATION = PHRVA,D79,074023;%%

%\cite{Martin:1998np}
\bibitem{Martin:1998np}
  A.~D.~Martin, R.~G.~Roberts, W.~J.~Stirling and R.~S.~Thorne,
  %``Scheme dependence, leading order and higher twist studies of MRST
  %partons,''
 {\it  Phys.\ Lett.\ } B {443} (1998) 301
  [arXiv:hep-ph/9808371]
  %%CITATION = PHLTA,B443,301;%%

%\cite{White:2005wm}
\bibitem{White:2005wm}
  C.~D.~White and R.~S.~Thorne,
  %``Comparison of NNLO DIS scheme splitting functions with results from  exact
  %gluon kinematics at small x,''
  {\it Eur.\ Phys.\ J.\ } C {45} (2006) 179
  [arXiv:hep-ph/0507244]
  %%CITATION = EPHJA,C45,179;%%

%\cite{Blumlein:1996rp}
\bibitem{Blumlein:1996rp}
  J.~Blumlein, S.~Riemersma, M.~Botje, C.~Pascaud, F.~Zomer, W.~L.~van Neerven and A.~Vogt,
  %``A detailed comparison of NLO QCD evolution codes,''
  arXiv:hep-ph/9609400
  %%CITATION = HEP-PH/9609400;%%


%\cite{Jung:2009eq}
\bibitem{Jung:2009eq}
  Z.~J.~Ajaltouni {\it et al.},
  %``Proceedings of the workshop: HERA and the LHC workshop series on the
  %implications of HERA for LHC physics,''
  arXiv:0903.3861 [hep-ph]
  %%CITATION = ARXIV:0903.3861;%%

%\cite{Vogt:2004ns}
\bibitem{Vogt:2004ns}
  A.~Vogt,
  %``Efficient evolution of unpolarized and polarized parton distributions  with
  %QCD-PEGASUS,''
  {\it Comput.\ Phys.\ Commun.\ } {170} (2005) 65
  [arXiv:hep-ph/0408244]
  %%CITATION = CPHCB,170,65;%%


%\cite{Salam:2008qg}
\bibitem{Salam:2008qg}
  G.~P.~Salam and J.~Rojo,
  %``A Higher Order Perturbative Parton Evolution Toolkit (HOPPET),''
  %\cite{Salam:2008qg}
  {\it Comput.\ Phys.\ Commun.\ } {180} (2009) 120
  [arXiv:0804.3755 [hep-ph]]
  %%CITATION = CPHCB,180,120;%%

%\cite{Ball:2008by}
\bibitem{Ball:2008by}
  R.~D.~Ball {\it et al.}  [NNPDF Collaboration],
  %``A determination of parton distributions with faithful uncertainty
  %estimation,''
 {\it  Nucl.\ Phys.\ } B {809} (2009) 1
  [{\it Erratum-ibid.\ } B {816} (2009) 293]
  [arXiv:0808.1231 [hep-ph]]
  %%CITATION = NUPHA,B809,1;%%

%\cite{Stump:2001gu}
\bibitem{Stump:2001gu}
  D.~Stump {\it et al.},
  %``Uncertainties of predictions from parton distribution functions. I:  The
  %Lagrange multiplier method,''
  {\it Phys.\ Rev.\ } D {65} (2002) 014012
  [arXiv:hep-ph/0101051]
  %%CITATION = PHRVA,D65,014012;%%

%\cite{Pumplin:2000vx}
\bibitem{Pumplin:2000vx}
  J.~Pumplin, D.~R.~Stump and W.~K.~Tung,
  %``Multivariate fitting and the error matrix in global analysis of data,''
  {\it Phys.\ Rev.\ } D {65} (2002) 014011
  [arXiv:hep-ph/0008191]
  %%CITATION = PHRVA,D65,014011;%%


%\cite{Pumplin:2001ct}
\bibitem{Pumplin:2001ct}
  J.~Pumplin {\it et al.},
  %``Uncertainties of predictions from parton distribution functions. II:  The
  %Hessian method,''
  {\it Phys.\ Rev.\ } D {65} (2002) 014013
  [arXiv:hep-ph/0101032]
  %%CITATION = PHRVA,D65,014013;%%

%\cite{Martin:2002aw}
\bibitem{Martin:2002aw}
  A.~D.~Martin, R.~G.~Roberts, W.~J.~Stirling and R.~S.~Thorne,
  %``Uncertainties of predictions from parton distributions. I: Experimental
  %errors. ((T)),''
  {\it Eur.\ Phys.\ J.\ }  C {28} (2003) 455
  [arXiv:hep-ph/0211080]
  %%CITATION = EPHJA,C28,455;%%

%\cite{Pumplin:2009sc}
\bibitem{Pumplin:2009sc}
  J.~Pumplin,
  %``Experimental consistency in parton distribution fitting,''
 {\it  Phys.\ Rev.\ } D {81} (2010) 074010
  [arXiv:0909.0268 [hep-ph]]
  %%CITATION = PHRVA,D81,074010;%%


%\cite{Pumplin:2009bb}
\bibitem{Pumplin:2009bb}
  J.~Pumplin,
  %``Parametrisation dependence and Delta Chi-squared in parton distribution
  %fitting,''
  arXiv:0909.5176 
  %%CITATION = ARXIV:0909.5176;%%

%\cite{Giele:1998gw}
\bibitem{Giele:1998gw}
  W.~T.~Giele and S.~Keller,
  %``Implications of hadron collider observables on parton distribution
  %function uncertainties,''
  {\it Phys.\ Rev.\ } D {58} (1998) 094023
  [arXiv:hep-ph/9803393]
  %%CITATION = PHRVA,D58,094023;%%

%\cite{Giele:2001mr}
\bibitem{Giele:2001mr}
  W.~T.~Giele, S.~A.~Keller and D.~A.~Kosower,
  %``Parton distribution function uncertainties,''
  arXiv:hep-ph/0104052.
  %%CITATION = HEP-PH/0104052;%%

%\cite{DelDebbio:2007ee}
\bibitem{DelDebbio:2007ee}
  L.~Del Debbio, S.~Forte, J.~I.~Latorre, A.~Piccione and J.~Rojo  [NNPDF Collaboration],
 %``Neural network determination of parton distributions: The nonsinglet case,''
 {\it JHEP} {0703}, 039 (2007) [arXiv:hep-ph/0701127]
 %%CITATION = JHEPA,0703,039;%%


%\cite{Ball:2009qv}
\bibitem{Ball:2009qv}
  R.~D.~Ball, L.~Del Debbio, S.~Forte, A.~Guffanti, J.~I.~Latorre, J.~Rojo and M.~Ubiali
                  [NNPDF Collaboration],
  %``Fitting Parton Distribution Data with Multiplicative Normalization
  %Uncertainties,''
  {\it JHEP} {1005} (2010) 075
  [arXiv:0912.2276 [hep-ph]]
  %%CITATION = JHEPA,1005,075;%%


%\cite{Ball:2009mk}
\bibitem{Ball:2009mk}
  R.~D.~Ball {\it et al.}  [The NNPDF Collaboration],
  %``Precision determination of electroweak parameters and the strange content
  %of the proton from neutrino deep-inelastic scattering,''
  {\it Nucl.\ Phys.\ } B {823} (2009) 195
  [arXiv:0906.1958 [hep-ph]]
  %%CITATION = NUPHA,B823,195;%%

%\cite{Dittmar:2005ed}
\bibitem{Dittmar:2005ed}
  M.~Dittmar {\it et al.},
  %``Parton distributions: Summary report for the HERA - LHC workshop,''
  arXiv:hep-ph/0511119
  %%CITATION = HEP-PH/0511119;%%

%\cite{Fortepdf4lhc}
\bibitem{Fortepdf4lhc}
S.~Forte {\it at PDF4LHC meeting}, DESY, Hamburg, Oct. 23 2009  



%\cite{Wattplots}
\bibitem{Wattplots}
G.~Watt, talk at PDF4LHC benchmarking meeting, CERN, March 26th, 2010, 
\href{http://indico.cern.ch/conferenceDisplay.py?confId=87871}{\tt http://indico.cern.ch/conferenceDisplay.py?confId=87871}; updated versions will appear at 
\href{http://projects.hepforge.org/mstwpdf/pdf4lhc/}{\tt http://projects.hepforge.org/mstwpdf/pdf4lhc/}


%\cite{Alekhin:2010dd}
\bibitem{Alekhin:2010dd}
  S.~Alekhin, J.~Blumlein, P.~Jimenez-Delgado, S.~Moch and E.~Reya,
  %``NNLO Benchmarks for Gauge and Higgs Boson Production at TeV Hadron
  %Colliders,''
  {\it Phys.\ Lett.\ } B {697} (2011) 127
  [arXiv:1011.6259 [hep-ph]]
  %%CITATION = PHLTA,B697,127;%%




%\cite{CooperSarkar:2007pj}
\bibitem{CooperSarkar:2007pj}
  A.~M.~Cooper-Sarkar,
  %``Impact of and constraints on PDFs at LHC,''
  arXiv:0707.1593 [hep-ph]
  %%CITATION = ARXIV:0707.1593;%%

%\cite{Laenen:1992zk}
\bibitem{Laenen:1992zk}
  E.~Laenen, S.~Riemersma, J.~Smith and W.~L.~van Neerven,
  %``Complete O (alpha-s) corrections to heavy flavor structure functions in
  %electroproduction,''
  {\it Nucl.\ Phys.\ } B {392}, 162 (1993)
  %%CITATION = NUPHA,B392,162;%%

%\cite{Harris:1995tu}
\bibitem{Harris:1995tu}
  B.~W.~Harris and J.~Smith,
  %``Heavy quark correlations in deep inelastic electroproduction,''
  {\it Nucl.\ Phys.\ } B {452} (1995) 109
  [arXiv:hep-ph/9503484]
  %%CITATION = NUPHA,B452,109;%%



%\cite{Bierenbaum:2009mv}
\bibitem{Bierenbaum:2009mv}
  I.~Bierenbaum, J.~Blumlein and S.~Klein,
  %``Mellin Moments of the {$O(\alpha_s^3$)} Heavy Flavor Contributions to
  %unpolarized Deep-Inelastic Scattering at $Q^2 \gg m^2$ and Anomalous
  %Dimensions,''
  {\it Nucl.\ Phys.\ } B {820} (2009) 417
  [arXiv:0904.3563 [hep-ph]]
  %%CITATION = NUPHA,B820,417;%%



%\cite{Buza:1996wv}
\bibitem{Buza:1996wv}
  M.~Buza, Y.~Matiounine, J.~Smith and W.~L.~van Neerven,
  %``Charm electroproduction viewed in the variable-flavour number scheme
  %versus fixed-order perturbation theory,''
  {\it Eur.\ Phys.\ J.\ } C {1} (1998) 301
  [arXiv:hep-ph/9612398]
  %%CITATION = EPHJA,C1,301;%%


%\cite{Aivazis:1993pi}
\bibitem{Aivazis:1993pi}
  M.~A.~G.~Aivazis, J.~C.~Collins, F.~I.~Olness and W.~K.~Tung,
  %``Leptoproduction of heavy quarks. 2. A Unified QCD formulation of charged
  %and neutral current processes from fixed target to collider energies,''
  {\it Phys.\ Rev.\ } D {50} (1994) 3102
  [arXiv:hep-ph/9312319]
  %%CITATION = PHRVA,D50,3102;%%

%\cite{Thorne:1997ga}
\bibitem{Thorne:1997ga}
  R.~S.~Thorne and R.~G.~Roberts,
  %``An ordered analysis of heavy flavour production in deep inelastic
  %scattering,''
  {\it Phys.\ Rev.\ } D {57} (1998) 6871
  [arXiv:hep-ph/9709442]
  %%CITATION = PHRVA,D57,6871;%%

%\cite{Chuvakin:1999nx}
\bibitem{Chuvakin:1999nx}
  A.~Chuvakin, J.~Smith and W.~L.~van Neerven,
  %``Comparison between variable flavor number schemes for charm quark
  %electroproduction,''
  {\it Phys.\ Rev.\ } D {61} (2000) 096004
  [arXiv:hep-ph/9910250]
  %%CITATION = PHRVA,D61,096004;%%

%\cite{Kramer:2000hn}
\bibitem{Kramer:2000hn}
  M.~Kramer, F.~I.~Olness and D.~E.~Soper,
  %``Treatment of heavy quarks in deeply inelastic scattering,''
  {\it Phys.\ Rev.\ } D {62} (2000) 096007
  [arXiv:hep-ph/0003035]
  %%CITATION = PHRVA,D62,096007;%%

%\cite{Tung:2001mv}
\bibitem{Tung:2001mv}
  W.~K.~Tung, S.~Kretzer and C.~Schmidt,
  %``Open heavy flavor production in QCD: Conceptual framework and
  %implementation issues,''
  {\it J.\ Phys.\ } G {28} (2002) 983
  [arXiv:hep-ph/0110247]
  %%CITATION = JPHGB,G28,983;%%


%\cite{Thorne:2006qt}
\bibitem{Thorne:2006qt}
  R.~S.~Thorne,
  %``A variable-flavour number scheme for NNLO,''
  {\it Phys.\ Rev.\ } D {73} (2006) 054019
  [arXiv:hep-ph/0601245]
  %%CITATION = PHRVA,D73,054019;%%

%\cite{Forte:2010ta}
\bibitem{Forte:2010ta}
  S.~Forte, E.~Laenen, P.~Nason and J.~Rojo,
  %``Heavy quarks in deep-inelastic scattering,''
  {\it Nucl.\ Phys.\ } B {834} (2010) 116
  [arXiv:1001.2312]
  %%CITATION = NUPHA,B834,116;%%


%\cite{Thorne:2008xf}
\bibitem{Thorne:2008xf}
  R.~S.~Thorne and W.~K.~Tung,
  %``PQCD Formulations with Heavy Quark Masses and Global Analysis,''
  arXiv:0809.0714 [hep-ph]
  %%CITATION = ARXIV:0809.0714;%%

%\cite{Binoth:2010ra}
\bibitem{Binoth:2010ra}
  J.~R.~Andersen {\it et al.}  [{\it SM and NLO Multileg Working Group}],
  %``The SM and NLO multileg working group: Summary report,''
  arXiv:1003.1241
  %%CITATION = ARXIV:1003.1241;%%

%\cite{Collins:1998rz}
\bibitem{Collins:1998rz}
  J.~C.~Collins,
  %``Hard-scattering factorisation with heavy quarks: A general treatment,''
  {\it Phys.\ Rev.\ } D {58} (1998) 094002
  [arXiv:hep-ph/9806259]
  %%CITATION = PHRVA,D58,094002;%%




%\cite{Nadolsky:2009ge}
\bibitem{Nadolsky:2009ge}
  P.~M.~Nadolsky and W.~K.~Tung,
  %``Improved Formulation of Global QCD Analysis with Zero-mass Matrix
  %Elements,''
 {\it  Phys.\ Rev.\ } D {79} (2009) 113014
  [arXiv:0903.2667 [hep-ph]]
  %%CITATION = PHRVA,D79,113014;%%

%\cite{Tung:2006tb}
\bibitem{Tung:2006tb}
  W.~K.~Tung, H.~L.~Lai, A.~Belyaev, J.~Pumplin, D.~Stump and C.~P.~Yuan,
  %``Heavy quark mass effects in deep inelastic scattering and global QCD
  %analysis,''
  {\it JHEP} {0702} (2007) 053
  [arXiv:hep-ph/0611254]
  %%CITATION = JHEPA,0702,053;%%

%\cite{Thorne:2010pa}
\bibitem{Thorne:2010pa}
  R.~S.~Thorne,
  %``The Effect of Variable Flavour Number Scheme Variations on PDFs and Cross
  %Sections,''
  {\it PoS} D {IS2010} (2010) 053
  [arXiv:1006.5925 [hep-ph]]
  %%CITATION = POSCI,DIS2010,053;%%

%\cite{Pumplin:2007wg}
\bibitem{Pumplin:2007wg}
  J.~Pumplin, H.~L.~Lai and W.~K.~Tung,
  %``The Charm Parton Content of the Nucleon,''
  {\it Phys.\ Rev.\ } D {75} (2007) 054029
  [arXiv:hep-ph/0701220]
  %%CITATION = PHRVA,D75,054029;%%


%\cite{Aubert:1982tt}
\bibitem{Aubert:1982tt}
  J.~J.~Aubert {\it et al.}  [European Muon Collaboration],
  %``Production of charmed particles in 250-GeV $\mu^+$ - iron interactions,''
  {\it Nucl.\ Phys.\ } B {213} (1983) 31
  %%CITATION = NUPHA,B213,31;%%



%\cite{Catani:1990eg}
\bibitem{Catani:1990eg}
  S.~Catani, M.~Ciafaloni and F.~Hautmann,
  %``High-energy factorisation and small x heavy flavor production,''
  {\it Nucl.\ Phys.\ } B {366} (1991) 135
  %%CITATION = NUPHA,B366,135;%%


%\cite{Laenen:1998kp}
\bibitem{Laenen:1998kp}
  E.~Laenen and S.~O.~Moch,
  %``Soft gluon resummation for heavy quark electroproduction,''
 {\it  Phys.\ Rev.\ } D {59} (1999) 034027
  [arXiv:hep-ph/9809550]
  %%CITATION = PHRVA,D59,034027;%%


%\cite{Martin:2006qz}
\bibitem{Martin:2006qz}
  A.~D.~Martin, W.~J.~Stirling and R.~S.~Thorne,
  %``MRST partons generated in a fixed-flavour scheme,''
  {\it Phys.\ Lett.\ } B {636} (2006) 259
  [arXiv:hep-ph/0603143]
  %%CITATION = PHLTA,B636,259;%%

%\cite{Martin:2010db}
\bibitem{Martin:2010db}
  A.~D.~Martin, W.~J.~Stirling, R.~S.~Thorne and G.~Watt,
  %``Heavy-quark mass dependence in global PDF analyses and 3- and 4-flavour
  %parton distributions,''
  {\it Eur.\ Phys.\ J.\ }  C {70} (2010) 51
  [arXiv:1007.2624 [hep-ph]]
  %%CITATION = EPHJA,C70,51;%%



%\cite{Kretzer:2003it}
\bibitem{Kretzer:2003it}
  S.~Kretzer, H.~L.~Lai, F.~I.~Olness and W.~K.~Tung,
  %``CTEQ6 parton distributions with heavy quark mass effects,''
  {\it Phys.\ Rev.\ } D {69} (2004) 114005
  [arXiv:hep-ph/0307022]
  %%CITATION = PHRVA,D69,114005;%%

%\cite{Gluck:2008gs}
\bibitem{Gluck:2008gs}
  M.~Gluck, P.~Jimenez-Delgado, E.~Reya and C.~Schuck,
  %``On the role of heavy flavor parton distributions at high energy
  %colliders,''
  {\it Phys.\ Lett.\ }  B {664} (2008) 133
  [arXiv:0801.3618 [hep-ph]]
  %%CITATION = PHLTA,B664,133;%%



%\cite{JimenezDelgado:2009tv}
\bibitem{JimenezDelgado:2009tv}
  P.~Jimenez-Delgado and E.~Reya,
  %``Variable Flavor Number Parton Distributions and Weak Gauge and Higgs Boson
  %Production at Hadron Colliders at NNLO of QCD,''
  {\it Phys.\ Rev.\ }  D {80} (2009) 114011
  [arXiv:0909.1711 [hep-ph]]
  %%CITATION = PHRVA,D80,114011;%%






%\cite{Guffanti:2010yu}
\bibitem{Guffanti:2010yu}
  A.~Guffanti and J.~Rojo,
  %``Top production at the LHC: the impact of PDF uncertainties and
  %correlations,''
  arXiv:1008.4671 [hep-ph]
  %%CITATION = ARXIV:1008.4671;%%



%\cite{Wattplotsalpha}
%\bibitem{Wattplotsalpha}
%G.~Watt, talk at PDF4LHC benchmarking meeting, CERN, March 26th, 2010, 
%\href{http://indico.cern.ch/conferenceDisplay.py?confId=87871}{\tt http://indico.cern.ch/conferenceDisplay.py?confId=87871}; updates at \href{http://projects.hepforge.org/mstwpdf/pdf4lhc/xsections7TeV.html}
%{\tt http://projects.hepforge.org/mstwpdf/pdf4lhc/xsections7TeV.html}

%\cite{Lai:2010nw}
\bibitem{Lai:2010nw}
  H.~L.~Lai, J.~Huston, Z.~Li, P.~Nadolsky, J.~Pumplin, D.~Stump and C.~P.~Yuan,
  %``Uncertainty induced by QCD coupling in the CTEQ-TEA global analysis of
  %parton densities,''
  arXiv:1004.4624 
  %%CITATION = ARXIV:1004.4624;%%

%\cite{Amsler:2008zz}
\bibitem{Amsler:2008zz}
  C.~Amsler {\it et al.}  [Particle Data Group],
  %``Review of particle physics,''
  {\it Phys.\ Lett.\ } B {667} (2008) 1
  %%CITATION = PHLTA,B667,1;%%

%\cite{Demartin:2010er}
\bibitem{Demartin:2010er}
  F.~Demartin, S.~Forte, E.~Mariani, J.~Rojo and A.~Vicini,
  %``The impact of PDF and alphas uncertainties on Higgs Production in gluon
  %fusion at hadron colliders,''
 {\it Phys.\ Rev.\ }  D {82} (2010) 014002
  [arXiv:1004.0962 [hep-ph]]
  %%CITATION = PHRVA,D82,014002;%%



%\cite{Ubiali:2010xc}
\bibitem{Ubiali:2010xc}
  M.~Ubiali, R.~D.~Ball, L.~Del Debbio, S.~Forte, A.~Guffanti, J.~I.~Latorre and J.~Rojo,
  %``Combined PDF and strong coupling uncertainties at the LHC with NNPDF2.0,''
  arXiv:1005.0397 
  %%CITATION = ARXIV:1005.0397;%%




%\cite{Catani:2001ic}
\bibitem{Catani:2001ic}
  S.~Catani, D.~de Florian and M.~Grazzini,
  %``Higgs production in hadron collisions: Soft and virtual QCD corrections  at
  %NNLO,''
  {\it JHEP} {0105} (2001) 025
  [arXiv:hep-ph/0102227]
  %%CITATION = JHEPA,0105,025;%%

%\cite{Harlander:2002wh}
\bibitem{Harlander:2002wh}
  R.~V.~Harlander and W.~B.~Kilgore,
  %``Next-to-next-to-leading order Higgs production at hadron colliders,''
  {\it Phys.\ Rev.\ Lett.\ }  {88} (2002) 201801
  [arXiv:hep-ph/0201206]
  %%CITATION = PRLTA,88,201801;%%

%\cite{Anastasiou:2002yz}
\bibitem{Anastasiou:2002yz}
  C.~Anastasiou and K.~Melnikov,
  %``Higgs boson production at hadron colliders in NNLO QCD,''
  {\it Nucl.\ Phys.\ }  B {646} (2002) 220
  [arXiv:hep-ph/0207004]
  %%CITATION = NUPHA,B646,220;%%

%\cite{Ravindran:2003um}
\bibitem{Ravindran:2003um}
  V.~Ravindran, J.~Smith and W.~L.~van Neerven,
  %``NNLO corrections to the total cross section for Higgs boson production  in
  %hadron hadron collisions,''
 {\it  Nucl.\ Phys.\ } B {665} (2003) 325
  [arXiv:hep-ph/0302135]
  %%CITATION = NUPHA,B665,325;%%


%\cite{Anastasiou:2004xq}
\bibitem{Anastasiou:2004xq}
  C.~Anastasiou, K.~Melnikov and F.~Petriello,
  %``Higgs boson production at hadron colliders: Differential cross sections
  %through next-to-next-to-leading order,''
  {\it Phys.\ Rev.\ Lett.\ } {93} (2004) 262002
  [arXiv:hep-ph/0409088]
  %%CITATION = PRLTA,93,262002;%%

%\cite{Anastasiou:2005qj}
\bibitem{Anastasiou:2005qj}
  C.~Anastasiou, K.~Melnikov and F.~Petriello,
  %``Fully differential Higgs boson production and the di-photon signal through
  %next-to-next-to-leading order,''
  {\it Nucl.\ Phys.\ } B {724} (2005) 197
  [arXiv:hep-ph/0501130]
  %%CITATION = NUPHA,B724,197;%%

%\cite{Brein:2003wg}
\bibitem{Brein:2003wg}
  O.~Brein, A.~Djouadi and R.~Harlander,
  %``NNLO QCD corrections to the Higgs-strahlung processes at hadron
  %colliders,''
  {\it Phys.\ Lett.\ } B {579} (2004) 149
  [arXiv:hep-ph/0307206]
  %%CITATION = PHLTA,B579,149;%%

%\cite{Abazov:2008ff}
\bibitem{Abazov:2008ff}
  V.~M.~Abazov {\it et al.}  [D0 Collaboration],
  %``Measurement of $\sigma(p\bar{p} \to Z + X)$ Br($Z \to \tau^+ \tau^-$) at
  %$\sqrt{s}$ = 1.96-TeV,''
  {\it Phys.\ Lett.\ } B {670} (2009) 292
  [arXiv:0808.1306 [hep-ex]]
  %%CITATION = PHLTA,B670,292;%%

%\cite{Abulencia:2005ix}
\bibitem{Abulencia:2005ix}
  A.~Abulencia {\it et al.}  [CDF Collaboration],
  %``Measurements of Inclusive W and Z Cross Sections in p-pbar Collisions at
  %sqrt{s} =1.96 TeV,''
  {\it J.\ Phys.\ } G {34} (2007) 2457
  [arXiv:hep-ex/0508029]
  %%CITATION = JPHGB,G34,2457;%%

\bibitem{D0wztot}
  D{\O} Collaboration: %``Measurement of the Cross Section for $W$ and $Z$ Production to Electron Final States with the D{\O} Detector at $\sqrt{s} = 1.96$ TeV''
  D{\O}Note 4403-CONF,
  %D{\O} Collaboration, ``Measurement of the Cross-Section for Inclusive $W$ Production in the Muon Channel at $\sqrt{s} = 1.96$ TeV Using the D{\O} detector''
  D{\O}note 4750,
  %D{\O} Collaboration, ``Measurement of the Cross section for Inclusive $Z$ Production in Di-Muon Final States at $\sqrt{s} = 1.96$ TeV''
  D{\O}note 4573

%\cite{Thorne:2008am}
\bibitem{Thorne:2008am}
  R.~S.~Thorne, A.~D.~Martin, W.~J.~Stirling and G.~Watt,
  %``Parton Distributions and QCD at LHCb,''
  arXiv:0808.1847 [hep-ph]
  %%CITATION = ARXIV:0808.1847;%%

%\cite{Blumlein:2006be}
\bibitem{Blumlein:2006be}
  J.~Blumlein, H.~Bottcher and A.~Guffanti,
  %``Non-singlet QCD analysis of deep inelastic world data at O(alpha(s)**3),''
  {\it Nucl.\ Phys.\ } B {774} (2007) 182
  [arXiv:hep-ph/0607200]
  %%CITATION = NUPHA,B774,182;%%


%\cite{Khorramian:2009xz}
\bibitem{Khorramian:2009xz}
  A.~N.~Khorramian, H.~Khanpour and S.~A.~Tehrani,
  %``Nonsinglet parton distribution functions from the precise
  %next-to-next-to-next-to leading order QCD fit,''
  {\it Phys.\ Rev.\ } D {81} (2010) 014013
  [arXiv:0909.2665 [hep-ph]]
  %%CITATION = PHRVA,D81,014013;%%


%\cite{Martin:2004dh}
\bibitem{Martin:2004dh}
  A.~D.~Martin, R.~G.~Roberts, W.~J.~Stirling and R.~S.~Thorne,
  %``Parton distributions incorporating QED contributions,''
  {\it Eur.\ Phys.\ J.\ } C {39} (2005) 155
  [arXiv:hep-ph/0411040]
  %%CITATION = EPHJA,C39,155;%%

%\cite{CarloniCalame:2007cd}
\bibitem{CarloniCalame:2007cd}
  C.~M.~Carloni Calame, G.~Montagna, O.~Nicrosini and A.~Vicini,
  %``Precision electroweak calculation of the production of a high
  %transverse-momentum lepton pair at hadron colliders,''
  {\it JHEP} {0710} (2007) 109
  [arXiv:0710.1722 [hep-ph]]
  %%CITATION = JHEPA,0710,109;%%

%\cite{Ciafaloni:2000df}
\bibitem{Ciafaloni:2000df}
  M.~Ciafaloni, P.~Ciafaloni and D.~Comelli,
  %``Bloch-Nordsieck violating electroweak corrections to inclusive TeV scale
  %hard processes,''
  {\it Phys.\ Rev.\ Lett.\ } {84} (2000) 4810
  [arXiv:hep-ph/0001142]
  %%CITATION = PRLTA,84,4810;%%


%\cite{Moretti:2006ea}
\bibitem{Moretti:2006ea}
  S.~Moretti, M.~R.~Nolten and D.~A.~Ross,
  %``Weak corrections to four-parton processes,''
  {\it Nucl.\ Phys.\ } B {759} (2006) 50
  [arXiv:hep-ph/0606201]
  %%CITATION = NUPHA,B759,50;%%

%\cite{Accomando:2004de}
\bibitem{Accomando:2004de}
  E.~Accomando, A.~Denner and A.~Kaiser,
  %``Logarithmic electroweak corrections to gauge-boson pair production at the
  %LHC,''
 {\it  Nucl.\ Phys.\ } B {706} (2005) 325
  [arXiv:hep-ph/0409247]
  %%CITATION = NUPHA,B706,325;%%

%\cite{Kuhn:2007cv}
\bibitem{Kuhn:2007cv}
  J.~H.~Kuhn, A.~Kulesza, S.~Pozzorini and M.~Schulze,
  %``Electroweak corrections to hadronic production of W bosons at large
  %transverse momenta,''
  {\it Nucl.\ Phys.\ } B {797} (2008) 27
  [arXiv:0708.0476 [hep-ph]]
  %%CITATION = NUPHA,B797,27;%%

%\cite{Baur:2006sn}
\bibitem{Baur:2006sn}
  U.~Baur,
  %``Weak Boson Emission in Hadron Collider Processes,''
  {\it Phys.\ Rev.\ }  D {75} (2007) 013005
  [arXiv:hep-ph/0611241]
  %%CITATION = PHRVA,D75,013005;%%


%\cite{Ciafaloni:2005fm}
\bibitem{Ciafaloni:2005fm}
  P.~Ciafaloni and D.~Comelli,
  %``Electroweak evolution equations,''
  {\it JHEP }{0511} (2005) 022
  [arXiv:hep-ph/0505047]
  %%CITATION = JHEPA,0511,022;%%

%\cite{Fadin:1998py}
\bibitem{Fadin:1998py}
  V.~S.~Fadin and L.~N.~Lipatov,
  %``BFKL pomeron in the next-to-leading approximation,''
  {\it Phys.\ Lett.\ }  B {429} (1998) 127
  [arXiv:hep-ph/9802290]
  %%CITATION = PHLTA,B429,127;%%

%\cite{Ciafaloni:1998gs}
\bibitem{Ciafaloni:1998gs}
  M.~Ciafaloni and G.~Camici,
  %``Energy scale(s) and next-to-leading BFKL equation,''
  {\it Phys.\ Lett.\ } B {430} (1998) 349
  [arXiv:hep-ph/9803389]
  %%CITATION = PHLTA,B430,349;%%


%\cite{Catani:1994sq}
\bibitem{Catani:1994sq}
  S.~Catani and F.~Hautmann,
  %``High-energy factorization and small x deep inelastic scattering beyond
  %leading order,''
  {\it Nucl.\ Phys.\ } B {427} (1994) 475
  [arXiv:hep-ph/9405388]
  %%CITATION = NUPHA,B427,475;%%

%\cite{Ball:2001pq}
\bibitem{Ball:2001pq}
  R.~D.~Ball and R.~K.~Ellis,
  %``Heavy quark production at high-energy,''
  {\it JHEP} {0105} (2001) 053
  [arXiv:hep-ph/0101199]
  %%CITATION = JHEPA,0105,053;%%

%\cite{Martin:2003sk}
\bibitem{Martin:2003sk}
  A.~D.~Martin, R.~G.~Roberts, W.~J.~Stirling and R.~S.~Thorne,
  %``Uncertainties of predictions from parton distributions. II: Theoretical
  %errors,''
 {\it  Eur.\ Phys.\ J.\ } C {35} (2004) 325
  [arXiv:hep-ph/0308087]
  %%CITATION = EPHJA,C35,325;%%

%\cite{White:2006yh}
\bibitem{White:2006yh}
  C.~D.~White and R.~S.~Thorne,
  %``A global fit to scattering data with NLL BFKL resummations,''
  {\it Phys.\ Rev.\ } D {75} (2007) 034005
  [arXiv:hep-ph/0611204]
  %%CITATION = PHRVA,D75,034005;%%

%\cite{Ciafaloni:2007gf}
\bibitem{Ciafaloni:2007gf}
  M.~Ciafaloni, D.~Colferai, G.~P.~Salam and A.~M.~Stasto,
  %``A matrix formulation for small-x singlet evolution,''
  {\it JHEP} {0708}, 046 (2007)
  [arXiv:0707.1453 [hep-ph]]
  %%CITATION = JHEPA,0708,046;%%

%\cite{Altarelli:2008aj}
\bibitem{Altarelli:2008aj}
  G.~Altarelli, R.~D.~Ball and S.~Forte,
  %``Small x Resummation with Quarks: Deep-Inelastic Scattering,''
  {\it Nucl.\ Phys.\ } B {799} (2008) 199
  [arXiv:0802.0032 [hep-ph]]
  %%CITATION = NUPHA,B799,199;%%

%\cite{Thorne:2001nr}
\bibitem{Thorne:2001nr}
  R.~S.~Thorne,
  %``The running coupling BFKL anomalous dimensions and splitting functions,''
  {\it Phys.\ Rev.\ }  D {64} (2001) 074005
  [arXiv:hep-ph/0103210]
  %%CITATION = PHRVA,D64,074005;%%

%\cite{Salam:1998tj}
\bibitem{Salam:1998tj}
  G.~P.~Salam,
  %``A resummation of large sub-leading corrections at small x,''
  {\it JHEP} {9807} (1998) 019
  [arXiv:hep-ph/9806482]
  %%CITATION = JHEPA,9807,019;%%


%\cite{GolecBiernat:2009be}
\bibitem{GolecBiernat:2009be}
  K.~Golec-Biernat and A.~M.~Stasto,
  %``F_L proton structure function from the unified DGLAP/BFKL approach,''
  {\it Phys.\ Rev.\ } D {80} (2009) 014006
  [arXiv:0905.1321 [hep-ph]]
  %%CITATION = PHRVA,D80,014006;%%

%\cite{Kowalski:2010ue}
\bibitem{Kowalski:2010ue}
  H.~Kowalski, L.~N.~Lipatov, D.~A.~Ross and G.~Watt,
  %``Using HERA Data to Determine the Infrared Behaviour of the BFKL
  %Amplitude,''
  {\it Eur.\ Phys.\ J.\ } C {70} (2010) 983
  [arXiv:1005.0355 [hep-ph]]
  %%CITATION = EPHJA,C70,983;%%




%\cite{Marzani:2008uh}
\bibitem{Marzani:2008uh}
  S.~Marzani and R.~D.~Ball,
  %``High Energy Resummation of Drell-Yan Processes,''
  {\it Nucl.\ Phys.\ } B {814} (2009) 246
  [arXiv:0812.3602 [hep-ph]]
  %%CITATION = NUPHA,B814,246;%%


%\cite{Sterman:1986aj}
\bibitem{Sterman:1986aj}
  G.~Sterman,
  %``Summation of Large Corrections to Short Distance Hadronic Cross-Sections,''
  {\it Nucl.\ Phys.\ } B {281} (1987) 310
  %%CITATION = NUPHA,B281,310;%%

%\cite{Appell:1988ie}
\bibitem{Appell:1988ie}
  D.~Appell, G.~Sterman and P.~B.~Mackenzie,
  %``SOFT GLUONS AND THE NORMALIZATION OF THE DRELL-YAN CROSS-SECTION,''
  {\it Nucl.\ Phys.\ } B {309} (1988) 259
  %%CITATION = NUPHA,B309,259;%%

%\cite{Catani:1989ne}
\bibitem{Catani:1989ne}
  S.~Catani and L.~Trentadue,
  %``Resummation Of The QCD Perturbative Series For Hard Processes,''
  {\it Nucl.\ Phys.\ } B {327} (1989) 323
  %%CITATION = NUPHA,B327,323;%%



%\cite{Kataev:1999bp}
\bibitem{Kataev:1999bp}
  A.~L.~Kataev, G.~Parente and A.~V.~Sidorov,
  %``Higher twists and alpha(s)(M(Z)) extractions from the NNLO {QCD} analysis
  %of the CCFR data for the xF3 structure function,''
  {\it Nucl.\ Phys.\ } B {573} (2000) 405
  [arXiv:hep-ph/9905310]
  %%CITATION = NUPHA,B573,405;%%

%\cite{Blumlein:2008kz}
\bibitem{Blumlein:2008kz}
  J.~Blumlein and H.~Bottcher,
  %``Higher Twist Contributions to the Structure Functions F_2^p(x,Q^2) and
  %F_2^d(x,Q^2) at Large x and Higher Orders,''
  {\it Phys.\ Lett.\ } B {662} (2008) 336
  [arXiv:0802.0408 [hep-ph]]
  %%CITATION = PHLTA,B662,336;%%

%\cite{Accardi:2009br}
\bibitem{Accardi:2009br}
  A.~Accardi, M.~E.~Christy, C.~E.~Keppel, P.~Monaghan, W.~Melnitchouk, J.~G.~Morfin and J.~F.~Owens,
  %``New parton distributions from large-x and low-Q^2 data,''
  {\it Phys.\ Rev.\ } D {81} (2010) 034016
  [arXiv:0911.2254]
  %%CITATION = PHRVA,D81,034016;%%

%\cite{Martin:2006qv}
\bibitem{Martin:2006qv}
  A.~D.~Martin, W.~J.~Stirling and R.~S.~Thorne,
  %``The role of F(L)(x,Q**2) in parton analyses,''
  {\it Phys.\ Lett.\ } B {635} (2006) 305
  [arXiv:hep-ph/0601247]
  %%CITATION = PHLTA,B635,305;%%

%\cite{Stein:1996wk}
\bibitem{Stein:1996wk}
  E.~Stein, M.~Meyer-Hermann, L.~Mankiewicz and A.~Schafer,
  %``IR-Renormalon Contribution to the Longitudinal Structure Function $F_L$,''
  {\it Phys.\ Lett.\ } B {376} (1996) 177
  [arXiv:hep-ph/9601356]
  %%CITATION = PHLTA,B376,177;%%

%\cite{Mueller:1985wy}
\bibitem{Mueller:1985wy}
  A.~H.~Mueller and J.~w.~Qiu,
  %``Gluon Recombination And Shadowing At Small Values Of X,''
  {\it Nucl.\ Phys.\ } B {268} (1986) 427
  %%CITATION = NUPHA,B268,427;%%

%\cite{Watt:2005iu}
\bibitem{Watt:2005iu}
  G.~Watt, A.~D.~Martin and M.~G.~Ryskin,
  %``Effect of absorptive corrections on inclusive parton distributions,''
  {\it Phys.\ Lett.\ }  B {627} (2005) 97
  [arXiv:hep-ph/0508093]
  %%CITATION = PHLTA,B627,97;%%

%\cite{Bartels:2009tu}
\bibitem{Bartels:2009tu}
  J.~Bartels, K.~Golec-Biernat and L.~Motyka,
  %``Twist expansion of the nucleon structure functions, F2 and FL, in the DGLAP
  %improved saturation model,''
  {\it Phys.\ Rev.\ } D {81} (2010) 054017
  [arXiv:0911.1935]
  %%CITATION = PHRVA,D81,054017;%%


%\cite{GolecBiernat:1998js}
\bibitem{GolecBiernat:1998js}
  K.~J.~Golec-Biernat and M.~Wusthoff,
  %``Saturation effects in deep inelastic scattering at low Q**2 and its
  %implications on diffraction,''
  {\it Phys.\ Rev.\ } D {59} (1998) 014017
  [arXiv:hep-ph/9807513]
  %%CITATION = PHRVA,D59,014017;%%

%\cite{Iancu:2000hn}
\bibitem{Iancu:2000hn}
  E.~Iancu, A.~Leonidov and L.~D.~McLerran,
  %``Nonlinear gluon evolution in the color glass condensate. I,''
  {\it Nucl.\ Phys.\ } A {692} (2001) 583
  [arXiv:hep-ph/0011241]
  %%CITATION = NUPHA,A692,583;%%

%\cite{Motyka:2008jk}
\bibitem{Motyka:2008jk}
  L.~Motyka, K.~Golec-Biernat and G.~Watt,
  %``Dipole models and parton saturation in ep scattering,''
  arXiv:0809.4191 [hep-ph]
  %%CITATION = ARXIV:0809.4191;%%

%\cite{Gelis:2010nm}
\bibitem{Gelis:2010nm}
  F.~Gelis, E.~Iancu, J.~Jalilian-Marian and R.~Venugopalan,
  %``The Color Glass Condensate,''
  arXiv:1002.0333
  %%CITATION = ARXIV:1002.0333;%%

%\cite{Watt:2007nr}
\bibitem{Watt:2007nr}
  G.~Watt and H.~Kowalski,
  %``Impact parameter dependent colour glass condensate dipole model,''
  {\it Phys.\ Rev.\ }  D {78} (2008) 014016
  [arXiv:0712.2670 [hep-ph]]
  %%CITATION = PHRVA,D78,014016;%%

%\cite{Thorne:2005kj}
\bibitem{Thorne:2005kj}
  R.~S.~Thorne,
  %``Gluon distributions and fits using dipole cross-sections,''
  {\it Phys.\ Rev.\ } D {71} (2005) 054024
  [arXiv:hep-ph/0501124]
  %%CITATION = PHRVA,D71,054024;%%

%\cite{Sherstnev:2007nd}
\bibitem{Sherstnev:2007nd}
  A.~Sherstnev and R.~S.~Thorne,
  %``Parton Distributions for LO Generators,''
  {\it Eur.\ Phys.\ J.\ } C {55} (2008) 553
  [arXiv:0711.2473 [hep-ph]]
  %%CITATION = EPHJA,C55,553;%%

%\cite{Sherstnev:2008dm}
\bibitem{Sherstnev:2008dm}
  A.~Sherstnev and R.~S.~Thorne,
  %``Different PDF approximations useful for LO Monte Carlo generators,''
  arXiv:0807.2132 [hep-ph]
  %%CITATION = ARXIV:0807.2132;%%


%\cite{Frixione:2002ik}
\bibitem{Frixione:2002ik}
  S.~Frixione and B.~R.~Webber,
  %``Matching NLO QCD computations and parton shower simulations,''
  {\it JHEP} {0206} (2002) 029
  [arXiv:hep-ph/0204244]
  %%CITATION = JHEPA,0206,029;%%


%\cite{Lai:2009ne}
\bibitem{Lai:2009ne}
  H.~L.~Lai, J.~Huston, S.~Mrenna, P.~Nadolsky, D.~Stump, W.~K.~Tung and C.~P.~Yuan,
  %``Parton Distributions for Event Generators,''
  {\it JHEP} {1004} (2010) 035
  [arXiv:0910.4183]
  %%CITATION = JHEPA,1004,035;%%

%\cite{Bacchetta:2010hh}
\bibitem{Bacchetta:2010hh}
  A.~Bacchetta, H.~Jung, A.~Knutsson, K.~Kutak and F.~Samson-Himmelstjerna,
  %``A method for tuning parameters of Monte Carlo generators and a its
  %application to the determination of the unintegrated gluon density,''
  {\it Eur.\ Phys.\ J.\ } C {70} (2010) 503
  [arXiv:1001.4675 [hep-ph]]
  %%CITATION = EPHJA,C70,503;%%




\bibitem{pdf4lhcweb}
The web page of the PDF4LHC forum can be found at \href{http://www.hep.ucl.ac.uk/pdf4lhc/}{\tt http://www.hep.ucl.ac.uk/pdf4lhc/}


\bibitem{Newman:2009mb}
P. Newman, Deep Inelastic Scattering at the TeV Energy Scale and the 
LHeC Project, 
{\it Nucl. Phys. Proc. Suppl.} 191 (2009) 307, [arXiv:0902.2292]

\bibitem{Deshpande:2008zz}
A Deshpande, Physics and challenges of the electron ion collider, 
AIP Conf.Proc.980 (2008) 343



%\cite{LHCHiggsCrossSectionWorkingGroup:2011ti}
\bibitem{LHCHiggsCrossSectionWorkingGroup:2011ti}
  LHC Higgs Cross Section Working Group {\it et al.},
  %``Handbook of LHC Higgs Cross Sections: 1. Inclusive Observables,''
  arXiv:1101.0593 [hep-ph]
  %%CITATION = ARXIV:1101.0593;%%

%\cite{Alekhin:2011sk}
\bibitem{Alekhin:2011sk}
  S.~Alekhin {\it et al.},
  ``The PDF4LHC Working Group Interim Report,''
  arXiv:1101.0536 [hep-ph]
  %%CITATION = ARXIV:1101.0536;%%

%\cite{Botje:2011sn}
\bibitem{Botje:2011sn}
  M.~Botje {\it et al.},
  ``The PDF4LHC Working Group Interim Recommendations,''
  arXiv:1101.0538 [hep-ph]
  %%CITATION = ARXIV:1101.0538;%%

%\cite{Aliev:2010zk}
\bibitem{Aliev:2010zk}
  M.~Aliev, H.~Lacker, U.~Langenfeld, S.~Moch, P.~Uwer and M.~Wiedermann,
  %``-- HATHOR -- HAdronic Top and Heavy quarks crOss section calculatoR,''
  {\it Comput.\ Phys.\ Commun.\ } {182} (2011) 1034
  [arXiv:1007.1327 [hep-ph]]
  %%CITATION = CPHCB,182,1034;%%




\end{thebibliography}
\end{document}